pdfoutput=1
\documentclass[goettingen, gauss, print]{thesis}
\title{Brightness Contrast of Solar Magnetic Elements Observed by Sunrise }
\author{Fatima Kahil}
\town{Nabat\^{i}y\'{e}, Libanon}

\thesisadvisorycommitteeb{Prof. Dr. Sami K. Solanki}
\instthesisadvisorycommitteeb{Max-Planck-Institut f\"ur Sonnensystemforschung, G\"ottingen}
\thesisadvisorycommitteec{Dr. Tino L. Riethm\"{u}ller}
\instthesisadvisorycommitteec{Max-Planck-Institut f\"ur Sonnensystemforschung, G\"ottingen}

\thesisadvisorycommitteea{Prof. Dr. Laurent Gizon}
\instthesisadvisorycommitteea{Institut f\"ur Astrophysik, Georg-August-Universit\"at G\"ottingen\\ Max-Planck-Institut f\"ur Sonnensystemforschung, G\"ottingen }

\refereea{Prof. Dr. Sami K. Solanki}
\instrefereea{Max-Planck-Institut f\"ur Sonnensystemforschung}
\refereeb{Prof. Dr. Wolfram Kollatschny}
\instrefereeb{Institut f\"ur Astrophysik, Georg-August-Universit\"at G\"ottingen}
\commissiona{Prof. Dr. Ariane Frey}
\instcommissiona{Physikalisches Institut, Georg-August-Universit\"at G\"ottingen }
\commissionb{Dr. Natalie A.~Krivova}
\instcommissionb{Max-Planck-Institut f\"ur Sonnensystemforschung}
\commissionc{Prof. Dr. Laurent Gizon}
\instcommissionc{Institut f\"ur Astrophysik, Georg-August-Universit\"at G\"ottingen \\ Max-Planck-Institut f\"ur Sonnensystemforschung }
\commissiond{Prof. Dr. Hardi Peter}
\instcommissiond{Max-Planck-Institut f\"ur Sonnensystemforschung}

\isbn{978-3-944072-68-5}
\examinationdate{29.01.2019}

\newcommand{\comment}[1]{}
\usepackage[english]{babel}
\usepackage{natbib}
\usepackage{printlen}
\usepackage{color}
\usepackage{hyperref}
\usepackage[bottom]{footmisc}
\usepackage{amsmath}
\usepackage{lscape}
\usepackage{changepage}

\begin{document}
\selectlanguage{english}
\maketitle


\tableofcontents

\newcommand{\sunrise}{\textsc{Sunrise}}
\newcommand{\carcsec}{$\mbox{.\hspace{-0.5ex}}^{\prime\prime}$}
\newcommand{\angstrom}{\textup{\AA}}
\newcommand{\psfpd}{\mathrm{PSF}_{\mathrm{PD}}}
\newcommand{\mtfpd}{\mathrm{MTF}_{\mathrm{PD}}}
\newcommand{\erfc}{\mathrm{erfc}}
\newcommand{\ud}{\,\mathrm{d}}
\graphicspath{{figures/}}


\chapter*{Summary\markboth{Summary}{Summary}}
\addcontentsline{toc}{chapter}{Summary}


Small-scale magnetic elements are part of the quiet-Sun network regions that form at the edges of supergranulation. They are also found in the vicinity of sunspots as part of active region plages. Studying the brightness of magnetic elements with respect to their magnetic field-free surroundings (i.e., their contrast) serves as an observational constraint for theoretical models of magnetic flux tubes. The brightness contrast is also an important input into models of solar irradiance variations.\\

The two types of magnetic elements (network and plage) form in two different magnetic regions, have different dynamics and evolve on different timescales. Plage are clusters of bright points, and affect granular convection in a different manner than small bright points found in the network. Hence there is a need for a separate evaluation of their contrast variation with the magnetic flux density. This evaluation serves as a constraint for spectral and total solar irradiance modelling which so far do not take this difference into account (e.g., the SATIRE model). In addition, brightness studies in the ultraviolet (UV) are crucial for climate modelling given that the emission in this part of the solar spectrum affects the chemistry in the Earth's atmosphere. However, high-resolution imaging in the UV is not so far possible with ground-based observations due to absorption by the Earth's atmosphere.
\textbf{Chapter~\ref{chapter_1}} is a brief introduction to the general properties and studies of small-scale magnetic elements, with an emphasis on observational and theoretical studies of their intensity contrast.\\

In this thesis, we revisit brightness studies of solar small-scale magnetic elements, but go beyond earlier studies in two ways. Firstly, we consider the UV between 200 and 400\,nm. Secondly, we use data of constant high spatial resolution. Both of these novelties are possible because the data employed here were obtained by the $\sunrise$ balloon-borne solar observatory.   

We carefully probe the photometric and magnetic properties of magnetic elements and the correlation between the two quantities. We use data acquired with the Imaging Magnetograph eXperiment (IMaX) and high-resolution imaging carried out by the Sunrise Filter Imager (SuFI), both onboard $\sunrise$. The solar observatory was launched two times to observe the Sun during two different magnetic activity periods. In 2009, $\sunrise$~I carried out observations of the Sun when it was mostly quiet, while $\sunrise$~II probed the active face of the Sun, 4 years after the first launch of the telescope.
With an aperture of 1 m, and at around 36 km above the terrestrial atmosphere, the telescope onboard $\sunrise$ provided nearly diffraction limited images in the visible and near UV, down to 200\,nm. 
Spectropolarimetric data at around 500\,nm are provided with IMaX. Simultaneously with IMaX, SuFI collected broad-band images in the UV of a smaller solar region embedded in the field of view of IMaX.
In \textbf{Chapter~\ref{chapter_2}} we describe in more detail these scientific instruments, along with the data acquisition and reduction steps. We also describe the sophisticated post-correction techniques applied to the data to correct for instrumental wavefront aberrations.\\

A number of studies based on $\sunrise$~I data of the properties of magnetic bright points (MBPs) at disk center were carried out, e.g., by \cite{riethmuller_bright_2010} who analysed the photometric contrasts of manually identified MBPs in the wavelengths imaged by IMaX and SuFI. 
In \textbf{Chapter~\ref{chapter_3}} we extend this study by analysing not only the individual bright points but also the surrounding granulation. We investigate the pixel-by-pixel correspondence between the visible/UV intensity contrast and magnetic flux density. To get the magnetic field vector, we perform Stokes inversions of IMaX data using the SPINOR code. In addition, the data employed in this study are corrected for instrumental stray light. We find that, unlike earlier lower spatial resolution studies, the brightness increases non-monotonically with the magnetic flux at all wavelengths, in qualitative agreement with radiative MHD simulations of the quiet Sun. In particular, at a continuum wavelength around 525\,nm, the averaged contrast in the quiet Sun does not show a turnover, i.e., a decrease of the contrast for large magnetic fields, indicative of having resolved the magnetic features.\\

In \textbf{Chapter~\ref{chapter_4}} we carry out the same study as in Chapter~\ref{chapter_3} but for active region plages observed by $\sunrise$ during its second flight. A comparison of the contrast in the visible and NUV of strong plage to that of the quiet-Sun network studied in Chapter~\ref{chapter_3} shows that, in accordance with observational studies and empirical models of flux tubes, the contrast in the quiet-Sun network
is higher than in active region plages, and the difference decreases with atmospheric height. We also find that in the UV the contrast of magnetic flux tubes in both regions increases with increasing field strength, independently of their size. Finally, whereas in the quiet Sun, magnetic elements stay brighter in the visible continuum than the average quiet Sun even at large field strength values, they become darker in plage regions. This is due to their dark magnetic cores as expected from larger features. 

\cite{hirzberger_quiet-sun_2010} computed the root-mean-square (RMS) intensity contrasts of 2D patches of quiet-Sun data at disk center, and found that they are lower than the output of radiative MHD simulations. A possible reason for this discrepancy is that the $\sunrise$~I data were not corrected for stray light. 
In \textbf{Chapter~\ref{chapter_5}} we extend the work of \cite{hirzberger_quiet-sun_2010} and model the total Point Spread Function (PSF) of the telescope in order to obtain the RMS contrasts of stray light free quiet-Sun data. The PSF is a combination of the core retrieved from phase-diversity measurements and of the wings obtained from the analysis of solar limb profiles. We describe the developed model, which allows for fitting the 1D solar limb profiles and for computing the telescope PSF. We also test the reliability of the stray-light correction based on solar limb profiles obtained during the first $\sunrise$ flight on the active region data recorded during the second $\sunrise$ flight.\\

In \textbf{Chapter~\ref{chapter_6}} we summarize the main results of the thesis. We then give a glimpse of on-going and future research which will provide more insights into the magneto-convective properties of small-scale flux tubes as seen by $\sunrise$.


\chapter{Introduction}
\label{chapter_1}

\section{Solar magnetic features}
\label{intro-features}


Stars, as observed from Earth, are unresolved point sources of light, hence the study of their surface dynamical properties is not possible.
The close proximity of the Sun to our planet offers a great opportunity to constrain theoretical models of stellar evolution: do other stars with the same mass and chemical composition as the Sun harbor the same surface magnetic features and phenomena?

A star is classified based on its observed spectrum and luminosity. The stellar spectrum reveals both, the chemical composition and the effective temperature of the stellar surface, while the luminosity is related to its mass and age~\footnote{ $\frac{L}{L_{\odot}} = (\frac{M}{M_{\odot}})^{3.5}$ for main sequence stars }.

The Sun is a main-sequence $G2V$ star, with nuclear fusion of Hydrogen to Helium occuring in its core for the last 5 billion years. In another 5 to 6 billion years, the Sun will expand to a red giant and will eventually end up as a cold white dwarf. 

Energy is carried from the Sun's center by \textit{radiation} up until a radius of 0.7$R_{\odot}$. Above this threshold, \textit{convection} takes over due to the large gradient in temperature, given that $dT/dr$ is proportional to opacity which increases with radius from the center~\footnote{$\frac{dT}{dr} = -\frac{3}{4ac}\frac{\kappa \rho}{T^3}\frac{L}{4\pi r^2}$, $\kappa$ is the absorption coefficient of the gas}. 
Due to the lower temperature at the top of the convection zone, free electrons recombine with atoms and radiation can escape to outer space~\footnote{In the core the temperature is so high that it ionises Hydrogen and scatters the photons with the free electrons. The latter become less present as the surface is approached and therefore the light can escape into free space.}. This layer is referred to as the solar surface or (bottom of the) \textit{photosphere} and, by convention, is defined where the optical depth at 500 nm is equal to unity. The most remarkable photospheric distribution in white light is the granulation. Hot plasma rises up from the solar interior to the surface, cools down then sinks back to form the intergranular lanes. The latter have widths of $\approx$300 km while the granules are of 1000\,km.
Convection also occurs at larger spatial scales (10 to 50 Mm) and called `supergranulation'.

\begin{figure}
\centering
\includegraphics[scale=0.4]{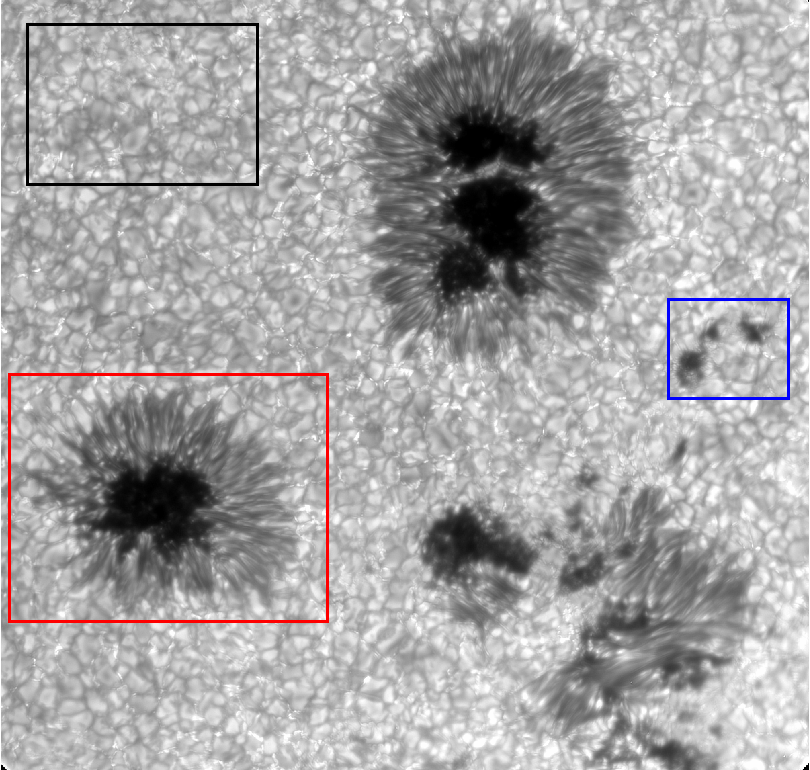}
\caption{An active region imaged in the G-band with the Swedish 1-m Solar Telescope \citep[SST,][]{scharmer_1-meter_2003}. Squares enclose the solar surface features explained in the text. The image is taken from The Instiute For Solar Physics, Stockholm University data center (www.isf.astro.su.se).} 
\label{G-band}
\end{figure}

Another distinctive feature of the solar photosphere are Sunspots (e.g. red box in Figure~\ref{G-band}). They are the darkest regions on the solar surface seen in white light and are characterized by their dark cores (umbrae) and their less dark filamentary surroundings (penumbrae). Pores are dark but smaller than sunspots. Pores resemble small umbrae (blue box in Figure~\ref{G-band}). In the vicinity of sunspots, plages, which have a filamentary shape, form in the intergranular lanes and are brighter than their surrounding granulation (mainly in the UV and cores of spectral lines). Parts of them are enclosed in the black box in Figure~\ref{G-band}, but as the figure shows, they are distributed over a large fraction of the respective solar region. 

Temperature above the photosphere declines to a minimum temperature of about 4400 K (\textit{temperature minumum}), then it increases again until 10 000 K to define the \textit{chromosphere}. Within 100\,km the temperature rises dramatically to reach about $10^5$ K (\textit{transition region}), then it increases slowly at about 2000 km above $\tau_{500} = 1$ defining the \textit{corona}. 

%
The interplay between the magnetic field with the solar plasma is the driver of the different features and phenomena seen in the solar atmosphere. This interplay, in terms of energy, is usually described by the plasma $\beta$, which is the ratio of the gas pressure to the magnetic pressure. It is `on average' larger than unity in the photosphere and decreases going to the upper chromosphere, where magnetic energy dominates. 

\begin{figure}[h!]
\centering
\includegraphics[scale=0.35]{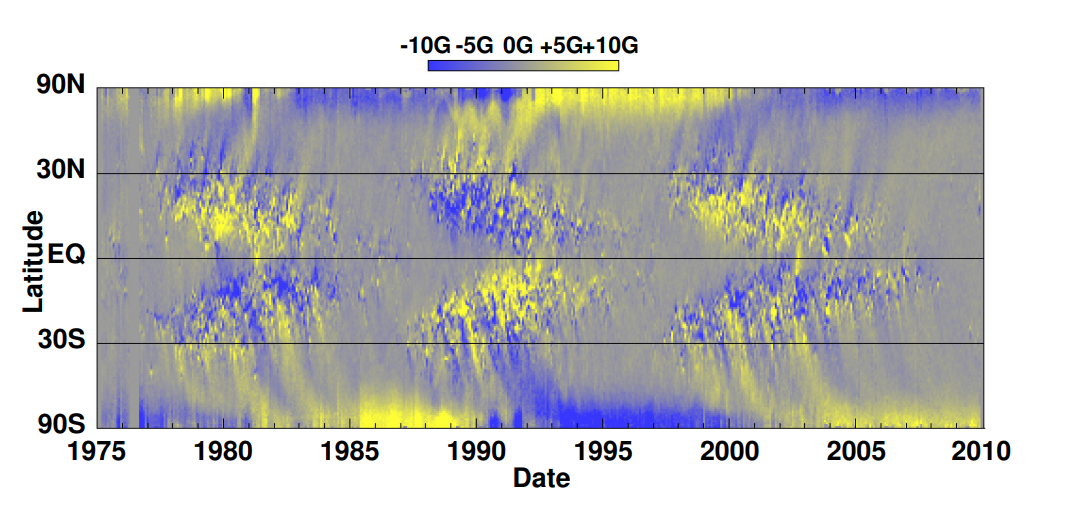}
\caption{The solar magnetic butterfly diagram. The evolution of the longitudinal magnetic flux distribution with time and latitude is shown in blue and yellow colors. From \cite{hathaway_solar_2010}.}
\label{solar_cycle}
\end{figure}

The first measurement of the magnetic field on the solar surface was that of a sunspot in 1908 by the American astronomer George Ellery Hale using a solar spectrograph. The discovery was based on the Zeeman splitting (see Section~\ref{intro-Zeeman}) of the $H\alpha$ line. Given their large magnetic fields (3000\,G in umbra and 800\,G in penumbra), sunspots are the drivers of many dynamic phenomena (solar flares, coronal mass ejections). The most remarkable discovery that emerged from sunspot observations is the solar cycle: the sunspots number would vary (decreasing and increasing) over the course of an 11 years. Not only that, the sunspots at the beginning of each cycle tend to lie at higher altitudes (at $\pm$30 deg from the equator), approaching the equator by the end of the cycle (\textbf{Sp\"{o}rer'law}) and defining the `activity belt'. In addition, in a given solar cycle, the polarities of the leading sunspots are opposite between the two hemispheres (also the polarities in the poles) (\textbf{Hale's law}), and in the next cycle they alternate, implying that the solar cycle is actually 22 years~\footnote{The polarity of the leading sunspot is usually the same as the respective one in the pole}. Another finding that emerged from observing the sunspots pair groups is that the tilt angle (the angle between the line joining the two sunspots and the equator) increases during the solar cycle (\textbf{Joy's law}). The variation of the solar magnetic flux with time and latitude is best visualized with the solar magnetic butterfly diagram in Figure~\ref{solar_cycle}. All of these observations led to the development of the solar global dynamo model, where the surface magnetic field evolution is explained in terms of its configuration as an axisymmetric dipole and the differential rotation of the Sun \citep{babcock_topology_1961}.

While the photospheric magnetic field is best observed by splitting of magnetically sensitive lines, the chromospheric magnetic field is hard to measure with the Zeeman effect (unless the splitting occurs at longer wavelengths, which is not convenient for a good spatial resolution). The reason for this is the weakening of the field at higher atmospheric layers. Proxies such as Ca\,{\sc ii} H and K, H$\alpha$ 6563 $\AA$, Ca\,{\sc ii} 8542 $\AA$ and Ly$\alpha$ 1216 $\AA$ spectral lines can be used to probe the bright and magnetic chromospheric features. These features also tend to be larger due to the expansion of the magnetic field with height (see Section~\ref{intro_flux_tubes}). These chromospheric emission lines are used to reveal other chromospheric structures, such as fibrils and spicules. Fibrils are a trace of closed magnetic field lines connecting opposite polarity flux features, while spicules are fibrils seen near the limb.

\newpage




\section{Small-scale magnetic elements}
\label{intro-ssme}
\begin{figure}
\centering
\hspace*{-1.3cm}\includegraphics[scale=0.37]{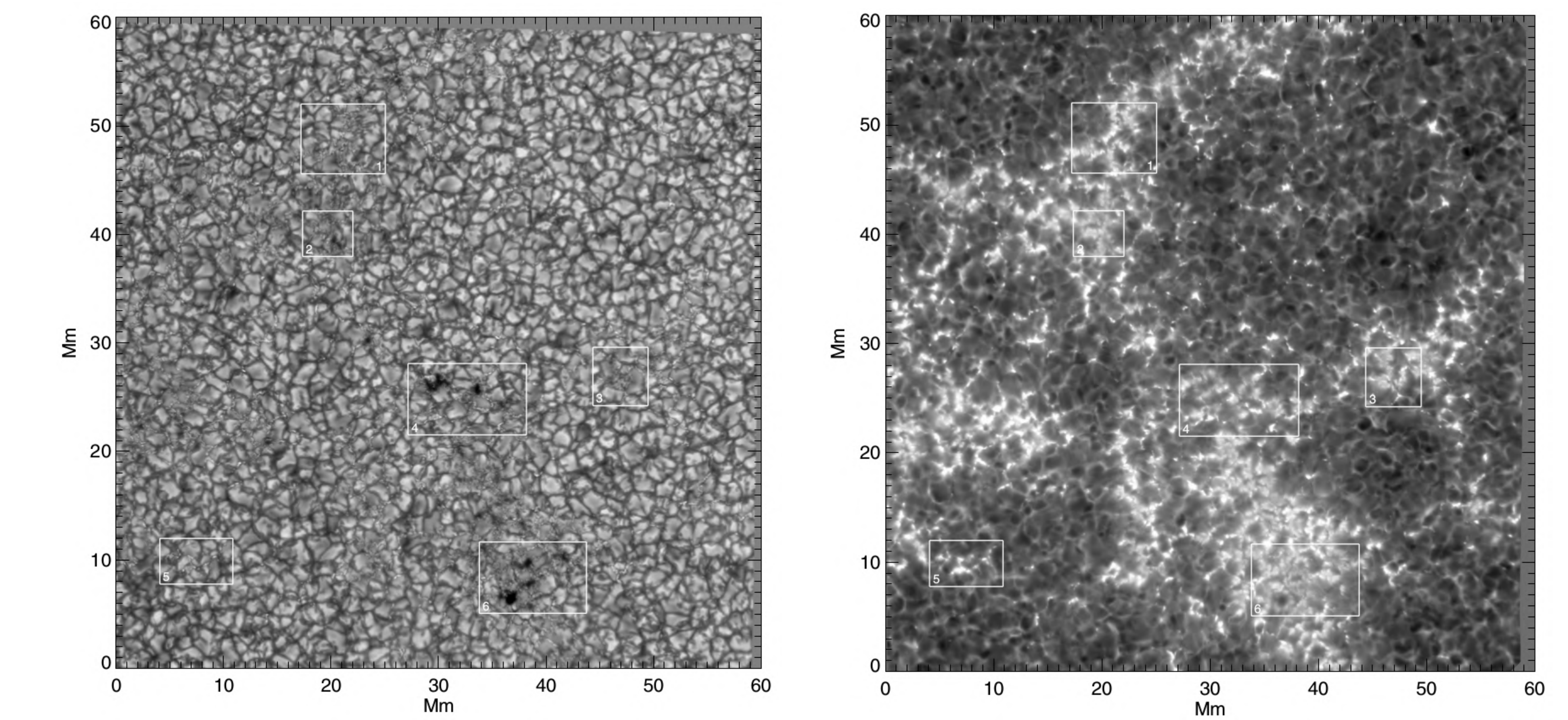}
\caption{Remnant active region plage at disk center in the G-band continuum bandpass (left panel) and core of Ca\,{\sc ii} H line at 369.88 nm (right panel). From \cite{berger_solar_2004}.  }
\label{G-band_calcium}
\end{figure}

With the advent of a photoelectric magnetograph, the Babcocks discovered in 1953 that the solar surface outside sunspots is permeated with magnetic fields. 
Now, we know that in the photosphere and besides sunspots and pores, the magnetic field is distributed at the boundaries of supergranules as quiet-Sun network or in active regions as plage (as seen in chromospheric spectral lines) or faculae (as seen in photospheric continuum radiation near the limb). This distribution is mostly visible in chromospheric emission due to the expansion of the field above these structures which makes them more diffuse. In addition, the shallower temperature gradient above magnetic flux tubes enhances their contrast compared to lower layers (see Section~\ref{intro_flux_tubes} and Figure~\ref{G-band_calcium}). Network and plage belong to \textit{small-scale magnetic elements} which are near the smaller end of the spectrum of magnetic features after sunspots and pores. They are characterized by kG magnetic fields and located in the dark intergranular lanes \citep[]{stenflo_magnetic-field_1973, schussler_small-scale_1992}. What distinguishes them from sunspots and pores is their excess brightness with respect to their magnetic field-free environement, and their sizes (see \citet{solanki_small-scale_1993} for a review on small-scale solar magnetic features). Their exact sizes were hard to measure because they lie below the achievable spatial resolution of solar telescopes. The high resolution observations of \cite{keller_resolution_1992} revealed that the smallest measurable flux tube sizes lie below 200\,km and that tubes larger than 300\,km tend to be darker. The $G$ band observations of magnetic bright points by \cite{dunn_solar_1973} inferred sizes of less than 200\,km.

In addition, the areas between network regions inside the supergranules are called internetwork. And only recently that kG magnetic fields were discovered in the internetwork with high spatial resolution observations \citep[e.g.][]{lagg_fully_2010}. The weak magnetic fields in the internetwork (close to the equipartition field strength) are eventually intensified by convective instability processes described later. That is why individual features have kG magnetic fields. 


The magnetic field generation in the quiet Sun internetwork is believed to be a result of a local dynamo driven by the turbulent motion of the convection \citep{vogler_solar_2007, danilovic_probing_2010, buehler_quiet_2013, lites_solar_2014}. \cite{buehler_quiet_2013} analysed QS regions observed by Hinode during half of a solar cycle and found no significant variation in the amount of both linear and circular polarization in the solar internetwork, eliminating the global dynamo effect. Plage, on the other hand, are the result of the decay of active regions which recycles the flux over time, and therefore related to sunspot activity linked to the global dynamo.
In the next sections we provide a description of the properties of magnetic elements in terms of their brightness in different wavelength bands and position on the solar disc, with an emphasis on the observational conditions under which they are observed.

To measure the magnetic field strength in the quiet Sun, \cite{stenflo_magnetic-field_1973} introduced the line-ratio technique which is valid in the weak field regime, where the broadening of the line due to Zeeman splitting is insignificant or on the order of the line width and Stokes~$V$ is proportional to $B \,\partial I/\partial \lambda$ (see Section~\ref{intro-Polarization}). 
Line inversion techniques are commonly used to compute the field strength along with other physical parameters, simultaneously. It is based on solving the polarized radiative transfer equations and fitting the synthesized Stokes profiles to the observed spectra (see Section~\ref{intro-RT}).

Detecting kG magnetic signals in the quiet Sun is challenged by detecting their circular polarization signals (from measuring Stokes~$V$, see Section~\ref{intro-Polarization}) which tend to cancel if the element is unresolved. Another challenge in observing small magnetic elements is the short timescale during which these elements evolve (around 5 minutes), which requires long time series in order to resolve their dynamical properties. In addition, short exposures are needed to overcome the effect of seeing (if observations are carried out from ground), but will limit the amount of photons received from the object of interest, lowering the signal-to-noise ratio which is related to the polarimetric sensitivity of the detector.
The spectral line used for quiet-Sun magnetic fields diagnosis also plays a major role, for instance there is a contradiction in the  measurements of the magnetic field from observations in the IR and the visible which turned out to be the result of the different sensitivities of the lines to the magnetic field \citep{sanchez_almeida_physical_2000}.

In addition to the magnetic field strength, there is still controversy on the inclination distribution of the magnetic field lines in the quiet Sun internetwork. While some observations detected horizontal magnetic fields \citep{orozco_suarez_strategy_2007,lites_horizontal_2008}, other studies suggest the tendency of internetwork fields and magnetic bright points to have a vertical distribution \citep{stenflo_distribution_2010, jafarzadeh_inclinations_2014} and that horizontal fields are mainly an artifact from the inversions that fit noise to Stokes~$Q$ and $U$ \citep{borrero_inferring_2011}, which are sensitive to transverse fields (Section~\ref{intro-inversions}). \cite{asensio_ramos_evidence_2009} came to the conclusion that inclinations in the QS internetwork have an isotropic distribution, meaning that the field of a given internetwork feature could have any orientation.

The interaction of the magnetic field with the convective motion in the photosphere (magneto-convection) on short time and spatial scales yields to many interesting phenomena. Observing and analysing these phenomena is now possible thanks to the advanced tools in spectropolarimetry allowing the detection of the polarized signals from Zeeman split components. High speed magnetized upflows is one of these interesting phenomena \citep{borrero_supersonic_2010}, it was discovered after detecting blueshifted circular polarization signals in the continuum of the IMaX Fe\,{\sc i} line (see Section~\ref{intro-imax}). These jets were investigated by \cite{borrero_is_2013} and \cite{quintero_noda_high_2014} who came to the conclusion that such jets are a result of magnetic reconnection \citep{parker_kinematical_1963}. This conclusion is based on detecting linear polarization signals between two opposite polarity features around the detected jets. Those events were also found by \cite{rubio_da_costa_centre--limb_2015} and \cite{martinez_pillet_ubiquitous_2011}. Another explanation is the rebound of downward flows caused by the convective instability when hitting dense layers \citep{grossmann-doerth_convective_1998,requerey_history_2014}.

Small-scale magnetic loops are detected in both QS and plage regions using Hinode data by \cite{centeno_emergence_2007} and \cite{ishikawa_comparison_2009}. Those loops are characterized by horizontal fields (inferred from Stokes~$Q$ and $U$ measurements) above granular features, accompanied by strong Stokes~$V$ signals (vertical field vectors) in the adjacent intergranular lanes, marking the footpoints of the loop. This discovery aims at explaining the origin of small-scale magnetic fields. The asymmetry detected in Stokes~$V$ profiles (which are supposed to be antisymmetric) is indicative of strong gradients in the line-of-sight velocity or magnetic field strength \citep{illing_broad-band_1975,grossmann-doerth_unshifted_1988, quintero_noda_photospheric_2014}.

Magnetic fields in the quiet Sun undergo a number of additional phenomena that alter the magnetic flux budget. \cite{anusha_statistical_2017} developped a feature tracking algorithm to track and study processes of merging, splitting, emerging, and cancellation of magnetic flux in a quiet-Sun region observed with $\sunrise$ during its first flight in 2009, updating earlier similar works \citep[e.g.,][]{deforest_solar_2007, zhou_solar_2010, lamb_solar_2013}.
 
Convective collapse is another interesting mechanism that nicely illustrates the interplay between convection and the magnetic field which leads to the formation of small bright points. This process will be described in more detail in the next section when addressing the flux tube model.

Magnetic fields in the quiet Sun emerge at the surface as small bipolar regions forming low-lying photospheric loops. The granulation motion moves the field lines of the footpoints until two opposite polarity features cancel each other's field. Hence their roles in heating the chromosphere and corona \citep[]{zhang_lifetime_1998, kubo_magnetic_2007,ishikawa_comparison_2009,zhou_solar_2010,gosic_chromospheric_2018}. However, the cancellation of the magnetic field lines is not solely responsible for the energy dissipation in higher layers. Other magnetic/non-magnetic mechanisms come into play, such as the transportation of energy by acoustic waves and dissipation by shocks, or the dissipation by electric currents (Joule heating) as a result of braiding of the field lines which gives rise to nanoflares \citep{parker_nanoflares_1988}. \cite{chitta_solar_2017} found after combining IMaX (see Section~\ref{intro-imax}) magnetograms with EUV observations from the Solar Dynamics Observatory (SDO) that the coronal footpoints in the photosphere are not unipolar but opposite polarity features exist at each footpoint, the cancellation of which supplies the energy necessary to heat the corona.
In the lower chromosphere, heating by oscillations and waves leads to an enhanced brightening where flux tubes are located. This is indicated by the correlation found in the brightness in the core of Ca\,{\sc ii} H and K lines and the magnetic field \citep[]{skumanich_statistical_1975,schrijver_relations_1989,ortiz_how_2005,rezaei_relation_2007,loukitcheva_relationship_2009,kahil_brightness_2017}.
In Chapters~\ref{chapter_3} and \ref{chapter_4} we will investigate the role of photospheric magnetic fields inferred from inverting IMaX data in the quiet Sun and AR plage in heating the lower chromosphere, using Ca\,{\sc ii} H as an indicator for chromospheric emission.\\

\begin{figure}
\centering
\hspace*{-1cm}\includegraphics[scale=0.35]{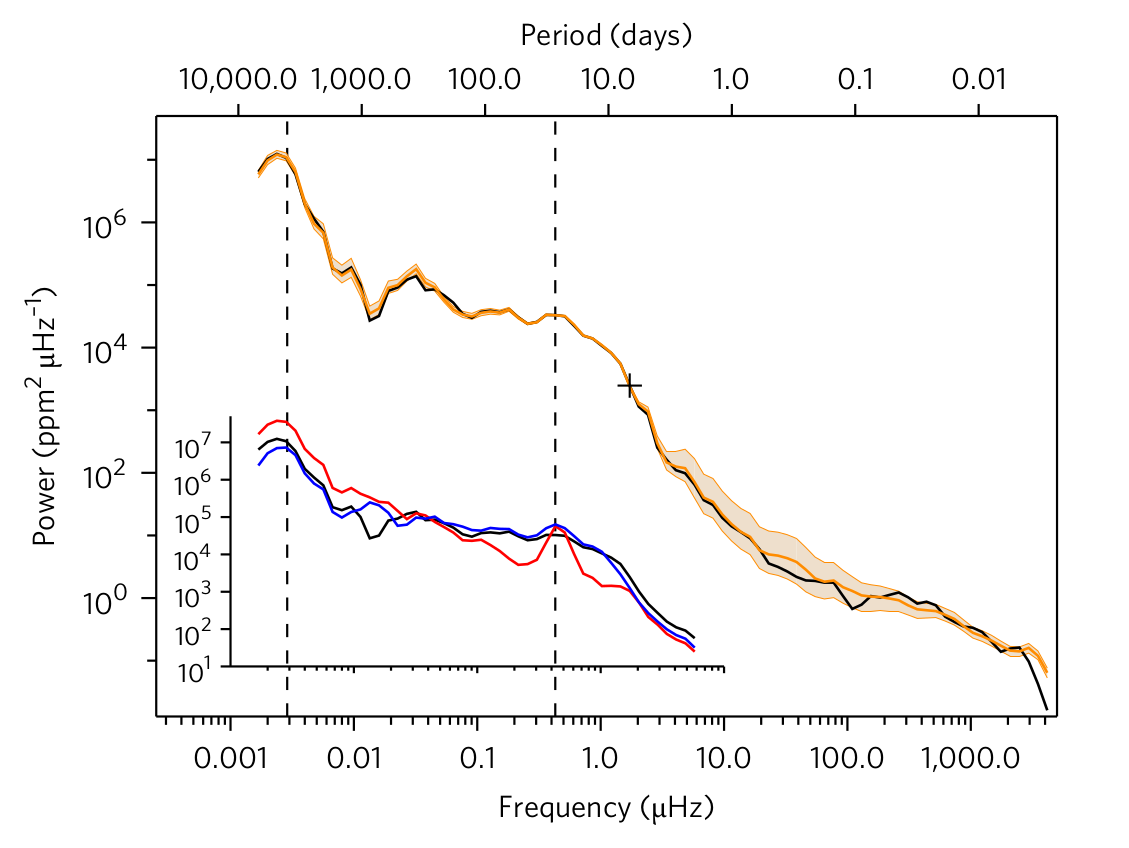}
\caption{The power spectra of modelled (black) and observed (orange) total solar irradiance variations over the course of 20 years. The short time variations caused by the granulation is modelled using radiative hydrodynamical simulations with the MURaM code. From \cite{shapiro_nature_2017}.}
\label{power_spectrum}
\end{figure}

Linked to the solar cycle variation (described in Section~\ref{intro-features}) is the variation of the total solar irradiance (TSI, energy flux integrated over wavelengths above the Earth's atmosphere). The Sun is on average brighter at activity maximum (when the number of the dark sunspots is maximum) compared to activity minimum. Now it is established that the deficit in the brightness due to the passage of sunspots (solar rotation) and the excess brightening from small-scale magnetic elements at the solar surface in the form of network and plage (or faculae) areas are the drivers of solar irradiance variations.
The brightness of small-scale magnetic elements overcompensates for the deficit in radiation caused by sunspots by 0.1\% between solar cycle activity minimum and maximum. High resolution observations of the quiet Sun with the Solar Optical Telescope (SOT) on-board Hinode revealed that the quiet Sun contributes by 0.15\% of excess brightening with respect to a hypothetical field-free Sun, which is larger than the 0.1\% TSI variation over a solar cycle \citep{schnerr_brightness_2011}. \cite{ermolli_measure_2003} identified network elements in full-disc observations using Ca\,{\sc ii} H emission maps and found an increase of TSI by a factor of $3-4\times 10^{-4}$ over a solar cycle.
Modelling the solar irradiance variations using the brightness dependence on wavelength and position on the solar disc, and distribution of magnetic features on the solar surface and their evolution reproduce the TSI observations very well \citep{krivova_reconstruction_2003, shapiro_nature_2017}.
Therefore, the magnetic flux distribution on the solar surface is believed to account for the solar irradiance variation on time scales from solar rotation to solar cycle.
At longer time scales (longer than the solar cycle), there is still a debate on the sources of the respective TSI variation (for a review, see \cite{solanki_solar_2013}). 
On very short timescales (minutes to a day), granulation and p-modes (5 minutes oscillation) contribute to the irradiance variation as shown in Figure~\ref{power_spectrum}.
The contribution from the convection to the TSI variations however is not well included  in solar irradiance modelling. This is mainly due to the lack of high resolution observations of such regions, especially in full disc observations needed for modelling such variations (like in SATIRE code, \cite{krivova_reconstruction_2003}). For example, the granulation contribution to the power spectrum of the TSI in Fig.~\ref{power_spectrum} is derived from solar hydrodynamic simulations.

\subsection{Brightness vs. magnetic flux density }
\label{intro-brightness}
The photospheric magnetic field is often measured from polarimetric data with high spectral resolution. Since magnetic flux tubes tend to have a vertical direction to the solar surface (due to their buoyancy), one is usually interested in the longitudinal component of the magnetic field, $B_{\rm LOS}$, which gives rise to a light that is circularly polarized within Zeeman-sensitive spectral lines (see Section~\ref{intro-Polarization}). The calibration of the net circular polarization signal to get $B_{\rm LOS}$ results in a longitudinal \textit{magnetogram}. Therefore, longitudinal magnetograms are sensitive to the line-of-sight component of the magnetic field.

Magnetograms require low noise, and therefore longer integration times, which affect the temporal resolution. Thus the brightness of small-scale magnetic elements is usually considered as a proxy for measuring the magnetic field outside sunspots. Since, simply put, magnetic flux concentrations which are smaller than sunspots and pores are, on average, bright. In additon to being an accessible proxy to magnetic field measurements, the contrast (see Section~\ref{intro-contrast}) of magnetic elements is an important physical quantity to measure since it is linked to the temperature variation between these elements and their field-free surroundings, and therefore to the radiative energy exchange between them. In addition, the motivation to asses the brightness of these elements stems from their significant contribution to the total solar irradiance variation between solar minimum and maximum magnetic activity, as pointed out in the previous section. The relationship between the intensity contrast of magnetic elements and the field strength however is not well defined since the (measured) brightness depends on many factors: 

\begin{enumerate}
\item[$\bullet$] The amount of magnetic flux in a specific region (and therefore the size of the element)
\item[$\bullet$] The wavelength of observation (since spectral lines form at specific atmospheric heights whereby thermal properties of solar features change, temperature sensitivity is wavelength dependent)
\item[$\bullet$] The position on the solar disc characterized by the heliocentric angle $\theta$, which is the angle between the local normal to the solar surface and the observer's line of sight. More commonly used in solar physics is $\mu = cos \,\theta$ ($\mu = 1$ at disk center and decreases approaching the limb)~\footnote{The study of the variation of any physical quantity or a relationship between two quantities with heliocentric angle is called Center-to-Limb Variation (CLV) study}
\item[$\bullet$] The spatial resolution of the observation (which depends on both, the imaging wavelength and the telescope configuration)
\item[$\bullet$] The instrumental/atmospheric degradation sources: stray (or scattered) light, telescope vibrations, pointing jitter, seeing
\item[$\bullet$] The criterion used to distinguish between the different magnetic elements from other solar features (pores and micropores): magnetic flux density, contrast (or both), location (proximity to sunspots)
\end{enumerate}

Quiet-Sun network features and plage differ mainly in the nature of the region they are embedded in, which is characterized by the amount of magnetic flux in the region.
Bright points are the smallest in the spectrum of magnetic elements and exist on the solar disk as roundish isolated bright features, mainly in network regions. While plages are found mainly in active regions and harbor a large number of bright points. At disk center they are sometimes called `filigree' \citep{dunn_solar_1973} after their elongated sheet-like appearence. 

The excess brightening of small bright points is caused by the evacuation of the gas inside the flux tube (see Section~\ref{intro_flux_tubes}), which leads to an opacity depression, allowing deeper and hotter layers to be seen. The convection becomes insufficient inside the magnetic structure to carry enough energy from the deeper layers. The gas within the magnetic feature is then cooled by radiation, which is balanced by the lateral radiative heating from the surrounding convection.
Magnetic bright points are particularly bright in chromospheric observations at the edges of supergranules because the chromospheric gas is heated more strongly, and the brightening there is more diffuse due to the expansion of the flux tubes at higher solar atmospheric levels \citep{jafarzadeh_structure_2013}. At high spatial resolution and at disk center, they exhibit large contrasts in the visible and near UV parts of the spectrum \citep{riethmuller_bright_2010}.

\begin{figure}
\centering
\includegraphics[scale=0.2]{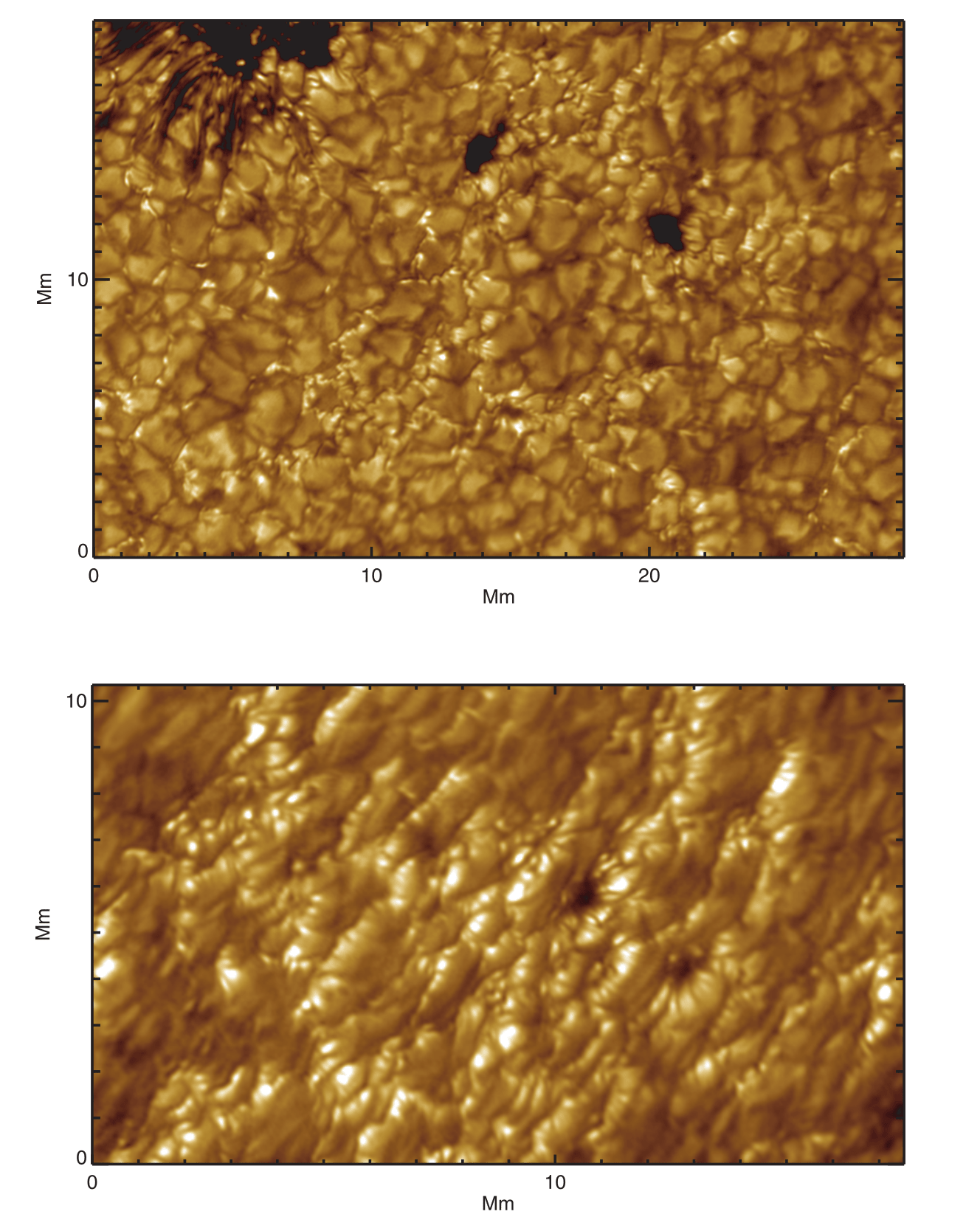}
\caption{G-band map of an active region imaged by the SST on May 2013 at disk center ($\mu=0.89$, upper panel) and near the limb ($\mu=0.42$, lower panel). The hot walls of the magnetic structures are more visible with decreasing $\mu$, which tend to be hotter then the floor of the flux tube (seen mostly at disk center). This results in the shape of faculae as bright sides of granules near the limb. From \cite{berger_contrast_2007}.}
\label{CLV_Gband}

\end{figure}

At a spatial resolution of 0.1$^{\prime\prime}$, the G-band \citep{muller_variability_1984} observations with the Swedich Solar Telescope \citep[SST,][]{scharmer_1-meter_2003} of plage regions at disk center revealed different appearences and morphologies of these structures \citep{berger_solar_2004}. The `ribbons' or `striations' are one type of the categorized features of plages, which have fluid-like shape.
The  striations are characterized by the `double humped' profile obtained upon plotting the line-of-sight (LOS) magnetic field and G-band contrast across them: 
both signals are higher at the boundaries of the ribbons and decrease towards their centers. Also in the G-band but at the limb, faculae appear as `illuminated granules' with one side brighter than the other \citep{hirzberger_solar_2005}. In the last proposed picture, faculae present a centerward `dark lane' and a limbward `bright tail', a result of seeing these structures from an oblique line of sight, making their brighter depressions visible against the limb-darkened surface (see Figure~\ref{CLV_Gband}).

This distribution is confirmed with numerical simulations of flux sheets (or tubes) embedded in a field-free environement and simulated for different sizes and at different heliocentric angles \citep{steiner_recent_2005}. For a narrow tube/sheet seen along a line of sight parallel to the tube's axis (simulating a disk-center observation), the intensity peaks in its center, while for a wider tube, the intensity distribution turns into double peaks at its boundaries and a dip in brightness towards its center (replicating the double-humped profiles of \cite{berger_solar_2004}).
For lines of sight inclined with respect to the solar normal, a dark lane on the disk center side of the sheet is obtained in accordancre with \cite{hirzberger_solar_2005}. The dark lanes of faculae near the limb were also modelled by \cite{keller_origin_2004} and \cite{carlsson_observational_2004}.

On average, in photospheric continuum radiation, faculae are bright at the limb (even with poor spatial resolution) and their contrast decreases approaching disk center (especially with poor spatial resolution). This behaviour has been explained with the `hot wall' picture introduced by \citep{spruit_pressure_1976}, and observed in many CLV studies of facular contrast: for large flux tubes, inclined lines of sight traverse the hot walls of a flux tube, which are brighter, while at disk center the cool dark floor is more visible \citep[]{topka_properties_1992,topka_properties_1997,ortiz_intensity_2002,kobel_discriminant_2009,yeo_intensity_2013}. 
\cite{topka_properties_1997} came to the conclusion that near the limb faculae are just micropores (as seen at disk center) seen from the sides. This points to the dependence of the brightness distribution and appearences of magnetic features on size and position on the solar disk.
One can naivly conclude that roundish bright points exist only at disk center while elongated faculae with their visible hot walls are seen only when approaching the limb, but both bright points and faculae can exist together at the same heliocentric angle \citep{kobel_discriminant_2009}. In \cite{berger_contrast_2007} bright points were found near the limb and therefore called facular grains. The contrast and the magnetogram signal of these facular grains peaked in their center, unlike the profiles seen in \cite{hirzberger_solar_2005}. In the last citation, faculae were found at disk center along with magnetic bright points. This coexistence was attributed to possible inclinations of faculae at disk center so that their bright walls become visible \citep{keller_origin_2004}.


Independently of the solar disk position, measured contrasts are higher when the same magnetic features are imaged in the cores of spectral lines which are formed higher in the solar atmosphere (mid-photosphere, lower chromosphere) than in their continuum radiation \citep{frazier_multi-channel_1970, title_differences_1992, yeo_intensity_2013}. That is due to the larger temperature difference with height with respect to the quiet surroundings (see Section~\ref{intro-contrast}).

Also, higher contrasts are observed when imaging in the blue part of the spectrum (eg. G-band observations). That is due to the contamination by spectral lines and because molecular lines have larger temperature sensitivity which leads to their depletion through molecular dissociation inside the hot flux tubes and decrease of their number density, and hence of the opacity of the lines which produces a line weakening \citep{berger_new_1995,shelyag_g-band_2004}.
Of particular importance to our work in Chapter~\ref{chapter_3}, the CN diatomic molecule at 388 nm  is used as a proxy for quiet-Sun magnetic fields \citep{zakharov_comparative_2005}.

\begin{figure}
\centering
\hspace*{-0.5cm}\includegraphics[scale=0.5]{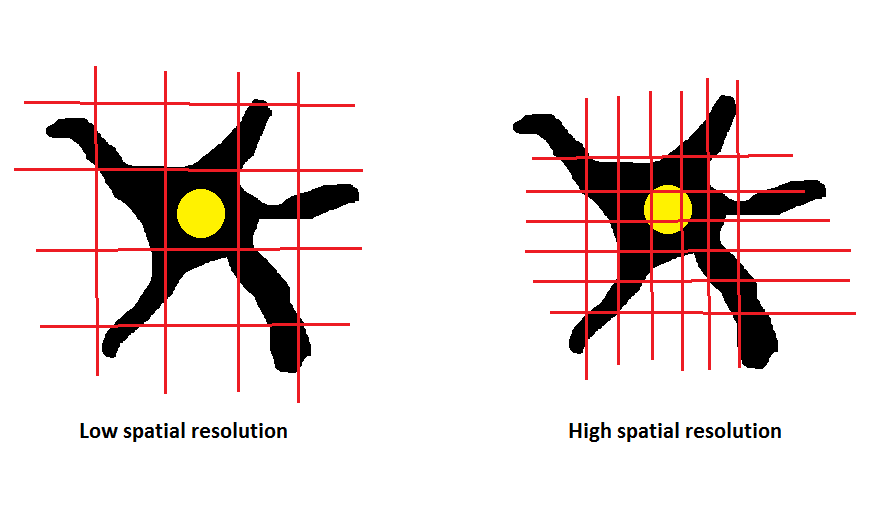}
\caption{A sktech of a bright point (yellow filled dot) embedded in the dark intergranular lanes and seen at both, low and high spatial resolution. The latter is represented by the size of the resolution element indicated by the squares in the red grid. When the resolution element size is bigger than the diameter of the bright point (left panel), more of the dark surroundings are included in the computation of the averaged brightness and field strength in the resolution element, which tend to lower their true values. While for smaller resolution element (right panel), the bright point contrast is less contaminated by its surroundings.}
\label{resolution}
\end{figure}

Due to their small sizes (diameters of order 100\,km), measuring the pixel brightness of magnetic features is hampered by the limited spatial resolution of the data. This is easy to understand as magnetic bright points are embedded in dark field-free intergranular lanes, and the poor spatial resolution blends the information from both solar components, leading to an underestimation of both, contrast and field strength of bright points averaged over the resolution element (see Fig.~\ref{resolution}). This explains why observations with low spatial resolution (when the resolution element is much larger than the intrinsic size of magnetic elements) and with considerable scattered light contamination returns negative contrasts for all magnetogram signals in active region plage \citep{topka_properties_1992, title_differences_1992,topka_properties_1997}. In the last citations, brightness below the quiet-Sun level is observed even for small magnetogram signals. This is due to the inclusion of unrelated pixels in the same magnetogram bin. For example the field-free surroundings of large features (like micropores) which after smearing have non-zero magnetic signal and are assigned to the same bin as magnetic features. This lowers the brightness in the corresponding bin right from small magnetic field values.  

\begin{figure}
\centering
\hspace*{-1cm}\includegraphics[scale=0.45]{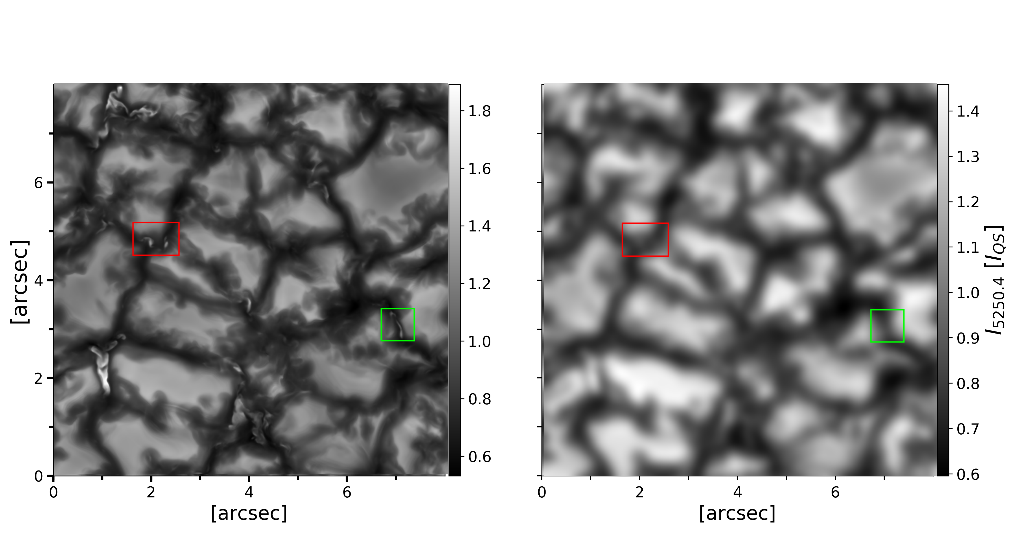}
\caption{MURaM simulated quiet-Sun region with an average vertical magnetic flux density of 30\,G. \textit{Left panel:} The synthesized continuum intensity map of the IMaX line. \textit{Right panel:} The continuum map degraded to a resolution of 0.23". }
\label{30G_MHD}

\end{figure}

The effect of poor spatial resolution on the computed contrast of magnetic bright points can be illustrated with the use of a radiative MHD simulation of a quiet-Sun region using the MURaM code \citep{vogler_simulations_2005}. Figure~\ref{30G_MHD} shows in the left panel the 2D continuum image taken from the synthesized Stokes profiles at a perfect spatial resolution of the line transition of IMaX (525.02 nm). The right panel is the same map convolved with a 2D Gaussian of FWHM of 0.23 arcsec, mimicing a real observation that is affected by a point spread function of a telescope. The red and green boxes enclose two bright magnetic flux concentrations at the original resolution of the simulation (left panel) which are darkened completely after the degradation in the right panel. This effect lowers the measured contrast by mixing the signal of the supposedly bright features with the background level, to which the contrast is usually defined \citep{schnerr_brightness_2011}. 

\cite{kobel_continuum_2011} investigated the contrast-$B_{\rm LOS}$ relationship at the wavelength of 630\,nm but with the Hinode spectropolarimetric data at a spatial resolution of 0.3 arcsec. Instead of decreasing monotonically with magnetogram signal, the contrast in plages peaked at intermediate field strengths and dropped below quiet-Sun level at higher field strengths. 
In fact, the monotonic decrease was reproduced upon degrading Hinode data and adding stray light, and following the same masking technique adopted with the SVST data of \cite{topka_properties_1992} and \cite{title_differences_1992}. However, the turnover of the averaged contrast at higher fields persisted in both, quiet-Sun network and AR plage.

\cite{danilovic_relation_2013} investigated the latter shape obtained in active region plage studies with the use of a MHD snapshot of a simulated plage (with an average vertical magnetic flux density of 200\,G) and degrading them to mimic the observational conditions of \cite{kobel_continuum_2011}.
The contrast vs. $B_{\rm LOS}$ plot turned from a monotonically increasing shape (at the perfect resolution of the simulations) into a peak and a turnover, comfirming the effect of spatial resolution on the shape of this relationship for small magnetic elements in plage. The qualitative influence of the finite spatial resolution on the relationship between the continuum contrast and longitudinal field strength in plage data was also investigated by \cite{rohrbein_is_2011} using the same snapshot as \cite{danilovic_relation_2013}. When data are degraded to a resolution of a 1-m telescope PSF at 630.2\,nm, the bolometric intensity contrast saturates at higher $B_{\rm LOS}$ (taken at log$\tau=1$).
However, these simulations, representative of plage regions on the solar surface, lack pores, the presence of which could lead to the formation of bigger magnetic elements which are supposed to be dark (and numerous if the simulated box is large enough) even at the perfect spatial resolution of the simulations. This questions the validity of comparing such simulations (with no pores) to observations in which pores are formed in order to interpret the physical reason for the peaking and negative contrasts obtained in observational studies of AR plages.
This point will be discussed in Chapter~\ref{chapter_4} in the light of our results of the dependence of the continuum contrast at 525.04 nm (lower photosphere) on $B_{\rm LOS}$ in the quiet Sun (Chapter~\ref{chapter_3}) and AR plages (Chapter~\ref{chapter_4}).

\subsection{Flux tube model}
\label{intro_flux_tubes}
In order to understand the brightness properties of magnetic elements and their dependence on the factors listed above, a comprehensive modelling of their structure is required.
The model describing the basic magnetic features in network and AR plage is that of an isolated flux tube. Under the assumption that the radius of the tube is negligible compared to the pressure scale height (which is around 100\,km in the photosphere), the thin flux tube approximation can be employed to describe the dynamics of flux tubes \citep{spruit_equations_1981}.

Due to the high conductivity of the plasma, the magnetic field is tied to the convective motion (frozen-in magnetic flux). In a first step, the field is swept by granular motion to the dark intergranular lanes and concentrated there by flux expulsion \citep{parker_kinematical_1963}. This process can concentrate magnetic fields up to the equipartition value \footnote{The magnetic field value that corresponds to an equilibrium between the magnetic energy and kinetic energy of the plasma, and is around 400\,G for typical solar values of the plasma}. The intensification of the magnetic field within intergranular lanes to kG values is believed to be driven by the convective instability, the displacement of material in an adiabatically stratified medium \citep{thomas_structure_1990}. The downward displacement of gas inside the tube leads to a reduction of gas pressure which increases the magnetic pressure to maintain total horizontal pressure equilibrium with the field-free environment \citep{parker_hydraulic_1978, spruit_convective_1979}. The evacuated tube tends to be vertical to the solar surface due to lower gas pressure inside the tube~\footnote{Gas pressure is related to the gas density via the equation of state, assuming LTE conditions}, which gives rise to a buoyancy force \citep{schussler_mhd_1986}.

\begin{figure}
\centering
\hspace*{-1cm}\includegraphics[scale=0.25]{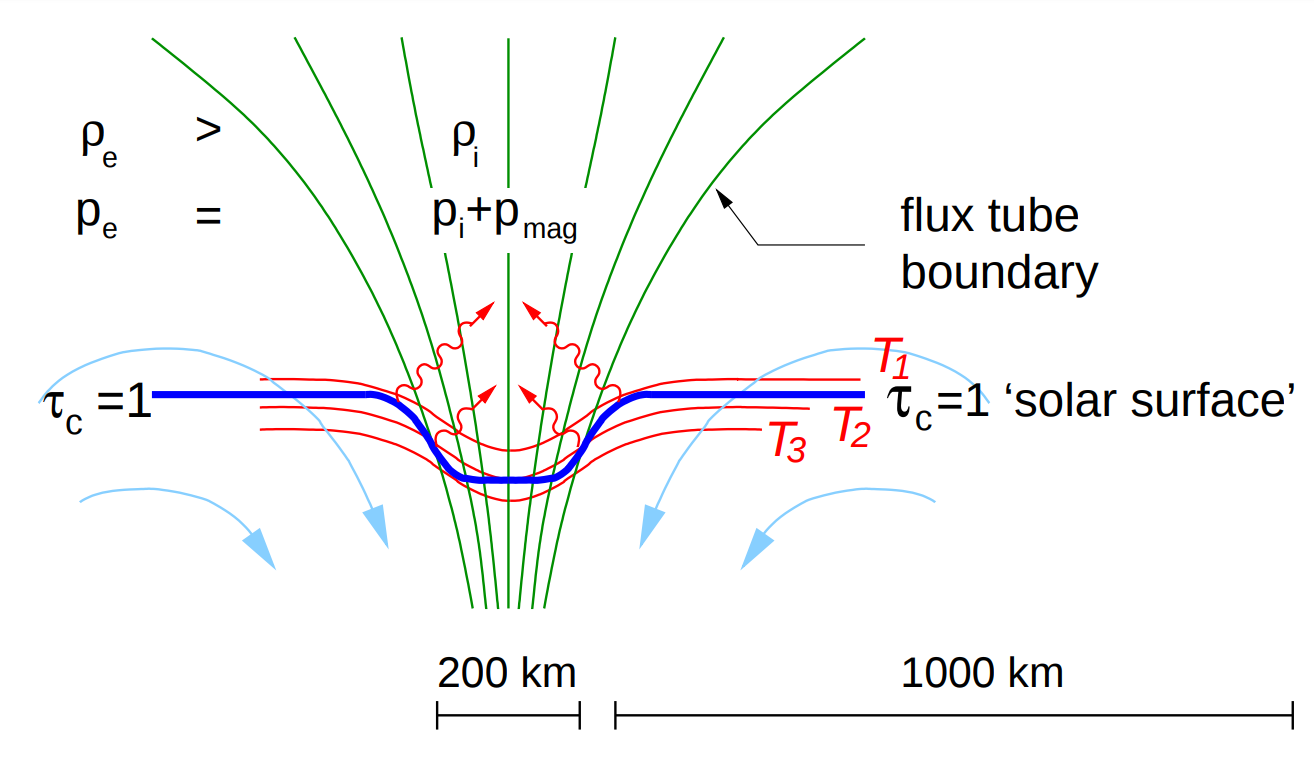}
\caption{Sketch of the geometry of a flux tube. The green lines are the magnetic field lines, the blue line is the optical depth unity level, which is depressed inside the tube. The red line are isotherms. From \cite{steiner_photospheric_2007}.}
\label{tft}
\end{figure}

The optical depth unity level (blue line in Figure~\ref{tft}) shifts downwards with respect to the surrounding convection (Wilson depression), and since temperature increases downwards, tubes appear brighter than their surroundings at the same optical depth level (see red lines in Figure~\ref{tft}). Inside the tube, the magnetic field is intensified and convection is inhibited due to the high Lorentz force, which renders the convection ineffective so that the interior cools down by radiation. However, the convection in the surrounding granules still transports nearly as much energy as without the magnetic features, and a horizontal radiative influx heats the interior due to its transparency to become hotter than the walls at equal optical depth level. This of course occurs if the size of the tube is small enough~\footnote{Size is comparable to the photon mean free path which in the solar photosphere is on the order of 100\,km}. Sunspots and pores are cooler than their surroundings at both equal geometrical and optical depths due to their very large sizes with respect to smaller tubes. The signatures of convective instability have been detected in several observational studies. The observational signatures of such a process are a combination of a decrease in gas pressure accompanied with magnetic field intensification (as inspected from magnetograms) and excess brigthening in the continuum, in addition to strong downflows as inspected from line-of-sight velocity maps \citep{nagata_formation_2008, fischer_statistics_2009}.

At higher average magnetic flux (active regions), flux tubes are concentrated at a larger number density, and they are clustered to form plage or faculae. Therefore they are better described by `flux sheets' instead of roundish flux tubes with circular cross sections \citep{yelles_chaouche_comparison_2009}.

The visible thermal and magnetic properties of flux tubes differ with atmospheric height, position on the solar disk, size and wavelength. At disk center and in the lower photosphere, small flux tubes are brighter than larger ones, owing to the efficiency of thermal heating by the convective surrounding walls. There, the brightness with respect to the average QS (contrast) is dependent on the size, while at higher layers, they are brighter than their field-free surroundings (lying at the same optical depth) regardless of their sizes. This implies that different mechanisms act on governing the thermal properties of magnetic elements in different atmospheric heights. In Chapter~\ref{chapter_4} we compare the contrast of magnetic elements in quiet-Sun network and plage in the photospheric continuum radiation and cores of spectral lines, making use of the high spatial and spectral resolution data imaged simultaneousely in spectral regions spanning the low photosphere to the low-mid chromosphere.

 
The magnetic field strength decreases exponentially with height, and since magnetic flux is conserved, the cross-sectional area increases (flux tube expansion). Viewed away from disk center, geometrical effects come into play in determining the brightness of magnetic elements. Inclined lines of sight expose the hotter walls of the magnetic elements, which are the sides of granulation while the dark floor of the flux tube is hidden in the background granulation. This is called the `hot wall' effect and is seen in many observations and numerical simulations \citep{spruit_pressure_1976, topka_properties_1997,keller_origin_2004, steiner_recent_2005}. The excess brightness is also a combination of the limb darkening effect, since the optical depth unity level will move upwards when looking closer to the limb, causing more darkening in the background granulation.

\chapter{Instrumentation, data, and analysis methods}
\label{chapter_2}

The data used throughout this thesis are provided with the balloon-borne observatory $\sunrise$, during its flights in June 2009 and June 2013. 
The 2009 observations occured during a period of minimum solar activity (minimum between solar cycle 23 and solar cycle 24) and targeted a quiet-Sun network region at disk center ($\mu=0.97$), while the 2013 observations (around maximum activity) probed an active region (AR) NOAA 11768 close to disk center ($\mu$ = 0.93) with the same capabilities as the first flight.

$\sunrise$ offers us the opportunity to have a closer look (close to the diffraction limit of the telescope) into the photometric and magnetic properties of small-scale magnetic elements at an unprecedented spatial and temporal resolution, and to image a given solar region in wavelengths ranging from the visible to the near UV, down to 200\,nm. This is possible thanks to the observatory's location above 99\% of the Earth's atmosphere, and to the clever light distribution to the post focus instruments.

In this chapter, I briefly describe the scientific instruments used to obtain the data relevant for our studies, which are presented in the following chapters, the reduction techniques, and the analysis methods followed to derive the relevant atmospheric quantities. 
The instrumentation is mainly unchanged between the two flights, except for minor updates in the 2013 flight that will be addressed in the relevant chapters. We will refer to the first flight of $\sunrise$ in 2009 as $\sunrise$~I and to the second flight in 2013 as $\sunrise$~II. For a detailed description of both flights and for an overview of the main results inferred from these data, the reader is referred to \cite{solanki_sunrise:_2010} and \cite{solanki_second_2017}.

\section{Sunrise}
\label{intro-sunrise}
\subsection{Telescope}
\label{intro-telescope}
The telescope onboard $\sunrise$ is the largest solar telescope that ever left the ground, with a Gregorian configuration~\footnote{A Gregorian telescope is one that uses two reflective surfaces (mirrors) to guide the light to the main focal plane. The light is reflected from the primary (usually parabolic) mirror to the secondary mirror back to a hole in the primary and to the subsequent detectors. A field stop is put in the location of the primary image between the two mirrors to block heat from outside the FOV that can reach the secondary mirror.}, an aperture size of 1 m and effective focal length of 24\,m. 

During the first (second) mission, the telescope along with the science instruments were mounted on a gandola and launched from Kiruna in northern Sweden on June 8, 2009 (June 12, 2013) to land in Somerset Island (Boothia Pennisula) in northern Canada on June 13, 2009 (June 17, 2013). The flight altitute for both missions was 36 km, remaining above 99\% of the Earth's atmosphere. This allowed for acquiring seeing-free data for the whole duration of the flight, which was 130 h (122 h).

$\sunrise$ consists of two focal-plane instruments fed from the telescope through a light distribution system. 
The two post-focus instruments onboard the $\sunrise$ observatory are: 
\begin{itemize}
\item A vector polarimeter operating in the visible regime: IMaX \citep[Imaging Magnetograph eXperiment;][]{martinezpillet_imaging_2011}
\item A UV imager: SuFI \cite[$\sunrise$ Filter Imager;][]{gandorfer_filter_2011}
\end{itemize}

IMaX and SuFI collected simultaneous observations of the same region but with SuFI having a smaller field of view (see Figs.~\ref{imax_sufi_2009} and \ref{imax_sufi_2013}) with the use of a dichroic beam splitter in the Image Stabilisation and Light Distribution unit \cite[ISLiD;][]{gandorfer_filter_2011}. ISLiD in addition contains a rapid tip-tilt mirror controlled by a Correlating Wavefront Sensor \cite[CWS;][]{berkefeld_wave-front_2011} to stabilize the images and correct them for any defocus and optical misalignments.

The corrections applied on the collected data are divided into two categories: in-flight and post-flight corrections. We describe the latter type of correction in the next sections.

\begin{figure}
\centering
\includegraphics[scale=0.5]{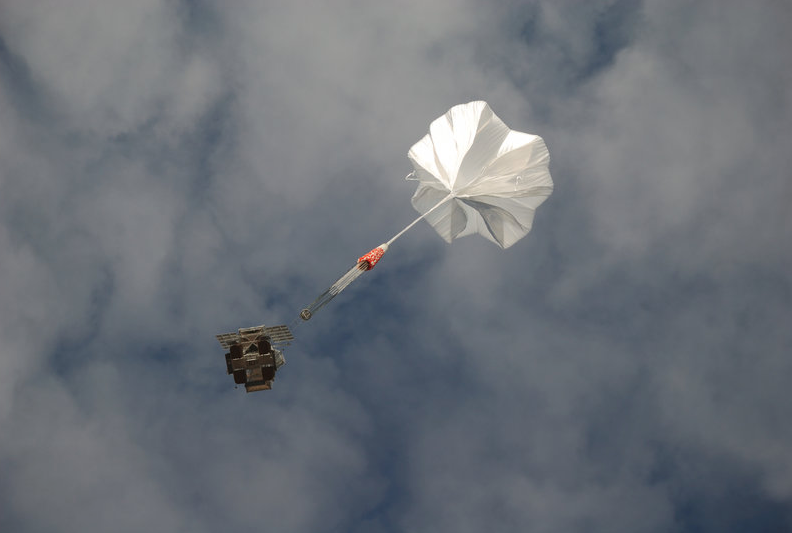}
\caption{$\sunrise$ telescope attached to the balloon and launched into space (2009). Image courtesy of $\sunrise$ team.}
\label{sunrise1}
\end{figure}

\begin{figure}
\centering
\includegraphics[scale=0.5]{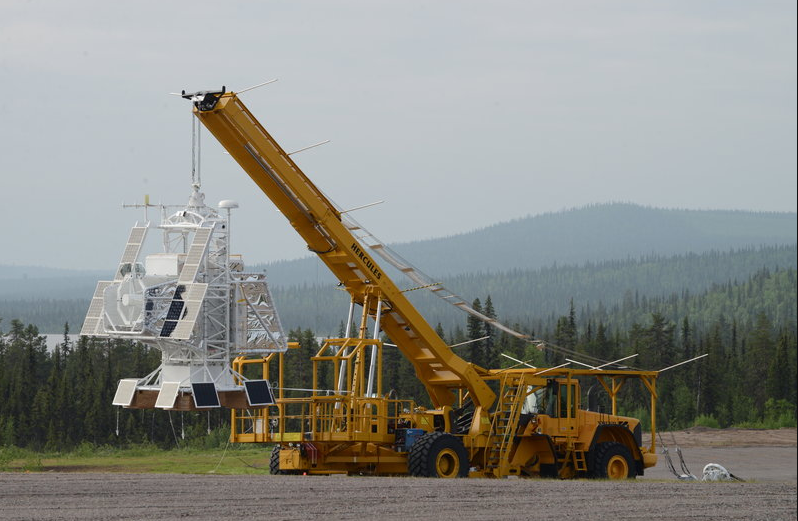}
\caption{$\sunrise$ telescope before launch, attached to the gondola (2013). Image courtesy of $\sunrise$ team.}
\label{sunrise2}
\end{figure}

\begin{figure}[h!]
\centering
\includegraphics[scale=0.3]{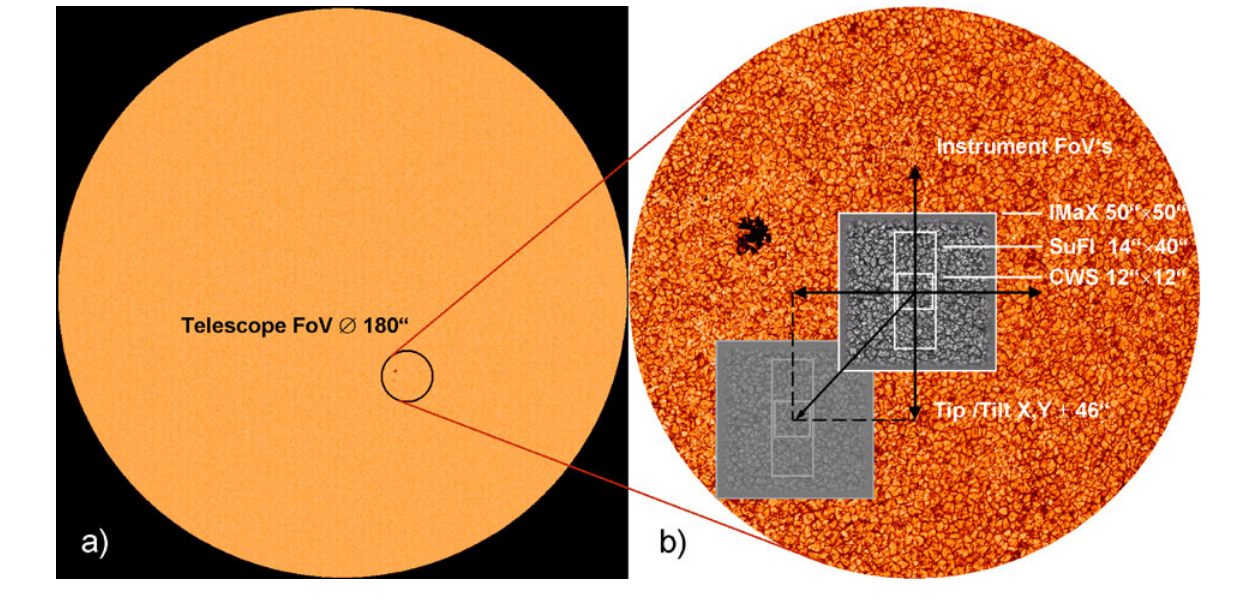}
\caption{The $\sunrise$ field-of-view in panel (a). The IMaX, SuFI, and CWS FOVs in panel (b). From \cite{barthol_sunrise_2011}.}
\label{imax_sufi_2009}
\end{figure}

\begin{figure}[h!]
\centering
\includegraphics[scale=0.3]{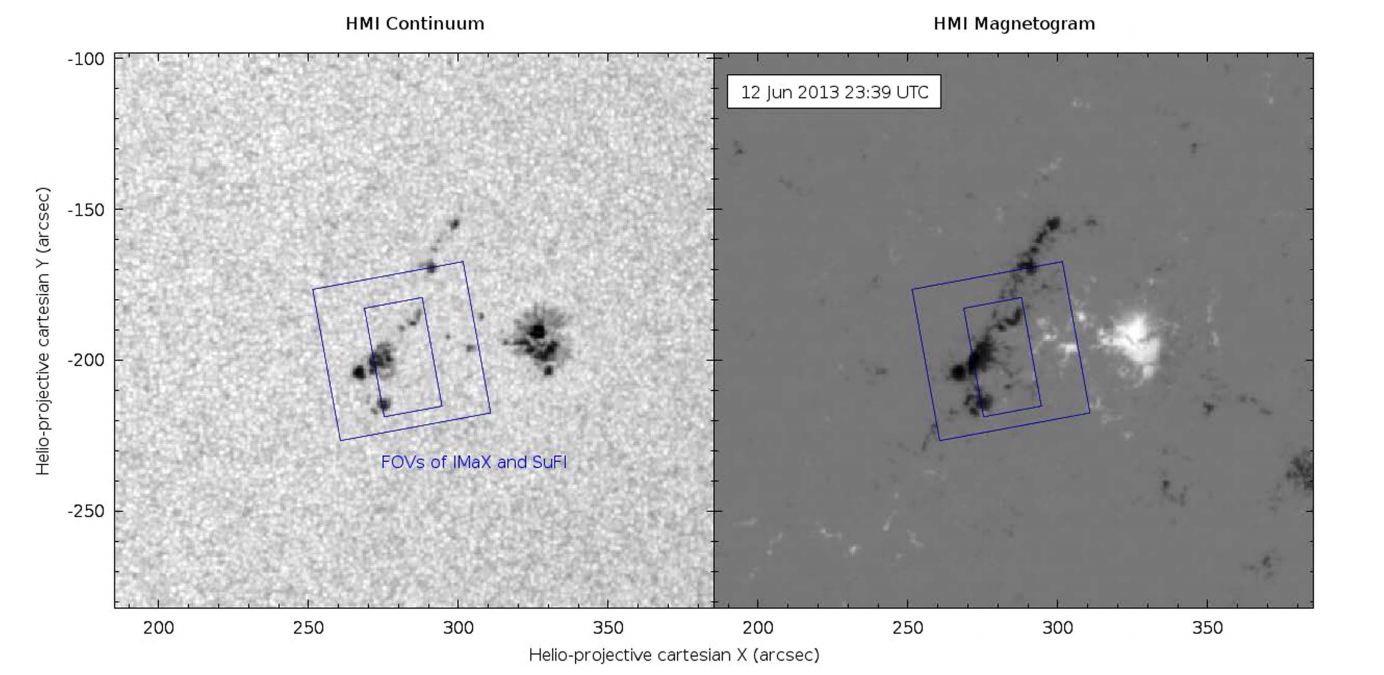}
\caption{IMaX and SuFI fields of view at the beginning of the $\sunrise$~II observations superimposed on HMI continuum (left panel) and magnetogram (right panel) images. From \cite{solanki_second_2017}.}
\label{imax_sufi_2013}
\end{figure}
\newpage

\subsection{IMaX}
\label{intro-imax}

Spectropolarimetry is the most suitable tool to analyse the dynamics of solar magnetic features, since simply put, the light passing through the solar plasma is partly polarized due to the presence of the magnetic field (see Section~\ref{intro-inversions}). However, the challenge in revealing any information from small-scale magnetic structures resides in building instruments that are capable of achieving simultaneously the following factors: high polarimetric sensitivity (for a precise magnetic field measurement), fast cadence (to track the fast evolution of magnetic elements), fine wavelength sampling (to restrict the information to a narrow atmospheric height range and avoid contamination from irrelevant spectral regions), and diffraction-limited observations (to resolve spatially the smallest magnetic structures).

IMaX is a spectropolarimeter which records two dimensional Stokes maps of a $50^{\prime\prime}\times50^{\prime\prime}$ solar region at multiple wavelength positions around the Fe\,{\sc i} line of rest wavelength $\lambda_0 =5250.2$\,\AA{}. 

\textit{For the spectral part}, IMaX as a Fabry-P\'{e}rot interferometer uses a double pass etalon through which the light passes twice, and a narrow band prefilter of FWHM = 1\,\AA{} to reduce the effect of secondary transmission peaks of the etalon. The spectral resolution achieved is about 85\,m\AA{} (see Fig.~1 of \cite{riethmuller_comparison_2014} for the IMaX spectral transmission profile).

\textit{As a polarimter}, IMaX measures the full Stokes vector, ${\bf{S}} = (I,Q,U,V)^{*}$, with the use of two Liquid Crystal Variable Retarders (LCVRs) followed by a linear polarizer (which is also a beam splitter). At the exit, two orthogonally polarized light beams are recorded on two CCD cameras, and four modulations are needed to record the full vector $\bf{S}$. The etalon is placed between the retarders and the linear polarizer~\footnote{Both the LCVRs and etalon were replaced with similar ones in $\sunrise$~II since they were damaged during the first landing}. To increase the signal-to-noise (S/N) ratio, a number of image accumulations is needed at each wavelength, all recorded in a time shorter than the timescale of evolution of granulation. We will refer to the above observing run after its completion (accumulated recorded Stokes vector in all wavelength positions) as a \textit{cycle}.

During the same flight, different IMaX observation modes were adopted, corresponding to a certain combination of number of wavelength scans, polarization states, and accumulations per wavelength point. These different modes provide observations with varying S/N ratio, cadence and spectral sampling, the choice of which depends on the scientific problem to be addressed with the corresponding data. There are mainly two polarization modes: {$\bf{L}$} and {$\bf{V}$}. {$\bf{L}$} corresponds to {$\bf{L}$}ongitudinal observations, recording only Stokes~$I$ and $V$, and {$\bf{V}$} for full Stokes {$\bf{V}$}ector measurements. These labels are followed by the number of spectral points and number of accumulations. For example, the standard mode recorded by IMaX in $\sunrise$~I and used in the analysis of the quiet-Sun data in Chapter~\ref{chapter_3} is $\bf{V}$5-4. This mode consists of measuring the full Stokes vector {$\bf{S}$} at 5 points within the Fe\,{\sc i} line (at $\Delta\lambda$ = $\pm80$, $\pm40$,$+227$ m\,\AA{} from $\lambda_0$) with 4 accumulations of 1 second exposure time per point, achieving a S/N ratio of $1000$ and a cadence (the time between two consecutive cycles) of $33$ seconds. The IMaX mode used for the analysed $\sunrise$~II AR data in Chapter~\ref{chapter_4} is $\bf{V}$8-4: full Stokes vector recorded at 8 wavelength positions within the line (at $\Delta\lambda$ = $\pm120$,$\pm80$, $\pm40$, $0$, $+227$ m\,\AA{} from $\lambda_0$) with 4 accumulations per wavelength point. The cadence achieved for this mode is 36.5 seconds, and the pixel size is unchanged between the two flights (0.054 arcsec/pixel). The noise levels as measured in Stokes~$V$ for the two flights are $3\times10^{-3}I_{QS}$ and $7\times 10^{-3}I_{QS}$, respectively. The noise is determined by fitting a Gaussian to Stokes~$V$ histograms in the continuum wavelength point since it is supposed to have zero $V$ signals (except in the presence of supersonic upflows as reported by \cite{borrero_supersonic_2010}).

IMaX/$\sunrise$~I data are corrected for fringes, dark current, and also flatfielded, at each wavelength and each polarization state. Next, they are corrected for instrumental polarization (demodulation) and for residual cross talk. To correct for wavefront aberrations introduced by the optical setup of the instrument, a thick glass is introduced into the light beam of one of the CCD cameras to record pairs of focused and defocused images (only in the continuum). The focused-defocused images are used for later post Phase-Diversity (PD) reconstruction to compute the aberrations and therefore the point-spread function, PSF, with which science images are deconvolved. 
Since the deconvolution is done in Fourier space (which is equivalent to a simple division by the Modulation Transfer Function, MTF~\footnote{The amplitude of the Optical Transfer Function (OTF, the Fourier transform of the PSF)}), an apodization~\footnote{Apodization is the smoothing of the boundaries down to zero intensity to avoid artifacts from applying discrete Fourier transform on the image} of the images was needed before performing the deconvolution, which led to the loss of information at the edges in the IMaX images. Since we had to trim these edges for furthur scientific analysis, the usable FOV is reduced to $40^{\prime\prime}\times40^{\prime\prime}$. For IMaX/$\sunrise$~II, these boundaries artifacts are avoided since instead of apodizing the images, the IMaX image was replicated around its edges before any deconvolution. In this way, the usable FOV has the original dimensions and no information is lost from any pixel at the boundaries. A detailed description of the phase-diversity method will follow in Section~\ref{intro-PD}. 
Furthermore, since spectral scans are not recorded simultaneously (because the etalon needs to be tuned), they were interpolated with respect to time to compensate for the solar evolution during the acquisition of each IMaX cycle. 

As a last correction step, instrumental stray light was removed by subtracting a certain percentage of the spatially averaged Stokes~$I$ profile from the individual Stokes~$I$ profiles. The subtracted amount corresponded to 12\% in IMaX/$\sunrise$~I data and 25\% in IMaX/$\sunrise$~II data.
The 12\% is obtained from the offset in the limb profiles at the IMaX continuum recorded by $\sunrise$~I from the zero level expected for non-contaminated profiles (see \cite{kahil_brightness_2017} and Chapter~\ref{chapter_5}). The 25\% global stray light is the value obtained by \cite{riethmuller_new_2017} after testing the strength of the magnetic field in the pores returned by the inversions of the stray-light corrected IMaX data. We summarize here the different levels corresponding to each of the correction steps explained above:
\begin{enumerate}

\item[$\bullet$] level~$0.0$: raw data in FITS format.
\item[$\bullet$] level~$1.0$: data corrected for dark current, flatfielded, fringes removed, demodulated, cross-talk corrected.
\item[$\bullet$] level~$2.0$: data deconvolved with the PD PSF obtained from PD measurements.
\item[$\bullet$] level~$2.2$:  level~$2.0$ data interpolated in time.
\item[$\bullet$] level~$2.3$: level~$2.2$ data corrected for stray light.
\end{enumerate}

The level~$2.3$ data are then used as an input for the SPINOR inversion code to fit synthesized Stokes profiles to the observed sepctra and compute the relevant atmospheric parameters for our study (see Section~\ref{intro-inversions} for more details on the inversion strategy).  
The accuracy of the physical parameters (especially the magnetic field vector) retrieval from IMaX data given the instrument configuration described earlier was investigated by \cite{orozco_suarez_retrieval_2010}. They concluded, using MURaM simulations, that a spectral resolution within $50-100$ m$\AA$ and 5 wavelength points in the Fe\,{\sc i} 5250.2\,\AA{} line are enough to retrieve reliable information on the solar magnetic field vector in the lower photosphere.

\subsection{SuFI}
\label{intro-sufi}
SuFI is a filter imager, collecting images of a part of the IMaX FOV at wavelengths in the Near Ultra-Violet (NUV) sampling heights from the lower photosphere to the lower chromosphere. The images are collected on a CCD camera with dimensions of $2048\times2048$ pixels, and with a filter wheel shifting through the wavelength channels. 
The various SuFI modes of observation differed mainly in the wavelengths observed.

The analysis of SuFI images in Chapter~\ref{chapter_3} is done in the following wavelength channels along with their bandwidths (in brackets): 214 nm (10 nm), 313 nm (OH-band; 1.2 nm), 388 nm (CN-band; 0.8 nm) and 397 nm (core of Ca\,{\sc ii} H line; 0.18 nm). The cadence at all wavelengths was 12 seconds (39 seconds) whenever observations in the 214 nm channel are excluded (included), since the photon flux at 214 nm is strongly attenuated by the Earth's atmosphere and needs longer exposures, even at high elevation angles. For SuFI/$\sunrise$~II data analysed in Chapter~\ref{chapter_4}, we use the images at 300 nm (4.4 nm bandwidth) and Ca\,{\sc ii} H narrowband (0.11 nm) and wide (0.18 nm) images, with a cadence of 7 sec. The plate scale is slightly wavelength dependent, but on average it is equal to 0.02 arcsec/pixel. 



The molecular bands (313 and 388 nm) provide high intensity contrast images of the magnetic structures due to the dissociation of the CN and OH molecules at high temperatures. The 214 nm channel samples the mid and upper photosphere. The 397 nm gets considerable contribution from the low-mid chromosphere, which allows for investigating the energy transport from lower layers to the upper atmosphere. 
Data are corrected for dark current and flatfielded. A PD image doubler is introduced in the focal plane, which consists of 2 parallel glass plates. Only half of the second plate is coated so that it transmits half of the incoming light to the detector without any reflection, while the other half undergoes two reflections. This creates a pair of focused-defocused images recorded simultaneously for later PD reconstruction described in Section~\ref{intro-PD}. The PD reconstruction was applied either to sub-boxes in each SuFI image, computing individual wavefront aberrations and deconvolving with independent PSFs in each box, or averaging those aberrations spatially over individual images and over time, i.e., over multiple images, and deconvolving the data with the averaged PSF.

The data used in the subsequent chapters are corrected for instrumental stray light. In Section~\ref{intro-SL}, we discuss in more details the origin of stray light and the different ways to correct for it. We only mention here that limb images recorded during the flight are used to derive the stray light contribution to the point spread function of the telescope optical system (see Chapter~\ref{chapter_5}). Images are then deconvolved with the computed PSF to obtain stray light-free data. Correcting for stray light is crucial for studying the contrast of magnetic elements and comparing with results from solar magneto-convection simulations, especially in the NUV. We summarize here the data reduction levels corresponding to the above corrections:
\begin{enumerate}
\item[$\bullet$] level~$0.0$: raw data in FITS format.
\item[$\bullet$] level~$1.0$: data corrected for dark current and flatfielded.
\item[$\bullet$] level~$2.0$: level~$1.0$ data deconvolved with individual PSFs 
\item[$\bullet$] level~$3.0$: level~$1.0$ data deconvolved with a PSF averaged over images and time
\item[$\bullet$] level~$3.1$: level~$3.0$ data corrected for stray light using limb profiles.
\end{enumerate}

The longest time-series of SuFI data in $\sunrise$~I lasted for 34 minutes (when 214 nm observations were included) and 19 minutes (without inclusion of the 214 nm channel), while in $\sunrise$~II it was 60 minutes (without 214 nm). However, since our study mainly focuses on the simultaneous magnetic and photometric information of magnetic elements, we only use the SuFI images from the time-series where IMaX data were also available. Accordingly, for the quiet-Sun data analysis in Chapter~\ref{chapter_3}, the time series lasted for 19 minutes while it was 17 minutes for the active region data in Chapter~\ref{chapter_4}. 

\newpage
\section{Image Restoration}
\label{intro-restoration}

\subsection{Phase-diversity reconstruction} 
\label{intro-PD}

The phase-diversity (PD) technique aims at estimating the aberrations caused by the Earth's atmosphere (for ground based instruments) and instrumental aberrations (for both, ground-based and space-based observations). Once those aberrations are found, deconvolving the recorded images with an appropriate PSF can return the original image affected only by diffraction effects (in addition to the noise).
For $\sunrise$ data, the PD technique consisted of recording two images of the same region, both are subject to an unknown aberration, but one with an additional known aberration introduced by the optical system. The recorded images are then divided into focused and defocused images used later for reconstruction.
As explained earlier, this was done for IMaX only in the continuum, and for SuFI, at all imaging wavelengths.\\

The usage of such methods on solar observations was described in \cite{lofdahl_wavefront_1994}. We will briefly describe the mathematical approach to this method, and the assumptions taken alongside.
We introduce the needed physical quantities in image space (Fourier space):\\ \\
$t$ ($T$) is the true image free from any aberrations.\\
$f$ ($F$) is the recorded image (the focused one).\\
$d$ ($D$) is the recorded defocused one.\\ 
$s$ ($S$) and $s_d$ ($S_d$) are the point spread functions of the two recorded images.\\ 
$\phi$ and $\phi_d$ are the unknown and known phase aberrations. \\
$n$ and $n_d$ are the amounts of noise in focused and defocused images, respectively. \\
The aberrations are usually expressed in terms of Zernike polynomials:
$\phi = \Sigma_{i=1}^{i_{max}} c_i Z_i$, with $i_{max}$ is the maximum number of Zernike polynomials $Z_i$ \citep{noll_zernike_1976}. \\ \\
$r$ is the two dimensional vector in the image plane. \\
$u$ is the two dimensional vector in the pupil plane~\footnote{Here the exit pupil, which is the image of the aperture as seen from the image plane.}. \\
$v$ is the two dimensional spatial frequency in Fourier space. \\

The point spread function and the aberrated phase of the focused image are related through the following expression:
\begin{equation}
s(r) = |F^{-1} \{ P(u) e^{i\phi(u)} \} |^2
\label{psf}
\end{equation}
And for the defocused image:
\begin{equation}
s_d(r) = |F^{-1} \{ P(u) e^{i(\phi(u)+\phi_d(u)} \} |^2
\end{equation}

$P(u)$ is the binary aperture function, it is equal to 1 inside the pupil and 0 outside. In the absence of any aberration (i.e., $\phi$ = 0), the point spread function takes the shape of an Airy pattern (diffraction-limited PSF).
In this problem, we have two unknowns, the point spread function $s$ and the true image $t$, which are estimated by minimizing an 'error metric`, $L$:

\begin{equation}
L = \Sigma_v |F-T \times S|^2 + \gamma|D-T \times S_d|^2
\label{Error}
\end{equation}

$L$ is the evaluated difference in mean-squared intensity between the collected images and those predicted by the optical system, which are related by (in Fourier space):

\begin{equation}
F(v) = T(v)\times S(v) + N(v)
\end{equation}  

And,

\begin{equation}
D(v) = T(v)\times S_d(v) + N_d(v)
\end{equation}           

The constant $\gamma=\sigma^2/\sigma_d^2$ is introduced if the noise variances $\sigma^2$ and $\sigma_d^2$ in the two images are not the same. 
An estimation of the optimum image $T_M$ is obtained first by minimizing $L$ with respect to the true image $T$:

\begin{equation}
T_M = Q^2(FS^{*}+\gamma DS_d^{*})
\label{real}
\end{equation}
With $Q^2 = (|S|^2 + \gamma.|S_d|^2)^{-1} $

Substituting $T_M$ into Eq.~\ref{Error} gives us:
\begin{equation}
L_M = \Sigma_v |E|^2
\end{equation}

With $E = Q (FS-DS_d)$.

Then the set of parameters $c_i$ are found such that $E$ (which is now a function of only the Zernicke coefficients) is optimum.
Once the problem is solved iteratively, the coefficients can be used to estimate the point spread function according to Eq.~\ref{psf} and therefore the optical transfer function, $S$ by a simple fourier transform. The real image can be restored using Eq.~\ref{real}.

As mentioned earlier in Sect.~\ref{intro-sufi} the wavefront aberrations can be derived for the whole image or for individual sub-apertures in the image. In the last case, the total wavefront can be averaged over all images, over the whole duration of the time-series. This will lead to the averaged wave-front errors for which data will be corrected and therefore called level~$3$ data \citep{hirzberger_quiet-sun_2010,hirzberger_performance_2011}.

Extensions to the phase diversity reconstruction method are the Multi-Frame Blind Deconvolution \citep[MFBD;][]{lofdahl_phase_1996} which makes use of multiple frames for the PD, and the Multi-Object Multi-Frame Blind Deconvolution \citep[MOMFBD;][]{van_noort_solar_2005} which makes use of multiple frames taken at different wavelengths (or different objects since wavelengths sample a certain atmospheric height where solar feature appearences change). This will analytically provide more information on the degraded image and lead to a better estimation of the wavefront aberrations and therefore the restored image.
\newpage
\subsection{Stray light correction}
\label{intro-SL}

The phase diversity technique is mainly sensitive to the aberrations that contribute to the core of the PSF, since only a limited number of Zernike polynomials can be fitted to avoid noise amplification. In contrast, the PSF wings are more affected by stray light, which dominates the higher spatial frequency small-scale wavefront aberrations. While large-scale wavefront aberrations affect the spatial resolution of the data, stray light affect the RMS contrast of the image, and it can vary across the FOV, i.e., it can be anisotropic.

In the case of satellite observations, instrumental stray light or scattered light (or the total point spread function resulting from both, optics and stray light) is usually estimated by observations of solar eclipses \citep{deubner_new_1975} or observing the edge of passing objects in front of the Sun like Mercury/Venus transits \citep{wedemeyer-bohm_point_2008, wedemeyer-bohm_continuum_2009, mathew_stray_2009, yeo_point_2014}, or, if data with transiting objects were not available, by using solar limb profiles \citep{mathew_properties_2007}. Other methods are based on the comparison of observed granulation contrast with MHD simulations \citep{danilovic_intensity_2008, afram_intensity_2011}. Once the PSF of the optical system is estimated, the observations are deconvolved with it to restore the `clean' images. Since no planetary or satellite transits were available during both $\sunrise$ flights, solar limb observations were used to estimate the wings of the PSF, while the core is succesfully modelled by the PD reconstruction (Sec.~\ref{intro-PD}). Note that the pointing jitter due to structural vibrations in the telescope also affects the shape of the PSF, and is usually hard to model.

In spectropolarimetric data, stray light has to be corrected for before inferring the physical parameters via an inversion technique, otherwise the false light could affect the computed value of the magnetic field strength/inclination angle, especially in the quiet Sun and in the dark pores. In some works, the stray light is considered to be a free parameter of the inversion \citep{orozco_suarez_quiet-sun_2007, borrero_inferring_2011}, in this case the pixel is divided into two parts, one that is magnetized (with a filling factor $\alpha$ and emergent Stokes vector [$I_m, Q_m, U_m, V_m$]) and another non-magnetized part filling a fraction of $1-\alpha$ of one pixel (with a signal $I_{nm}$). The Stokes vector is then defined by:

\begin{align}
I = \alpha \times & I_{m}+(1-\alpha)\times I_{nm} \\
& Q = \alpha \times Q_m \\
& U = \alpha \times U_m \\
& V = \alpha \times V_{m} \\
\end{align}

(see \citet{stenflo_magnetic-field_1973}; \cite{stenflo_dependence_1985}; \cite{solanki_small-scale_1993}). The Stokes signals, $Q$, $U$, and $V$ are then reduced by a factor of $\alpha$ due to instrumental defocus or stray light. This is called the local stray light approach, since the signal coming from stray light $I_{nm}$ is taken as the average of the profiles of the inverted surrounding pixels in one resolution element.
As mentioned in Sect.~\ref{intro-imax}, IMaX data in Chapters~\ref{chapter_3} and ~\ref{chapter_4} were corrected for stray light by subtracting a certain percentage of the averaged mean Stokes~$I$ across the image. This global stray light approach is based on the approximation that stray light is constant across the image, and not varying locally. For the SuFI data used in the subsequent chapters, solar limb profiles recorded during the flights were used to asses the stray light contribution to the total PSF of the optical system.  
     
\section{Inversions} 
\label{intro-inversions}

Electromagnetic radiation from the Sun interacts with the medium in which it travels before reaching us. Information about the atmospheric parameters controlling this interaction is embedded in the detected intensity profiles. The magnetic field is one major atmospheric quantity which leaves a distinctive imprint on the spectral line profiles. Below I give a brief overview of the effect of introducing a magnetic field $\textbf{B}$ on the emergent spectra (Zeeman effect), one distinctive property of the radiation escaping a $\textbf{B}$-filled medium (polarization), and the equation relating the atmospheric properties to the emergent intensity spectra in a magnetized medium (the polarized radiative transfer equation). I end this section by describing how solar physicists use the radiative transfer equation to infer the physics of a magnetized medium observed through a spectropolarimeter (Inversions). \textbf{The contents of this Section are based on readings from \cite{solanki_photospheric_1987}, \cite{del_toro_iniesta_introduction_2003} and \cite{stix_sun_2004}, so the reader is refered to such works and to references therein for a more in-depth derivation of the equations.}  

\subsection{Zeeman effect}
\label{intro-Zeeman}
In the presence of a magnetic field $\bold{B}$, and in the Zeeman regime (i.e., the L-S coupling dominates over the $\bold{B}$-electron magnetic moment interaction), the degeneracy of the energy levels is removed, i.e., the energy of a transition is not only determined by the total angular momentum $\bold{J = L + S}$ of the electron but also on the projection $\bold{M}$ of $\bold{J}$ along the direction $\bold{B}$.
The energy shift from the central energy (corresponding to $\bold{B}$=0) is given by:

\begin{equation}
E = M\frac{e h}{2mc} g B
\end{equation}

M takes $2J+1$ values ranging from $-J$ to $J$. $g$ is the Land\'{e} factor characterizing the sensitivity of the energy level of total angular momentum $\textbf{J}$ to $\textbf{B}$, and is given by:
\begin{equation}
g = \frac{3}{2} +\frac{s(s+1)-l(l+1)}{2j(j+1)}
\label{g}
\end{equation}

An atomic transition from a lower level \textit{l} ($J_l, M_l, g_l$) to an upper level \textit{u} ($J_u, M_u, g_u$) occurs if $\Delta J=J_u-J_l=0,\pm 1$ and $\Delta M = M_u-M_l=0,\pm 1$. The shift $\Delta\lambda$ from the central wavelength $\lambda_0$ (in the absence of any magnetic field) is:
\begin{equation}
\Delta \lambda = \frac{e\lambda_0^2B}{4\pi m_e c}(M_lg_l - M_ug_u)
\label{del_lam}
\end{equation}

Transitions with $\Delta M = 0$ are called $\pi$ components, while those with $\Delta M = \pm 1$ are called $\pm \sigma$ components. The Zeeman splitting corresponds to the wavelength shift of one of the $\sigma$ components with respect to $\pi$.
\textbf{Normal Zeeman effect} corresponds only to three transitions, and according to Eq.~\ref{del_lam} this happens when both levels have the same Land\'{e} factors ($g_l = g_u$), or when the transition occurs between $J=0$ (and therefore $g$ = 0) and $J=1$ levels.  
In the other cases where more than 3 components are formed (\textbf{anomalous Zeeman effect}), an effective Land\'{e} factor is used, and the corresponding wavelength shift is given by:

\begin{equation}
\Delta \lambda = \frac{e\lambda_0^2B}{4\pi m_e c}g_{eff}
\label{del_lam_2}
\end{equation}
$g_{eff}$ corresponds to a separation $\Delta \lambda$ between the centers of gravity of the shifted $\sigma$ components from the unshifted $\pi$, and is given by:

\begin{equation}
g_{eff} = \frac{1}{2}(g_u+g_l)+\frac{1}{4}(g_u-g_l)[j_u(j_u+1)-j_l(j_l+1)]
\label{g_eff}
\end{equation}

\subsection{Polarization}
\label{intro-Polarization}
One distinctive property of the three components described earlier is that they are polarized. The state of polarization of each is dependent on the direction of the magnetic field with respect to the line-of-sight of the observer.
In the local frame of the atom, the unshifted $\pi$ component is always directed along the magnetic field vector, while the two components $\sigma_{+}$ and $\sigma_{-}$ are precessing around it. As a result, if signals are seen along the magnetic field line, the $\sigma$ components are right and left circularly polarized, while the $\pi$ component disappears. On the other hand, if the observer's line-of-sight is perpendicular to the magnetic field, the $\pi$ component is linearly polarized while $\sigma_{+}$ and $\sigma_{-}$ are linearly polarized perpendicularly to it.
For a random orientation of the field, the light in the $\sigma$ components is elliptically polarized.

Since Stokes~$V$ is a measure of circular polarization, and Stokes~$Q$ and $U$ of linear polarization (see \cite{stenflo_solar_2013} for a review on solar spectropolarimetry), one can come to the conclusion that Stokes~$V$ is sensitive to the longitudinal (vertical to the solar surface at $\mu=1$) magnetic field, while Stokes $Q$ and $U$ are a measure of the transverse (horizontal to the solar surface) component of the field.

\begin{figure}
\centering
\includegraphics[scale=0.1]{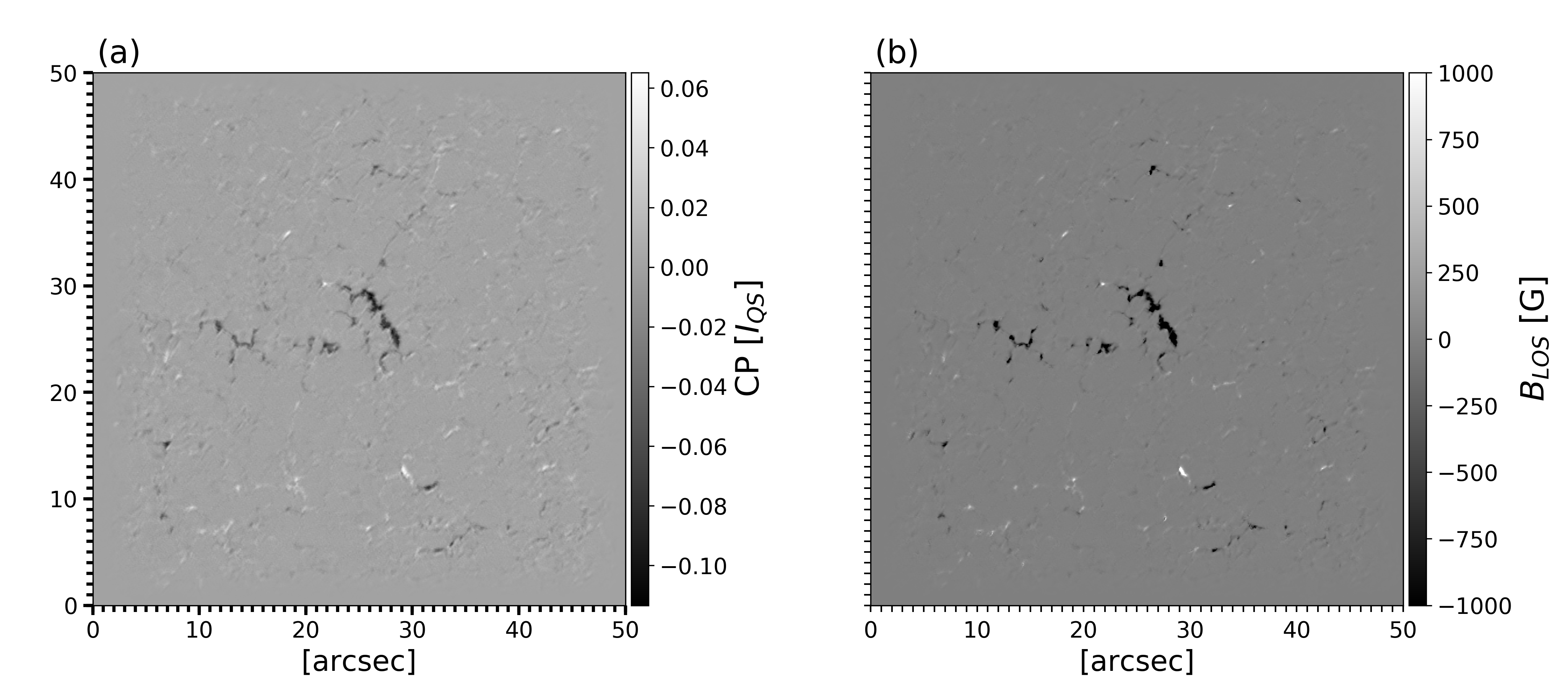}
\caption{(a) The circular polarization map computed from recorded Stokes vector by IMaX in $\sunrise$~I. (b) The longitudinal magnetic field map derived from inverting the same Stokes data. }
\label{CP}
\end{figure}

Figure~\ref{CP}~(a) shows the circular polarization (CP) map which is computed by averaging Stokes~$V$ signals over the 4 wavelengths (excluding the continuum point) recorded by IMaX during the $\sunrise$~I flight (according to the definition of CP in \cite{riethmuller_bright_2010} and \cite{jafarzadeh_structure_2013}. Figure~\ref{CP}~(b) is the corresponding LOS magnetic field map derived from the inversions of IMaX data described in the next section. This shows that circular polarization is proportional to and can be used as a proxy for measuring $B_{\rm LOS}$ (Eq.~\ref{weak_field_eq }). Note that this is only valid in the weak-field regime, where Zeeman splitting $\Delta \lambda$ (Eq.~\ref{del_lam}) is too small compared to the width of the line in the absence of any magnetic field ($\Delta \lambda_D$, introduced in the next subsection), and where first order perturbation to the Stokes~$I$ yields to:

\begin{equation}
V = -\Delta \lambda \,g \,\cos\theta \,\frac{\partial I}{\partial \lambda} 
\label{weak_field_eq }
\end{equation}
(see \cite{landi_deglinnocenti_polarization_2004}), $g$ is the Lande factor, and $\theta$ is the inclination angle of the magnetic field vector with respect to the line of sight.

\newpage
\subsection{Radiative Transfer Equation}
\label{intro-RT}

\subsubsection{Synthesis mode}
The equation relating the emergent intensity to the atmospheric parameters along the light path is referred to as the \textbf{R}adiative \textbf{T}ransfer \textbf{E}quation (RTE). In the presence of a magnetic field, splitting of energy levels and polarization of the corresponding transitions occur. The emergent intensity takes then the form of a Stokes vector, $\bold{I}=(I,Q,U,V)^\intercal$:

\begin{equation}
\frac{d\bold{I}}{ds} = -\bold{K}I + \bold{J}
\end{equation}

The RTE describes the change in Stokes vector $d\bold{I}$ at a specific wavelength  over a distance $ds$. This change is caused by absorption and emission processes along the light path. All the physical parameters of the medium creating these two processes are included in the $4\times 4$ propagation matrix $\bold{K}$ and in the emission vector $\bold{J}$.

In continuum optical depth scale ($\tau_c$), and with $d\tau_c = - \bold{k}\, ds$, the RTE can be written as:

\begin{equation}
\frac{d\bold{I}}{d\tau_c} = \bold{K} (\bold{I}-\bold{S} )
\end{equation}

$\bold{S}$ is the source function, which in LTE translates to $(B(T),0,0,0)^\intercal$. 
$\bold{K}$ is the sum of the continuum and line opacities:
\begin{equation}
\bold{K} = \kappa_{cont} 1+ \kappa_{line} \Phi = \kappa_{cont} (1 + \eta_0 \Phi )
\end{equation}

$\kappa_{cont}$ is the continuum absorption coefficient, $1$ is the unity matrix (of dimensions $4\times 4$), $\kappa_{line}$ is the line absorption coefficient, $\eta_0$ is the line-to-continuum absorption coefficient ratio, and $\Phi$ is the absorption matrix. For a given bound-bound transition from a lower level $l$ with an Einstein coefficient $B_{lu}$~\footnote{The probability per unit time per unit energy density that an electron jumps from level $l$ to $u$ upon absorbing a photon} and number density $n_l$ of atoms populating that level, to an upper level $u$ ($B_{ul}$, $n_u$), $\kappa_{line}$ is given by:

\begin{equation}
\kappa_{line} = \frac{h\nu}{4\pi} (n_l\,B_{lu} - n_u\,B_{ul})
\end{equation}

This expression of $\kappa_{line}$ is derived from a \textbf{quantum} mechanical approach. Treating the electrons as damped oscillators in the presence of a radiation field (Lorentz \textbf{classical} model), $\kappa_{line}$ is given by:

\begin{equation}
\kappa_{line} = \frac{n_l\,\pi \,e^2}{m\,c}
\end{equation}
 
$m$ and $e$ are the mass and charge of the electron, respectively. In addition to the absorption matrix which describes the reduction of the intensity, $\bold{K}$ includes two other matrices. The dichroism matrix which describes the production/destruction of polarisation, and the dispersion matrix which describes the conversion of one type of polarisation into another. The propagation matrix is then given by:

\begin{equation}
\bold{K} = 
\begin{pmatrix}
\eta_I & \eta_Q & \eta_U & \eta_V \\
\eta_Q & \eta_I & \rho_V & -\rho_U\\
\eta_U & -\rho_V & \eta_I & \rho_Q \\
\eta_V & \rho_U & -\rho_Q & \eta_I \\
\end{pmatrix}
\end{equation}

$\rho$ and $\eta$ are the matrix elements of $\bold{K}$ and contain all the information on the geometry of the problem (inclination ($\gamma$) and azimuthal ($\chi$) angles of the magnetic field vector with respect to the line of sight) and properties of the medium, and are given by:

\begin{align*}
\eta_I &= 1+\frac{\eta_0}{2} \left(\phi_p \sin^2\gamma + \frac{\phi_b+\phi_r}{2}(1+\cos^2 \gamma)\right) \\
\eta_Q &= \frac{\eta_0}{2} \left(\phi_p - \frac{\phi_b + \phi_r}{2}    \right) \sin^2 \gamma \cos 2\chi \\
\eta_U &= \frac{\eta_0}{2} \left(\phi_p - \frac{\phi_b + \phi_r}{2}    \right) \sin^2 \gamma \sin 2\chi \\
\eta_V &= \frac{\eta_0}{2} \left(\frac{\phi_r - \phi_b}{2}    \right) \cos \gamma \\
\rho_Q &=  \frac{\eta_0}{2} \left(\psi_p - \frac{\psi_b + \psi_r}{2}    \right) \sin^2 \gamma \cos 2\chi \\
\rho_U &=  \frac{\eta_0}{2} \left(\psi_p - \frac{\psi_b + \psi_r}{2}    \right) \sin^2 \gamma \sin 2\chi \\
\rho_V &= \frac{\eta_0}{2} \left(\psi_r - \psi_b  \right) \cos \gamma 
\end{align*}

$\phi$ and $\psi$ are the absorption and dispersion profiles for all Zeeman components, $i=(p,b,r)$ (or $\pi,\, \sigma_{+}, \,\sigma_{-}$ components introduced in Section~\ref{intro-Polarization}), of the atomic transition under study:

\begin{equation}
\phi_i = \sum_j S_{i_{j}} \, H(a,v) \hspace*{0.3 in} \psi_i = \sum_j S_{i_{j}} \, F(a,v) 
\end{equation}

$H$ and $F$ are the Voigt and Faraday-Voigt functions, describing the absorption and dispersion profiles, respectively. The summation runs over all transitions $j$ in each component $i$ in case of anomalous Zeeman splitting (see Section~\ref{intro-Zeeman}). 
In the expression of these profiles enters the wavelength shift of the Zeeman split profiles, $v$. This shift is a summation of the splitting caused by the magnetic field (Eq.~\ref{del_lam_2}), and the shift caused by the presence of a macroscopic velocity $v_z$ along the line of sight ($\frac{v_z \lambda_0}{c \Delta \lambda_D}$), in addition to shift caused by the Doppler effect $\frac{\lambda - \lambda_0}{\Delta \lambda_D}$. These velocities are expressed in terms of the natural width of the line ($\Delta \lambda_D = \frac{\lambda_0}{c}\,\sqrt{\frac{2\,K\,T}{m}+v_{mic}}$), with $v_{mic}$ is the microturbulence velocity (which results from turbulent motions occuring on scales smaller than the photon mean free path). Broadening due to collisions and the radiative lifetime of the energy level enters in the expression of the damping parameter, $a$, which tweaks the shape of the wings of the absorption and dispersion profiles.

Knowledge of the quantum properties of the specific transition (e.g., oscillator strength, Einstein coefficients, ionization energy) allows for the computation of $\eta_0$, $\phi$, and $\psi$ which enter in the expression of the absorption matrix $\bold{K}$. Saha and Boltzmann equations are used, assuming LTE conditions, to find the level population of atoms and ionization levels. This is where thermodynamical (temperature and gas density) and atomic properties come into play. The Planck function is also estimated from the temperature.

\subsubsection{Inversion mode}
If one knows the physical parameters representing a certain region on the solar surface, and their dependence on the optical depth, the RTE can be used to synthesize Stokes vector, $\bold{I}$. The equation is then used in its \textbf{forward} mode.

In solar physics applications, one is interested in retrieving the physical conditions under which the observed Stokes vector is obtained by the detector. The RTE is then used in its \textbf{inverse} mode, and the problem is called \textit{Inversion}. For reviews on inversions techniques, check \cite{del_toro_iniesta_stokes_1996} and \cite{bellot_rubio_stokes_2006}.

The inversion process starts with computing Stokes vector $\bold{I_{syn}}$ of the line transition under study from an initial estimate of the physical parameters: temperature, magnetic field vector, gas density, microturbulence, line-of-sight velocity. Usually, these parameters are inferred from a model atmosphere corresponding to the solar region under study (quiet Sun, plage, sunspot) or from the output of magnetohydrodynamical (MHD) simulations with average magnetic flux corresponding to these regions.

The observed Stokes vector $\bold{I_{obs}}$ is then fitted iteratively with the synthesized Stokes vector $\bold{I_{syn}}$ until an optimized solution is found. Different optimization schemes can be used, the most common is the non-linear least square minimization based on the Levenberg-Marquardt algorithm \citep{press_numerical_2007} \citep[e.g.][]{ruiz_cobo_inversion_1992,ruiz_cobo_sensitivity_1994}. Other codes refer to global optimization algorithms like genetic algorithms \cite[PIKAIA;][]{lagg_retrieval_2004}. Global optimization algorithms are used to look for a solution in a wider parameter space and therefore converging to the global minimum, but it is computationally expensive. 

In terms of approximations, Milne-Eddington inversions are based on the approximation that physical parameters do not change along the line of sight (the absorption matrix $\bold{K}$ is constant) and the source function varies linearly with optical depth \citep[e.g.,][]{lagg_retrieval_2004, borrero_inferring_2011}. This allows for solving the RTE analytically (see \cite{landi_deglinnocenti_magnetic_1992} and \cite{del_toro_iniesta_introduction_2003} for the numerical solution of the RTE in this approximation).

SPINOR (The {\bf S}tokes-{\bf P}rofiles-{\bf IN}version-{\bf O}-{\bf R}outines) inversion code works with gradients of atmospheric parameters and can synthesize atomic and molecular lines at any given disc position \citep{frutiger_properties_2000}. It is also possible to set a number of model atmosphere components (for example when the observed features are not resolved) and compute the emergent Stokes vector for each component. SPINOR uses the STOPRO routines \citep{solanki_photospheric_1987} for solving the RTE under LTE conditions, and INVERT routine for fitting observed Stokes profiles using the Levenberg-Marquardt algorithm.
In Chapters~\ref{chapter_3} and \ref{chapter_4}, the inversion code SPINOR is used to retrieve the atmospheric parameters. I will not repeat here the strategy for inverting the spectropolarimetric IMaX data employed in this thesis, as it will be described in Chapters~\ref{chapter_3} and \ref{chapter_4}.

\section{Centre-Of-Gravity (COG) method} 
\label{intro-COG}

As an alternative to Stokes inversions, other proxies can be used to derive the LOS magnetic field, $B_{\rm LOS}$. The Centre-of-Gravity method \citep{semel_contribution_1967,rees_line_1979} computes the $B_{\rm LOS}$ values by measuring the separation $\Delta\lambda_{CG}$ between the two circularly polarized components $I+V$ and $I-V$ profiles. It is valid only in the weak-field regime. Another assumption is that the magnetic field is assumed to be constant along the line of sight.

If $I_c$ is the continuum intensity at each pixel, $\Delta \lambda_{\pm}$ are the centers of gravity of the residual intensities $(I_c-I\pm V)/2$ profiles and are given by:
\begin{equation}
\Delta \lambda_{\pm} = \frac{\int (I_c - (I \pm V))\Delta \lambda \,d\Delta \lambda}{\int(I_c - (I \pm V)) \,d\Delta \lambda}
\end{equation}

$\Delta\lambda_{CG}$ can then be computed by subtracting the $\Delta \lambda_{\pm}$  values:

\begin{equation}
\Delta \lambda_{CG} = \frac{\Delta \lambda_{+} - \Delta \lambda_{-}}{2} =  \frac{\int V\, \Delta \lambda\, d \Delta \lambda}{\int (I_c-I) \,d\Delta \lambda}
\end{equation}

For a normal Zeeman splitting, $B_{\rm LOS}$ can be computed via: 
\begin{equation}
B_{\rm LOS} = \frac{\Delta \lambda_{CG}}{Cg\lambda_0^2}
\end{equation} $\lambda_0$ is the central wavelength of the line (measured in the laboratory), $C$ is a constant equal to $e/4\pi mc^2$ = 4.67$\times10^{-13}$ \AA{}$^{-1}$ G$^{-1}$, $g$ is the Land\'{e} factor for normal Zeeman splitting and should be replaced with the effective Land\'{e} factor $g_{eff}$ for anomalous Zeeman splitting.

\begin{figure}[h!]
\centering
\includegraphics[scale=0.5]{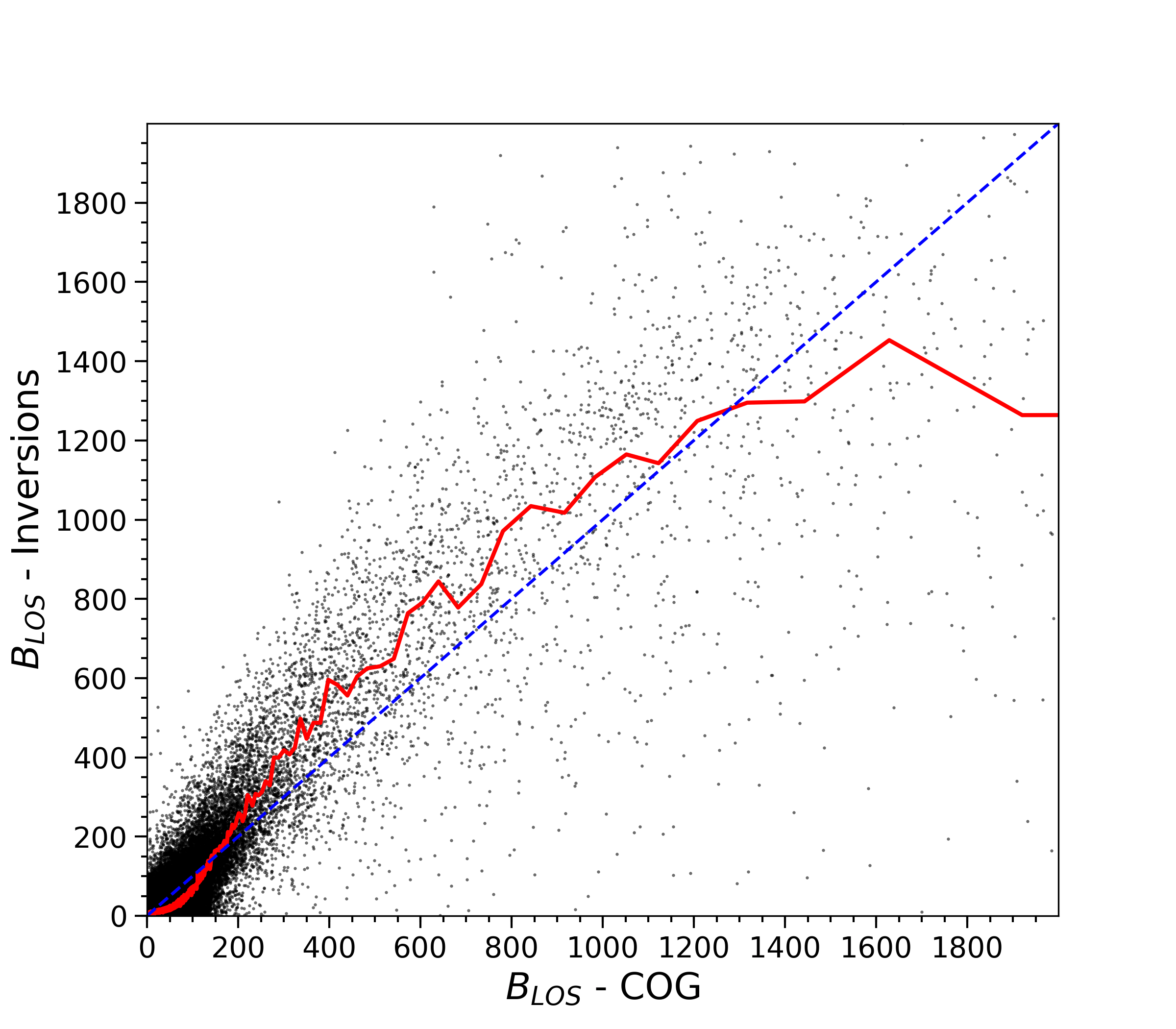}
\caption{Pixel-by-pixel scatterplot of the $B_{\rm LOS}$ values computed from the inversions of IMaX Stokes profiles vs. those obtained from the center-of-gravity method applied to the 5 discrete wavelength points. The red curve joins the averaged pixels in bins of 200 data points each. The blue line has a slope of unity.}
\label{cog_inv}
\end{figure}

Below I will be discussing the difference in the computed $B_{\rm LOS}$ values with both methods (inversions and COG) applied to IMaX Stokes profiles. Both methods are applied to V5-6 mode data described earlier in Section~\ref{intro-imax}, and restricted only to one quiet-Sun cycle taken from the time series used in Chapter~\ref{chapter_3}.
The difference between the computed values is best illustrated with a pixel-by-pixel scatterplot. Figure~\ref{cog_inv} shows the scatterplot of the $B_{\rm LOS}$ values computed from COG (and called $B_{COG}$) and those computed from the inversions described in Section~\ref{intro-inversions}, and are called $B_{inv}$.

\begin{figure}[h!]
\centering
\includegraphics[scale=0.5]{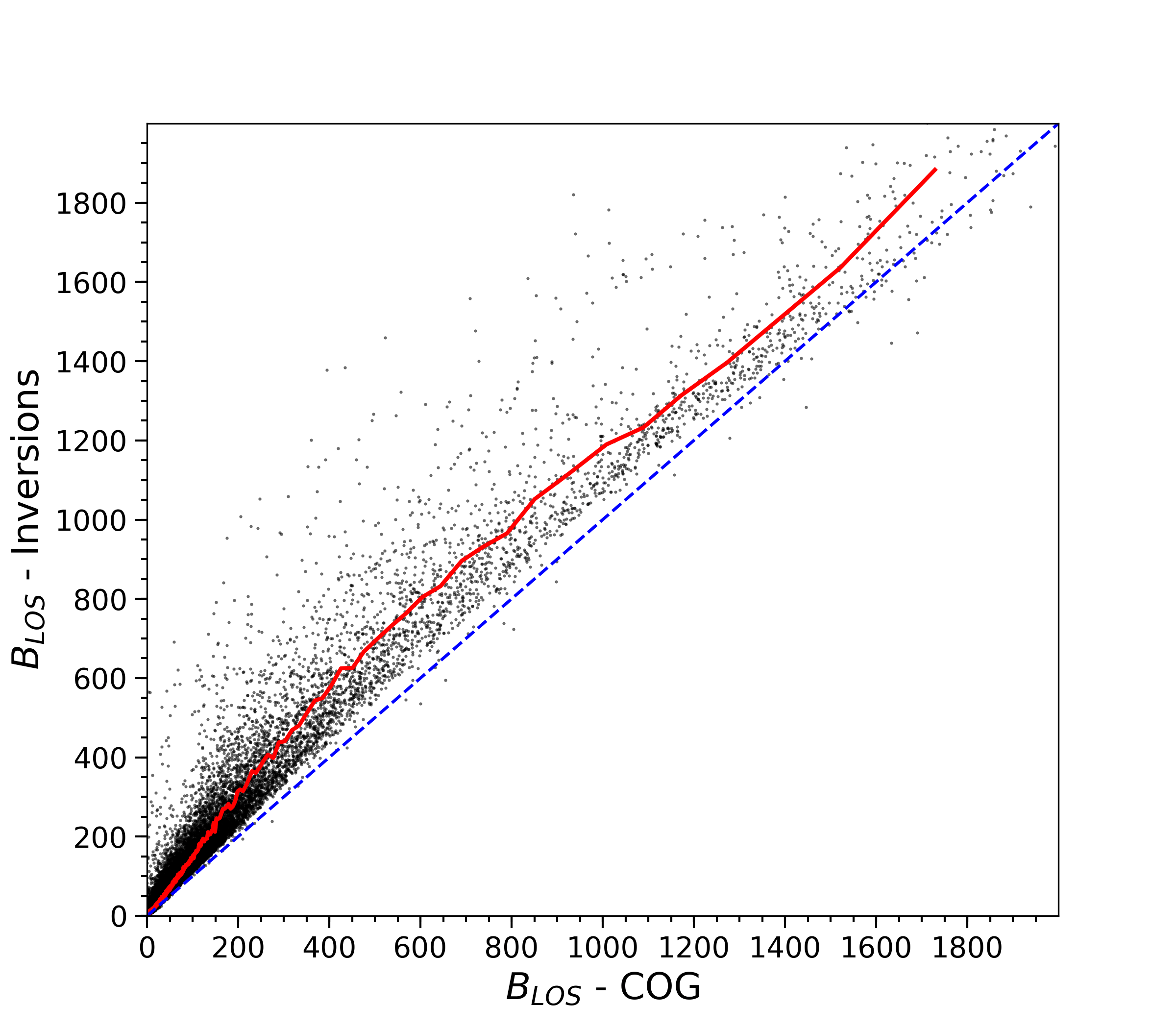}
\caption{Same as Figure~\ref{cog_inv} but with the $B_{\rm LOS}$ on the x-axis computed from the C-O-G method on the inverted Stokes profiles instead of the 5 data points.}
\label{cog_on_inv}
\end{figure}

For equal values of $B_{\rm LOS}$ returned from both methods, the red curve should follow the blue line (with a slope of unity), but as Figure~\ref{cog_inv} shows, the COG seems to underestimate the magnetic field value for small and intermediate field strengths (compared with the Inversion results) until about 1000\,G, but overestimate it for higher fields, due to the few scattered points in the right side of the plot. 
The small number of data points in this mode (5 wavelength points) could be the reason for this scatter. A few number of data points is not enough to compute the exact centroids of the $(I_c-I\pm V)/2$ profiles. This computation is even prone to failure for strongly Zeeman split Stokes~$I$ profiles. This leads to either an over or underestimation of the magnetic field strength.
As a test, I applied the COG method on the best-fit profiles returned by the SPINOR inversions, having now more data points sampling the line (51 data points), and therefore better statistics.
Figure~\ref{cog_on_inv} shows the corresponding scatterplot with the x-axis now corresponding to $B_{\rm LOS}$ values from the COG method applied to the inverted Stokes profiles. The scatter is less pronounced, but the COG is still underestimating the $B_{\rm LOS}$ values for all field strengths.

\begin{figure}
\centering
\hspace*{-1.5cm}\includegraphics[scale=0.4]{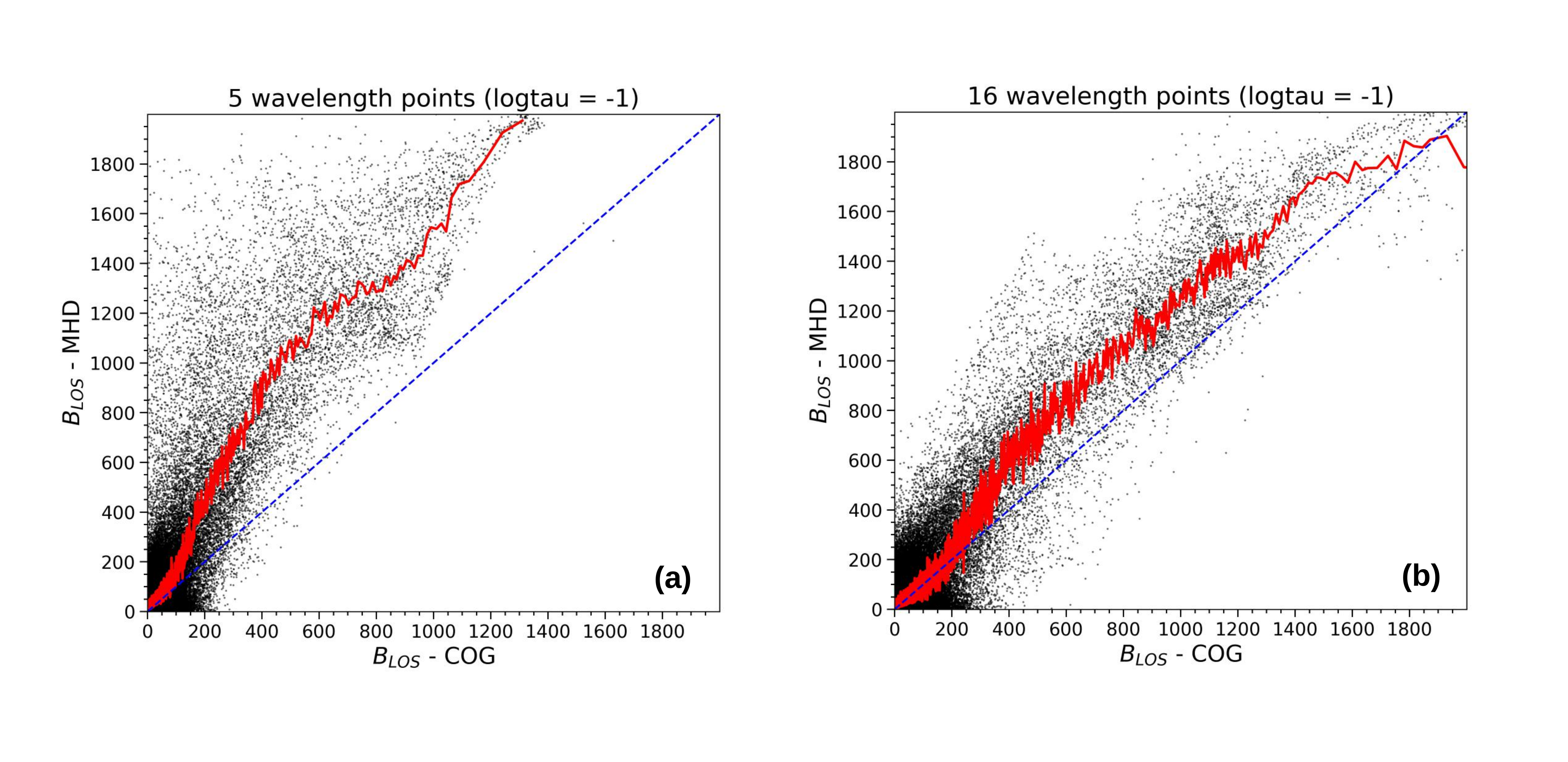}
\caption{(a): Scatterplot of the $B_{\rm LOS}$ of a QS region simulation at $\log \tau=-1$ at the original MURaM resolution vs. $B_{\rm LOS}$ returned by the COG method on the 5 wavelength points (as sampled by IMaX) of the synthesized Stokes profiles. (b) same as panel (a) but with the x-axis corresponding to $B_{\rm LOS}$ values computed with the COG method on 16 wavelength points instead of 5. }
\label{cog_inv_mhd}
\end{figure}

As an additional test, I have applied the COG method on synthesized Stokes profiles of the Fe\,{\sc i} line at 5250.2\, \AA{}. The Stokes profiles are synthesized with SPINOR in its forward mode using MHD data cubes from MURaM simulations of a quiet-Sun region at disk center, with an average vertical magnetic flux density of $30$\,G. I use here the Stokes data which are convolved with the the IMaX spectral PSF to account for the limited spectral resolution \citep{riethmuller_comparison_2014}. The sampled wavelengths lie at 16 positions from the centre of the Fe\,{\sc i} line, I choose from those the 5 wavelength points sampled by IMaX in the V5-6 mode, and compute the $B_{\rm LOS}$ values with the COG method. 
I compare these values with the $B_{\rm LOS}$ map returned by MURaM at log$\tau$ = -1 via a scatterplot that is shown in Figure~\ref{cog_inv_mhd}~a.

The COG method underestimates the $B_{\rm LOS}$ values with respect to the simulated values, just like it does with the IMaX data. To test the effect of the number of data points on the scatter, I apply the COG method, this time on 16 data points within the synthesized Stokes profiles. The corresponding scatterplot is shown in Figure~\ref{cog_inv_mhd}~b. The scatter now is less pronounced with more wavelength points, but the COG still returns lower values. Inversions of the synthesized profiles after taking into account other degradation effects (noise and jitter) are needed to compare the reliability of the COG compared to the inversions in retrieving the  $B_{\rm LOS}$ in simulated QS regions, but our point here is to show that using a geometrical method to find the centroids of the circularly polarized components is only reliable with enough spectral data points. In Chapters~\ref{chapter_3} and \ref{chapter_4}, we refer to the inversions of IMaX data using SPINOR to compute the line-of-sight component of the magnetic field.

\section{Contrast}
\label{intro-contrast}
\subsection{Why is it important?}

In studying small-scale magnetic features, one is generally interested in measuring the excess brightening of a flux tube with respect to its magnetic field-free surroundings. This quantity is of continuous interest since in the valid LTE condition in the solar photosphere, the brightness scales with the temperature. Hence, it is important in learning about the thermal structure of flux tubes, but also about their geometry. Moreover, it serves as a proxy for solar magnetic fields, especially when magnetograms are hard to record. As mentioned in Section~\ref{intro_flux_tubes}, the brightness is dependent, among other factors, on the amount of magnetic flux wherein, thus, the relationship between intensity contrast and magnetic field strength is a constraint for well known theoretical models of flux tubes and an important input for constructing spectral solar irradiance models when contrast is measured in different wavelength bands. 

Moreover, the brightness is highly dependent on the spatial resolution of the observations, since the brightness of a magnetic element could be affected by the dark lanes surrounding it in the same resolution element. This leads to an underestimation of its contrast value (see Sec.~\ref{intro-brightness}), making it hard to learn about the internal structure of a flux tube. The contrast value is also dependent on the stray light handeling, be it instrumental or atmospheric. As explained before in Sections~\ref{intro-imax} and ~\ref{intro-sufi}, our $\sunrise$ data are corrected for wavefront aberrations and for stray light, achieving a spatial resolution of 0.15 arcsec. Thus, it is tempting, at this high spatial resolution, to revisit earlier studies of the brightness of small-scale magnetic elements. 

\subsection{Mean quiet-Sun computation}

The term "contrast" at a specific wavelength band that will be used frequently in Chapters~\ref{chapter_3} and \ref{chapter_4} is the brightness at each pixel $I_{i}$ in the science image normalized to the mean quiet-Sun intensity $<I_{qs}>$. In some works, the contrast is defined as the relative brightness minus unity ($\frac{I_i}{<I_{qs}>}-1$), but we will use the relative intensity  definition of contrast here and refer to bright pixels (with respect to the mean QS intensity) as having contrast > 1 and dark pixels with contrast < 1.

Since the solar surface during $\sunrise$~I observations was quiet, the mean quiet-Sun intensity for each wavelength is taken as the spatial average over all pixels in each individual image.
In the continuum and core of the IMaX Fe\,{\sc i} line, the full usable IMaX field of view of $40^{\prime\prime}\times40^{\prime\prime}$ (excluding the apodized edges resulting from phase-diversity reconstruction) is used. The same is applied to the SuFI images but for the smaller FOV of $15^{\prime\prime}\times40^{\prime\prime}$. In $\sunrise$~II observations, IMaX and SuFI fields of view contained a wide variety of magnetic features that are both active and quiet. Therefore, the normalization in the visible was done by taking a quiet-Sun sub-region containing a fair amount of granulation in the IMaX FOV away from strong magnetic activity. These regions are available in the large field of view of IMaX, but for the smaller FOV of SuFI, finding the quiet Sun is not trivial. In Appendix~\ref{appendix_B} I explain how the normalization is done for the SuFI images in $\sunrise$~II observations.

Fluctuations on day-night timescale of the mean quiet-Sun intensity during the studied time series have to be checked, especially in the near UV, since wavelengths in this range are absorbed by the Earth's atmosphere~\footnote{$\sunrise$ observations were carried out only above 99\% of the Earth's atmosphere.} differently with time depending on the elevation angle and hence the airmass (see Figure~1 in Appendix~\ref{appendix_B}). Another oscillation pattern that is superimposed on the day-night cycle at these wavelengths are the 5 minutes oscillations. Normalizing each image of the time series by its mean quiet-Sun value is done to correct for this factor.

\subsection{IMaX continuum and line-core intensity computation} 

\begin{figure}
\begin{center}

\hspace*{-1.5cm}\includegraphics[scale=0.3]{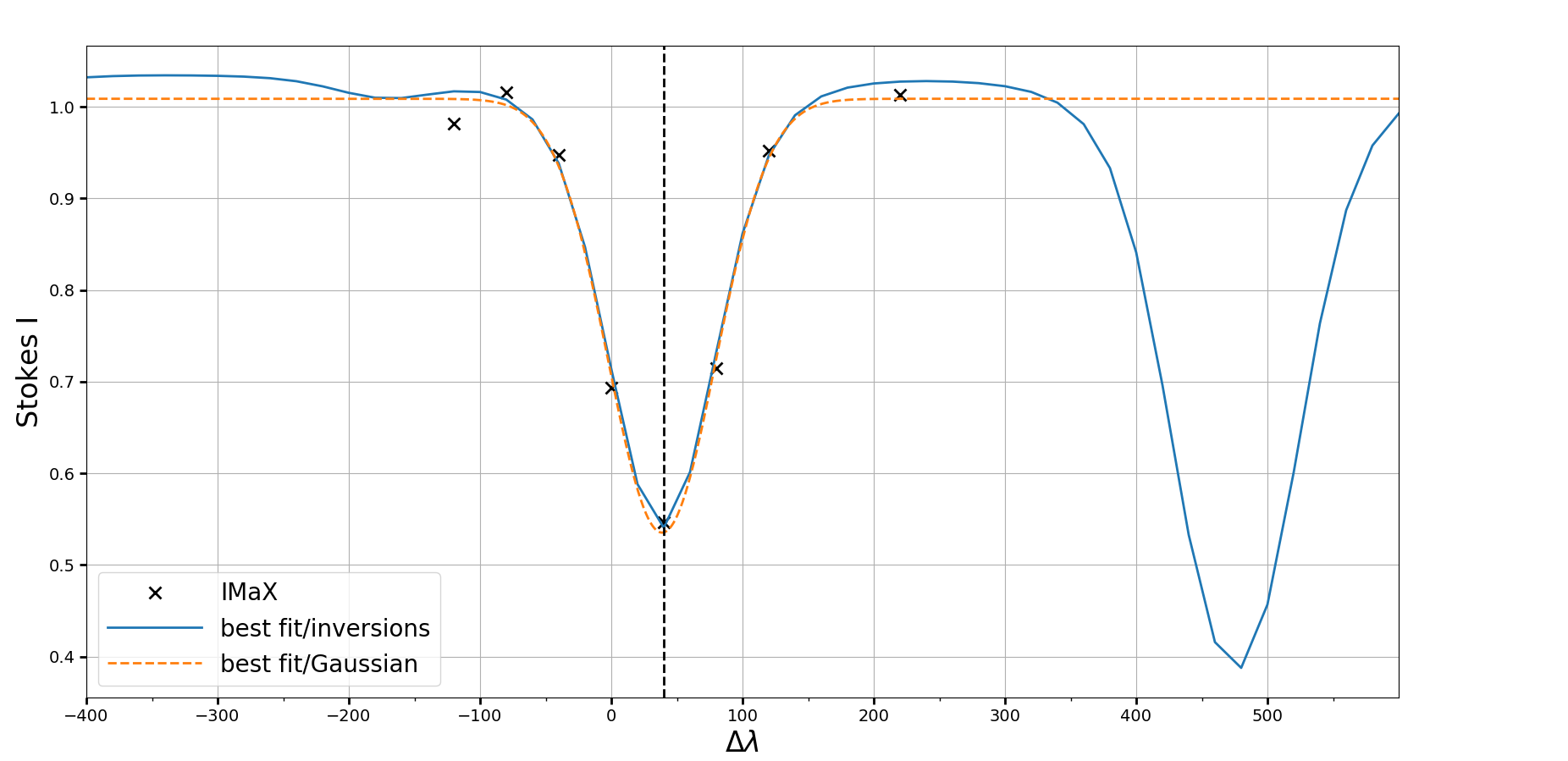}
\caption{Stokes~$I$ profile at 8 wavelength points in the IMaX Fe\,{\sc i} line for a typical granular pixel. The blue curve is the best-fit profile returned by the SPINOR inversions of the corresponding Stokes vector. The dashed orange curve is the Gaussian fit to the IMaX Stokes~$I$ data points. The dashed black vertical line marks the position of the central wavelength of the line relative to the reference wavelength ($0$ on the horizontal axis).}
\label{fits_weak}
\end{center}

\end{figure}

To compute the continuum and line-core intensities in the IMaX data, I have tested different methods. In scientific publications making use of $\sunrise$/IMaX data, the IMaX continuum intensity is taken as the one recorded at $\Delta \lambda$ = + 227\,m\AA{} from Fe\,{\sc i} line center, but this value might not be the true continuum if the line is strongly blue/redshifted or broadened. The problem when the line is blueshifted lies in the contamination of the signal at the continuum point from the nearby Fe\,{\sc i} line at 5250.645\,\AA{} (see Figure~\ref{fits_weak}).
Moreover, the IMaX line core intensity is usually computed by averaging the Stokes~$I$ signals at $\Delta \lambda = \pm 40$\,m\AA{} positions, which might not hold if the line is shifted.

Below are the alternatives for computing both intensities:
\begin{enumerate}
\item From the inverted profiles: the continuum value is taken as the maximum intensity between $-400$ m\AA{} and $-220$ m\AA{} from the central wavelength (see Figure~\ref{fits_weak})
\end{enumerate}
For computing the line-core intensity:
\begin{enumerate}
\item Gaussian fits are made to the recorded Stokes~$I$ data points (5 points in $\sunrise$~I and 8 points in $\sunrise$~II)

\item From the inverted profiles: the line-core intensity is taken as the minimum intensity (if the line is not strongly Zeeman split)

\item If the Zeeman splitting is important, I compute the total wavelength shift and evaluate the Stokes~$I$ inverted profile at that wavelength shift value (see Chapter~\ref{chapter_4} for more details)
\end{enumerate}

\begin{figure}
\hspace*{-1.5cm}\includegraphics[scale=0.1]{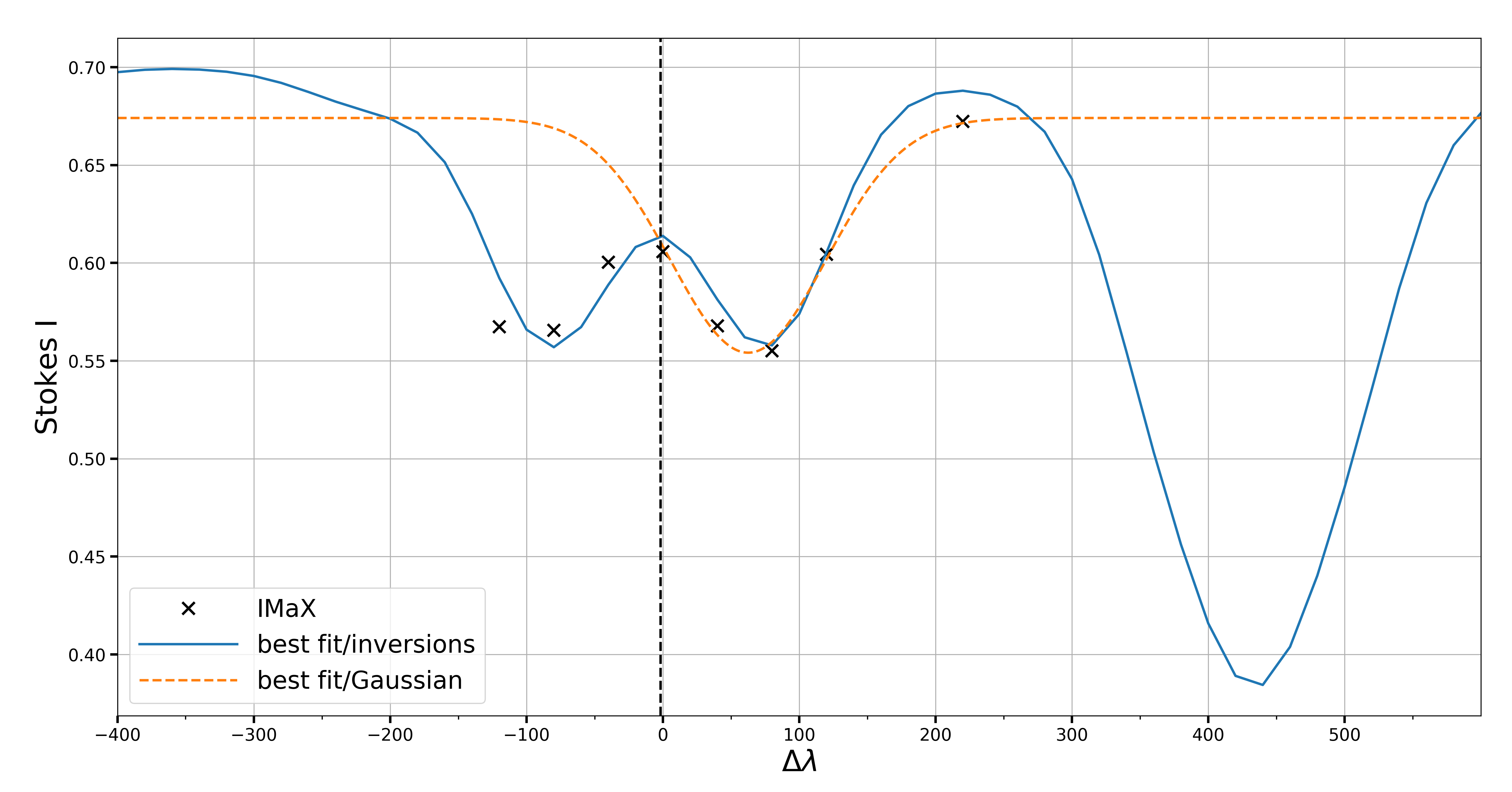}
\caption{Same as Fig.~\ref{fits_weak} but for a pore pixel.}
\label{fits_strong}
\end{figure}

In Figures~\ref{fits_weak} and \ref{fits_strong} I show examples of Stokes~$I$ profiles resulting from inverting the corresponding Stokes data, in addition to the corresponding Gaussian fits. The figures correspond to IMaX data points in two pixels belonging to a granule and a pore taken from the time series analysed in Chapter~\ref{chapter_4}.
Figure~\ref{fits_weak} shows that the Gaussian fit is able to compute a reasonable value for the line-core intensity, similar to the one returned by the inverted profiles (dashed black line), but it also shows the higher continuum value in the continuum region next to the Co\,{\sc i} line (5249.997 \AA{}) than the one recorded at +227\,m\AA{} from $\lambda_0$.

The invalidity of using the Gaussian fits to derive the line-core intensity in the case of strong Zeeman splitting is illustrated in Figure~\ref{fits_strong} where the Gaussian fit fails in representing the shape of the line, leading to a miscomputation of the IMaX line-core value. In contrast, inferring it from the inverted profiles after computing the total wavelength shift from $\lambda_0$ (dashed black line) is clearly a better way to do it since it seems to capture the real line center.

\newpage

\subsection{RMS contrast}
In Chapter~\ref{chapter_5} we study the Center-to-Limb Variation (CLV) of the root mean square (RMS) of the intensity fluctuations in the quiet-Sun granulation in the visible and NUV wavelengths.
The RMS contrast is also called `granulation contrast' \citep{danilovic_intensity_2008}, and it is not to be confused with the intensity contrast defined earlier in this section. It is defined as \citep[e.g.,][]{abramenko_detection_2012}:

\begin{equation}
RMS = \frac{1}{\bar{I}} \sqrt{ \frac{\Sigma^{N}_{i=0} (I_i - \bar{I})}{N}}
\end{equation} where $N$ is the number of pixels in an image with mean intensity $\bar{I}$.

The RMS contrast is a measure of the temperature difference between hot (bright) granules and cold (dark) intergranular lanes and is related to the net energy flux transported to the Sun by the granulation. It is dependent on both the amount of magnetic flux in the solar region and the solar heliocentric angle. For example, at disk center, going to high average magnetic flux density region, bright magnetic points fill the intergranular lanes and reduce the RMS contrast compared to a completely quiet region. Moving to the solar limb, the optical depth unity moves upward where the granulation pattern is reversed and temperature gradients are flatter, leading to a decrease of the RMS contrast above a quiet-Sun granulation region. Measuring the CLV of the continuum intensity fluctuations also constrains the 3D structure of the granulation overshooting, since lines of sight traverse more granular structures approaching the limb.  

\subsection{Wavelengths of study}

\begin{figure}[h]
\centering
\includegraphics[scale=0.18]{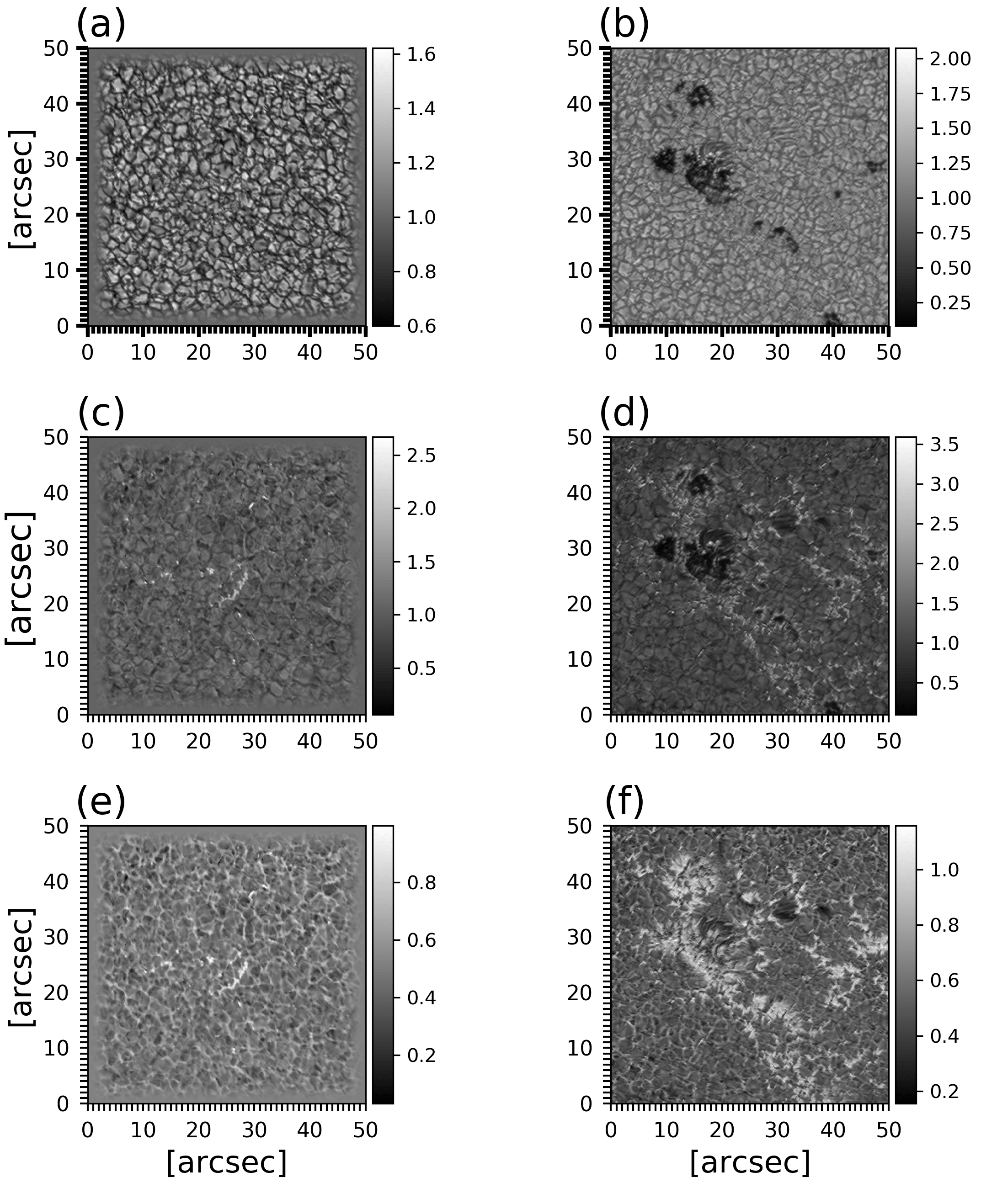}
\caption{\textit{First column:} Example IMaX image set (i.e., images belonging to a single wavelength scan) of the observed QS region during $\sunrise$~I. \textit{Second column:} Example IMaX image set of the observed AR during $\sunrise$~II. \textit{First row}: IMaX continuum normalized to the mean QS continuum intensity. \textit{Second row:} IMaX line core intensity normalized to the mean QS line core intensity. \textit{Third row:} IMaX line core intensity normalized to the local IMaX continuum intensity. }
\label{contrasts_imax}
\end{figure}

Here I will briefly describe the wavelengths used in my study, the information they provide on the properties of solar magnetic features, and the appearences of the latter. Examples of the continuum and core of the Fe\,{\sc i} line intensity images from both flights are shown in Figure~\ref{contrasts_imax}.
In the continuum images of IMaX from both flights (panels (a) and (b)), the normal granulation pattern (away from pores in panel (b)) is clearly seen: bright (hot) granules and dark (cold) intergranular lanes. In the line-core images (panels (e) and (f)), this pattern is reversed. The reversed granulation pattern is explained by the shallower temperature gradient within intergranular lanes than in the granules \citep{cheung_origin_2007}. As a result, at a larger height, granules will be cooler (and therefore darker) than their surroundings lanes. This pattern is seen better when normalizing the line-core intensities to the local continuum (panels (e) and (f)).

Bright points and plage elements are mainly located in the intergranular lanes away from sunspots, appearing brighter in the line-core images (panels (c) and (d)). The formation height difference between the two wavelength bands is about 380\,km and 300\,km as computed in \cite{jafarzadeh_inclinations_2014} for a quiet-Sun and plage atmospheric model, respectively. 

In LTE conditions the contrast translates into a difference in temperature. Due to the shallower temperature gradient above the strong kG features embedded in intergranular lanes with respect to weaker field regions at the same optical depth level \citep{lagg_fully_2010}, and the larger formation height of the core of the Fe\,{\sc i} line at 5250.2\, \AA{} compared to its continuum, the temperature difference (and therefore contrast) measured in the line core is higher than the one measured in the continuum for the same field strength. This conclusion is confirmed with $\sunrise$ data analysed in Chapters~\ref{chapter_3} and \ref{chapter_4}. 

\begin{figure}[h]
\centering
\hspace*{-2cm}\includegraphics[scale=0.051]{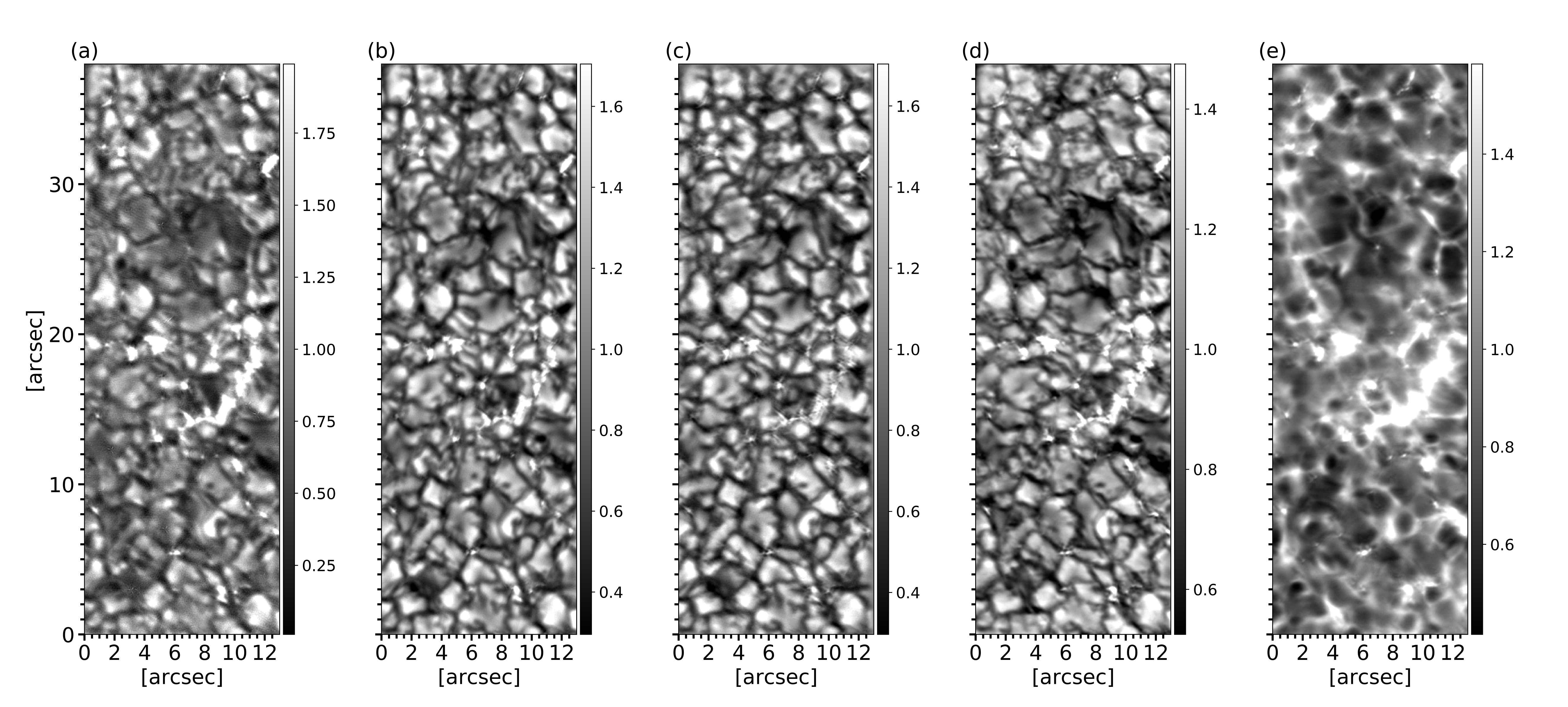}
\caption{Example SuFI/$\sunrise$~I images of the a quiet-Sun region at disk center ($\mu=0.97$) imaged in (a) 214 nm, (b) 300 nm, (c) 313 nm, (d) 388 nm, (e) 397 nm. The gray scale is set to cover two times the rms range of each image individually. }
\label{sufis_2009}
\end{figure}

\begin{figure}[h]
\centering
\includegraphics[scale=0.1]{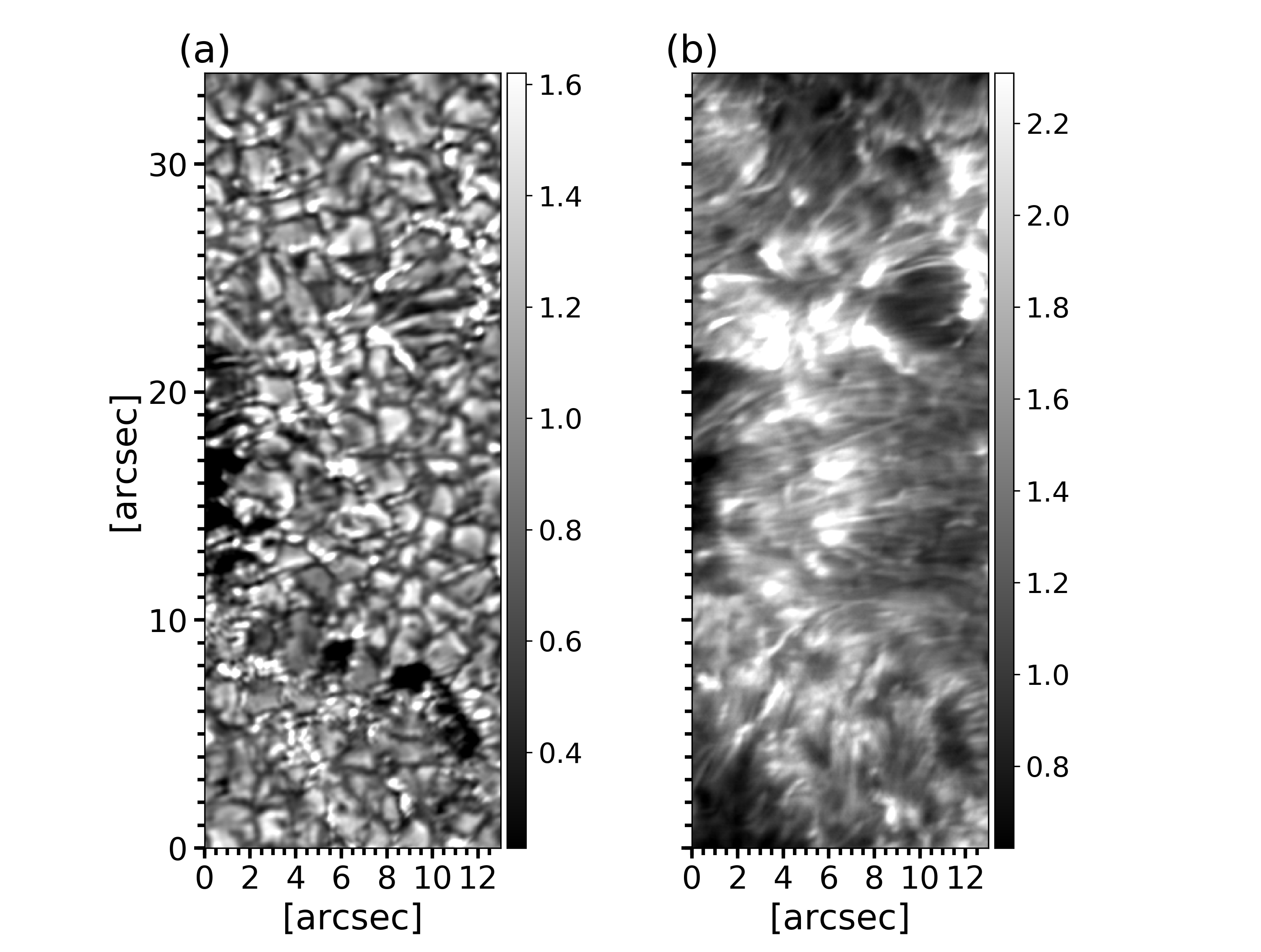}
\caption{Example SuFI/$\sunrise$~II images of an active region near disk center ($\mu = 0.93$) imaged at (a) 300 nm, (b) 397 nm. The gray scale is set to cover two times the rms range of each image individually. }
\label{sufis_2013}
\end{figure}

One thing that makes $\sunrise$ data special is that the same solar region, although with a narrower FOV (Sec.~\ref{intro-sunrise}), is simultaneously imaged in different wavelength bands that show different faces of the magnetic elements. This difference is either due to the wavelengths probing different ranges of atmospheric heights, or to their different temperature sensitivity. In Section~\ref{intro_flux_tubes} I mentioned some of the molecular bands used in imaging magnetic elements (OH and CN bands), that in addition to their short wavelengths (which enhances the spatial resolution), they are also sensitive to the temperature variations through their molecular dissociation, yielding excess brightening of these elements at heights not very far from the continuum formation level. These bands are proved to be excellent tracers of magnetic elements brightenings in the solar photosphere \citep{berdyugina_molecular_2003}.
I show in Figures~\ref{sufis_2009} and ~\ref{sufis_2013} the contrast images of the smaller FOV of SuFI in the wavelengths of 214 nm, 300 nm, 313 nm, 388 nm, 397 nm imaged by $\sunrise$~I and in 300 nm and 397 nm imaged by $\sunrise$~II. Displayed is the common field of view between SuFI images due to differential offsets. It is $13^{\prime\prime}\times38^{\prime\prime}$ and $13^{\prime\prime}\times34^{\prime\prime}$ in $\sunrise$~I and $\sunrise$~II, respectively.

Small-scale magnetic elements in both, quiet-Sun network and AR plage are bright in the NUV. They are particularly bright in calcium images where they appear more diffuse compared to lower layers. As mentioned in Chapter~\ref{chapter_1}, this is due to the expansion of the magnetic field lines with height. The reversed granulation is clearly seen just like in the line core of IMaX, except that at the latter wavelength, magnetic elements are less diffuse. The contrasts of small-scale magnetic elements in the quiet Sun network and AR plage in the different wavelength bands imaged by $\sunrise$ are compared in Chapters~\ref{chapter_3} and \ref{chapter_4}, respectively.

Information about the formation height of observed spectral lines in a specific wavelength band is usually dependent on the used filter and the physical conditions of the observed medium, and is usually difficult to derive in wavelength bands with high density of spectral lines \citep{solanki_sunrise:_2010}. The contribution function which informs us on how the different layers \textit{contribute} to the observed spectrum can be used to infer the averaged formation height of spectral lines, and is given by: 

\begin{equation}
C(\tau_c) = e^{-\int_0^{\tau_c} k(t) dt} \,k(\tau_c)\,S(\tau_c)
\end{equation}

(see \cite{grossmann-doerth_unshifted_1988}). This function gives the emission at each atmospheric level (optical depth). The atmospheric conditions under which spectral lines are synthesized correspond to whether lines are formed in LTE or NLTE regimes, and on the atmospheric model (corresponding to quiet Sun, bright quiet Sun, or plage regions) used to get the average geometrical height. Below are the works in which formation heights of the wavelength bands under study are computed. The heights are given with respect to the optical depth unity level at 500\,nm, and they correspond to the mean of the contribution function over the whole range of geometrical heights, after convolving the CFs by the filter transmission profiles at each height and integrating the filter convolved CFs over wavelength:

\begin{enumerate}
\item[$\bullet$] Calcium emission as seen by the Ca\,{\sc ii} H filter in $\sunrise$~I: 437\,km, 456\,km, 500\,km for FALC (QS), FALF (bright points in QS) and FALP (plage) model atmospheres \citep{jafarzadeh_structure_2013}
\item[$\bullet$] IMaX continuum : 20\,km for FALC and FALP, IMaX line core: 400\,km (FALC) and 320\,km (FALP) \citep{jafarzadeh_inclinations_2014} 
\item[$\bullet$] 300 nm (OH band): 50\,km (FALP) and 10\,km (FALC) \citep{jafarzadeh_high-frequency_2017}
\item[$\bullet$] Calcium emission as seen by the Ca\,{\sc ii} H filter in $\sunrise$~II: 550\,km \citep{jafarzadeh_slender_2017}
\end{enumerate}

\newpage


\chapter{Brightness of solar magnetic elements as a function of magnetic flux at high spatial resolution}
\label{chapter_3}
\textbf{*This chapter is based on a published paper to ApJ \citep{kahil_brightness_2017}, with the permission of the journal} \\ \\

\textbf{Abstract} \\
We investigate the relationship between the photospheric magnetic field of small-scale magnetic elements in the quiet Sun (QS) at disc centre, and the brightness at 214 nm, 300 nm, 313 nm, 388 nm, 397 nm, and at 525.02 nm. To this end we analysed spectropolarimetric and imaging time series acquired simultaneously by the IMaX magnetograph and the SuFI filter imager on-board the balloon-borne observatory $\sunrise$ during its first science flight in 2009, with high spatial and temporal resolution.

We find a clear dependence of the contrast in the near ultraviolet (NUV) and the visible on the line-of-sight component of the magnetic field, $B_{\rm LOS}$, which is best described by a logarithmic model. This function represents well the relationship between the Ca\,{\sc ii} H-line emission and $B_{\rm LOS}$, and works better than a power-law fit adopted by previous studies. This, along with the high contrast reached at these wavelengths, will help with determining the contribution of small-scale elements in the QS to the irradiance changes for wavelengths below 388 nm. At all wavelengths including the continuum at 525.40 nm the intensity contrast does not decrease with increasing $B_{\rm LOS}$. This result also strongly supports that $\sunrise$ has resolved small strong magnetic field elements in the internetwork, resulting in constant contrasts for large magnetic fields in our continuum contrast at 525.40 nm vs. $B_{\rm LOS}$  scatterplot, unlike the turnover obtained in previous observational studies. This turnover is due to the intermixing of the bright magnetic features with the dark intergranular lanes surrounding them. 
\clearpage

\section{Introduction} 
Small-scale magnetic elements or magnetic flux concentrations are described by flux tubes, often with kG field strengths, located in intergranular downflow lanes \citep{solanki_small-scale_1993}. 
Studying the intensity contrast of magnetic elements relative to the QS in the continuum, and line core of spectral lines, is of importance, because it provides information about their thermal structure. Due to their enhanced brightness, particularly in the cores of spectral lines \citep[]{title_differences_1992, yeo_intensity_2013} and in the UV \citep{riethmuller_bright_2010}, these elements are believed to contribute to the variation of the solar irradiance, especially on solar cycle time scales \citep{foukal_magnetic_1988,fligge_modelling_2000, krivova_reconstruction_2003, yeo_solar_2014}. The contrast in the visible, and in the UV spectral ranges contributes by 30\% and 60\% respectively, to the variation of the total solar irradiance (TSI) between minimum and maximum activity \citep{krivova_reconstruction_2006} and spectral lines contribute to a large part of this variation \citep[]{livingston_spectrum_1988, shapiro_role_2015}.
The magnetic flux in magnetic elements is also believed to be responsible for the structuring and heating of the chromosphere and corona. The relationship with chromospheric heating is indicated by the strong relationship between excess brightening in the core of the Ca\,{\sc ii} K-line and the photospheric magnetic flux \citep[e.g.,][]{skumanich_statistical_1975,schrijver_relations_1989,loukitcheva_relationship_2009}.  

Several studies have been carried out to investigate how the brightness at given wavelength bands depends on the photospheric magnetic field at disc centre, and different results have been obtained. \citet{title_differences_1992} and \cite{topka_properties_1992} made a pixel-by-pixel comparison of the longitudinal magnetogram signal and the continumm intensity at 676.8, 525, 557.6, and 630.2 nm, using simultaneous continuum filtergrams, in addition to line centre filtergrams at 676.8 nm in \cite{title_differences_1992}, and magnetograms of magnetic features in active regions at disc centre, acquired by the 50 cm Swedish Solar Vacuum Telescope with a 0.3$^{\prime\prime}$ resolution. They obtained negative contrasts in the continuum (darker than the average QS) for all magnetogram signals (as long as the observations were carried out almost exactly at solar disc centre), and an increase in the line core brightness with the magnetogram signal until 600\,G followed by a monotonic decline.

\cite{lawrence_contrast_1993} applied the same method at roughly the same spatial resolution as \cite{topka_properties_1992} to quiet-Sun network data. Their scatterplot showed positive contrast values (brighter than the average QS) for magnetic fields larger than 200\,G, until nearly 500\,G, and negative contrasts for higher magnetic fields.
\cite{ortiz_intensity_2002} used low resolution (4$^{\prime\prime}$), but simultaneous magnetograms and full-disc continuum intensity images of the Ni\,{\sc i} 6768\, \AA{} absorption line, recorded by the Michelson Doppler Imager (MDI) on-board the Solar and Heliospheric Observatory (SoHO). They found that the contrast close to disc centre (at $\mu \approx 0.96$) initially increases slightly with the magnetic field, before decreasing again for larger fields.
\cite{kobel_continuum_2011} made the same pixel-by-pixel study of the continuum contrast at 630.2 nm with the longitudinal magnetic field in the quiet-Sun network and in active region (AR) plage near disc centre, using data from the Solar Optical Telecope on-board Hinode (0.3$^{\prime\prime}$ spatial resolution). They found that for both the QS and AR, the contrast initially decreases and then increases for weak fields, until reaching a peak at $\approx$ 700\,G, to decrease again for stronger fields, even when the pores in their AR fields of view (FOVs) were expilicitly removed from their analysis. They explained the initial, rapid decrease to have a convective cause, with the bright granules harbouring weaker fields than the intergranular lanes. The following increase in contrast is due to the magnetic elements being brighter than the average QS. To explain the final decrease at high field strengths, they argued that due to their limited spatial resolution (0.3$^{\prime\prime}$ for Hinode/SP), many flux tubes were not resolved, and therefore, the field strength of small bright elements is attenuated, while those of bigger structures (such as micropores), which are darker than the mean QS, were weakly affected by the finite resolution, so that the average contrast showed a peak at intermediate field strengths, and decreased at higher strengths.

At a constant spatial resolution of 1$^{\prime\prime}$ achieved by the Helioseismic and Magnetic Imager (HMI) on-board the Solar Dynamics Observatory (SDO), \cite{yeo_intensity_2013} studied the dependence of the continuum and line core contrasts in  the Fe\,{\sc i} line at 6173\, \AA{}, of network and faculae regions on disc position and magnetogram signal, using simultaneous full-disc magnetograms and intensity images. For a quiet-Sun region at disc centre ($\mu > 0.94$), their scatterplots of the continuum intensity against $<B>/\mu$ exhibited a peak at $<B>/\mu \sim$ 200\,G.  

\cite{rohrbein_is_2011} simulated a plage region using the MURaM code \citep{vogler_simulations_2005}. They studied the relation between the continuum contrast at 630.2 nm and the vertical magnetic field both, at the original MURaM resolution and at the resolution of telescopes with 1.0 m and 0.5 m apertures. For the original resolution, the contrast monotonically increased with increasing field strength, confirming the expectations of a thin flux-tube model \citep{spruit_pressure_1976}, which predicts that for higher field strengths, the flux tubes get more evacuated, which leads to lateral inflow of heat from the hot walls of the evacuated flux concentrations, and therefore an increase in brightness as a result of the optical depth surface depression, which allows deeper layers to be seen. According to \cite{rohrbein_is_2011}, at the resolution of a  1.0 m telescope the average simulated contrast saturates for stronger fields, while it shows a turnover at the resolution corresponding to 0.5 m. This points to a non-trivial effect of finite spatial resolution on the relation between continuum brightness, and photospheric magnetic field.

In contrast to continuum radiation, spectral lines display a relatively monotonic increase in rest intensity with magnetic flux. Of particular importance is the Ca\,{\sc ii} H line, due to its formation in the chromosphere. \cite{frazier_multi-channel_1971} showed by using simultaneous observations of the calcium network and photospheric magnetic field that the line core of Fe\,{\sc i} at 5250.2 \AA{} and the Ca\,{\sc ii} K emission increases with the magnetic field until his limit of about 500\,G. \cite{schrijver_relations_1989} carried out a quantitative study of the relationship between the Ca\,{\sc ii} K emission and magnetic flux density in an active region (outside sunspots). After subtracting the basal flux (the non-magnetic contribution of the chromospheric emission) they found that the relation follows a power law, with an exponent of 0.6.
\cite{ortiz_how_2005} found the same relation with a power-law exponent of 0.66, for a quiet-Sun region at disc centre, while \cite{rezaei_relation_2007} reported a value of 0.2 (including the internetwork), and 0.4--0.5 for the network, with a strong dependence of the power exponent on the magnetic field threshold.
\cite{loukitcheva_relationship_2009} in their study of the correlation between emissions at different chromospheric heights with the photospheric magnetic field in a quiet-Sun region close to the disc centre, found an exponent of 0.31 for Ca\,{\sc ii} K.

Here, we study the contrast at a number of wavelengths in the NUV (between 214 nm and 397 nm) and the visible (around 525 nm) in the QS close to solar disc centre. We employ high-resolution, seeing-free measurements of both, the intensity and the magnetic field obtained with the $\sunrise$ balloon-borne observatory. The structure of the rest of the chapter is as follows. In Sec.~\ref{chapter_3_data}, we describe the data used for this analysis, in addition to presenting the detailed data reduction steps. In Sec.~\ref{plots}, we present our results, and compare them to the literature. In Sec.~\ref{chapter3_conc}, we summarize and discuss the results.

\section{Observations and Data Preparation}
\label{chapter_3_data}
\subsection{IMaX and SuFI data}
We use a time series recorded during the first $\sunrise$  flight on 2009 June 9, between 14:22 and 15:00 UT. $\sunrise$ is composed of a telescope with a 1.0 m diameter main mirror mounted on a gondola with two post-focus instruments \citep{solanki_sunrise:_2010,barthol_sunrise_2011}. It carries out its observations hanging from a stratospheric balloon. The images are stabilized against small-scale motions by a tip-tilt mirror placed in the light distribution unit connected to a correlation tracker and wavefront sensor \citep{berkefeld_wave-front_2011, gandorfer_filter_2011}. $\sunrise$ carried two instruments, a UV imager (SuFI) and a magnetograph (IMaX).

The Imaging Magnetograph eXperiment \citep[IMaX;][]{martinezpillet_imaging_2011} acquired the anaylsed spectropolarimetric data by scanning the photospheric Fe\,{\sc i} line at 5250.2\, \AA{} (Land{\'e} factor $g$ = 3) at five wavelengths positions (4 within the line at $-80$, $-40$, $+40$, $+80$\, m\AA{} and one in the continuum at $+227$\, m\AA{} from the line centre), with a spectral resolution of 85\, m\AA{}, and measuring the full Stokes vector $(I,Q,U,V)$ at each wavelength position. The total cadence for the V5-6 observing mode (V for vector mode with 5 scan positions and 6 accumulations of 250 ms each) was 33\,s. We use the phase-diversity reconstructed data with a noise level of 3$\times$10$^{-3} I_c$, and achieving a spatial resolution of $0.15^{\prime\prime}-0.18^{\prime\prime}$ (see \cite{martinezpillet_imaging_2011} for more details on the instrument, data reduction and data properties).

For the NUV observations, we use the data acquired by the $\sunrise$ Filter Imager \citep[SuFI;][]{gandorfer_filter_2011} quasi-simultaneously with IMaX, in the spectral regions 214 nm, 300 nm, 313 (OH-band) nm, 388 nm (CN-band), and 397 nm (core of Ca\,{\sc ii} H), at a bandwidth of 10 nm, 5 nm, 1.2 nm, 0.8 nm, and 0.18 nm, respectively. The cadence of the SuFI data for a given wavelength was 39\,s. We analyse data that were reconstructed using wave-front errors obtained from the in-flight phase-diversity measurements, via an image doubler in front of the CCD camera \citep[level 3 data, see][]{hirzberger_quiet-sun_2010, hirzberger_performance_2011}. The SuFI data were corrected for stray light by deconvolving them with the stray light modulation transfer function (MTFs) derived from comparing the limb intensity profiles recorded for the different wavelengths, with those from the literature (A. Feller et al. 2018, in preparation)

\begin{landscape}
\begin{figure*}
\centering
\includegraphics[width=\linewidth]{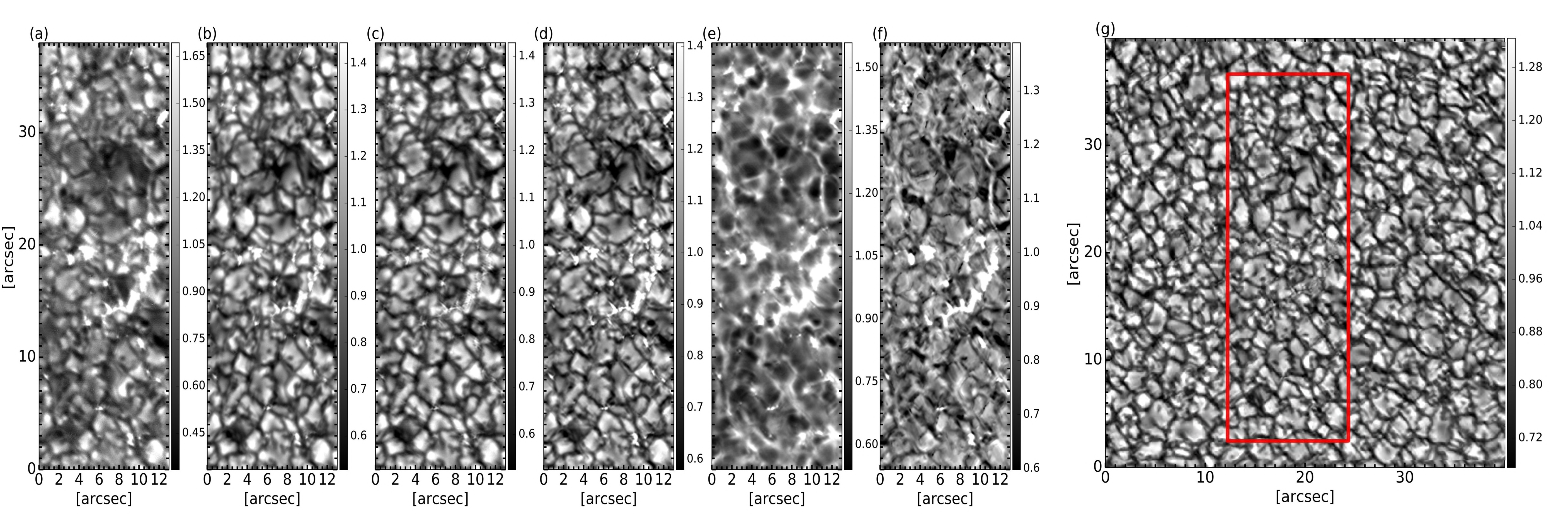}
\caption{Example of aligned SuFI and IMaX contrast images, a) SuFI at 214 nm, b) SuFI at 300 nm,  c) SuFI at 313 nm (OH-band), d) SuFI at 388 nm (CN-band), e) SuFI  Ca\,{\sc ii} H line core at 397.0 nm, f) IMaX line core, g) full FOV of IMaX Stokes~$I$ continuum at 5250.4\, \AA{}. The red box overlaid on the IMaX FOV is the common FOV ($13^{\prime\prime}\times38^{\prime\prime}$), to which the other images in this figure are trimmed. The gray scale is set to cover two times the rms range of each image.
}
\label{images}
\end{figure*}
\end{landscape}

\begin{figure}
\centering
\includegraphics[scale=0.1]{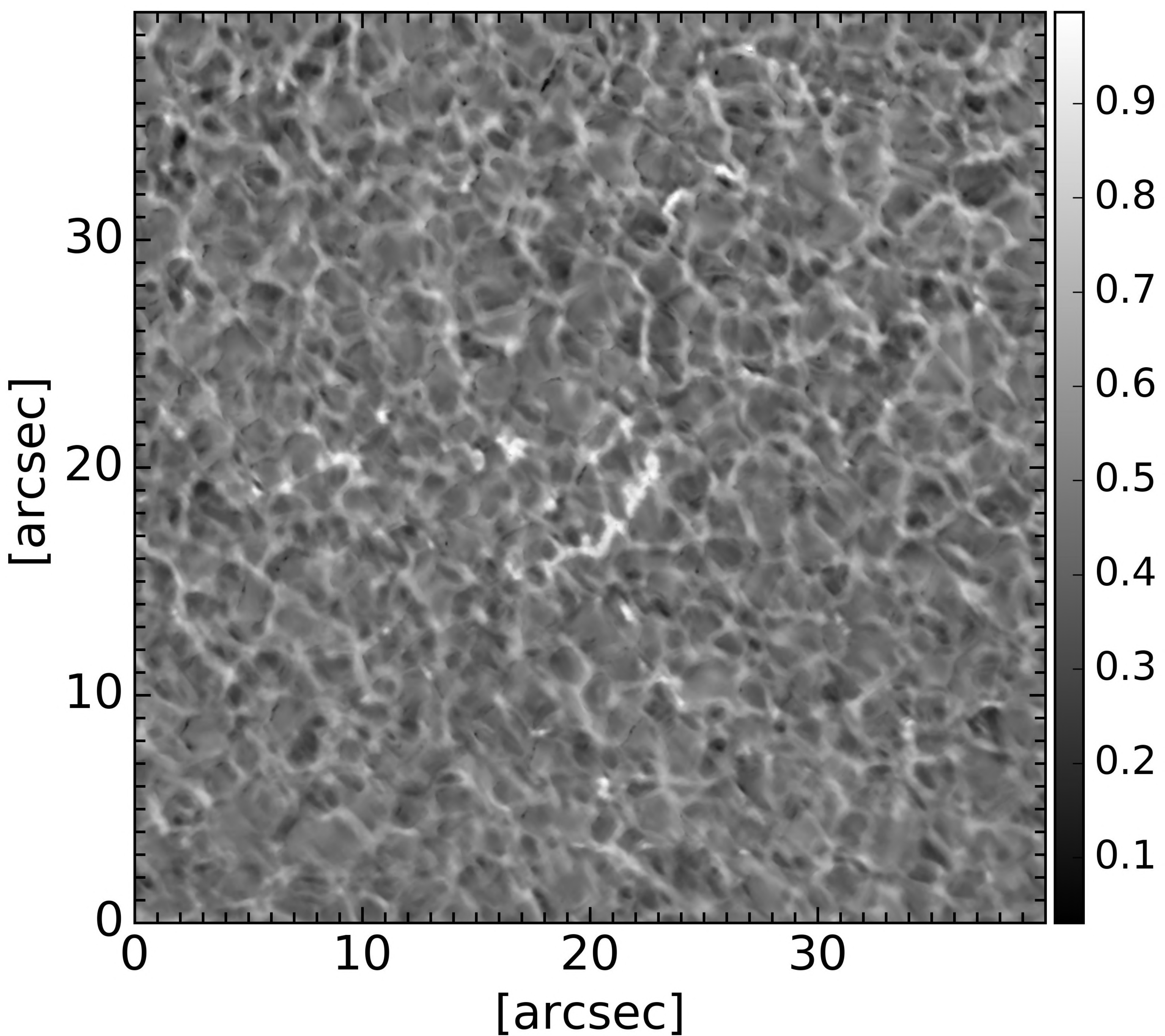}                                                                                                                                       
\caption{Same IMaX FOV shown in Fig.~\ref{images} but for Stokes~$I$ line core normalized to the local continuum.}
\label{LC_qs}
\end{figure}

\subsection{Stokes inversions}
In this study we use the line-of-sight (LOS) component of the magnetic field vector $B_{\mathrm{LOS}}$, which was retrieved from the
reconstructed Stokes images by applying the SPINOR\footnote{The {\bf S}tokes-{\bf P}rofiles-{\bf IN}version-{\bf O}-{\bf R}outines.}
inversion code \citep{solanki_photospheric_1987,frutiger_properties_2000}. Such an inversion code assumes that the five spectral positions within the
525.02~nm Fe\,{\sc i} line are recorded simultaneously. This is not the case for the IMaX instrument that scans through the positions sequentially
with a total acquisition time of 33\,s. To compensate for the solar evolution during this cycle time we interpolated the spectral scans with
respect to time.

We also corrected the data for stray light because we expect that the stray-light contamination has a serious effect on the inversion results,
in particular on the magnetic field results in the darker regions (intergranular lanes, micropores). Unfortunately, Feller et al. (in preparation)
could only determine the stray-light MTF for the Stokes~$I$ continuum images but not for the other spectral positions. Since a severe
wavelength dependence of the stray light cannot be ruled out, we decided to apply a simplistic global stray-light correction to the IMaX
data by subtracting 12\% (this value corresponds to the far off-limb offset determined in the continuum by Feller et al., in preparation) of the spatial mean
Stokes profile from the individual profiles.

After applying the time interpolation and the stray-light correction, the cleaned data were inverted with the traditional version of the SPINOR code.
In order to get robust results, a simple one-component atmospheric model was applied that consists of three optical depth nodes for the temperature
(at $\log\tau=-2.5, -0.9, 0$) and a height-independent magnetic field vector, line-of-sight velocity and micro-turbulence. The spectral resolution
of the instrument was considered by convolving the synthetic spectra with the spectral point-spread function of IMaX \citep[see bottom panel of
Fig.~1 in][]{riethmuller_comparison_2014}.

The SPINOR inversion code was run five times in a row with ten iterations each. The output of a run was smoothed and given as initial atmosphere
to the following run. The strength of the smoothing was gradually decreased which lowered the spatial discontinuities in the physical quantities caused
by local minima in the merit function. The final LOS velocity map was then corrected by the etalon blueshift which is an unavoidable instrumental effect
of a collimated setup \citep[see][]{martinezpillet_imaging_2011} and a constant velocity was removed from the map so that the spatially averaged velocity is zero.

The inversion strategy used for the 2009 IMaX data analysed in this chapter is identical to the one applied to the 2013 data in Chapter~\ref{chapter_4} which is described in more detail
by \cite{solanki_second_2017}.

\subsection{Image Alignment}

When comparing the SuFI and IMaX data, we need to align the two data sets with each other and transform them to the same pixel scale. Because some of the SuFI wavelengths show granulation (at 300 nm, 313 nm, 388 nm, and 214 nm), these were aligned with IMaX Stokes~$I$ continuum images. The SuFI Ca\,{\sc ii} H images at 397 nm were aligned with IMaX Stokes~$I$ line-core images (see Sect.~\ref{contrast_def} for the derivation of the line-core intensity), since both wavelength bands sample higher layers in the photosphere and display reversed granulation patterns.

In a first step, the plate scale of SuFI images of roughly 0\carcsec{}02 pixel$^{-1}$ were resampled via bi-linear interpolation to the plate scale of IMaX (0\carcsec{}05 pixel$^{-1}$). After setting all the images to the same pixel size, IMaX images with $50^{\prime\prime}\times50^{\prime\prime}$ FOV were first trimmed to exclude the edges lost by apodisation. The usable IMaX FOVs of $40^{\prime\prime}\times40^{\prime\prime}$ were then flipped upside down, and trimmed to roughly match the FOV of the corresponding SuFI images of $15^{\prime\prime}\times40^{\prime\prime}$. A cross-correlation technique was used to compute the horizontal and vertical shifts with sub-pixel accuracy. After shifting, the IMaX and SuFI FOVs  were trimmed to the common FOV (CFOV) of all data sets of $13^{\prime\prime}\times38^{\prime\prime}$. This value is smaller than the FOV of individual SuFI data sets since the images taken in the different SuFI filters are slightly shifted with respect to each other due to different widths and tilt angles of the used interference filters.
Figure~\ref{images} shows from right to left, an IMaX Stokes~$I$ continuum contrast image with an effective FOV of $40^{\prime\prime}\times40^{\prime\prime}$, with the CFOV overlaid in red, an IMaX Stokes~$I$ line core contrast image trimmed to the CFOV,  and the corresponding resampled and aligned SuFI contrast images at 397 nm, 388 nm, 313 nm, 300 nm, and 214 nm (see Sect.~\ref{contrast_def} for the definition of intensity contrast).

The reversed granulation pattern in the line-core image is more visible if normalized to the local continuum intensity as shown in Fig.~\ref{LC_qs}.

\subsection{Contrast}
\label{contrast_def}
The relative intensity (hereafter referred to as contrast), $C_{\rm WB}$ at each pixel for each wavelength band, $\rm WB=\{\rm CONT, \rm LC, 214, 300, 313, 388, 397 \}$ is computed as follows:

\begin{equation}
C_{\rm WB} = \frac{I_{\rm WB}}{I_{\rm WB, QS}}
\end{equation}

Where $C_{\rm CONT}$ and $C_{\rm LC}$ are the IMaX Stokes~$I$ continuum and line-core intensity contrasts, respectively. $C_{\rm 214}$, $C_{\rm 300}$, $C_{\rm 313}$, $C_{\rm 388}$ and $C_{\rm 397}$ are the SuFI intensity contrasts at 214 nm, 300 nm, 313 nm, 388 nm and 397 nm, respectively.

$I_{\rm WB, QS}$ is the mean quiet-Sun intensity averaged over the entire common FOV. When comparing SuFI contrast with IMaX-based magnetic field parameters this common FOV is $13^{\prime\prime}\times38^{\prime\prime}$, when comparing IMaX continuum or line-core intensity with IMaX magnetic field, the full usable IMaX FOV of $40^{\prime\prime}\times40^{\prime\prime}$ is employed.

The line-core intensity, $I_{\rm LC}$ at each pixel was computed from a Gaussian fit to the 4 inner wavelength points of the Stokes~$I$ profile. For comparison later in Sect.~\ref{scatter_ca}, we also compute the line core  by averaging the IMaX Stokes~$I$ intensity at $-40$ m${\angstrom}$ and $+40$ m${\angstrom}$ from the line centre:
\begin{equation}
I_{\rm LC,\pm 40} = \frac{I_{+40}+ I_{-40}}{2}
\label{LC_40}
\end{equation}

Each scatterplot in Sect.~\ref{plots} contains data from all the 40 available images in the time series.

\section{Results}
\label{plots}
\subsection{Scatterplots of IMaX continuum and line core contrasts vs. $B_{\rm LOS}$}
\label{plots_b_c}
\begin{figure}[!h]
\centering
\includegraphics[scale=0.2]{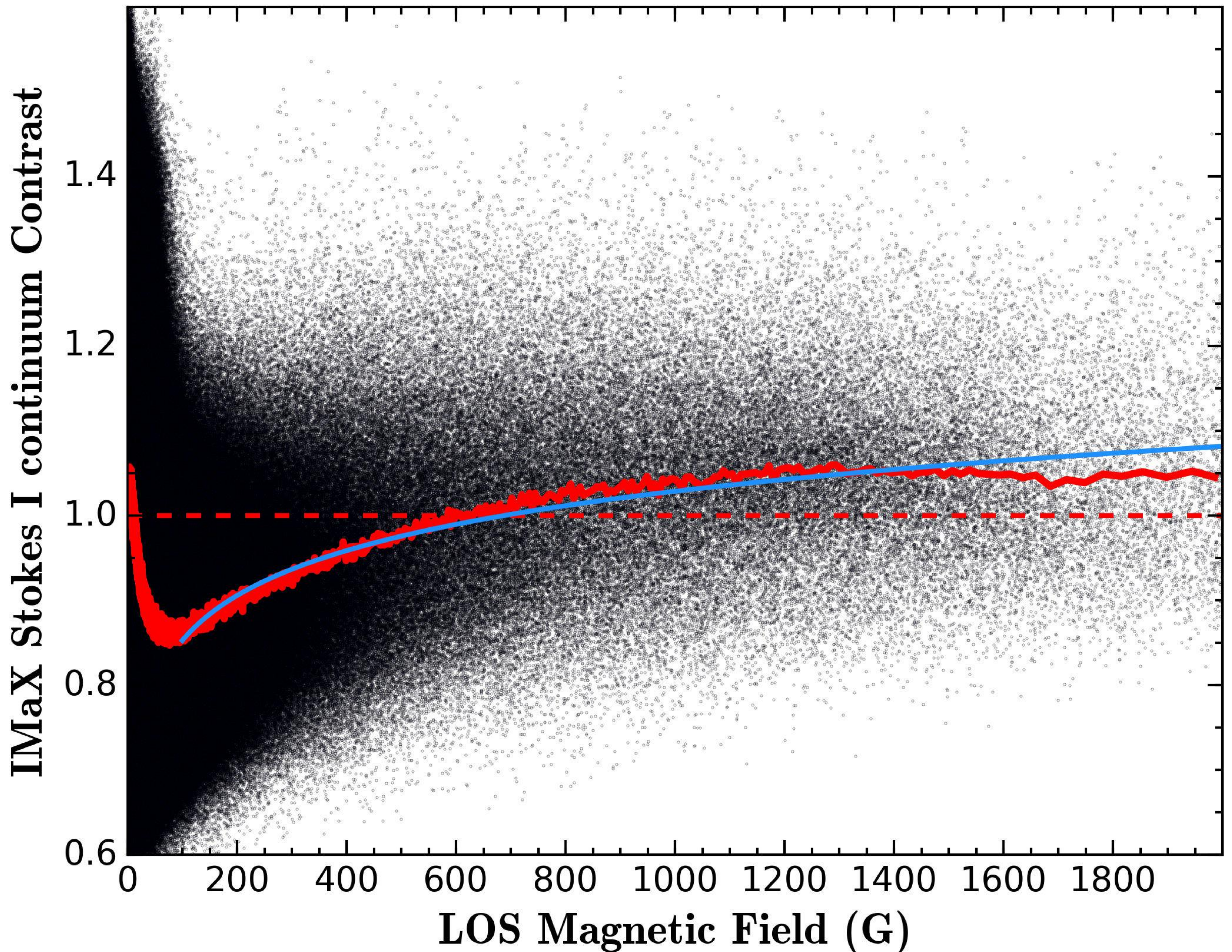}
\caption{Scatterplot of the IMaX continuum contrast at 5250.4\, \AA{} vs. the LOS component of the magnetic field in the QS at disc centre. The horizontal dashed red line indicates the mean quiet-Sun continuum intensity level, i.e., a contrast of unity. The red curve is composed of the binned values of the contrast, with each bin containing 500 data points. The blue curve is the logarithmic fit to the binned values starting at 90\,G . }
\label{cont_con}
\end{figure}

\begin{figure}
\centering
\includegraphics[scale=0.2]{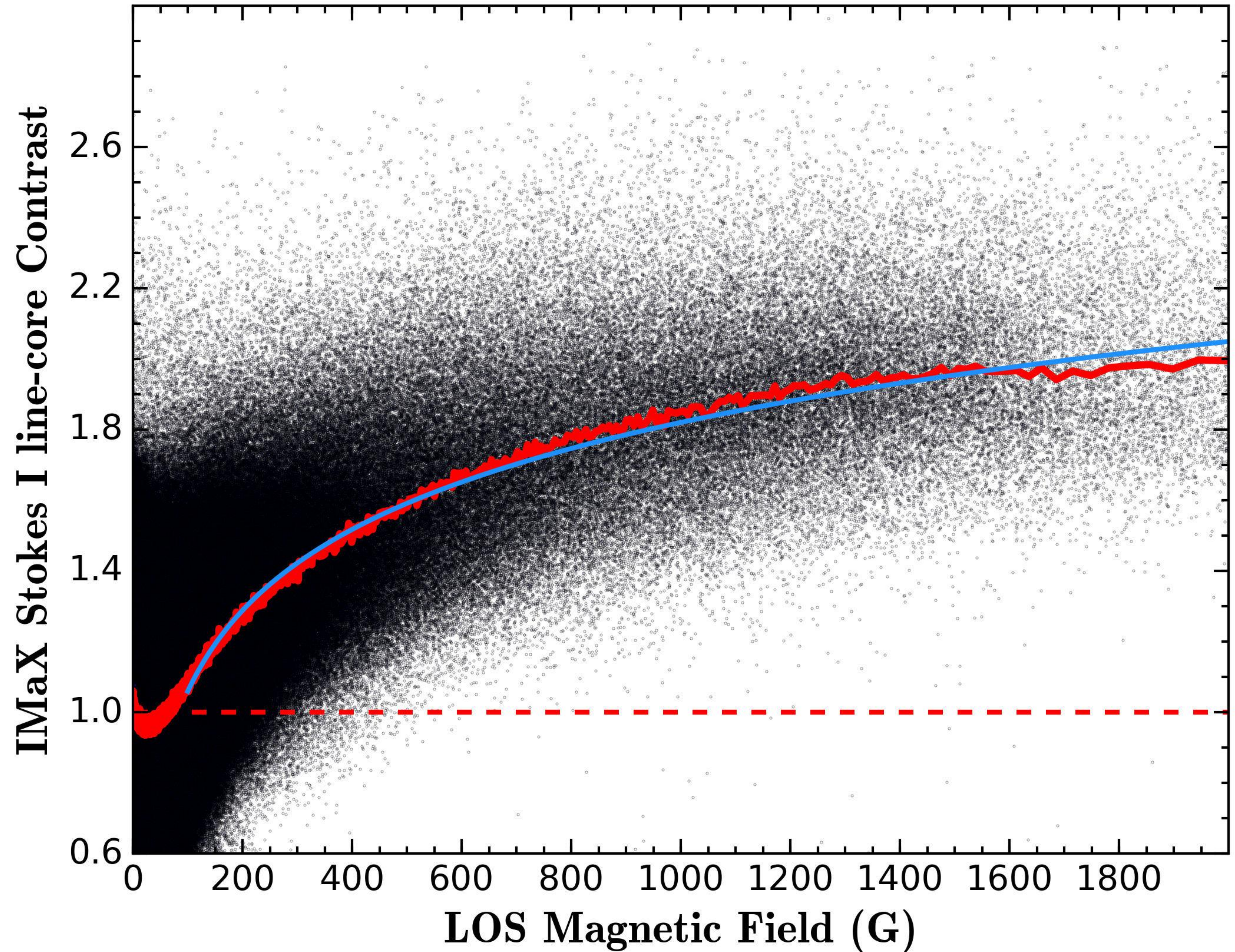}
\caption{Same as Fig.~\ref{cont_con}, but for the IMaX line-core contrast derived from Gaussian fits to the Stokes~$I$ line profiles.}
\label{lc_con}
\end{figure}

\begin{figure}
\centering
\includegraphics[scale=0.2]{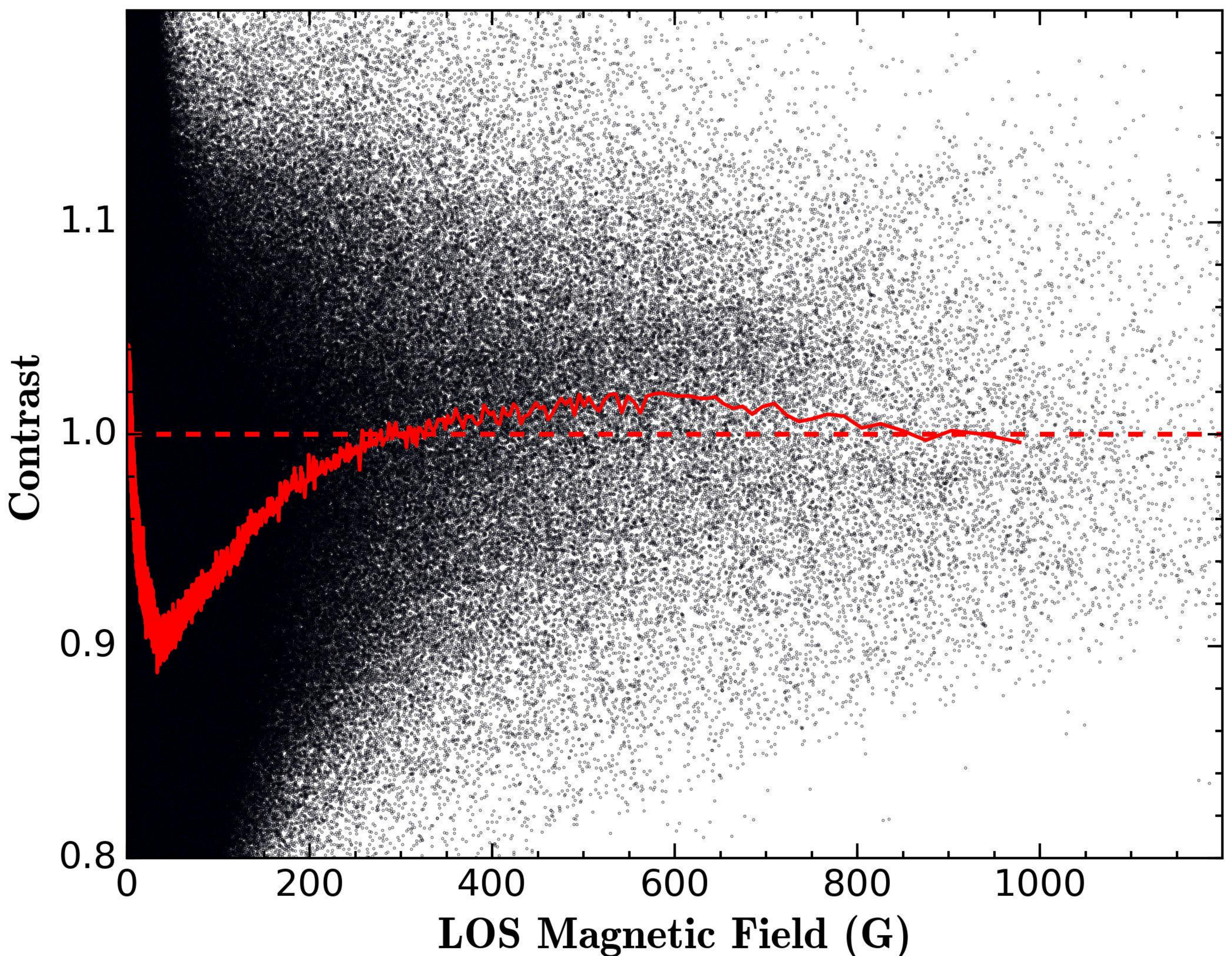}
\caption{Same as Fig.~\ref{cont_con}, but after degrading the underlying continuum Stokes images with a Gaussian of FWHM 0\carcsec{}32 to mimic the spatial resolution of the Hinode Spectropolarimter.}
\label{b_c_hinode}
\end{figure}

Pixel-by-pixel scatterplots of the IMaX continuum and line core contrasts are plotted vs. the longitudinal component of the magnetic field, $B_{\rm LOS}$ in Figs.~\ref{cont_con} and \ref{lc_con}. The contrast values are averaged into bins, each containing 500 data points, which are overplotted in red on Figs.~\ref{cont_con} and \ref{lc_con}, as well as on Figs.~\ref{b_c_hinode} and \ref{chap_3_sufi_UV} that are discussed later.

For the continuum contrast (Fig.~\ref{cont_con}), the large scatter around $B_{\rm LOS} \sim 0$ is due to the granulation. At very weak fields, the average contrast decreases with increasing field strength, because only the weakest fields are present in granules, while slightly stronger fields, close to the equipartition value, are concentrated by flux expulsion in the dark intergranular lanes \citep{parker_kinematical_1963}. These weak fields are typically at or below the equipartition field strength of around 200--400\,G at the solar surface \citep[e.g.,][]{solanki_solar_1996}, which corresponds roughly to 120--240\,km about 1 scale height above the solar surface, a very rough estimate of the height at which Fe\,{\sc i} 525.02 nm senses the magnetic field. Fields of this strength have little effect on the contrast, so that these pixels are darker than the mean quiet-Sun intensity (shown as the horizontal dashed red line). The contrast reaches a minimum at approximately 80\,G, then increases with increasing field strength, as the pressure in the flux tubes decreases, and these brighten, becoming brighter than the mean QS at around 600\,G. Together, these various effects give rise to the ``fishhook" shape of the continuum contrast curve, as described by \cite{schnerr_brightness_2011}. The contrast then saturates at larger field strengths.

\begin{table}[h!]
\centering
\begin{tabular}{||cccc||}
\hline
Threshold (\,G) & $\alpha$  & $\beta$ & $\chi^2$ \\
\hline\hline
90 &0.17 $\pm$0.001 & 0.51$\pm$0.002 & 7.62 \\
130 & 0.18 $\pm$0.001&0.47$\pm$0.002&4.04\\
170 & 0.19$\pm$0.001&0.46$\pm$0.003&3.45\\
210 & 0.19$\pm$ 0.002&0.46$\pm$0.004&3.26\\
250 & 0.18$\pm$0.002&0.47$\pm$0.006&3.05\\
 \hline
\end{tabular}
\caption{Parameters of logarithmic fits according to Eq.~\ref{log} to the continuum contrast at 5250.4\, \AA{} vs. the $B_{\rm LOS}$.}
\label{cont_log}
\end{table}

\begin{table}[h!]
\centering
\begin{tabular}{||cccc||}
\hline
Threshold (\,G) & $\alpha$ &$\beta$ &$\chi^2$ \\
\hline\hline
100 &0.76$\pm$0.002 & -0.47$\pm$0.004 & 6.3 \\
140 & 0.80 $\pm$0.002&-0.55$\pm$0.002&3.15\\
180 & 0.80$\pm$0.002&-0.58$\pm$0.007&2.55\\
220 & 0.81$\pm$ 0.003&-0.60$\pm$0.009&2.28\\
240 & 0.81$\pm$0.004&-0.60$\pm$0.01&2.22\\
\hline
\end{tabular}
\caption{Parameters of logarithmic fits according to Eq.~\ref{log} to the IMaX line core contrast vs. the $B_{\rm LOS}$.}
\label{lc_log}

\end{table}

The contrast values reached in the line core data (Fig.~\ref{lc_con}) are much higher than those in the continuum, in agreement with \cite{title_differences_1992} and \cite{yeo_intensity_2013}, and are on average larger than the mean QS intensity for $B_{\rm LOS}>$ 50\,G. 

The high average contrast values (both in the continuum and line core) with respect to the mean QS for strong magnetic fields proves the enhanced brightness property of small scale magnetic elements present in our data. Moreover, the average continuum contrast values reported here, are higher than the ones measured by \cite{kobel_continuum_2011}, partly due to our higher spatial resolution, but partly also due to the shorter wavelength of 525 nm vs. 630 nm of the Hinode data employed by \cite{kobel_continuum_2011}. In addition, our plots do not show any peak in the contrast at intermediate field strengths, nor a downturn at higher values as reported by \cite{kobel_continuum_2011} and \cite{lawrence_contrast_1993}, or a monotonic decrease as obtained by \cite{topka_properties_1992}.

To reproduce the scatterplot obtained by \cite{kobel_continuum_2011}, and to show the effect of spatial resolution on the shape of the $C_{\rm CONT}$ vs. $B_{\rm LOS}$ scatterplot, we degrade our data (Stokes~$I$ and $V$ images) to the spatial resolution of Hinode,  with a Gaussian of 0\carcsec{}32 \textit{FWHM}. Then, the magnetic field is computed at each pixel, using the centre of gravity (COG) technique \citep{rees_line_1979} for the degraded Stokes images.
Figure~\ref{b_c_hinode} shows the corresponding scatterplot, based on the degraded contrast and magnetic field images, with the data points binned in the same manner as for the undegraded data. One can clearly see both, a decrease in the contrast values and a leftward shift of data points towards lower magnetogram signals. Upon averaging, the binned contrast peaks at intermediate field stengths, and turns downwards at higher values.\\

To derive a quantitative relationship between the continuum contrast and the quiet-Sun magnetic field, we tried fitting the scatterplot with a power-law function of the form: 

\begin{equation}
I(B) = I_0 + a B^b
\label{pl}
\end{equation}

This function could not represent our scatterplots since the log--log plots (not shown) did not display a straight line in the range of $B_{\rm LOS}$ where we expected the fit to work, i.e., above the minimum point of the fishhook shape described earlier. Consequently, we looked for other fitting functions, and we found that the scatterplots could be succesfully fitted with a logarithmic function, which is the first time that it is used to describe such contrast curves:
\begin{equation}
I(B) = \beta + \alpha \log B 
\label{log}
\end{equation}

The fit represents the data quite well (the lin--log plots show a straight line) for data points lying above 90\,G for the continuum contrast vs $B_{\rm LOS}$, and from 100\,G for the line-core contrast vs. $B_{\rm LOS}$.
To investigate how the quality of the fit and the best-fit parameters depend on the magnetic field threshold below which all the data points are ignored, we list in Tables~\ref{cont_log} and \ref{lc_log} the best-fit parameters for the continuum and line core contrast variation with $B_{\rm LOS}$, respectively, along with the magnetic flux threshold, and the corresponding $\chi^2$ values. Fitting the original data points or the binned values returns similar results. We have plotted and tabulated the curves obtained by fitting the binned values. The $\chi^2$ values are large for smaller thresholds, and decrease with increasing threshold where less points are fitted. In contrast to this, the best-fit parameters show only a rather small variation with the threshold used, which is not the case with the power-law fit used later in Sec.~\ref{scatter_ca} when describing the relationship between the Ca\,{\sc ii} H emission and $B_{\rm LOS}$, a relationship that has traditionally been described with a power-law function.

\begin{figure}
\centering
\includegraphics[width=\linewidth]{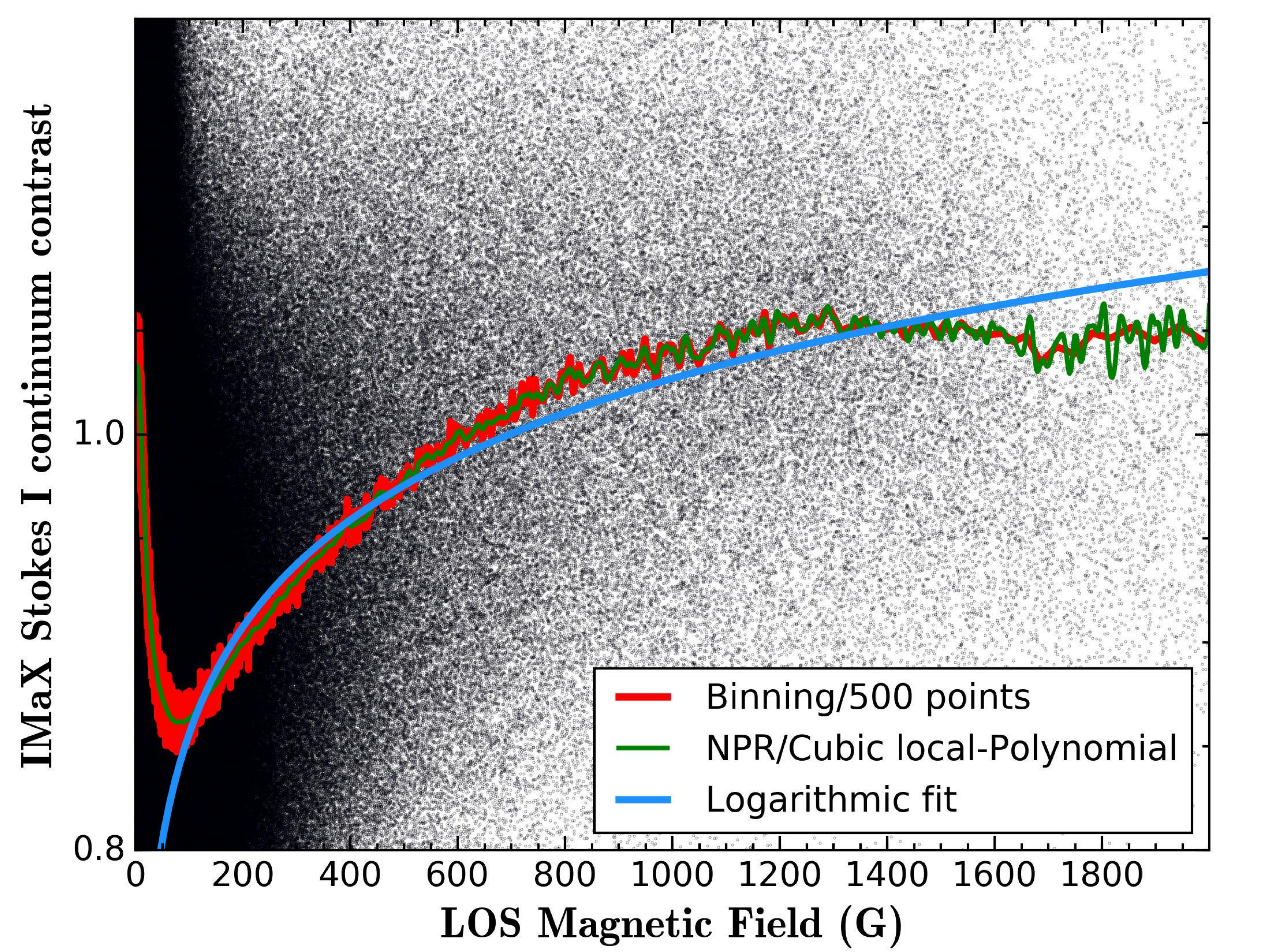}
\caption{Same scatterplot shown in Fig.~\ref{cont_con} of the IMaX continuum vs. $B_{\rm LOS}$. The red curve is the binned values of the contrast. The green curve the non-parametric regression curve obtained from applying the Kernel smoothing technique, with a local-polynomial of order $q=3$ as a regression method (check Appendix~\ref{appendix_A} for more details). The blue curve is the logarithmic fit applied to data points starting at 90\,G and extrapolated to smaller $B_{\rm LOS}$ values.}
\label{bc_smooth}
\end{figure}

In order to test the validity of the binning method used to represent the trend of the scatterplots throughout the paper, and to which the parametric models described above (logarithmic and power-law models) are fitted, we also apply non-paramteric regression (NPR) methods to the data points. These methods do not require specific assumptions about how the data should behave, and are used to find a non-linear relationship between the contrast and magnetic field by estimating locally the contrast value at each $B_{\rm LOS}$, depending on the neighbouring data points. We show in Fig.~\ref{bc_smooth} the scatterplot of the IMaX continuum contrast vs. $B_{\rm LOS}$ discussed earlier in this section and depicted in Fig.~\ref{cont_con}. We plot in blue the logarithmic fit extrapolated to small $B_{\rm LOS}$ values. The red curve is the graph joining the binned contrast values and the green curve  is the regression curve obtained after applying one of the NPR techniques that are described in detail in  Appendix~\ref{appendix_A}.

The NPR curve fits the data, including the fishhook shape at small values of $B_{\rm LOS}$, and lies extremely close to the curve produced by binning contrast values. This agreement gives us considerable confidence in the binned values we have used to compare with the simple analytical model functions.
The most relevant conclusion that can be drawn from Fig.~\ref{bc_smooth} is that our binning method is appropriate to represent the behaviour of the data points, and that it is valid to fit the logarithmic or power-law models to the binned values of the data points.

This test was repeated also for the scatterplots analysed in the next sections and the results are discussed in  Appendix~\ref{appendix_A}.

\subsection{Scatterplots of SuFI UV brightness vs. $B_{\rm LOS}$}
\label{uv_vs_B}
\begin{figure*}
\centering
\includegraphics[width=\textwidth]{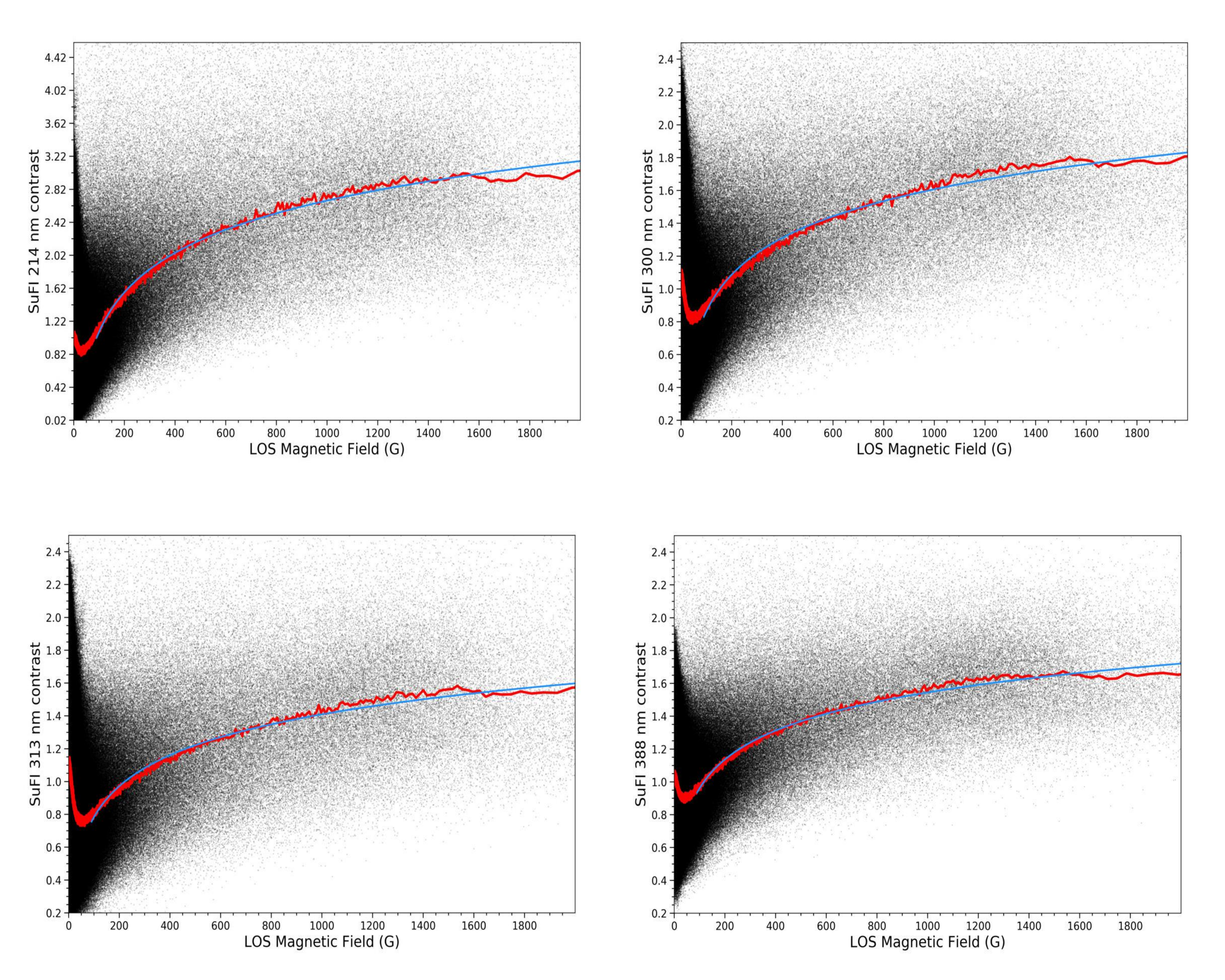}
\caption{Scatterplots of the intensity contrast relative to the average QS in four NUV wavelength bands sampled by SUFI vs. $B_{\rm LOS}$. The red curves are the values binned over 500 data points each, the blue curves are the logarithmic fits to the binned curves, starting from 90\,G,  with the fitting parameters listed in Table.~\ref{uv_log}.}
\label{chap_3_sufi_UV}
\end{figure*}

The pixel-by-pixel scatterplots of the contrast at 214 nm, 300 nm, 313 nm, and 388 nm vs. $B_{\rm LOS}$ are shown in Fig.~\ref{chap_3_sufi_UV}. The data points are binned following the procedure described in section~\ref{plots_b_c}.

For all wavelengths in this range, the contrasts are much larger than in the visible, especially at 214 nm where the contrast is greatly enhanced \citep[see][]{riethmuller_bright_2010}. The averaged contrast increases with increasing field strength, even for higher $B_{\rm LOS}$. Also, the fishhook shape is well visible at all UV wavelengths, and the minimum in the contrast occurs at similar $B_{\rm LOS}$ values (30\,G -- 50\,G), while contrast $>$ 1 is reached at somewhat different $B_{\rm LOS}$ values, ranging from 65\,G to 220\,G for the different UV wavelengths. 

The data points are fitted with a logarithmic function (Eq.~\ref{log}), since it describes the fitted relation better than a power-law function. Table \ref{uv_log} lists the best-fit parameters for the different spectral regions, from a threshold of 90\,G, at which the fits start to work, along with the corresponding $\chi^2$ values. 
The logarithmic function represents well the contrast vs. $B_{\rm LOS}$ relationship for all wavelengths in the NUV. A test-wise fit for different magnetic field thresholds showed that the fit results are quite insensitive to the threshold.

\begin{table}[h!]
\centering
\caption{Parameters of logarithmic fits according to Eq.~\ref{log} to the observed NUV contrast vs. the $B_{\rm LOS}$. The wavelengths sampled by SuFI are shown in the first column.}
\begin{tabular}{||cccc||}
\hline
Wavelength (nm)& $\alpha$ & $\beta$ & $\chi^2$ \\
\hline\hline
214 & 1.57$\pm$0.006 & -2.04$\pm$0.01&2.48 \\
300 & 0.74$\pm$0.004&-0.61$\pm$0.009&3.75 \\
313&0.62$\pm$0.003&-0.46$\pm$0.007&2.98 \\
388&0.58$\pm$0.002&-0.21$\pm$0.005&2.61 \\
\hline
\end{tabular}
\label{uv_log}
\end{table}

\subsection{Scatterplot of chromospheric emission vs. $B_{\rm LOS}$}
\label{scatter_ca}
\begin{figure}
\centering
\includegraphics[width=\linewidth]{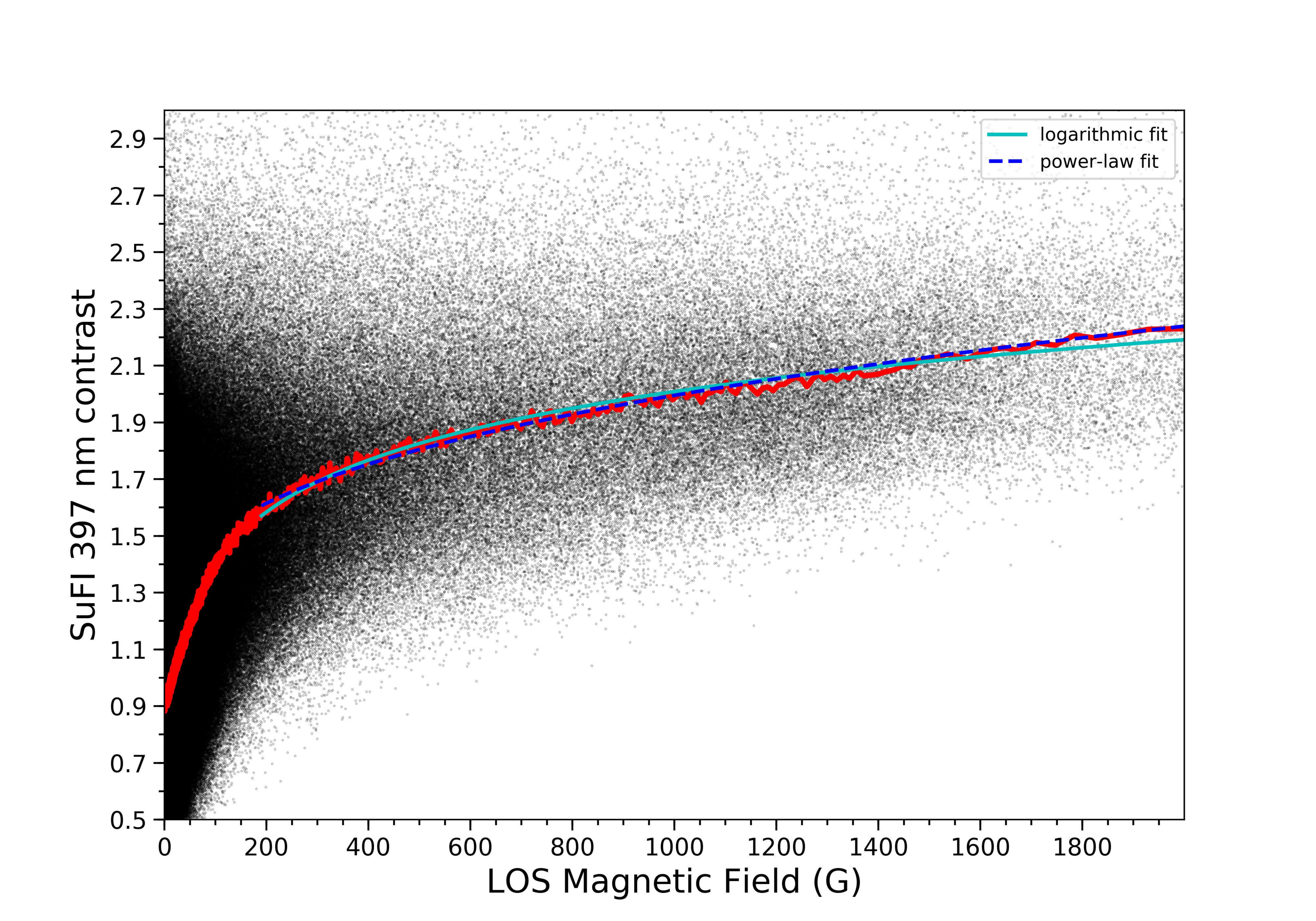}
\caption{Scatterplot of the Ca\,{\sc ii} H intensity vs. the longitudinal component of the magnetic field. The red curve represents the binned data points, the solid and dashed blue curves are the logarithmic and power-law fits to the binned data points, starting from 190\,G.}
\label{b_ca}
\end{figure}

A scatterplot of the contrast in the SuFI 397 nm Ca\,{\sc ii} H band vs. $B_{\rm LOS}$ is shown in Fig.~\ref{b_ca}.\\
The Ca\,{\sc ii} H spectral line gets considerable contribution from the lower chromosphere. This was shown in \citet{jafarzadeh_structure_2013}(see their Fig.~2c), and \cite{danilovic_comparison_2014} (their Fig.~1) who determined the average formation heights of this line as seen through the wide and the narrow $\sunrise$/SuFI Ca\,{\sc ii} H filter, respectively, by convolving the spectra with the corresponding transmission profile and computing the contribution function for different atmospheric models. The model corresponding to an averaged quiet-Sun area returned an average formation height of 437\,km.

As pointed out in Sects.~\ref{plots_b_c} and \ref{uv_vs_B} a logarithmic function fits the contrast vs. $B_{\rm LOS}$ relationship better than a power-law function. Nonetheless, we have fitted the  Ca\,{\sc ii} H contrast vs. $B_{\rm LOS}$ relation with both, a power law (described by Eq.~\ref{pl}) and a logarithmic function (described by Eq.~\ref{log}). We first discuss the power law fits, as these have been widely used in the literature \citep[e.g.,][]{schrijver_relations_1989,ortiz_how_2005,rezaei_relation_2007,loukitcheva_relationship_2009}. 
As pointed out by \cite{rezaei_relation_2007} the best-fit parameters of the power law depend significantly on the $B_{\rm LOS}$ threshold below which the fit is not applied. To investigate this dependence, we fit the data points exceeding different threshold values, which are listed in Table~\ref{ca_pl} along with the corresponding best-fit parameters and $\chi^2$ values. The data are well-represented by a power-law function, for points lying above 190\,G~\footnote{The log-log plot of the data shows a straight line starting from this value, defining where the data follows a power law function.}. However, have strongly different best-fit parameters, depending on the threshold in $B_{\rm LOS}$ applied prior to the fit.

The data were also succesfully fitted by a logarithmic function (Eq.~\ref{log}). The fit works from a lower threshold (50\,G) than the power-law fit, and the best-fit parameters vary only slightly with the threshold as can be seen in Table~\ref{ca_log}. Although the log function produces a reasonable fit starting already from 50\,G, the $\chi^2$ is rather large for this threshold and drops to values close to unity only for a $B_{\rm LOS}$ threshold $>$ 190\,G, although the fit parameters remain rather stable. 
\\

\begin{table}[h!]
\centering
\caption{Parameters of power-law fits according to Eq.~\ref{pl} to the Ca\,{\sc ii} H emission vs. the $B_{\rm LOS}$. The first column is the threshold for the magnetic field strength, the third column is the power-law index, the fourth column is the offset, and the last one is the corresponding $\chi^2$ value.}
\begin{tabular}{||ccccc||}
\hline
Threshold (\,G) & $a$  &$b$ &$I_0$ & $\chi^2$ \\
\hline\hline
190 & 0.49 $\pm$0.13&0.17$\pm$0.02&0.37$\pm$0.19&0.86\\
210 & 0.38$\pm$0.11&0.19$\pm$0.03&0.55$\pm$0.17&0.81\\
230 & 0.25$\pm$ 0.08&0.23$\pm$0.03&0.76$\pm$0.14&077\\
250 & 0.12$\pm$0.04&0.29$\pm$0.03&1.03$\pm$0.09&0.67\\
\hline
\end{tabular}
\label{ca_pl}

\end{table}

\begin{table}[h!]
\centering
\caption{Parameters of logarithmic fits according to Eq.~\ref{log} to the Ca\,{\sc ii} H emission vs. the $B_{\rm LOS}$}
\begin{tabular}{||cccc||}
\hline
Threshold (\,G) & $\alpha$ & $\beta$ & $\chi^2$ \\
\hline\hline
50 & 0.62$\pm$0.001&0.14$\pm$0.004&3.45\\
90 &0.59$\pm$0.002&0.22$\pm$0.005&1.58\\
170&0.58$\pm$0.003&0.25$\pm$0.009&1.08\\
190&0.58$\pm$0.004&0.25$\pm$0.01&1.04\\
210&0.59$\pm$0.004&0.24$\pm$0.01&0.99\\
230&0.59$\pm$0.005&0.23$\pm$0.01&0.97\\
250&0.59$\pm$0.005&0.24$\pm$0.01&0.94\\
\hline
\end{tabular}
\label{ca_log}
\end{table}

\begin{landscape}
\begin{figure*}
\centering
\includegraphics[width=\linewidth]{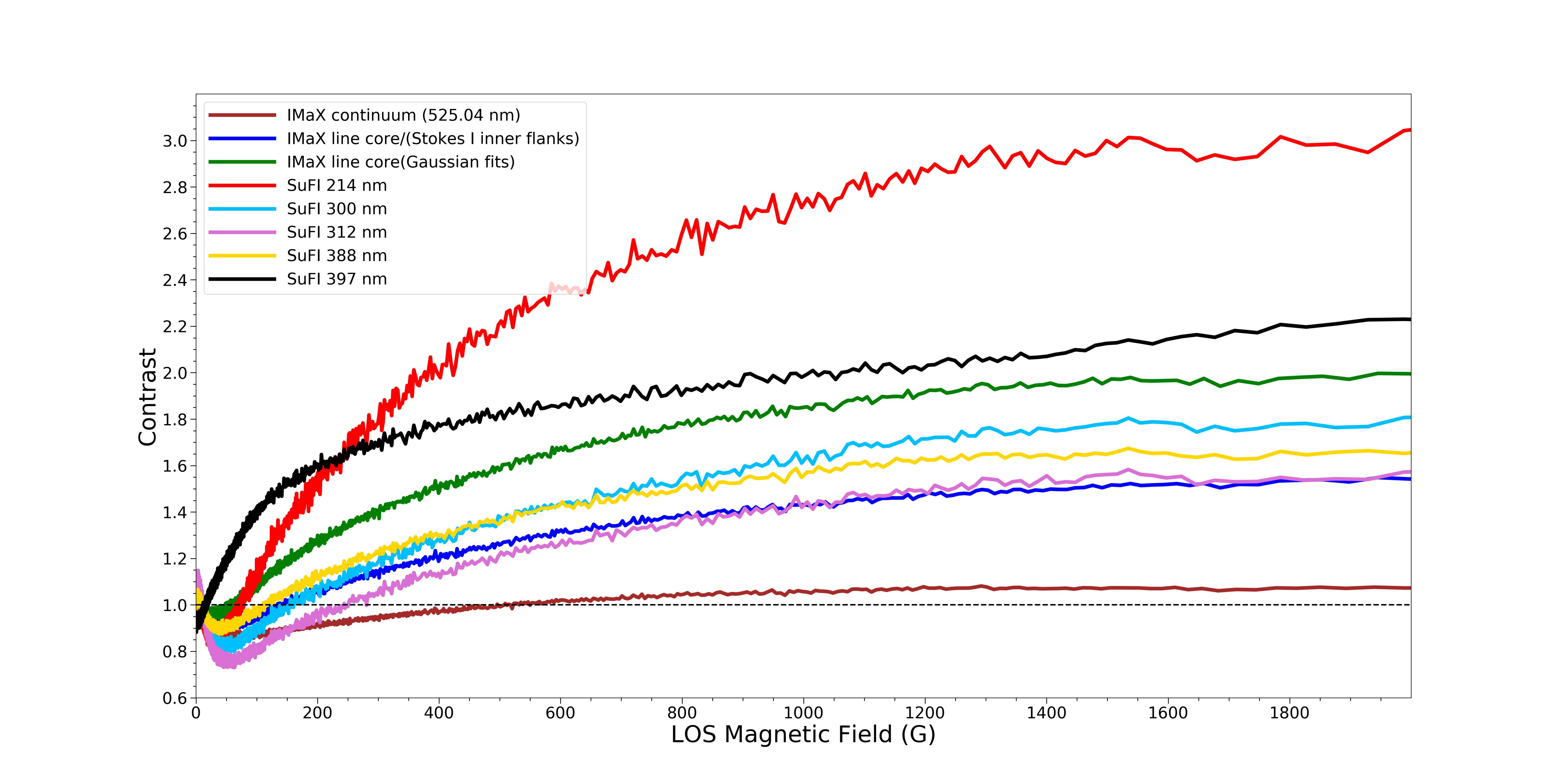}
\caption{ All binned contrast vs. $B_{\rm LOS}$ curves from Figs.~\ref{cont_con}, \ref{lc_con}, \ref{chap_3_sufi_UV} and \ref{b_ca} plotted together. Also plotted is the contrast of the 5250\, \AA{} line core obtained by averaging the intensities at the wavelength positions $+40$ and $-40$ m${\angstrom}$ apart. The curves are identified by their colour in the upper left part of the figure. The black dashed line marks the mean quiet-Sun intensity level, i.e., contrast of unity.}
\label{bins}
\end{figure*}
\end{landscape}

To better compare the magnitudes of the contrast at the different wavelengths studied here, we plot in Fig.~\ref{bins} all the scatterplots composed of the binned points obtained in the previous sections for the NUV wavelengths at 214 nm, 300 nm, 313 nm, 388 nm, and 397 nm (from Figs.~\ref{chap_3_sufi_UV} and \ref{b_ca}), and at the visible wavelengths, i.e., the continuum at 525 nm (from Fig.~\ref{cont_con}) and line core derived from the Gaussian fit to the IMaX Stokes~$I$ profiles (from Fig.~\ref{lc_con}). Also plotted is the approximate line-core value obtained from the average of the $-40$ m${\angstrom}$ and $+40$ m${\angstrom}$ line positions of the Stokes~$I$ line profile using Eq.~\ref{LC_40}.

Several qualitative observations can be made from this graph. Firstly, the contrast reached at 214 nm is higher than the Ca\,{\sc ii} H contrast. We could not find an instrumental reason for this, e.g. the calcium images were not overexposed and the response of the detector was quite linear. Therefore, we believe that the larger contrast seen in the 214 nm wavelength band is intrinsic. We expect that it is due to the very large density of lines at 214 nm and to the relatively broad Ca\,{\sc ii} H filter of 1.8 \AA{}, which has considerable contributions from the photosphere. The much higher temperature sensitivity of the Planck function at short wavelengths also plays a role.

Secondly, the large difference between the line core contrasts derived from the two different methods, with the Gaussian fits to the line profile giving almost twice the contrast (and almost reaching that of the Ca\,{\sc ii} H line core at 397 nm for higher fields) suggests that the sum of intensities at $\pm 40$ m${\angstrom}$ from the line centre is not a good approximation of the line core itself. At the other NUV wavelengths the contrast depends mainly on the number and the temperature sensitivity of the molecular lines in the passbands \citep[see][]{schussler_why_2003}. Thus the 388 nm has a larger contrast than 312 nm due to the large density of CN lines in the former. 

\section{Discussion and conclusions}
\label{chapter3_conc}
\subsection{Brightness in the visible vs. $B_{\rm LOS}$}

The constant continuum contrast reached in our scatterplots for field strengths  higher than 1000\,G (see Fig.~\ref{cont_con}), confirms the previous results of \cite{rohrbein_is_2011}. They compared the relation between the bolometric intensity contrast and magnetic field strength for MHD simulations degraded to various spatial resolutions. For field strengths higher than 300--400\,G they found a monotonic increase at the original full resolution, a saturation at the spatial resolution of a 1-m telescope, while at the spatial resolution of a 50 cm telescope a turnover at around 1000\,G followed by a contrast decrease. Our results also agree with the analysis presented by \cite{kobel_continuum_2011}, who found that even at Hinode/SP resolution, the strong magnetic features are not resolved, leading to a turnover in their scatterplots for higher fields, similar to the behaviour obtained by \cite{lawrence_contrast_1993}, for a quiet-Sun region. The $\sunrise$/IMaX data display a saturation of the contrast at its maximum value in the visible continuum, in qualitative agreement with the work of \cite{rohrbein_is_2011}, although the wavelength range of their simulated contrast is different, which indicates the need to repeat the study of \cite{rohrbein_is_2011}, but for the actual measured wavelength bands and to compare these results quantitavely with the $\sunrise$ data. Interestingly, after degrading the $\sunrise$/IMaX data to Hinode's spatial resolution, a peak and downturn of the contrast were reproduced (see Fig.~\ref{b_c_hinode}).

The higher spatial resolution reached by IMaX allowed us to constrain the effect of spatial resolution on the relation between continuum brightness at visible wavelengths, and the LOS component of the photospheric magnetic field. At a resolution of 0.15$^{\prime\prime}$ (twice that of Hinode/SP), magnetic elements in the quiet-Sun internetwork start to be spatially resolved \citep[see][]{lagg_fully_2010}, leading to a constant and high contrast becoming visible in strong magnetic features.

\subsection{Brightness in the NUV vs. $B_{\rm LOS}$ }
The relationship between the intensity and the photopsheric magnetic field for several wavelengths in the NUV provided new insights into the quantitative relation between the two parameters. The wavelength range between 200 and 400 nm is of particular importance for the variable Sun's influence on the Earth's lower atmosphere, as the radiation at these wavelengths affects the stratospheric ozone concentration \citep[e.g.,][]{gray_solar_2010,ermolli_recent_2013,solanki_solar_2013}. Although there is some convergence towards the level of variability of the solar irradiance at these wavelengths \citep{yeo_uv_2015}, there is a great need for independent tests of the employed modelled spectra in the UV. Such UV data are also expected to serve as sensitive tests of MHD simulations.

Of the UV wavelengths imaged by $\sunrise$ only the Ca\,{\sc ii} H line (discussed below) and the CN band head at 388 nm have been observed at high resolution earlier. E.g. \citet{zakharov_comparative_2005} obtain a contrast of 1.48 in bright points with the SST, which is close to the mean contrast of 1.5 we find for large field strengths.   

Most of the studies made so far on the relationship between the Ca\,{\sc ii} H emission and the photospheric magnetic field were carried out using ground-based data and different results have been obtained concerning the form of this relation. As mentioned in the introduction, several authors fitted their data with a power-law function, obtaining power-law exponents that varied from 0.2 \citep{rezaei_relation_2007} to 0.66 \citep{ortiz_how_2005}. 

We were also able to fit our Ca\,{\sc ii} H data with a power-law function, obtaining different exponents for different thresholds of the magnetic field strength from which the fit started (see Table~\ref{ca_pl}). A nearly equally good fit was provided by a logarithmic function, starting from lower field strengths, and showing no strong variations of best-fit parameters with the threshold (see Table~\ref{ca_log}).
Other advantages of the logarithmic fit are that it has a free parameter less than the power law fit and that it also works well and equally independently of the threshold for the other observed wavelengths (see Tables~\ref{cont_log}--\ref{uv_log}), whereas the power-law fit did not lead to reasonable results.

As mentioned earlier in this paper, irradiance changes from below 400 nm are the main contributors to the TSI variations over the solar cycle. The magnetic flux from small-scale magnetic elements in the QS is believed to contribute considerably to not just these changes \citep{krivova_reconstruction_2003}, but likely lie at the heart of any secular trend in irradiance variations \citep[e.g.,][]{krivova_reconstruction_2007, dasi-espuig_reconstruction_2016}, which are particularly uncertain, but also particularly important for the solar influence on our climate.

This contribution depends on the size and position on the solar disc of these elements and possibly on their surroundings. Here we studied the intensity contrast of a quiet-Sun region near disc centre. Next steps include repeating such a study for MHD simulations, carrying out the same study for different heliocentric angles and extending it to active region plage, so that the results can be used to test and constrain the atmosphere models used to construct spectral solar irradiance models.

\chapter{Intensity contrast of solar plage as a function of magnetic flux at high spatial resolution}

\textbf{*This Chapter is based on a published paper to Astronomy and Astrophysics \citep{kahil_intensity_2019}, with the permission of the journal.}\\ \\
\label{chapter_4}

\textbf{Abstract} \\

The contrast of magnetic elements depends on the type of region they are located in (e.g. quiet Sun, or active region plage). Observed values also depend on the spatial resolution of the data. Here we investigate the contrast-magnetic field dependence in active region plage observed near disk center with $\sunrise$ during its second flight in 2013. The wavelengths under study range from the visible at 525\,nm to the Near-Ultra-Violet (NUV) at 300\,nm and 397\,nm. We use quasi-simultaneous spectropolarimetric and photometric data from the Imaging Magnetograph eXperiment (IMaX) and the Sunrise Filter Imager (SuFI), respectively. We find that in all wavelength bands, the contrast represents a qualitatively similar dependence on the line-of-sight magnetic field, $B_{\rm LOS}$, as found in the quiet Sun, with the exception of the continuum at 525\,nm. There, the contrast of plage magnetic elements peaks for intermediate values of $B_{\rm LOS}$ and decreases at higher field strengths, whereas the contrast of quiet-Sun magnetic elements saturates at its maximum value at large $B_{\rm LOS}$. We find that the explanation of the turnover in contrast in terms of the effect of finite spatial resolution of the data is incorrect in the $\sunrise$ data, as the plage magnetic elements are larger than the quiet Sun magnetic elements and are well-resolved. The turnover comes from the fact that the core pixels of these larger magnetic elements are darker than the quiet Sun. We find that plages reach lower contrast than the quiet Sun at disk center at wavelength bands formed deep in the photosphere, such as the visible continuum and the 300\,nm band. This difference decreases with height and disappears in the Ca\,{\sc ii} H core, in agreement with empirical models of magnetic element atmospheres.
  
  \newpage
\section{Introduction}
In addition to quiet-Sun (QS) network, small-scale magnetic elements on the solar surface manifest themselves in the form of plage regions or faculae. 
Faculae are seen near the solar limb in white light as bright structures often surrounding sunspots and pores. The same structures seen in chromospheric emission are called plages. Here, we use the latter nomenclature to describe such magnetic flux concentrations seen at wavelengths sampling heights ranging from the low photosphere to the lower chromosphere. The network is seen at the edges of supergranules, while plages are part of active regions (ARs). Network and plage both harbor kG magnetic fields \citep{stenflo_magnetic-field_1973}.

At the photospheric level, magnetic elements are located within the intergranular lanes and have been successfully described by flux tubes \citep{spruit_pressure_1976, solanki_small-scale_1993}. The magnetic field is swept by granular motion to the dark intergranular lanes and concentrated there by the magnetic flux expulsion mechanism \citep{parker_kinematical_1963}. The intensification of the magnetic field within these elements causes the internal gas pressure at a given height to drop to maintain total horizontal pressure equilibrium with the non-magnetic surroundings. This leads to an opacity depression, so that one can see into deeper layers of the photosphere within such elements. Their brightness is dependent on their size, the extent of the opacity depression (i.e., effectively the magnetic field strength), which together determine the effective heating from the surrounding hot walls, which correspond to the sides of the surrounding granules.

Small-scale magnetic elements are also linked to the heating of the chromopshere and corona. For the chromosphere, this is indicated by the correlation found between the brightness in the cores of Ca II H and K lines and the magnetic field \citep[]{skumanich_statistical_1975,schrijver_relations_1989,ortiz_how_2005,rezaei_relation_2007,loukitcheva_relationship_2009,kahil_brightness_2017}, while the importance of small magnetic features for coronal heating has been pointed out by \cite{zhang_lifetime_1998,ishikawa_comparison_2009, zhou_solar_2010,chitta_solar_2017}. Together, the network and faculae are responsible for the brightening of the Sun with higher magnetic activity on time scales longer than solar rotation \citep{krivova_reconstruction_2003,yeo_solar_2017}. 

The reported brightness of magnetic elements with respect to the quiet surrounding (i.e., their contrast) in the photosphere, along with their morphological structure depends on the wavelength band in which these structures are seen, their position on the solar disk, the selection method used to identify them, and the spatial resolution at which they are observed. Consequently, different observations sometimes gave seemingly contradictory results regarding their photometric and magnetic properties.
Below we give an overview of some of the studies that have looked into the contrast-magnetic field relationship, concentrating on the (near) disk center observations.\\

At a spatial resolution of 4$^{\prime\prime}$, \citet{ortiz_intensity_2002} used simultaneous full-disk magnetograms and continuum images (at 676.8 nm) acquired by the Michelson Doppler Imager (MDI) on-board the Solar and Heliospheric Observatory (SoHO). At disk center, the contrast of active-region faculae (identified by their higher magnetic field values) is negative while it is positive for smaller magnetic elements (the network). They attributed the contrast difference to the size of the features, whereby the radiative heating from the surrounding walls is more important and effective when the cross-section of a flux tube is smaller. This study was redone at higher resolution by \cite{yeo_intensity_2013} using HMI data. They obtained essentially the same qualitative results, with the differences that the contrast values at weak magnetogram signals in their identified network are negative.

\citet{frazier_multi-channel_1971} used simultaneous observations of photospheric magnetic field inferred from the Fe\,{\sc i} 5233.0\, \AA{} line and continuum filtergrams at Fe\,{\sc i} 5250.2\, \AA{} in addition to line-core data in Fe\,{\sc i} 5250.2\, \AA{} and Ca\,{\sc ii} K of 2 active regions at disk center. Whereas continuum contrast increased up until 200\,G followed by a decrease at higher fields, the line-core and Ca\,{\sc ii} K contrasts kept on increasing with increasing magnetogram signal. For the core of another photospheric line, Fe\,{\sc i} 6173\, \AA{}, \cite{yeo_intensity_2013} obtained that the line contrast remains positive at all studied field strengths, but it does decrease somewhat at higher field strengths. For Ca\,{\sc ii} H and K this qualitative behavior was confirmed by, e.g., \cite{skumanich_statistical_1975}, \cite{schrijver_relations_1989}, \cite{loukitcheva_relationship_2009}, \cite{kahil_brightness_2017}, although the exact dependence varied. 

\citet{title_differences_1992}, \citet{topka_properties_1992} and \citet{lawrence_contrast_1993} used AR data from the 50\,cm Swedish Solar Vacuum Telescope (SVST), while \citet{lawrence_contrast_1993} also analyzed the quiet-Sun network.
At disk center, and at a spatial resolution of 0.3$^{\prime\prime}$ their measured continuum brightness at visible wavelengths (676.8, 525, 557.6, and 630.2 nm) is negative (less than the mean QS intensity) for all magnetogram signals. The line core brightness at 676.8 nm increases until 600\,G, then drops monotonically at higher fields in active regions.
They reported higher (more positive) contrast in the quiet Sun \citep{lawrence_contrast_1993}.

\citet{kobel_continuum_2011} looked into the brightness relationship with the magnetogram signal of both, QS and AR plages. They used spectropolarimetric data at 630.2\,nm  from the Solar Optical Telecope on-board Hinode (spatial resolution of 0.3$^{\prime\prime}$). At disk center, the contrast of both regions showed the same behaviour when plotted against the longitudinal field strength: contrast peaking at 700\,G, and turning over at higher fields. This partly contradicts the results of \cite{title_differences_1992}; \cite{topka_properties_1992} and \cite{lawrence_contrast_1993}. \citet{kobel_continuum_2011} also obtained higher contrast in the QS compared to AR plages, which confirmed the results of \cite{title_differences_1992} and \cite{lawrence_contrast_1993}.

At a spatial resolution of 0.1$^{\prime\prime}$, \citet{berger_contrast_2007} studied the relationship between the G-band contrast of solar magnetic elements and the magnetic field at different heliocentric angles using the Swedich Solar Telescope \citep[SST,][]{scharmer_1-meter_2003}. For faculae close to disk center ($\mu=0.97$), the G-band contrast is positive for fields up until 700\,G and becomes negative for higher fields, with sunspots and pores excluded from their study.
Also using the SST, \citet{narayan_small-scale_2010} analysed the dynamical properties of solar magnetic elements close to disk center, using spectropolarimetric data collected by the CRisp Imaging SPectropolarimter (CRISP) in the Fe\,{\sc i} line at 630.25\, nm, with the magnetic field obtained via Milne-Eddington inversions. After masking the dark pores in their images, their scatterplot of the continuum contrast against the longitudinal component of the magnetic field showed an initial decrease of the contrast for fields less than 200\, G, then an increase until 600\,G followed by a drop  below the QS reference for higher fields.

So far the observational results show a strong disagreement with the predictions of radiative magnetohydrodynamic simulations (MHD) at their original resolution. According to such simulations, in agreement with  the flux tube model, the continuum contrast of small-scale magnetic elements should increase monotonically with the magnetic field strength \citep[][their Fig.~13]{vogler_simulations_2005}. In all observational studies the continuum contrast (in visible wavelengths) of quiet-Sun network and faculae decreases at higher fields, even when excluding pores in the case of faculae. This decrease was attributed to the limited spatial resolution of observations (even if they are carried out using space-borne instruments) compared to the higher resolution of MHD simulations \citep[]{rohrbein_is_2011,danilovic_relation_2013}. At the spatial resolution of 0.15$^{\prime\prime}$ achieved by the Imaging Magnetograph eXperiment (IMaX) onboard $\sunrise$, \cite{kahil_brightness_2017} confirmed this hypothesis by demonstrating that the continuum contrast at 525\,nm of a quiet-Sun region saturated at the strongest fields. Inspecting this relationship in the larger plage features at the spatial resolution of IMaX is one of the primary aims of the current work. \\

A qualitative and quantitative study has been carried out in Chapter~\ref{chapter_3} for the pixel-by-pixel relationship of contrast to magnetic field in quiet-Sun network at disk center. We used observations recorded by the balloon-borne observatory $\sunrise$ during its first flight in 2009 ($\sunrise$ I), in the UV and visible wavelength ranges. Here, we look into the same relationship, but for plage regions observed by $\sunrise$ during its second flight in 2013 ($\sunrise$ II).  
The data and analysis techniques are described in Section \ref{observations}. In Section \ref{results} we show our results and compare them to similar studies and to those obtained in Chapter~\ref{chapter_3} for quiet-Sun data. In Section~\ref{conclusion} we discuss our findings and come up with conclusions and insights for future investigations.

\section{Observations and data reduction}
\label{observations}
The time series used in this study is recorded with the 1-m telescope on the balloon-borne observatory $\sunrise$ \citep{solanki_sunrise:_2010,barthol_sunrise_2011,berkefeld_wave-front_2011} during its second flight on 2013 June 12 \citep{solanki_second_2017}. We use simultaneous spectropolarimetric and imaging data collected by the two scientific instruments onboard, an imaging spectropolarimeter \citep[IMaX;][]{martinezpillet_imaging_2011} and a UV filter imager \citep[SuFI;][]{gandorfer_filter_2011}. The observations were made close to disk center (cosine of heliocentric angle $\mu$ = 0.93), targeting the active region (AR) NOAA 11768 (see Section~\ref{inversions} for more details on the AR field of view). 
\subsection{IMaX data}
\label{imax}
The spectropolarimetric data were acquired by the Imaging Magnetograph eXperiment \citep[IMaX;][]{martinezpillet_imaging_2011} between 23:39 and 23:55 UT in their V8-4 mode. This mode consists of measuring the full Stokes vector $(I,Q,U,V)$ at 8 wavelength positions around the center of the photospheric Fe\,{\sc i} 5250.2\, \AA{} line, with 4 accumulations at each wavelength. The spectral positions sampled by IMaX were located at $\Delta \lambda$ = $-120$, $-80$, $-40$, $0$, $+40$, $+80$, $+120$\, m\AA{} from the line center of rest wavelength $\lambda_0$ = 5250.2\, \AA{} with an additional one in the red continuum (at $\Delta \lambda$ = $+227$\, m\AA{}).
With an exposure time of $250$ ms for each of the 4 polarization measurements at each of the 8 spectral positions, the total cadence achieved for this mode was $36.5$ s. The 17 minutes time series consists of 28 sets of observations, i.e., images at 8 wavelengths in all 4 Stokes profiles, each image covering $51^{\prime\prime}\times51^{\prime\prime}$ on the solar surface with a platescale of 0\carcsec{}0545 pixel$^{-1}$.

During the flight, images were stabilized, by the use of a tip-tilt mirror controlled by the Correlating-Wave Front Sensor \citep[CWS;][]{berkefeld_wave-front_2011}.
Since the spectral scans were not recorded simultaneously, they were interpolated with respect to time to compensate for the solar evolution time during the acquisition of each observation cycle. In addition, they were corrected for instrumental polarization, and for low-order wavefront aberrations to improve the spatial resolution \citep{martinezpillet_imaging_2011}. At this stage, data are called phase-diversity (PD) reconstructed (level~$2.2$), with a noise level of 7$\times$10$^{-3} I_c$ in Stokes~$V$ ($I_c$ is the Stokes~$I$ continuum intensity) and a spatial resolution of $0.15^{\prime\prime}-0.18^{\prime\prime}$. 

Finally, a very simple approach was taken to correct for the instrumental straylight by subtracting 25\% of the spatial mean Stokes~$I$ profiles from the individual profiles at each of the 8 sampled wavelengths. The Stokes~$Q,U$ and $V$ data were not corrected for straylight since it is considered to be non-polarized (see \cite{riethmuller_new_2017} for more details on the stray-light correction). After stray-light correction, data are called level~$2.3$ data.

\subsection{Inversions}
\label{inversions}
After applying the above corrections, IMaX data were inverted to retrieve the physical parameters needed for our study. The Stokes inversion code SPINOR \citep{frutiger_properties_2000}, which uses the STOPRO routines for the radiative transfer \citep{solanki_photospheric_1987} was employed. The main retrieved parameters of relevance for this study are the magnetic field strength ($B$), the line-of-sight velocity ($V_{\rm LOS}$), and field inclination ($\gamma$). 
A simple one-component atmospheric model was considered, with three optical depth nodes at $\log\tau=-2.5, -0.9, 0$ for the temperature and a height-independent magnetic field vector, line-of-sight velocity and micro-turbulence. More details on the inversion strategy can be found in \cite{solanki_second_2017} and Chapter~\ref{chapter_3}. The line-of-sight velocity maps obtained from the inversions were corrected at each pixel for the wavelength blueshift caused by the Etalon used for the spectral analysis, and a constant velocity (0.6\, km/s) was removed so that the spatially averaged velocity across the FOV is zero.

\begin{figure*}
\centering
\includegraphics[width=\textwidth]{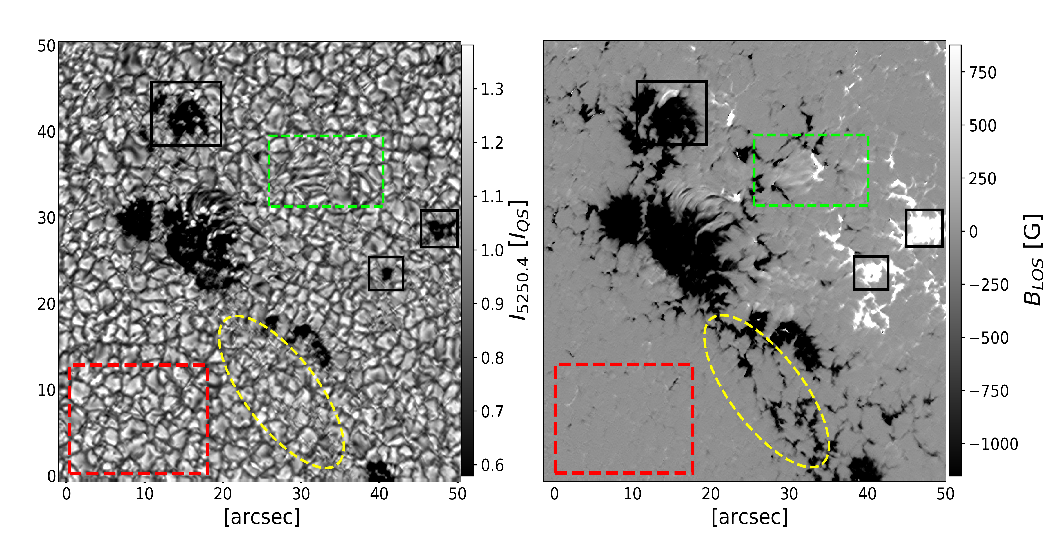}
\caption{\textit{Left panel:}
Continuum contrast at 525 nm. \textit{Right panel:} its co-spatial and co-temporal longitudinal magnetic field map retrieved from the inversions. The black boxes (solid lines) contain small pores characterized by contrast below unity and kG magnetic fields. The dashed green box encloses an area with emerging flux. The dashed yellow ellipse outlines a region of plage composed of magnetic elements embedded in intergranular lanes. The dashed red box contains a quiet-Sun internetwork region with weak fields (average of 20\,G) and mean contrast of unity (by definition).}
\label{fig1}
\end{figure*}

Figure~\ref{fig1} shows an example IMaX continuum map (at $\Delta \lambda=+227$\, m\AA{} from $\lambda_0$, left panel) and the corresponding magnetogram (longitudinal magnetic field, $B_{\rm LOS}$ map) returned by the inversions. Figure~\ref{fig1} displays a photospheric zoo of different structures with various photometric  and magnetic properties: a large pore, which has some penumbral structure attached to it, smaller pores with kG magnetic fields (black boxes), a flux emergence (FE) region (enclosed in the dashed green box), a small QS internetwork region (dashed red box) and plage (partly enclosed by the dashed yellow ellipse). The latter is located mainly in the vicinity of the pores and shows elongated structures or `ribbons' in the intergranular lanes as described in \citet{berger_solar_2004}. Such structures (wherever present in the FOV; i.e., not just in the yellow ellipse) will be analyzed in Section~\ref{turnover}.

\subsection{SuFI}
\label{sufi}

Simultaneously with IMaX, the Sunrise Filter Imager \citep[SuFI;][]{gandorfer_filter_2011} acquired high resolution images sampling the low-mid photosphere (300\,nm, bandwidth of 4.4\,nm) and the lower-chromosphere (397\,nm; core of Ca\,{\sc ii} H, bandwidth of 0.11\,nm) of the same region but with a narrower FOV of $15^{\prime\prime}\times38^{\prime\prime}$. The SuFI FOV covered a part of the leading polarity flux and flux-emergence regions, in addition to small pores and plage regions (see yellow and red boxes overlaid on IMaX FOV in Figs.~\ref{fig2}c and d). The integration time for both 300\,nm and 397\,nm images is 500 ms, with a cadence of 7\,s, and the corresponding  plate scale is 0\carcsec{}02 pixel$^{-1}$.
The spatial correction for wave-front aberrations was done using the Multi-Frame Blind Deconvolution technique \citep[MFBD;][]{van_noort_solar_2005}. The spatial resolution achieved at 397 nm after reconstruction is approximately 70\,km.
The reconstructed data are then corrected for instrumental straylight using the solar limb profiles recorded during the $\sunrise$~I flight. A stray-light modulation transfer function (MTF), with which images are deconvolved, is obtained upon comparing the observed limb profiles to those in the literature at each of the observed wavelength bands (see Chapter~\ref{chapter_5}).

\subsection{Alignment}
\label{alignment}
\begin{landscape}
\begin{figure*}
\centering
\includegraphics[width=\linewidth]{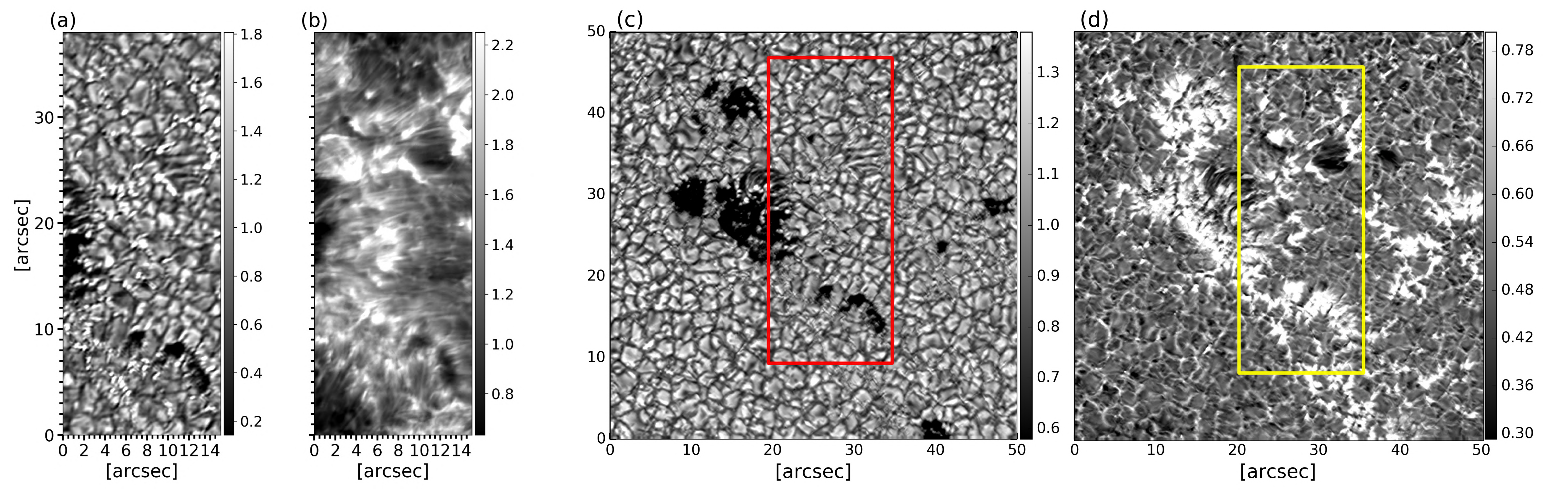}
\caption{The SuFI images coaligned to the IMaX FOV shown in Figure~\ref{fig1}. (a) SuFI 300 nm. (b) SuFI Ca\,{\sc ii} H. (c) IMaX continuum intensity (normalized to the mean quiet-Sun intensity) to which SuFI 300 nm (overlaid in red) is aligned. (d) IMaX line-core intensity derived from the inversions (see Section~\ref{contrast}) and normalized to the local continuum ($I_{+227}$), and to which calcium images (overlaid in yellow) are aligned. The yellow and red boxes are slightly misaligned due to the differential offsets in the SuFI wavelegnth channels. The gray scale is set to cover two times the rms range of each image.} 
\label{fig2}
\end{figure*}
\end{landscape}

Since we are comparing images recorded by SuFI with magnetograms obtained by IMaX, a precise alignment procedure has to be followed.
We apply here the same alignment procedure described in Chapter~\ref{chapter_3} for the quiet-Sun data, mainly resampling SuFI images to the pixel size of IMaX images, and using a cross-correlation technique to compute the horizontal and vertical shifts. 
The SuFI images at 300 nm are aligned with IMaX Stokes~$I$ continuum images, since both of these spectral regions form in the low photosphere and show a normal granulation pattern (although bright points are more prominent at 300 nm, see Figure~\ref{fig2} and \cite{riethmuller_bright_2010}). The Ca\,{\sc ii} H data are aligned with IMaX Stokes~$I$ line-core images.
Both wavelength bands share similar spatial structures, such as the reversed granulation pattern (dark granules, bright intergranular lanes), a result of forming higher in the atmosphere (see Figs.~\ref{fig2}b and d). In the calcium images also other features besides reversed granulation are used (as the latter is best seen outside active regions). These include brightenings associated with strong magnetic fields, although these are somewhat diffuser in Ca\,{\sc ii} H  than in the line core of IMaX due to the expansion of the magnetic field, which forms a low-lying magnetic canopy in plage \citep{jafarzadeh_high-frequency_2017}. Note that the IMaX line-core image shows a reversed granulation pattern only if normalized by the local continuum (see Fig.~\ref{fig2}d). 

\subsection{Contrast}
\label{contrast}
At each pixel, the Stokes~$I$ continuum intensity is determined as the highest intensity of the synthetic best-fit profile retrieved from the inversion of IMaX data. We will consider these values instead of the ones observed at $\Delta \lambda = +227$\,m\AA{}, which can deviate from the true continuum if the neighboring line (Fe\,{\sc i} 5250.6\, \AA{}) influences the 227 m\AA{} intensity (due to a wavelength shift or broadening of the line).

We also refer to the inverted profiles to compute the IMaX line-core intensity as follows: we use the $V_{LOS}$ values returned by the inversions to derive the wavelength shift from the rest wavelength of the Fe\,{\sc i} line ($\lambda_0=5250.2$\, \AA{}) due to the Doppler effect. We add to it the shifts caused by the etalon blueshift and the convective blueshift that were already subtracted from the inverted $V_{LOS}$ maps, this gives the total shift of the line with respect to $\lambda_0$. The line-core intensity is the Stokes~$I$ value at the total computed wavelength shift. This approach was adopted since at pixels with higher magnetic fields, where Zeeman splitting is important, a simple Gaussian fit to Stokes~$I$ profiles was found to underestimate the line-core intensity (see Chapter~\ref{chapter_2}). Moreover, the profiles fitted to IMaX data points by the inversions are far superior to Gaussian fits, since the latter do not take into account the various effects that contribute to the shape of the spectral line (temperature, density, magnetic field, etc..).

The relative intensity (or contrast) at each wavelength band was computed at every pixel by normalizing to the mean quiet-Sun intensity of a very quiet region (of weak magnetic fields). For the IMaX continuum and line core, the QS region is outlined by the dashed red rectangle in both panels of Figure~\ref{fig1}.

Due to the narrower field of view of SuFI, finding a quiet-Sun region in these data is more challenging, particularly as the QS region has to contain a large number of granules. Locating this region in the calcium images is even harder since one should look for dark regions with low fields, but due to the long fibrils found almost everywhere in the images (also protruding into neighbouring quiet regions), most of the low field regions are rather bright in calcium emission, affecting the computed mean quiet-Sun value. 

Luckily, quiet-Sun images are available as flatfields, which are sufficient for the purpose of obtaining the average quiet-Sun intensity. Obviously flatfields were recorded before and after the observations, but not during them. This is an issue as the photon flux at 300 nm varies with elevation angle (and hence time) due to absorption by the Earth's atmosphere. We use this day-to-night cycle variation to derive the mean quiet-Sun intensities at the times of our data acquisition. At 397 nm, this cyclic variation is negligible, so that the average quiet-Sun intensity does not vary with time. In Appendix~\ref{appendix_B} we provide more details on the computation of the mean quiet-Sun intensities in SuFI data.

\subsection{Masking}
\label{masking}
We are mainly interested in studying the relationship between contrast and magnetic field of magnetic elements in plages. These features are distributed everywhere in the FOV around the large and small pores. Hence, pores are masked by using an intensity threshold in the IMaX continuum maps. We compute the mean and standard deviation ($\sigma$) of each continuum image. After applying a low-pass filter consisting of the running mean over $33\times33$ pixels (to smooth the image), we discard pixels with intensities in the smoothed image lower than the mean minus $\sigma$ value. The smoothing greatly reduces the granular contrast. In particular intergranular lanes are much less dark, so that they lie above the threshold of mean minus $\sigma$ in the smoothed image. However, pores remain darker and the smoothing also leads to the exclusion of their immediate surroundings which are darkened by the pores after smoothing. To ignore pixels around the pores, which do not belong to magnetic elements (see Section~\ref{turnover} for a further discussion on the masking technique), the surroundings of pores in addition to the flux-emergence area (black and green boxes in Figure~\ref{fig1}) were masked manually. The flux-emergence region can be identified more easily when examining movies of Stokes~$I$ and $V$ images at the inner wavelength points, i.e.,  at $\Delta \lambda$ = $-40$, $0$, $+40$ m\AA{} from line center.

All unmasked pixels of the 28 images composing the IMaX time series are included in the pixel-by-pixel scatterplots analyzed in the next sections. For studying the contrast at 525 nm vs. $B_{\rm LOS}$, the pixels in the whole IMaX FOV of $51^{\prime\prime}\times51^{\prime\prime}$ are taken, while in the UV wavelengths we consider the pixels in the smaller FOV of $15^{\prime\prime}\times38^{\prime\prime}$ with the corresponding cropped $B_{\rm LOS}$ maps from the inverted IMaX data.
The masks computed from IMaX continuum maps were also used to mask the magnetic features in the corresponding IMaX magnetograms and SuFI images at 300 nm and 397 nm. 

\section{Results}
\label{results}

\subsection{Scatterplots of IMaX continuum and line-core constrasts vs. $B_{\rm LOS}$}
\label{results_imax}
\begin{figure}
\centering
\includegraphics[width=\hsize]{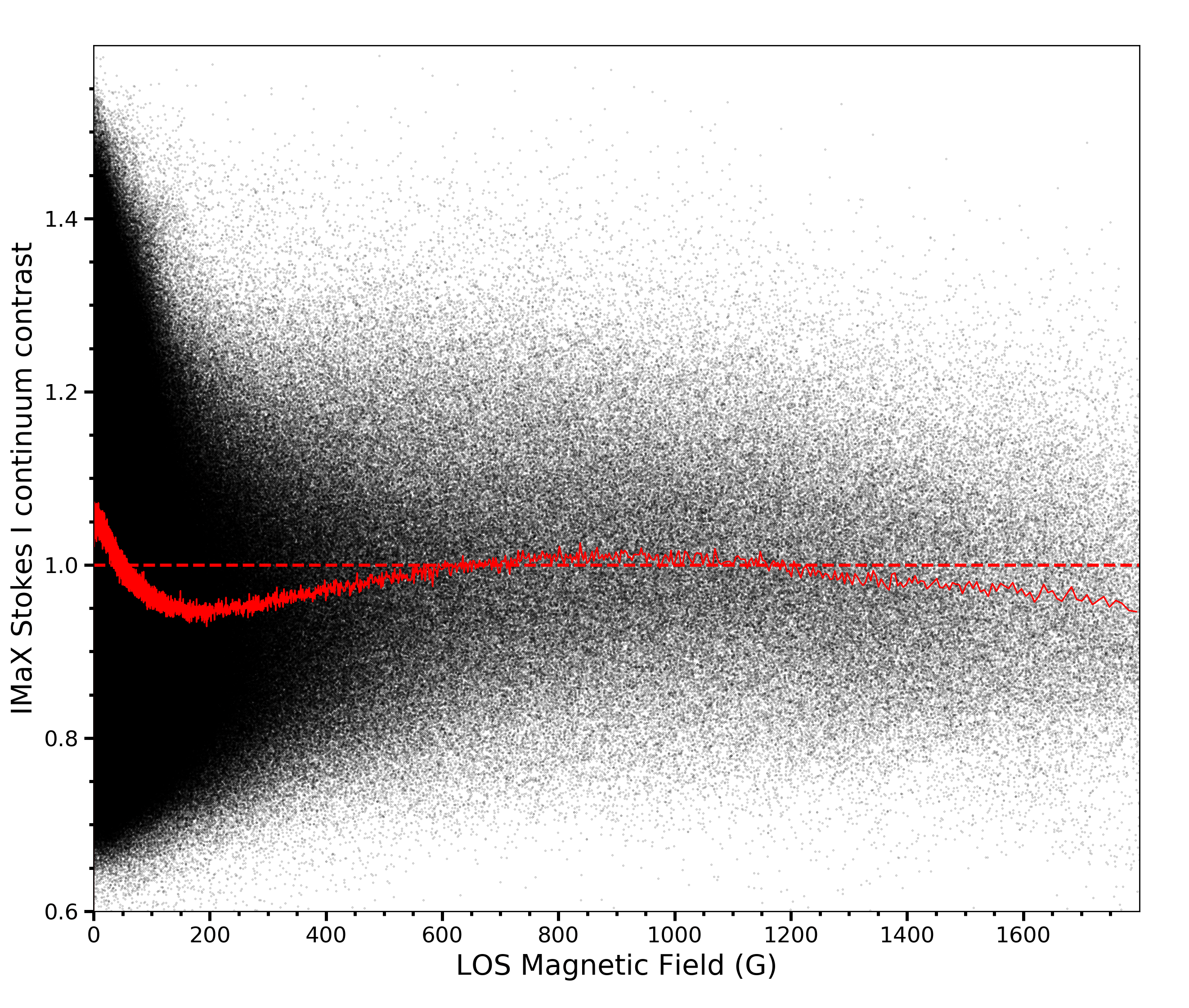}
\caption{Scatterplot of the continuum contrast  of Fe\,{\sc i} 5250.4\, \AA{} vs. the LOS component of the magnetic field, $B_{\rm LOS}$, for the AR plage. Pores and the FE region are excluded. The horizontal dashed red line is the mean QS continuum intensity level (the mean contrast of the pixels in the dashed red box in Figure~\ref{fig1}). The red curve joins the average contrast values in bins of 500 data points each.}
\label{imax_cont_sp}
\end{figure}

\begin{figure}
\centering
 \includegraphics[width=\hsize]{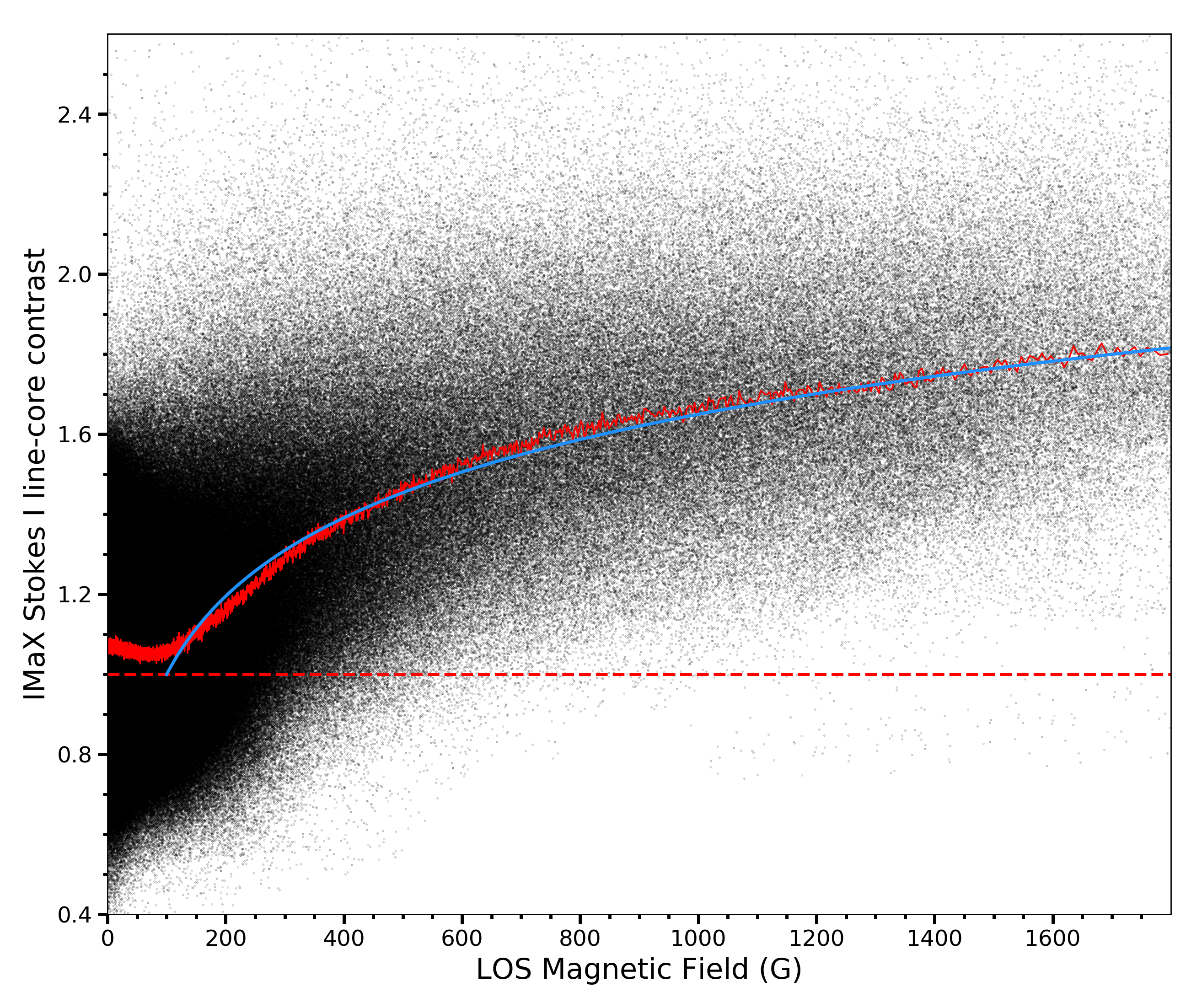}
\caption{Same as Figure~\ref{imax_cont_sp} but for the line core of Fe\,{\sc i} 5250.4\, \AA{}. The blue curve is the logarithmic fit to the averaged line-core contrast values starting at 120\,G.  }
\label{imax_lc_sp}
\end{figure}

After masking the pores, we plot in Figures~\ref{imax_cont_sp} and \ref{imax_lc_sp} pixel-by-pixel scatterplots of the contrast in the IMaX continuum and line-core of Fe\,{\sc i} 5250.2\, \AA{} versus the line-of-sight component of the magnetic field, $B_{\rm LOS}$. To show the trend of data points in the scatterplots, the contrast values are averaged into bins, each containing 500 data points. These binned values are overplotted in red in both figures.\\

The large scatter around $B_{\rm LOS}=0$ in Figure~\ref{imax_cont_sp} is due to the distribution of the magnetic field in the granulation \citep{schnerr_brightness_2011}. At low magnetic field values ($B_{\rm LOS}$ < 150\,G), the contrast in the visible continuum decreases with increasing $B_{\rm LOS}$ since those fields are mostly concentrated in the intergranular lanes by flux expulsion \citep{parker_kinematical_1963}. In an image, these pixels (not shown here) are located in the interiors of bright granules and dark intergranular lanes with weak magnetic fields. For fields above 200\,G, which correspond to magnetic features located in the intergranular lanes, the contrast increases with $B_{\rm LOS}$ to reach positive values (above unity) at 600\,G. The resulting shape of the curve at low fields is referred to as the fishhook \citep{schnerr_brightness_2011}.

Above 600\,G, the average continuum contrast is positive and continues to increase until it reaches a maximum at $B_{max}=855$\,G, then decreases again, dropping below unity at around 1200\,G. The pixels belonging to this range (around the peak and after the turnover) are mainly located in the intergranular lanes, as is typical of small-scale magnetic elements \citep{solanki_small-scale_1993}. 
The numerical values of the longitudinal magnetic field obtained above (at the minimum of the fishhook and at the highest contrast, $B_{max}$) are computed by modelling the scattered data points using one of the non-parametric regression techniques described in Appendix~\ref{appendix_A}.
We prefer the non-parametric regression over a simple parametric model, such as the polynomial fit used in \citet{kobel_continuum_2011}, to find the contrast and $B_{\rm LOS}$ at the peak of their scatterplots, for two reasons. Firstly, the polynomial fit is highly dependent on the lower and upper limits of $B_{\rm LOS}$ within which the averaged contrast values are fitted. Secondly, it is also dependent on the degree of the fitting polynomial.

The IMaX continuum contrast in plage exhibits the same qualitative behaviour as the observations of \citet{frazier_multi-channel_1971}, \citet{narayan_small-scale_2010}, \citet{kobel_continuum_2011}, \citet{kostik_properties_2012}, and the degraded simulations of \citet{danilovic_relation_2013}: a peak in the contrast at intermediate field strengths followed by a turnover at higher fields.  The main difference is that the magnetic field value at which the contrast reaches its maximum (850\,G) is higher than the values of 700\,G and 650\,G reported by \citet{kobel_continuum_2011} at their spatial resolution of 0.3" and by \citet{narayan_small-scale_2010} at 0.15", respectively. These values support the correlation between the position of the peak and the corresponding spatial resolution of the observations, as pointed out by \cite{yeo_intensity_2013}. 

The scatterplot of the line-core contrast versus $B_{\rm LOS}$ in Figure~\ref{imax_lc_sp} shows a weaker fishhook shape compared to the visible continuum. The monotonic increase of the line-core contrast with $B_{\rm LOS}$ was not seen in similar studies where line-core data were available \citep[]{title_differences_1992, yeo_intensity_2013}. Their scatterplots of the line-core contrast dropped at high magnetogram signals, even after masking out the pores.
Following paper I we model this relationship with a logarithmic function of the form $I(B) = \beta + \alpha.log B$, which is overplotted in blue in Figure~\ref{imax_lc_sp}. The IMaX continuum contrast could not be modeled with a logarithmic function given the non-monotonic increasing shape. This behaviour will be analysed in Section~\ref{turnover}.

\subsection{Scatterplots of UV contrast vs. $B_{\rm LOS}$}
\label{results_sufi}
\begin{figure}
\centering
\includegraphics[scale=0.1]{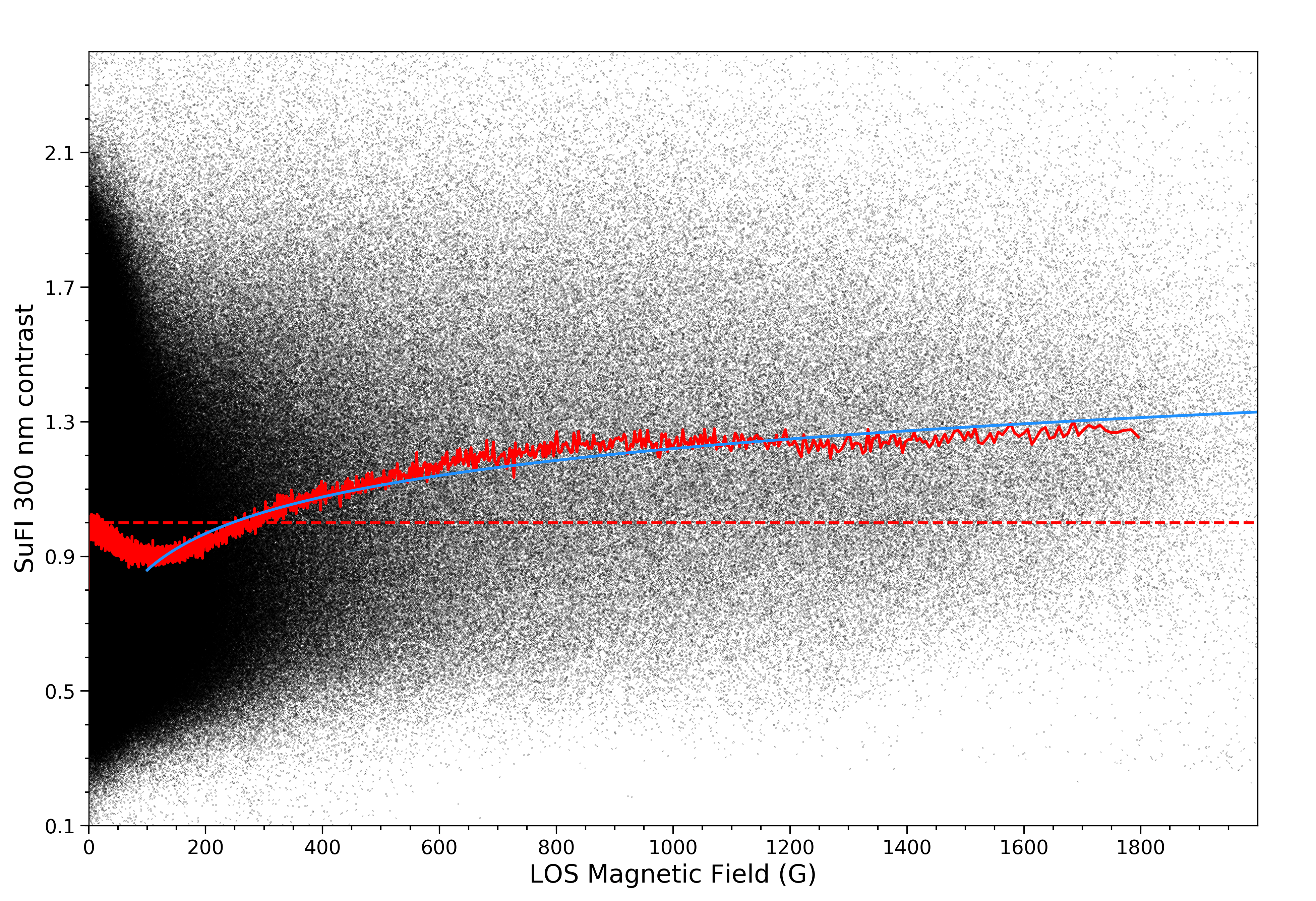}
\caption{Scatterplot of the SuFI intensity contrast at 300\,nm vs. $B_{\rm LOS}$. The red curve joins the averaged contrast values inside each bin, the blue curve is the logarithmic fit to the red curve starting from 100\,G.}
\label{sufi_300_sp}
\end{figure}

\begin{figure}
\centering
\includegraphics[scale=0.1]{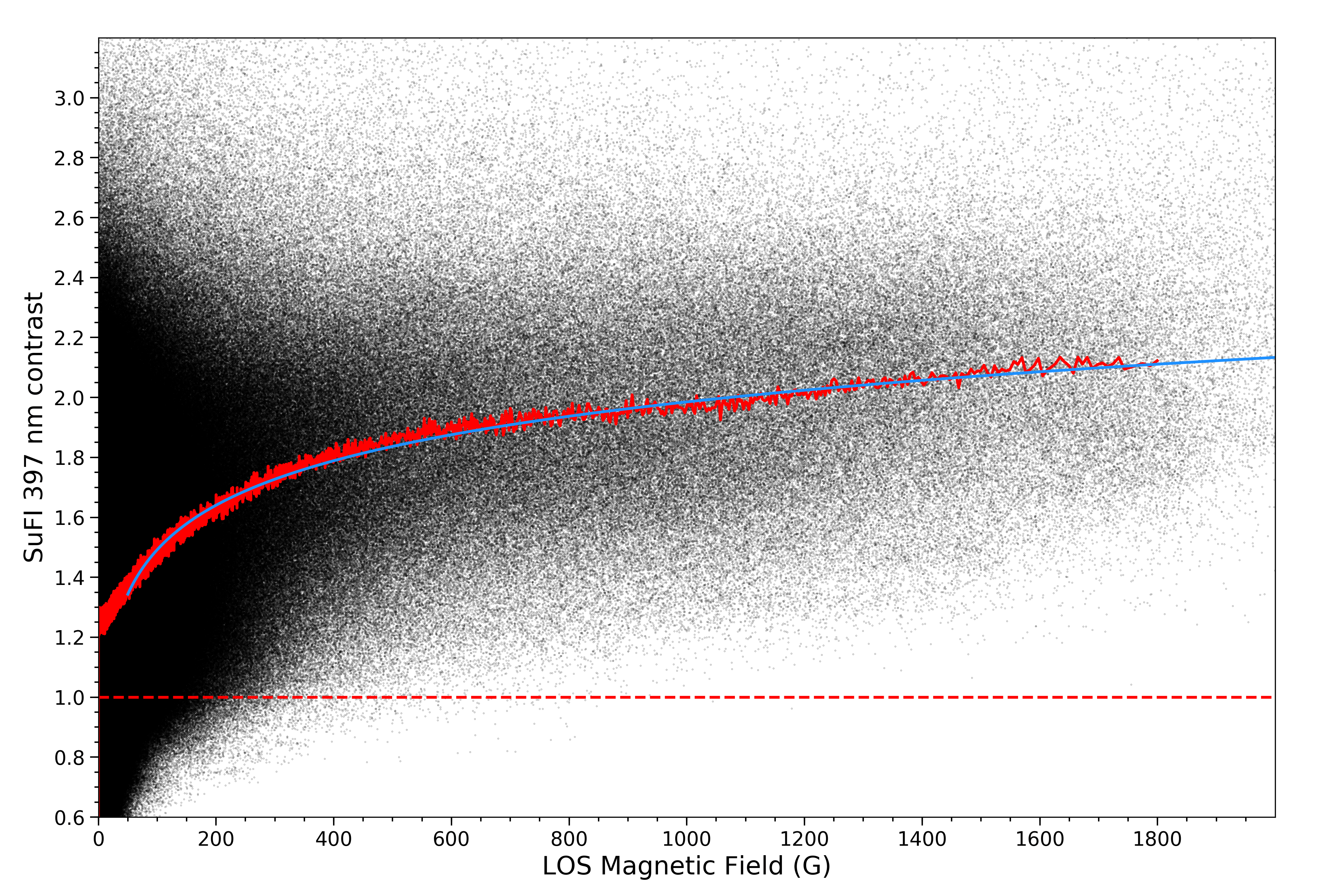}
\caption{Same as Figure~\ref{sufi_300_sp} but for the line core of Ca\,{\sc ii} H. The blue curve is the logarithmic fit to the red curve starting from 50\,G.}
\label{sufi_397_sp}
\end{figure}

Scatterplots of the contrast at 300 and 397\,nm against $B_{\rm LOS}$ are shown in Figures~\ref{sufi_300_sp} and \ref{sufi_397_sp}, respectively. Contrast values are averaged following the method described in Section~\ref{results_imax}.

At 300\,nm, the fishhook shape is still visible due to the relatively low formation height of this wavelength band of around 50\,km, as computed by  \cite{jafarzadeh_high-frequency_2017}. After an initial decrease with $B_{\rm LOS}$, the contrast at 300\,nm reaches its minimum  at around 100\,G. Beyond that point, the contrast increases until it saturates at higher field strengths.

The contribution function of the Ca\,{\sc ii} H bandpass, computed from an atmospheric model corresponding to AR plages and showing a mean formation height of 550\,km \citep{jafarzadeh_slender_2017}, implies that this line has a contribution from the photosphere, but also from the low-to-mid chromosphere.
The fishhook shape in the scatterplot of Figure~\ref{sufi_397_sp} is gone, due to the absence of granulation at the larger formation height of this line compared to the 300\,nm bandpass and the 525\,nm continuum. The features are on average brighter than the QS even at very weak fields, due to the presence of bright fibrils that are prominent even in weak magnetic field regions. Weak field regions that are dark can be seen mainly in the upper middle and lower left regions of the calcium images, while fibrils are found nearly everywhere (see Fig.~\ref{fig2}). The contrast then increases with $B_{\rm LOS}$ until it saturates at higher fields. Pixels in this region of the scatterplot belong to plages as identified in the IMaX continuum images. 

In the past, a power-law function has been used to describe the relation between Ca\,{\sc ii} H brightness and magnetic flux density for network and IN features both, within limited fields of view \cite[]
{rezaei_relation_2007,loukitcheva_relationship_2009}  
and for magnetic features in full-disk observations \citep[]{ortiz_how_2005, chatzistergos_analysis_2017}. Here, however, we follow the analysis in Chapter~\ref{chapter_3}, where we found that a logarithmic function provides a superior representation to the data at the high spatial resolution reached by $\sunrise$. The fit parameters are shown in Table~\ref{tab1}.

\begin{table}
\caption{Parameters of logarithmic fits to the contrast in the core of Fe\,{\sc i} line at 5250.2\, \AA{} (starting from 120\,G) and for 300 nm and 397 nm (starting from 100\,G) vs. $B_{\rm LOS}$. }
\label{tab1}
\centering
\begin{tabular}{c c c c}
\hline \hline

Wavelength band & $\alpha$ & $\beta$ & $\chi^2$ \\
\hline
IMaX line core & 0.673$\pm$ 0.002 & -0.365$\pm$ 0.004 & 9.232 \\
\hline
300 & 0.353 $\pm$ 0.002&0.156$\pm$ 0.005&9.213\\
\hline
397 & 0.512$\pm$ 0.001&0.456$\pm$ 0.003&5.55\\
\hline
\hline
\end{tabular}

\end{table}

\subsection{Comparison between wavelengths}
\label{all_wav}

\begin{figure*}
\centering
\includegraphics[width=\hsize]{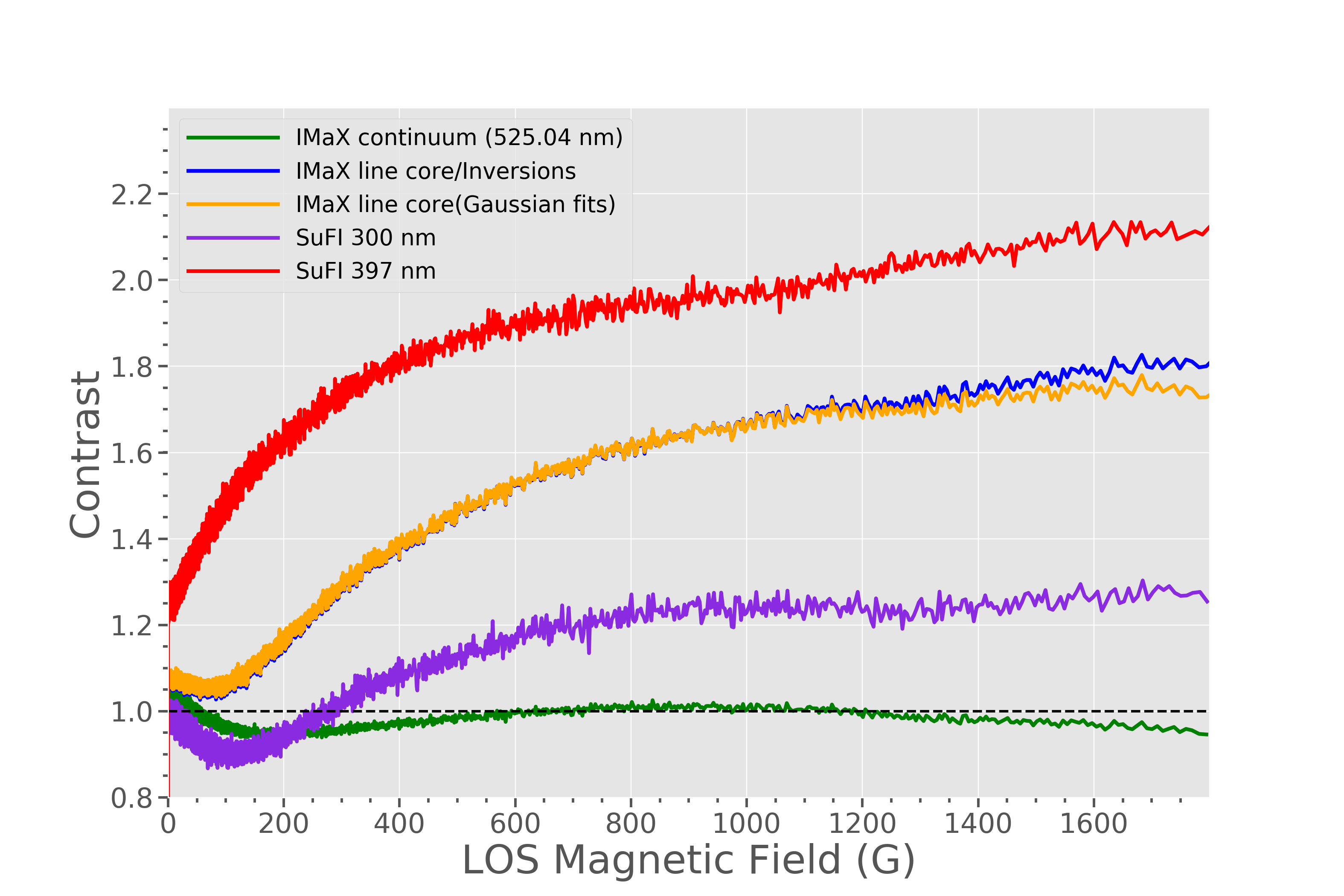}
\caption{Intensity contrasts of plage vs. $B_{\rm LOS}$ in the wavelengths considered in the current study (see the legend in the figure). Plotted are the contrast values (plotted in red in Figures~\ref{imax_cont_sp}, \ref{imax_lc_sp}, \ref{sufi_300_sp} and \ref{sufi_397_sp}). For comparison, also plotted is the binned line-core contrast at 525.02\,nm derived from Gaussian fits to individual Stokes~$I$ profiles.}
\label{all}
\end{figure*}

\begin{figure*}[ht!]
\centering
\hspace*{-2.5cm}\includegraphics[scale=0.45]{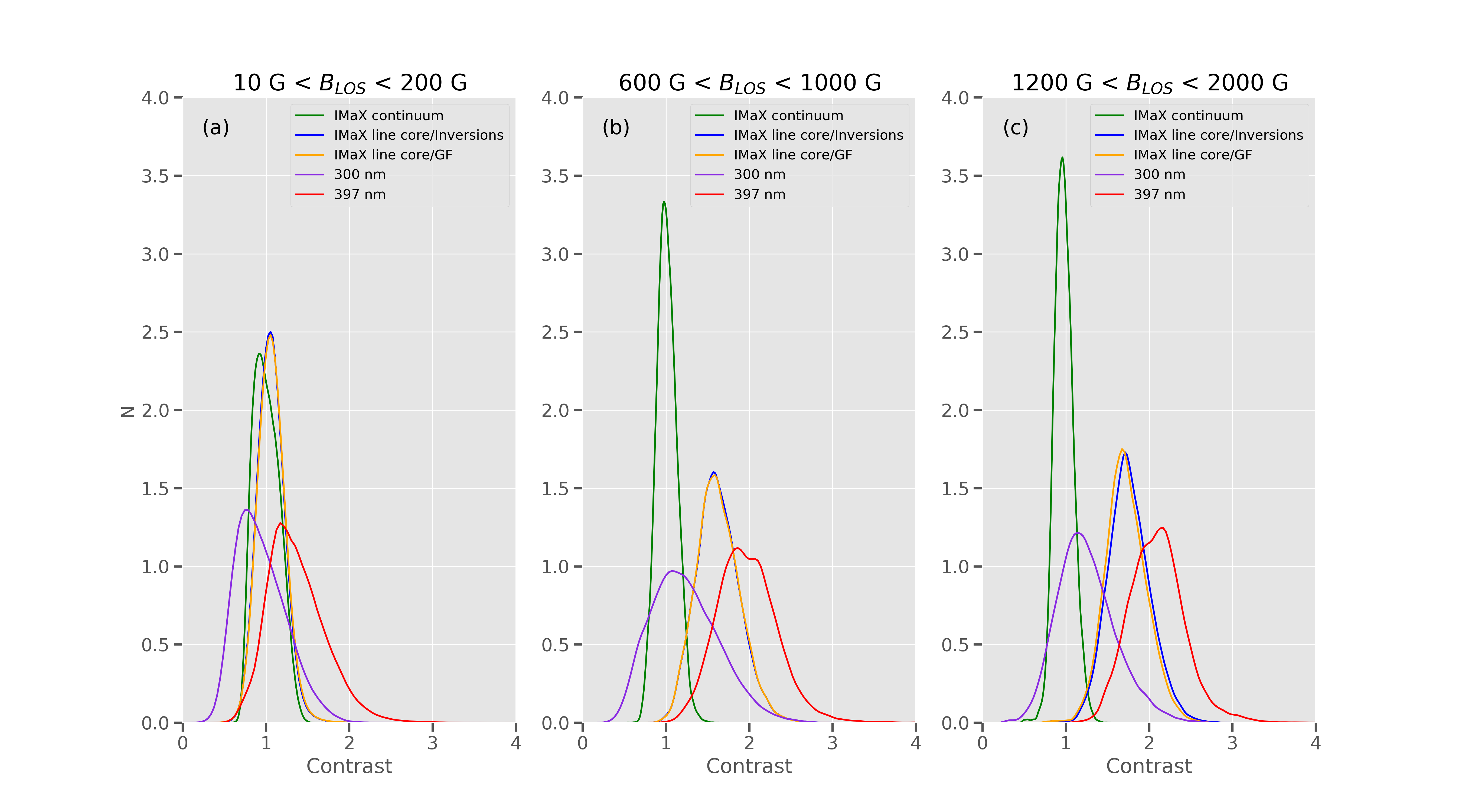}
\caption{Contrast histograms of pixels belonging to the common FOV of SuFI and IMaX after masking out pores. Colors correspond to different wavelengths, and panels correspond to different ranges of $B_{\rm LOS}$ values (see text for a detailed explanation). The histograms are normalized such that the integral over each is equal to one.  }
\label{hist}
\end{figure*}

We plot in Figure~\ref{all} the binned contrast curves for the wavelengths in the UV at 300 and 397\,nm, and in the visible continuum at 525.04\,nm plus line core of Fe\,{\sc i} at 525.02\,nm derived from the inversions as explained in Section~\ref{contrast}. We plot for comparison the curve corresponding to the line-core contrasts derived from the Gaussian fits to individual IMaX Stokes~$I$ profiles. 
Histograms of the measured contrasts at the studied wavelengths are displayed in Figure~\ref{hist}. These histograms are for the common FOV ($13^{\prime\prime}\times34^{\prime\prime}$) of IMaX and SuFI data sets for a one to one correspondence of the contrasts. All the pixels of the time series except for those belonging to pores   contributed to these histograms. They are plotted for different ranges of $B_{\rm LOS}$, these ranges correspond to magnetic field values where the scatterplot of the contrast at 525.04 nm vs. $B_{\rm LOS}$ shows different behaviour: the weak field region (10\,G < $B_{\rm LOS}$ < 200\,G), the range where features have an average contrast above unity (600 G < $B_{\rm LOS}$ < 1000\,G), and the range where features start to be darker than the mean quiet-Sun intensity ($B_{\rm LOS}$ > 1200\,G).

For weak fields (Fig.~\ref{hist}a), 525\,nm continuum (green) and 300 nm (violet) have averaged contrasts < 1. Both bands exhibit lower contrasts than the rest of the wavelengths at all field strengths, with the contrast at 300 nm becoming larger than 525 nm with increasing $B_{\rm LOS}$ (Figs.~\ref{hist}b and c).

The higher contrast measured in the UV (at 300 and 397 nm) compared to the visible continuum at 525.04\,nm is explained by the stronger response of the Planck function to the temperature variations at shorter wavelengths plus their greater height of formation in the case of the Ca\,{\sc ii} H line core . For the Fe\,{\sc i} line core, the reason for the larger contrast, is this line's great temperature sensitivity and its greater formation height. For all values of $B_{\rm LOS}$, Ca\,{\sc ii} H (red) exhibits the highest contrasts.

Figures~\ref{all} and \ref{hist} also show the difference in the computed line-core intensity from the inversions (blue) and the Gaussian fits (orange), which is more notable at higher fields (Fig.~\ref{hist}c). As explained in Section~\ref{contrast}, this difference is due to the underestimation of the line-core intensity at higher fields if derived from the Gaussian fits. The curves obtained by the two methods agree rather well, except at the very highest field strengths, suggesting that the Zeeman splitting plays a minor role at the spectral resolution of IMaX.

\subsection{Why does the continuum contrast display a turnover?}
\label{turnover}

\begin{figure*}[ht!]
\centering
\includegraphics[scale=.6]{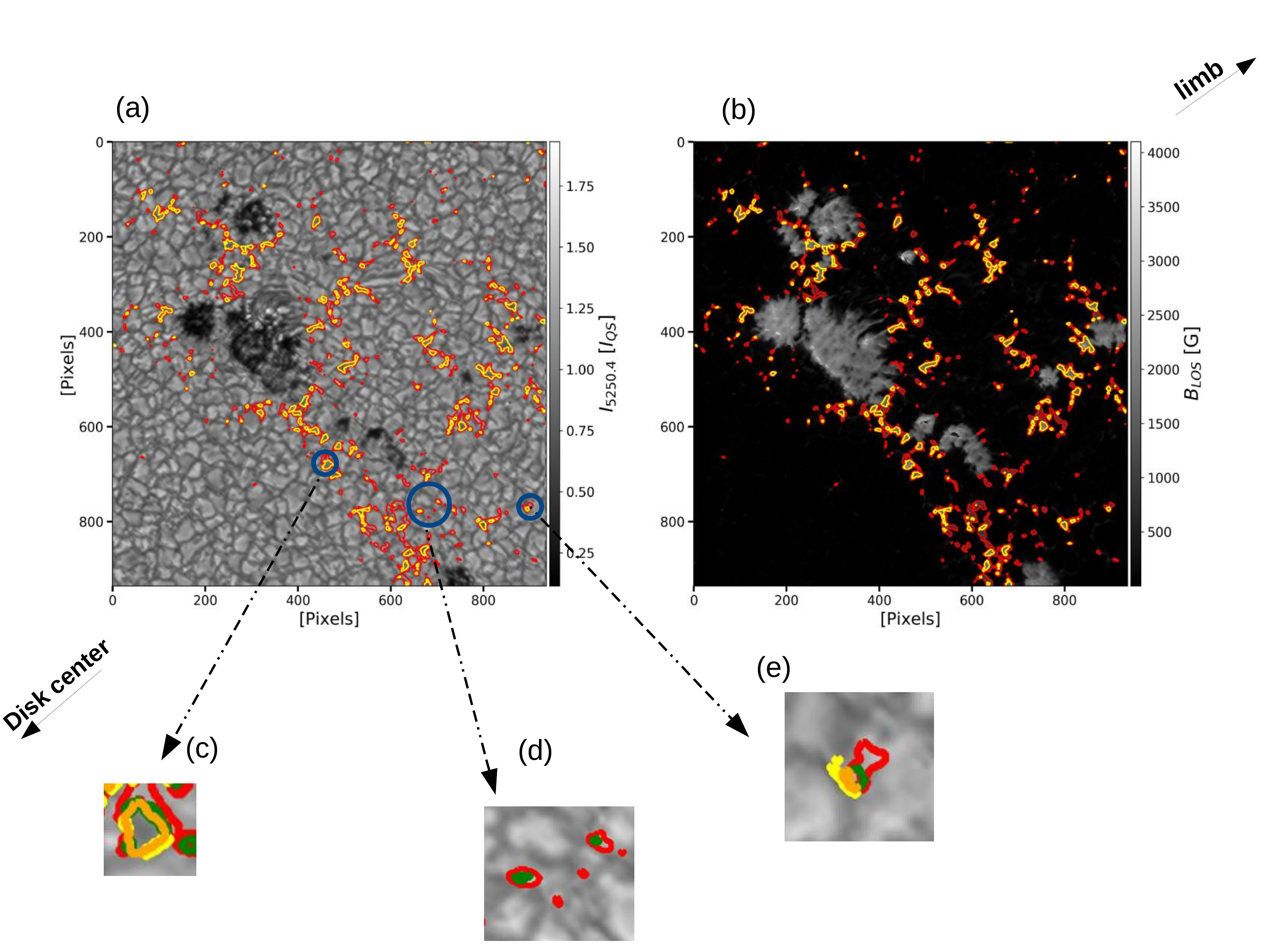}
\caption{(a), (b): IMaX data for the same time as in Figure~\ref{fig1}, now with the absolute value of $B_{\rm LOS}$ plotted in the right panel. The red contours enclose the `peak' pixels with positive averaged contrast in the continuum of 525 nm and 600\,G < $B_{\rm LOS}$ < 1000\,G, while the yellow contours enclose the `turnover' pixels with negative averaged contrasts and $B_{\rm LOS}$ larger than 1200\,G. (c), (d), (e): Blowups of three magnetic features of different sizes. Red and green contours enclose the bright (contrast > 1) parts of both families of pixels, while orange and yellow contours enclose the dark (contrast < 1) parts.
}
\label{contour}
\end{figure*}

In this subsection we will try to understand why the averaged continuum contrast of plage features at 525\,nm peaks at intermediate field strengths and decreases at higher fields when observed by IMaX near disk center. 
\subsubsection{Treatment of magnetic signals around pores} 
The dependence of the contrast on $B_{\rm LOS}$ near disk center is believed to vary according to the treatment of pixels around pores and sunspots \citep[]{kobel_continuum_2011,yeo_intensity_2013}. Therefore, care has to be taken when removing pores since plages are often located close to their boundaries, and inclusion of pixels which might belong to pores could affect the contrast-$B_{\rm LOS}$ relationship in the plage areas.
We test this by extending our masks around the pores, described in Section~\ref{masking}, to exclude more of their surroundings. We then again produce the scatterplot to see if the shape described above persists or not. The shape of the scatterplot (not shown here) and in particular the position of the contrast peak were hardly affected. 

\subsubsection{Masking technique}
The criterion we use here to exclude pores from our data is based solely on the IMaX continuum intensity (see Sec.~\ref{masking}).
We also try a mask based on a contrast-$B_{\rm LOS}$ threshold, such as that employed by \cite{kobel_continuum_2011}. The threshold used for creating their masks (contrast < 0.85 and $B_{\rm LOS}$ > 900\,G) aimed to remove the inner dark and magnetic parts of pores. In addition, they extended their masks around pores to exclude the pixels darkened by telescope diffraction. However, in our data this threshold corresponds to micropores that are part of the plage, and also to the dark magnetic pixels found in the ribbon-like features which result from their possible inclination with respect to the solar normal (see Sec.~\ref{morphology}).
Thus, adopting the same contrast-magnetic field threshold in our data will remove the pixels contributing to the downturn in the bottom right part of our scatterplot in Figure~\ref{imax_cont_sp} in addition to a few brighter pixels if those masks are extended, since micropores in our data tend to have brighter edges (see Figs.~\ref{profiles}a and b). This points to the dependence of the photospheric continuum contrast-$B_{\rm LOS}$ relationship on the criterion used to distinguish between magnetic elements and pores. This also implies that a careful examination of the magnetic/photometric distribution of small-scale magnetic elements has to be done before adopting any masking criterion, which is only possible if features are well resolved, as is the case with IMaX (see Sects.~\ref{morphology} and \ref{QS_2009_2013}).  

\subsubsection{Spatial resolution}
\label{morphology}
In lower spatial resolution studies \citep[]{title_differences_1992,topka_properties_1992,lawrence_contrast_1993} using the SVST, the continuum contrast monotonically decreased with magnetogram signal (see Introduction). The inability to resolve magnetic regions, or the fact that they are partly blended with the non-magnetic surroundings, primarily with the dark intergranular lanes, were considered the reason for the negative contrasts at higher field strengths. 
This was confirmed by \cite{danilovic_relation_2013} using MHD snapshots of a plage region. After degrading the simulations to the spatial resolution of Hinode/SP, their scatterplots turned from a monotonic relationship at the original MuRAM resolution to a peak and a turnover, in agreement with the findings of \cite{kobel_continuum_2011}.

\cite{danilovic_relation_2013} showed that smearing isolated bright magnetic structures embedded in a dark, nearly field-free environment shifts the data points with high contrast and $B_{\rm LOS}$ at the original resolution to a region of the scatterplot with lower contrasts and intermediate field strengths. The accumulation of data points there gives rise to a peak upon averaging the contrast values. Bigger and less bright magnetic structures are only slightly affected by the smearing and remain dark with higher field strengths, causing the turnover in the right side of the scatterplot. 
This interpretation of the effect of finite spatial resolution is valid for regions with low average magnetic field such as the quiet-Sun network, where the magnetic features have little internal structure. Thus, it works well for the data described in Chapter~\ref{chapter_3}. However, The plage features in our data are larger than those found in the quiet Sun, so that they should be affected less by spatial smearing, as test have confirmed. Instead the magnetic features in the plage show more complex photometric and magnetic distributions than the simple picture of magnetic bright points surrounded by dark field-free lanes. 

To isolate these structures, we locate the pixels with $B_{\rm LOS}$ > 1200\,G and binned contrast < 1 as seen in Figure~\ref{imax_cont_sp}. We call them the `turnover' pixels. We also locate the pixels centered at $\pm$ 200\,G from $B_{\rm max}$ (i.e., 600\,G < $B_{\rm LOS}$ < 1000\,G) with an averaged contrast > 1, and call them the `peak' pixels. Figures~\ref{contour}a and b show the contours enclosing both families of pixels. We find that together they form magnetic structures of different sizes and appearences embedded in intergranular lanes and spread mainly around the large and small pores, with the turnover pixels (yellow contours) typically being surrounded by the peak pixels (red contours). Small structures have the shape of circular bright points, while slightly larger ones resemble the so called `striation' or `ribbon' in the G-band observations of \cite{berger_solar_2004}. Figs.~\ref{contour}c, d, and e are blowups of three features, picked to illustrate three distinctive types: a micropore (Fig.~\ref{contour}c), bright points (Fig.~\ref{contour}d), and a ribbon-like feature (Fig.~\ref{contour}e). The red and yellow contours now enclose the bright (contrast > 1) and dark (contrast < 1) parts of the peak pixels, while the green and orange contours enclose the bright and dark parts of the turnover pixels. Clearly, the smallest features, the bright points in Fig.~\ref{contour}d, contain only bright pixels, while the larger features display an increasing fraction of dark pixels as their area increases. This behaviour is found to be typical. 

\begin{figure*}[ht!]
\centering
\hspace*{-1cm}\includegraphics[scale=0.43]{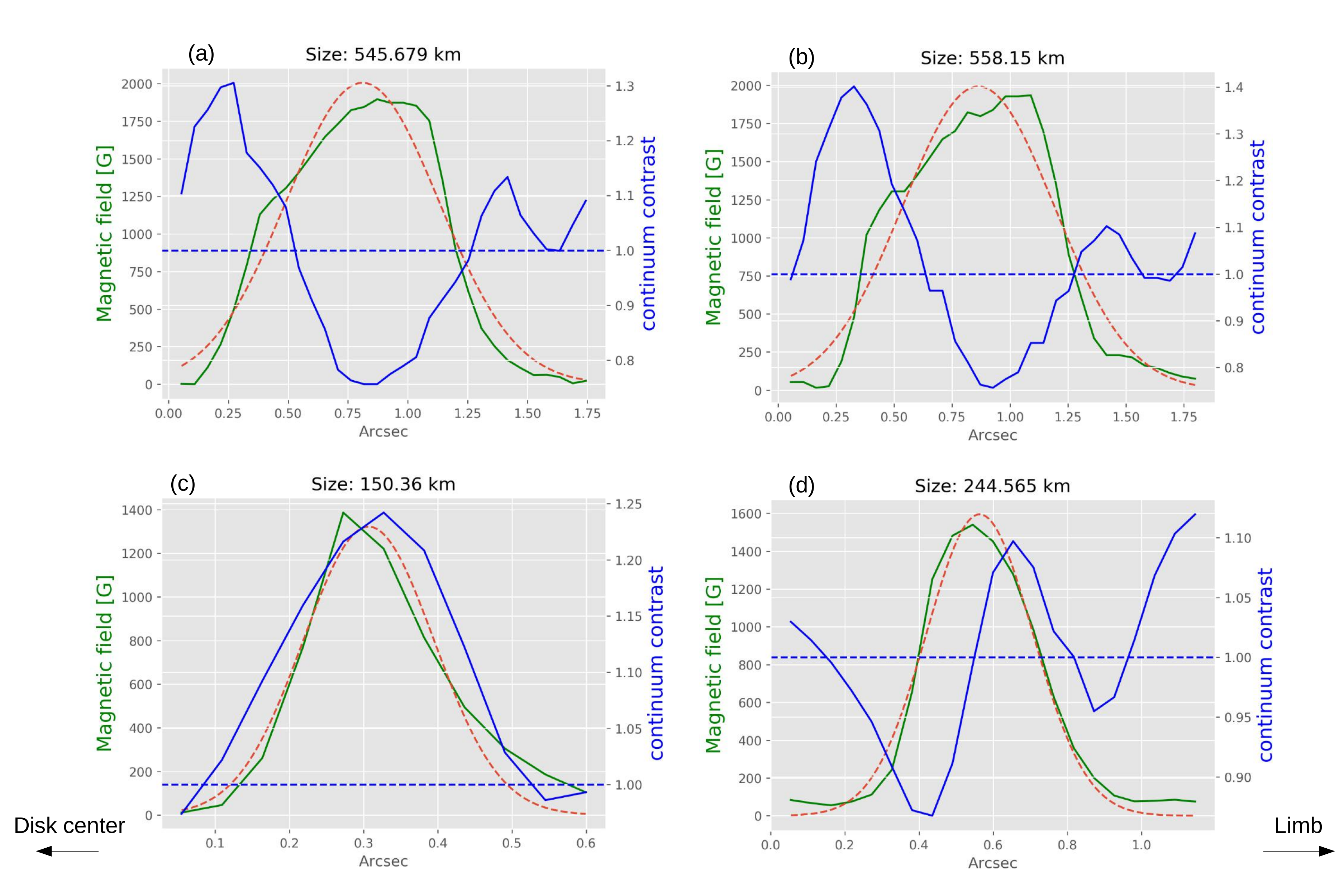}
\caption{Profiles of $B_{\rm LOS}$ (green curve) and continuum intensity contrast at 525\,nm (blue curve) along cuts through 4 different magnetic structures in an example IMaX image (see text for details). The red dashed curves are the Gaussian fits to the $B_{\rm LOS}$ profiles. The FWHM of these Gaussian fits are given at the top of each frame. The horizontal dashed blue line is where the contrast is equal to unity. The limb is to the right of the figure, while disk center is to the left.}
\label{profiles}
\end{figure*}

We plot in Figure~\ref{profiles} horizontal cuts through 4 features belonging to the three types of magnetic elements shown in Figure~\ref{contour}. Displayed are the continuum intensity at 525 nm and the LOS magnetic field. The cut is perpendicular to the limb, with the nearest limb being to the right of the figure, and disk center to the left.

In all 4 features the LOS component of the magnetic field peaks in their centers and diminishes towards their edges. The continuum contrast at 525\,nm, however, shows three distinctive patterns according to the sizes of the features (given above each frame): for large features (horizontal extent of 500\,km-600\,km) it is > 1 at the edges and < 1 at the center where $B_{\rm LOS}$ peaks (Figs.~\ref{profiles}a and b). For small bright points (100\,km-200\,km), it is > 1 in their centers and the brightness peak coincides with the position where $B_{\rm LOS}$ peaks (Fig.~\ref{profiles}c). In the elongated features with intermediate sizes (200\,km-400\,km), the continuum contrast is > 1 in the limb side of the feature, and < 1 in the disk center side, with a neutral contrast in the center (Fig.~\ref{profiles}d). The latter pattern is seen in most of the features in the  IMaX data.
The feature size is determined from a Gaussian fit (dashed red curves in Fig.~\ref{profiles}) to the $B_{\rm LOS}$ profile.

The profiles of the continuum intensity in Figs.~\ref{profiles}a and b resemble the symmetric double-humped profiles at disk center in both G-band observations and numerical simulations \citep{berger_solar_2004, steiner_recent_2005}. While the profile in Fig.~\ref{profiles}d resembles the asymmetric pattern of contrast in magnetic elements near the limb \citep{hirzberger_solar_2005,steiner_recent_2005}. Given that our observations are not carried out exactly at disk center ($\mu$ = 0.93), it could be that the shape of the intensity profiles seen across the mid-sized features are due to the inclined line of sight (the hot wall effect). But it could also be a result of inclined magnetic fields with respect to the solar normal, as proposed in \cite{keller_origin_2004}.

Variations of the intensity contrast and magnetic field distributions across IMaX features with different sizes explain the scatter we see in the plot of IMaX continuum contrast vs. $B_{\rm LOS}$ in the range 600\,G-2000\,G, which corresponds to magnetic elements belonging to plage. We conclude that the scatter around the mean QS level is real and is not a consequence of poor spatial resolution or noise in the data. This point is strengthened later when inspecting the brightness-magnetic field relationship in quiet-Sun areas located within the FOV of the AR observations (Section~\ref{QS_2009_2013}).

\subsection{Comparison with 2009 quiet-Sun data}
\label{QS_2009_AR}

In this section we present a qualitative and quantitative comparison of the contrast-magnetic field relationship near disk center in AR plage with that of quiet Sun, as obtained in Chapter~\ref{chapter_3}. This comparison is obviously restricted to the wavelengths that were recorded during both flights (IMaX continuum and line core, 300\,nm continuum, and Ca\,{\sc ii} H line core at 397\,nm). The reason we don't compare with the quiet regions extracted from the active-regions scans is that the quiet Sun is located mainly outside the SuFI FOV. In the next section we will analyse the IMaX properties of quiet-Sun features from 2013 scans, and compare them to the 2009 results to test for consistency. 



\begin{figure*}[ht!]

\centering
\hspace*{-1cm}\includegraphics[scale=0.11]{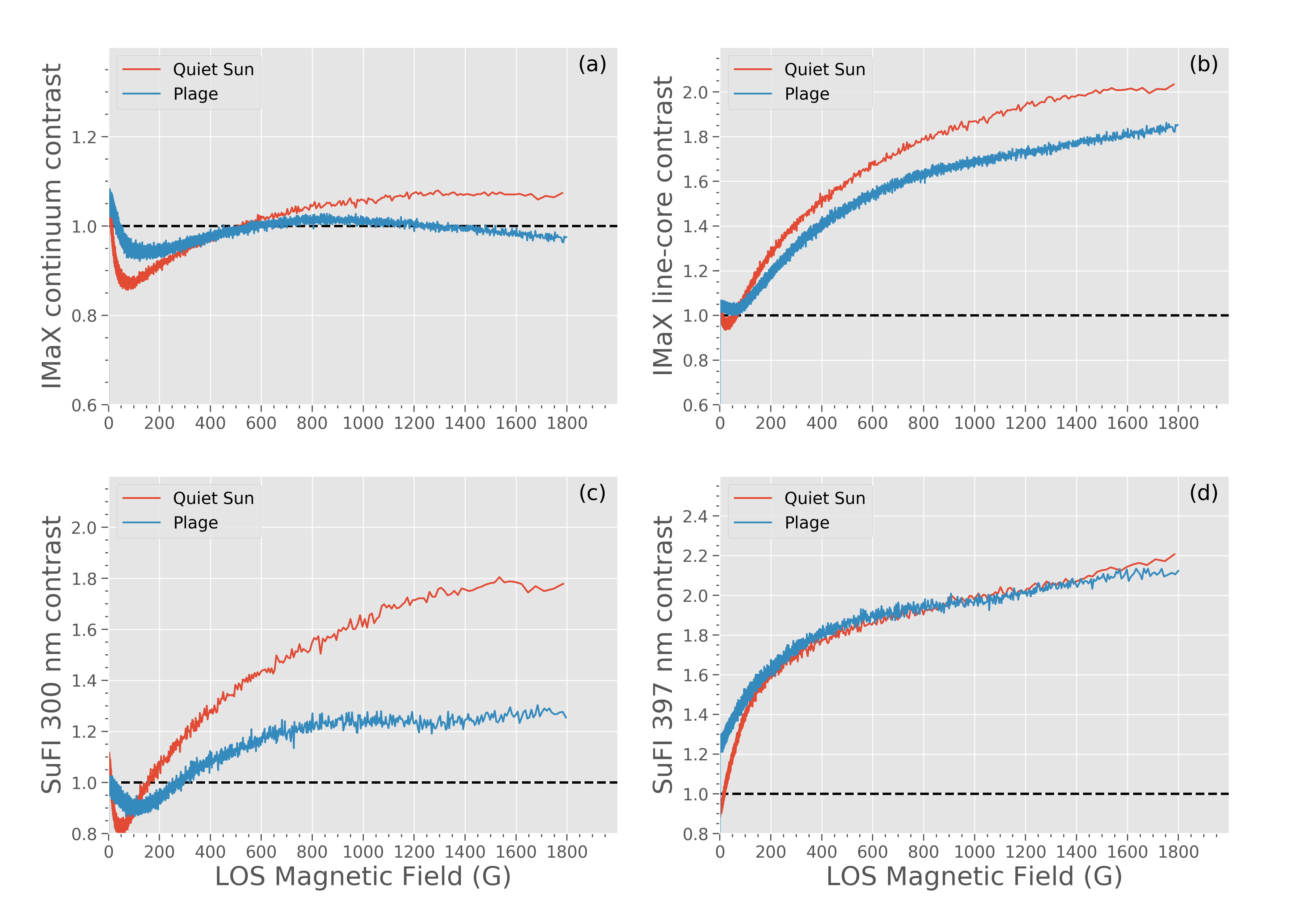}
\caption{Averaged contrast of (a) IMaX continuum, (b) IMaX line core, (c) SuFI 300 nm, (d) SuFI 397 nm against $B_{\rm LOS}$ for the quiet Sun observed by $\sunrise$ in 2009 (blue curves) and for plage (red curves) in 2013.}
\label{qs_plage}
\end{figure*}

In Figure~\ref{qs_plage}, we overplot the averaged contrasts for the quiet Sun (red curves) observed with $\sunrise$~I and analysed in Chapter~\ref{chapter_3} and for the AR plage (blue curves) analysed in the current chapter, of the IMaX continuum (panel (a)) and line core (panel (b)) at 525 nm, the continuum of 300 nm (panel (c)), and the line core of Ca\,{\sc ii} H at 397 nm (panel (d)) against the photospheric magnetic field. The black dashed line corresponds to a contrast of 1.

In all 4 panels, the plage curves lie generally higher than the QS ones at low $B_{\rm LOS}$, but are lower at larger $B_{\rm LOS}$, with the latter effect being marginal for Ca\,{\sc ii} H. Very close to $B_{\rm LOS}=0$, however, the QS curves are higher again for SuFI 300\,nm and for IMaX continuum.
For SuFI 300\,nm and IMaX continuum the larger variation of the contrast curve in QS at small $B_{\rm LOS}$ is likely due to the brighter granules and darker intergranular lanes than in the abnormal granulation present in the plage region \citep{narayan_small-scale_2010}. This `fishhook shape' \citep{schnerr_brightness_2011} is extended to larger $B_{\rm LOS}$ in plage due to the filling of intergranular lanes (and to some extent the sides of granules) with weak field. The greater brightness of the IMaX continuum in the QS is in accordance with the findings of \cite{title_differences_1992}, \cite{lawrence_contrast_1993} and \cite{kobel_continuum_2011}. Whereas the transition to higher QS contrast in magnetic features occurs at 500\,G for the IMaX continuum, it occurs already at around 100\,G for the IMaX line core and the SuFI 300\,nm channel. In addition, plage magnetic elements at 300\,nm reach a contrast >1 at around 250\,G, while QS magnetic elements make this transition already at 150\,G. The brightness in the Ca\,{\sc ii} H line core in both, quiet-Sun network and plage is similar for $B_{\rm LOS}$ above 1000\,G. The slightly higher QS contrast above approximately 1400\,G is based on too few QS points to be truly significant.
 
In the literature, the higher contrasts at larger $B_{\rm LOS}$ values reached by the visible continuum in the QS than in AR plages was attributed to the higher efficiency of convective energy transport in the quiet Sun. Going from QS to AR, i.e., with increasing magnetic flux averaged over the region, convection is suppressed in the surroundings of the stronger and larger magnetic features (ARs) leading to less efficient heating from the convective walls, and therefore less brightening with respect to their counterparts in the quiet Sun \citep[]{vogler_effect_2005,morinaga_suppression_2008,kobel_continuum_2012,criscuoli_comparison_2013,riethmuller_comparison_2014}.
The same reasoning may also explain the significantly lower contrast in plage at larger $B_{\rm LOS}$ seen at 300\,nm, given the similar formation height to the IMaX continuum.

In the cores of spectral lines (Figs.~\ref{qs_plage}b and d), this difference is less significant compared to lower layers. This result is in accordance with the temperature models of network and plage flux tubes derived from the FTS observations by \cite{solanki_solar_1996}, cf.~\cite{solanki_photospheric_1987} and \cite{solanki_continuum_1992}. In these works, flux tubes in the network were found to be hotter than in plage, especially in lower regions of the photosphere (see Chapter~\ref{chapter_6}).
\subsection{Comparison between 2009 and 2013 quiet-Sun data}
\label{QS_2009_2013}

As mentioned in Section~\ref{inversions}, we can identify some quiet-Sun areas in IMaX scans (see for example red box in Figure~\ref{fig1}). We also delimit a couple of quiet-Sun areas in the upper right side of the FOV. We produce contrast-$B_{\rm LOS}$ scatterplots of the pixels embedded in these boxes to test if we can reproduce the relationship obtained from observations of the solar surface when the Sun was mostly quiet, i.e., $\sunrise$ observations from 2009. For brevity, we refer to QS data from 2009 as QS-2009 and QS boxes from scans recorded in 2013 as QS-2013. Since the quiet-Sun boxes in 2013 data are outside the SuFI field of view, we restrict this comparison to the IMaX continuum and line-core contrasts vs. $B_{\rm LOS}$.

\begin{figure*}[ht!]
\centering
\hspace*{-1.05cm}\includegraphics[scale=0.11]{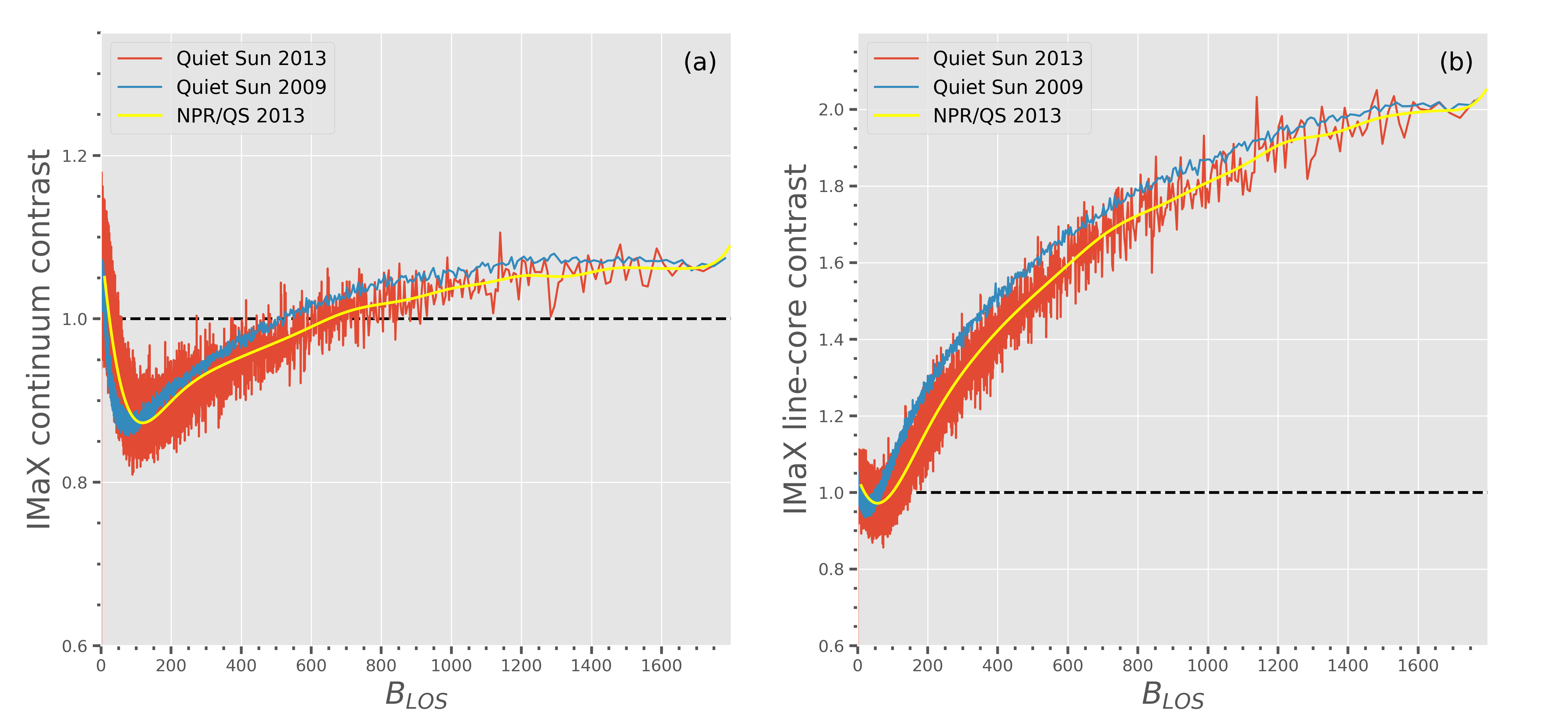}
\caption{IMaX continuum (panel a) and line-core (panel b) contrasts vs. $B_{\rm LOS}$. The red curves are the binned contrasts of quiet-Sun areas extracted from the 2013 ARs scans, and the blue curves are taken from Paper~I for quiet-Sun observations in 2009. The yellow curves are the result of non-parametric smoothing applied to the scattered data points of quiet-Sun pixels of 2013. The black dashed line is the quiet-Sun level where contrast is equal to unity.}
\label{qs_2009_2013}
\end{figure*}

In Figure \ref{qs_2009_2013} we show the scatterplots of the binned IMaX continuum (panel (a)) and line-core (panel (b)) contrasts, in blue for QS-2009 and in red for QS-2013, vs. $B_{\rm LOS}$. The black dashed horizontal line is the mean QS intensity level. Due to the smaller number of data points in QS-2013, the corresponding curves of averaged contrasts look noisier. Hence, we use non-parametric smoothing on the scattered data points of QS-2013 to make the trend easy for the eye to follow. The smoothed curves are overplotted in yellow in Figure~\ref{qs_2009_2013}.

According to Fig.~\ref{qs_2009_2013}, contrast-$B_{\rm LOS}$ relationships in both, IMaX continuum and line core agree qualitatively in the quiet-Sun regions from both flights.
In Chapter~\ref{chapter_3} we came to the conclusion that magnetic fields in the quiet-Sun network at disk center are resolved by IMaX since the contrast in the visible continuum saturated at higher $B_{\rm LOS}$. The fact that the quiet-Sun features embedded in AR scans exhibit the same behaviour implies that such fields are also resolved in $\sunrise$ II observations of active regions.
Not surprisingly, the behaviour of the averaged contrast variation with $B_{\rm LOS}$ of magnetic elements in the quiet Sun is independent of solar magnetic activity level \citep{ortiz_intensity_2006}, so that any difference between the two flights would reflect instrumental differences.

Quantitatively, in panel (a) of Figure~\ref{qs_2009_2013}, the minumum contrast of the fishhook in QS-2013 occurs at around 150\,G, a value higer than the 80\,G obtained for QS-2009. In addition, the features in QS-2013 start to become brighter than the mean QS at 650\,G, i.e., at fields higher than the 500\,G for QS-2009. \\ 
The averaged contrast of magnetic features in the QS-2009 data is higher than in the QS-2013 in both wavelength bands. Considering the basic picture of the `hot wall' and that observations in 2013 were done further from disk center ($\mu=0.93$) than in 2009 ($\mu = 0.98$), we expect that magnetic elements are brighter (in the visible continuum) with increasing heliocentric angle, but the opposite behaviour is seen in Figure~\ref{qs_2009_2013}a. However, the difference between the two data sets is small. The line core of IMaX (Figure~\ref{qs_2009_2013}b) displays the same behavior, which in this case is consistent with the findings of \cite{yeo_intensity_2013}.
The differences in the contrast could be interpeted as the effect of the higher average magnetic flux density in AR scans of 2013 data (50\,G) compared to the quiet Sun observed by $\sunrise$ in 2009 (30\,G). The convection is more hampered during the period of higher activity (2013), which leads to less effective heating inside the magnetic elements from their surroundings, and therefore less radiative energy emitted by these elements.

\section{Discussion and conclusions}
\label{conclusion}
We have analyzed the properties of solar AR plage near disk center, as observed by $\sunrise$ during its second flight on June 2013. In particular, we have qualitatively and quantitatively described how the brightness in the visible and UV of the different components of magnetic features depends on the longitudinal field strength, $B_{\rm LOS}$. The latter is computed among other physical parameters from SPINOR inversions of IMaX Stokes profiles corrected for wavefront aberrations and straylight.

At the spatial resolution of 0.15$^{\prime\prime}$ achieved by IMaX, the scatterplot of the continuum contrast at 525 nm vs. $B_{\rm LOS}$ (Fig.~\ref{imax_cont_sp}) shows that the contrast peaks at around 850\,G beyond which it decreases with increasing $B_{\rm LOS}$, a behaviour that agrees qualitatively with the findings of \cite{berger_contrast_2007}, \cite{narayan_small-scale_2010} and \cite{kobel_continuum_2011}. The findings of the latter citation were interpreted to be an effect of the finite spatial resolution of the data \citep{danilovic_relation_2013}.

After locating the pixels in the $B_{\rm LOS}$ range of 600\,G-2000\,G, we found that these pixels belong to the same features, which are well resolved by $\sunrise$ (see Fig.~\ref{contour}). The $B_{\rm LOS}$ is always high in their centers and decreases towards their edges, while the contrast displays a more complex behaviour that can be different from one feature to another, depending on its size. In most of the features, however, the contrast is >1 in the limb direction, <1 towards disk center, and neutral near their centers (Fig.~\ref{profiles}d), whereas for larger size features, the centers are dark and brighten towards their edges (Figs.~\ref{profiles}a, b). These complex profiles are attributed to the fact that plage features are large and they are located slightly off-disk-center in the analyzed data \citep{steiner_recent_2005}.

Consequently, one and the same plage magnetic element provides pixels in different parts of the contrast vs. $B_{\rm LOS}$ plot, including bright/dark intermediate field pixels, and dark strong-field pixels. As a result, the turnover of brightness with $B_{\rm LOS}$ is not an effect of image smearing induced by the limited spatial resolution but rather of the internal morphology of individual plage magnetic features. This conclusion is strengthened by the saturation of the continuun contrast at higher $B_{\rm LOS}$ in quiet-Sun magnetic features (Chapter~\ref{chapter_3} and Sect.~\ref{QS_2009_2013}), which are smaller than the features in the plage, so that their contrasts are more likely to suffer from insufficient spatial resolution. The network features also show much less internal structure.

We think that the simulations used so far to interpret the downturn of the contrast in observations as an artifact of the poor spatial resolution \citep{rohrbein_is_2011, danilovic_relation_2013}, lack the larger magnetic features found in strong plage that dominate their continuum contrasts. By introducing higher average magnetic flux ($\approx$ 400\,G) and by extending the vertical depth, we expect the boxes to be deep enough to allow convection to produce larger features. In addition, the current simulations for plage regions are analyzed for lines-of-sight parallel to the solar normal (i.e., $\mu=1$). Therefore, we propose that for such simulations to reproduce our data, bigger and deeper simulation boxes are needed, in addition to performing spectral line synthesis with such simulations for lines-of-sight that are inclined to the local normal ($\mu=0.93$).

In the UV at 397\,nm (core of the chromospheric Ca\,{\sc ii} H line) and in the core of the IMaX line, the averaged contrasts of plage features increase with increasing $B_{\rm LOS}$. Given the larger formation heights of these wavelengths compared to the IMaX continuum, this implies that the brightness at larger atmospheric heights is independent of the feature size, while lower in the photosphere, the size of the magnetic features plays a decisive role in determining their contrast \citep{solanki_small-scale_2001}.  

Comparison of the contrast-$B_{\rm LOS}$ relationship in plage with quiet-Sun observations presented in Chapter~\ref{chapter_3} confirms the findings of earlier studies: at the photospheric level and for larger field strengths, the contrast in the quiet Sun network is higher than active region plages \citep{title_differences_1992, morinaga_suppression_2008, kobel_continuum_2011} due to the less efficient convective energy transport in the latter \citep{morinaga_suppression_2008, kobel_continuum_2012,criscuoli_comparison_2013}. In the cores of spectral lines, this difference is found to be smaller. At the atmospheric layers sampled by these wavelengths (upper photosphere and lower chromosphere), radiative heating from convection becomes less efficient in determining the contrast of these features, and other processes dominate in transporting and dissipating energy (e.g. oscillations and waves).

\chapter{Stray light determination of Sunrise data and application to CLV of quiet-Sun intensity contrasts}

\label{chapter_5}

\textbf{Abstract}\\

We investigate the center-to-limb variation of quiet-Sun intensity contrasts, based on observations at high spatial resolution, covering multiple wavelengths in the near UV (300 nm, 312 nm, 388 nm) and visible (525 nm) spectral range. The data have been obtained during the first science flight of the $\sunrise$ balloon-borne solar observatory in 2009. To estimate and correct for the effect of the point spread function on the observed intensity contrasts, we combine an analysis of solar limb profiles with phase diversity wavefront sensing. The results are compared to theoretical predictions inferred from magnetohydrodynamic simulations and radiative transfer computations. In the near UV we find a significant discrepancy between our observed intensity contrasts and the theoretical predictions. In particular, the theoretical center-to-limb variations of the intensity contrast in the near UV reach a local maximum at $\mu$ values between 0.5 and 0.6, whereas our observations show a monotonic decrease from disk center towards the limb. The applicability of the point spread function derived from the analysis of limb images obtained during the 2009 flight of $\sunrise$ is tested on the active region data observed by $\sunrise$ during its second science flight in 2013.      

\newpage

\section{Introduction} \label{sec:intro}

Observations of spatial intensity fluctuations in the quiet-Sun (QS) continuum provide important information on the temperature distribution in the lower solar photosphere and hence gives insight into convective energy transfer, as well as into the energetics of small-scale magnetic features. The sub-arcsecond spatial resolution of a 1m telescope allows proper sampling of the intensity differences between the bright granules and the darker intergranular lanes, which together represent the key structures of solar surface convection. 

In the near-UV (NUV) the photospheric thermal radiation is more sensitive to temperature fluctuations than at longer wavelengths. As an example, a temperature difference of 100 K results in a relative change of the Planck function~\footnote{for a typical photospheric temperature of about 5800 K} of 14\% at 300 nm, compared to only 8\% at 525 nm.  

The root-mean-square of the spatial continuum intensity fluctuations in the QS, normalized to the average intensity, is a meaningful statistical quantity to describe the intensity fluctuations. This quantitiy (defined explicitly in Section~2.5 of Chapter~\ref{chapter_2}), which we denote in the following by \emph{QS intensity contrast}, has been extensively studied in the past. For a recent overview of such studies, limited to disk center observations, and to the visible spectral range, see e.g. \cite{sanchez_cuberes_center--limb_2000}, \cite{danilovic_intensity_2008}, \cite{mathew_stray_2009}, and references therein. \cite{hirzberger_quiet-sun_2010} finally extended the disk center contrast studies to the NUV. 

An evaluation of the center-to-limb variation (CLV) of QS intensity contrasts provides additional information on the height dependence of the temperature inhomogeneities in the photosphere and on the undulation of the height of the solar surface. Such observations have also been described by many authors, starting with the results from US and Soviet balloon-borne stratospheric solar observatories \citep{edmonds_statistical_1962, pravdyuk_distribution_1974}, and followed by studies relying on ground-based observations \citep{keil_new_1977, schmidt_center_1979, druesne_speckle_1989, wilken_speckle_1997, sanchez_almeida_physical_2000, sanchez_cuberes_centre--limb_2003}, on observations with the Spektrostratoskop balloon-borne telescope \citep{schmidt_center_1979, durrant_balloon-borne_1983} and on observations from space \citep{zakharov_diagnostic_2006, wedemeyer-bohm_continuum_2009, afram_intensity_2011, yeo_point_2014}. These CLV studies are limited to the visible and near-infrared spectral range with a reduced photospheric temperature sensitivity, with the exception of \cite{afram_intensity_2011} who present, as their shortest observed wavelength, Hinode/SOT results obtained at the CN band head (388 nm).

This work can be considered as a follow-up of the study by \cite{hirzberger_performance_2011}, who have discussed disk-center quiet-Sun intensity contrasts in the NUV as observed with the $\sunrise$ Filter Imager \citep[SuFI; see][]{gandorfer_filter_2011} onboard the $\sunrise$ balloon-borne solar observatory \citep{barthol_sunrise_2011, solanki_sunrise:_2010, solanki_second_2017}. Like this earlier work, the present study is based on observations obtained during the first $\sunrise$ science flight, but the evaluation of quiet-Sun intensity contrasts is extended to a study of their CLV, using both the SuFI NUV observations as well as observations in the visible spectral range, obtained with the second scientific instrument onboard \textsc{Sunrise}, the Imaging Magnetograph eXperiment \citep[IMaX; see][]{martinezpillet_imaging_2011}. 

In Section \ref{sec:psf} we describe the relevant data reduction steps, in particular the correction for the effect of the extended point spread function (PSF). In Section~\ref{SL_up} the stray light model developed in Section~\ref{sec:psf} is tested on the active region data observed by $\sunrise$ during its second flight on June 2013. In Section~\ref{conclusions} we summarize and discuss our results.

\section{Estimate of an extended point spread function} \label{sec:psf}

\begin{figure}
\centering
\includegraphics[scale=0.3]{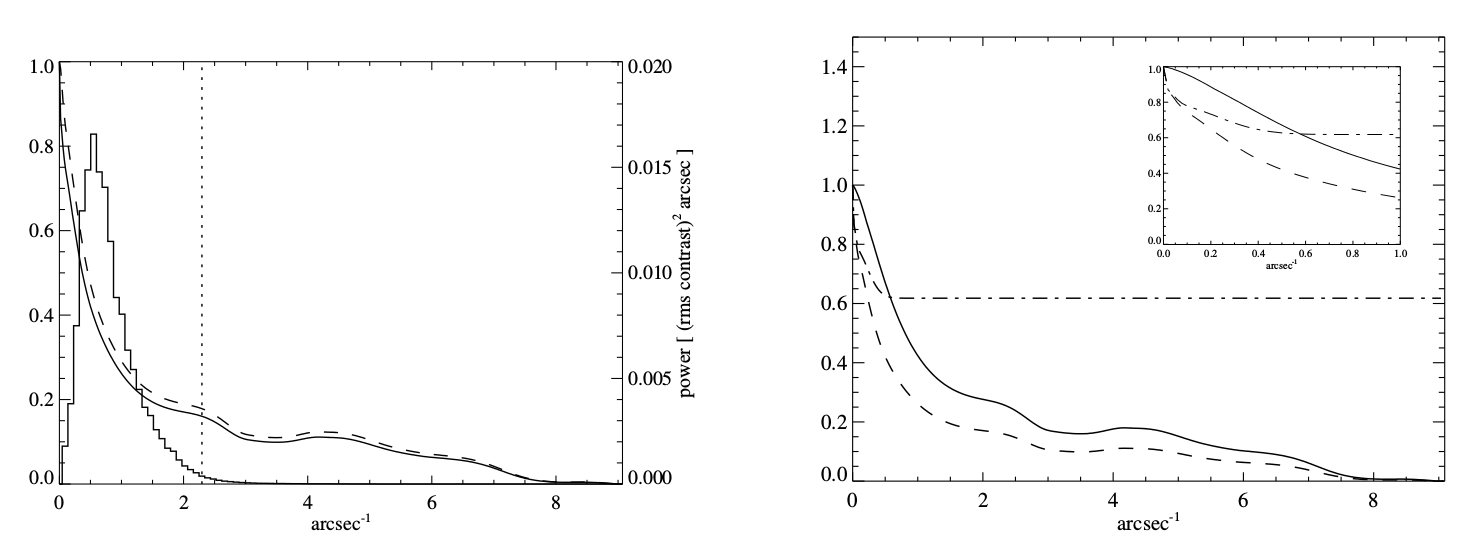}
\caption{\textbf{Left panel:} Spatial frequencies relevant to the QS intensity contrast. The histogram plot shows the azimuthally averaged power spectrum of a typical IMaX disk-center image at reduction level-2 (i.e., PD restored). The vertical dotted line denotes the upper limit of the frequency range encompassing 99\% of the total power. The solid and dashed curves represent the extended MTF based on the local and large-FOV stray light assumptions respectively. \textbf{Right panel:} Comparison of different MTFs using the example of IMaX: MTF derived from PD wavefront sensing (solid line), extended MTF including the 2 Gaussian stray light terms (dashed line), and reduced MTF (dash-dotted line) used for the actual stray light deconvolution of PD restored images. The inlay shows a blow-up for the region 0 to 1 arcsec$^{-1}$. See text for a detailed description of the MTFs.\label{fig:mtf}}
\end{figure}

The QS intensity contrast is significantly affected by image degradation, which reduces the amplitude of the power spectrum of the observed solar images in the frequency range dominated by solar granulation, between 0 and about 2.5 arcsec$^{-1}$ (cf. Fig. \ref{fig:mtf}). In the case of $\sunrise$ the relevant contributions to image degradation in this frequency range are optical aberrations and residual instrument jitter~\footnote{We assume here that the effect of the Earth's atmosphere on image degradation can be neglected at the $\sunrise$ flight altitude}. For both SuFI and IMaX the core of the PSF (on spatial scales of order 0\farcs1) is estimated from PD wavefront sensing (cf. section \ref{sec:psf_pd}). An estimate of the wings of the PSF (on spatial scales of order arcsec), corresponding to higher spatial frequencies in the pupil wavefront is beyond the PD sensitivity range. As mentioned in the introduction of this thesis (Section~2.2 in Chapter~\ref{chapter_2}), planetary transits or eclipses are usually used to estimate the stray light contribution to the total PSF. But since none occured during the $\sunrise$ observations, we will take recourse to the analysis of solar limb images recorded during the flight. In short, we process the recorded 2D limb images to derive the 1D solar limb profile. We develop a model that can quantify the contribution of the different degradation sources that led to the broadening of the observed limb profile compared to a theoretical profile. An expression of the PSF is then derived based on the best-fit parameters. In the following, we denote the PSF including both contributions, PD core and stray light wings, as \emph{extended PSF}.

\subsection{Phase diversity wavefront sensing} \label{sec:psf_pd}

For a description of the phase diversity reconstruction method on SuFI data, see \cite{hirzberger_quiet-sun_2010, hirzberger_performance_2011} and Section~2.1 of Chapter~\ref{chapter_2}. The correction process on the spectropolarimetric data of IMaX is described in \cite{martinezpillet_imaging_2011}.

\subsection{Processing of solar limb images} \label{process}
\begin{figure}
\centering
\hspace*{-0.8cm}\includegraphics[scale=0.6]{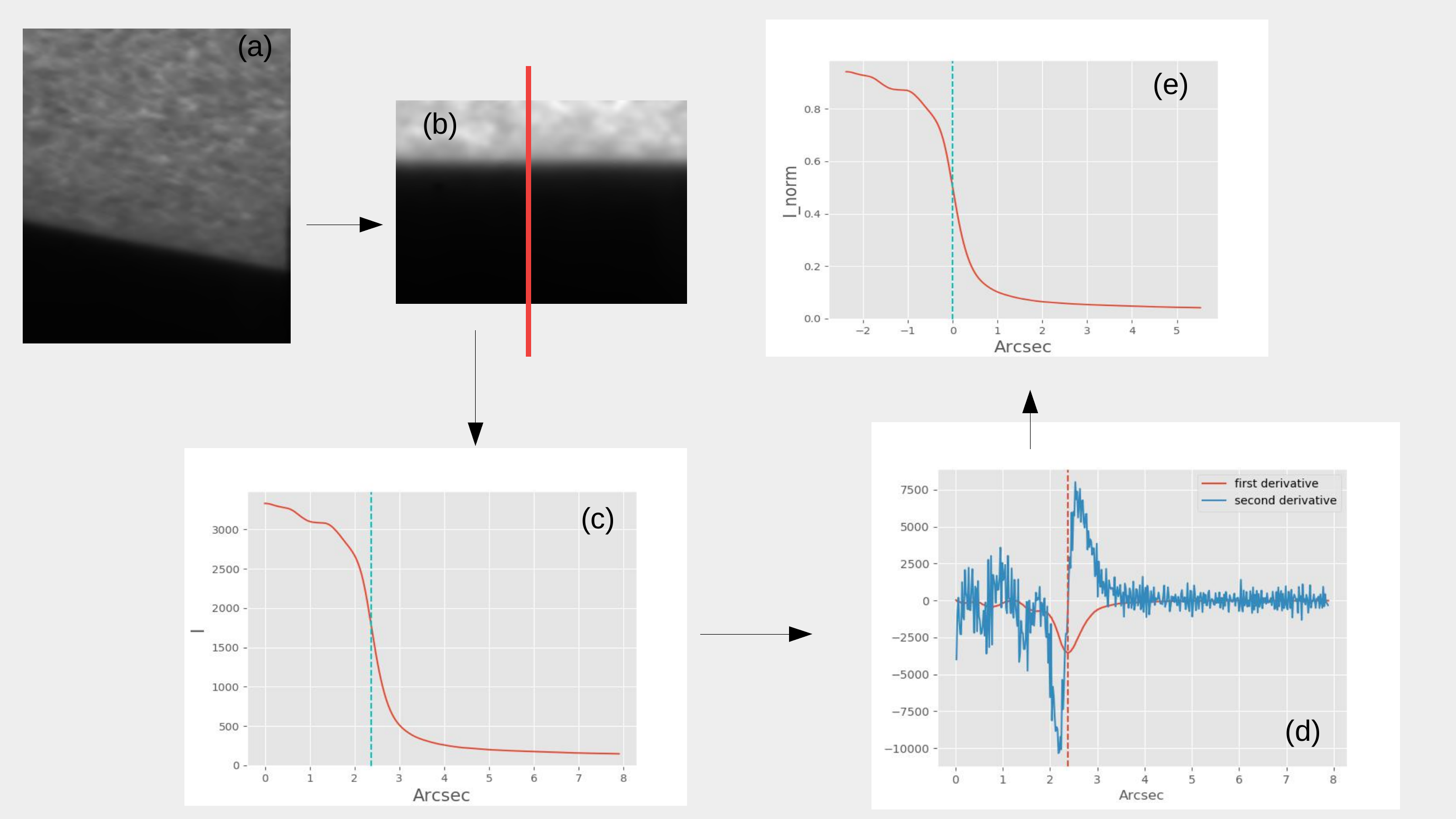}
\caption{A short description of the different limb processing steps which led to the derivation of the 1D averaged limb profile used later in the analysis. 
(a) The original level-1 limb image (the focused image collected on the SuFI camera). (b) The stretched limb image, the red line is perpendicular to the limb. (c) The 1D limb profile resulting from averaging the profiles of the red lines along an axis parallel to the limb. (d) Evaluation of the position of the inflection point (where the first derivative $dl/ dx$ is minimum or the second derivative $d^2l/ d^2x$ is equal to zero). (e) Normalizing the limb profile such that the intensity at the inflection point is equal to 0.5. The example here is for a 300\,nm limb image. $x$ is given in arcsecond (platescale of SuFI is 0.02 arcsec/pixel). }
\label{limb_process}
\end{figure}

As the PD restored images of SuFI and IMaX often suffer from artefacts at the limb, we only use unrestored level-1 images for our evaluation (definitions of the SuFI reduction levels are given in Section~2.1 of Chapter~\ref{chapter_2}). These images come in pairs of focused and manually defocused images (for phase diversity reconstruction), we only use the focused images (see Fig.~\ref{limb_process}~a). The 2D limb images are numerically rotated and destretched to get rid of the limb curvature. Let us consider a cut through the straightened solar limb image, perpendicular to the limb (red line in Fig.~\ref{limb_process}~b). Alternatively an average across such an image along an axis parallel to the limb can be considered. The resulting one-dimensional profile is denoted henceforth as \emph{limb profile} (Fig.~\ref{limb_process}~c).  
To reduce the number of degrees of freedom in the fit described in the next section, the inflection point can be accurately determined from a numerical derivative (where $d^2 l/ d x^2 = 0$) since the observed limb profiles are sampled with a large number of pixels (Fig.~\ref{limb_process}~d). 

\begin{figure}
\centering
\includegraphics[scale=0.8]{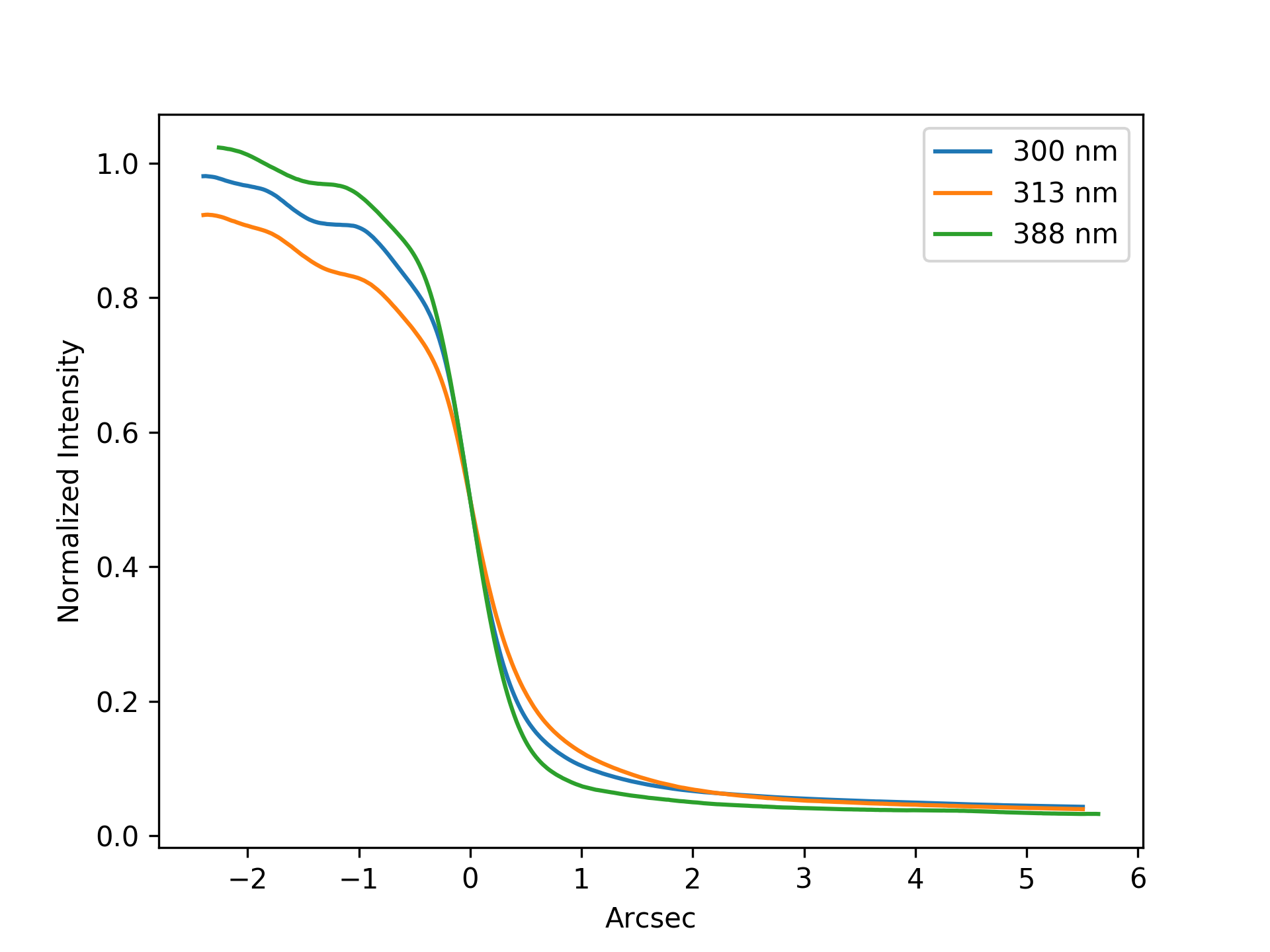}
\caption{Averaged observed solar limb profiles for the different spectral regions sampled by SuFI.}
\label{av_limb_pro}
\end{figure}

For reasons that we will show later, prior to fitting the observed limb profile, we shift it such that the inflection point is located at $x=0$ and normalize it such that the intensity at the inflection point is equal to $1/2$ (Fig.~\ref{limb_process}~e). These steps are applied to several limb images. These are then averaged to obtain the averaged limb profiles plotted in Figure~\ref{av_limb_pro} for each spectral window.

\subsection{Evaluation of observed solar limb profiles} \label{sec:psf_limb}

The observed solar limb profile is not a rectangular (i.e., Heaviside) function. There is an intrinsic broadening, of order 0.5 arcsec (corresponding to roughly 360\,km on the Sun), which results from the extent of the solar atmosphere and from radiative transfer. Furthermore, the observed limb suffers from broadening due to instrumental effects, as described above. The instrumental broadening can influence the PSF on scales ranging from fractions of an arcsec up to wings of order 10\arcsec or more. The main difficulty of a limb-based PSF evaluation is to properly disentangle the different contributions.

The relationship between the observed limb profile $l(x)$ and the intrinsic limb profile $l_{t}(x)$ is modeled as follows:
\begin{equation}
  l(x) = l_{t}(x) \ast \underbrace{ \left(w_{1} \, \psfpd(x) + \sum_{i=2}^{3} w_{i} \, g_{s,i}(x) \right) }_{\mathrm{PSF}(x)}
\label{eq_strayx}
\end{equation} 
$\psfpd$ is inferred from PD wavefront sensing and affects the central part of the extended PSF. The Gaussian terms
\begin{equation}
\label{eq_gauss}
g_{s,i} = \frac{1}{\sqrt{2 \pi} \sigma_{s,i}} \, \exp \left(-\frac{x^2}{2
  \sigma_{s,i}^2} \right)
\end{equation} 
are used to model the wings of the extended PSF, which are not properly captured by PD wavefront sensing. Both the $\psfpd$ and the Gaussian terms are normalized to an integrated value of 1, and weighted with the factors $w_i$, with
\begin{equation}
\label{eq_weights}
  \sum_{i=1}^{3} w_{i} = 1. 
\end{equation}
The $\ast$ operator denotes a convolution.
In Fourier space eq. \ref{eq_strayx} becomes:
\begin{equation}
L(k) = L_{t}(k) \cdot \underbrace{ \left(w_{1} \, \mtfpd(k) + \sum_{i=2}^{3}
  w_{i} \, G_{s,i}(k) \right) }_{\mathrm{MTF}(k)},
\label{eq_strayk}
\end{equation}
where $L$, $L_{t}$ and $G_{s,i}$ are the Fourier transforms (FTs) of the terms denoted by the corresponding lower-case symbols in eq. \ref{eq_strayx}, and where $k$ is the spatial frequency. The FT of the Gaussian terms is given by
\begin{equation}
\label{eq_ftgauss}
  G_{s,i}(k) = \exp(-2 \pi^2 \sigma_{s,i}^2 k^2).
\end{equation}
In the context of this convolution problem we can discard the phases and only consider the modulation transfer function (MTF), i.e., the modulus of the FT of the extended PSF. \\

The intrinsic (i.e., solar) limb profile can be sufficiently well approximated by
\begin{equation}
l_{t}(x) = \frac{1}{2} \, \erfc \left( \frac{x}{\sqrt{2} \sigma} \right),  
\end{equation}
where $\erfc$ is the conjugated Gaussian error function (CGEF)
\begin{eqnarray}
 \frac{1}{2} \, \erfc\left( \frac{x}{\sqrt{2} \sigma} \right) 
 & = & (h \ast g) (x) \nonumber \\
 & = & \frac{1}{\sqrt{2 \pi} \sigma}
 \int_{x}^{\infty} \exp \left(-\frac{y^2}{2 \, \sigma^2} \right) \ud y, 
\end{eqnarray}
which results from the convolution of the Heaviside function 
\begin{equation}
\label{eq_heaviside}
h(x) = \left\{ \begin{array}{rl}
0 & \mathrm{if} \, x < 0 \\
1 & \mathrm{if} \, x \ge 0
\end{array} \right.
\end{equation}
with a Gaussian $g$ in the normalized form of Eq.~ \ref{eq_gauss}. The origin $x=0$ is defined as the position of the inflection point $d^2 l_t / d x^2 = 0$, and $l_t$ is normalized such that $l_t = 1/2$ for $x=0$. For the purpose of fitting the solar limb profile, the term $\mtfpd$ from PD wavefront sensing is approximated with a Gaussian as well (see the alternative for this approximation in Section~\ref{SL_up}):
\begin{equation}
L(k) = H(k) \, G_{t}(k) \cdot \left(w_{1} \, G_{\mathrm{PD}}(k) +
\sum_{i=2}^{3} w_{i} \, G_{s,i}(k) \right)
\label{eq_strayappr}
\end{equation}
The function $H(k) = \frac{1}{2} \left( \frac{i}{\pi k} + \delta(k) \right)$ is the FT of the Heaviside function, as defined in Eq.~ \ref{eq_heaviside}, and $\delta(k)$ is the Dirac delta function. After multiplying out Eq.~ \ref{eq_strayappr} we finally end up with
\begin{equation}
L(k) = H(k) \cdot \sum_{i=1}^{3} w_{i} \, G_{i}(k),
\label{eq_ffitfunc}
\end{equation}
where the $G_{i}(k)$ are all Gaussian terms in the normalized form of Eq.~ \ref{eq_ftgauss}. Back in image space, this is equivalent to a sum of CGEFs:
\begin{equation}
l(x) = \frac{1}{2} \, \sum_{i=1}^{3} w_{i} \, \erfc \left(
\frac{x}{\sqrt{2} \sigma_i} \right)
\label{eq_fitfunc} 
\end{equation}

Eq.~ \ref{eq_fitfunc} is used as fitted function for the observed limb profiles, with 5 free parameters $w_{1}$, $w_{2}$, $\sigma_{1}$, $\sigma_{2}$, $\sigma_{3}$. The third weighting parameter $w_3$ is constrained by Eq.~ \ref{eq_weights}. 

\subsection{Derivation of an extended PSF}

Eqs. \ref{eq_strayappr} and \ref{eq_ffitfunc}, define the relations between the widths of the different Gaussian terms:
\begin{eqnarray}
\sigma_{t}^{2} & = & \sigma_{1}^{2} - \sigma_{\mathrm{PD}}^2 \nonumber, \\
\sigma_{s,i}^{2} & = & \sigma_{i}^{2} - \sigma_{t}^{2}, \quad i=2,3,
\label{sigmas}
\end{eqnarray} 
where $\sigma_{\mathrm{PD}}$ is obtained independently from PD wavefront sensing (fitting a Gaussian to the 1D radial profile of the PSF returns a $\sigma_{\mathrm{PD}}$ value in the order of 0.02 arcsec for all wavelengths, see Figure~\ref{rad_psf}), and where $\sigma_{1\ldots3}$ are free parameters of the limb fitting. 
With this interpretation of the observed limb profiles, we obtain the following estimate of the MTF:
\begin{equation}
  \mathrm{MTF}(k) = w_{1} \, \mtfpd(k) + \sum_{i=2}^{3} w_{i} \, \exp(-2 \pi^2
  \sigma_{s,i}^2 k^2).
  \label{eq_mtf_stray}
\end{equation}

Apart from the broadening, the observed limb profiles asymptotically tend to different, wavelength-dependent offsets (cf. Fig. \ref{av_limb_pro}). The offsets are handeled according to the \textbf{local stray light assumption}. 
The basis of this assumption is that the stray light which produces the offsets is generated only within the FOV of the respective instrument. In this case the offset can be handled as an additional term in the MTF
\begin{equation}
  \mathrm{MTF}(k) = w_{1} \, \mtfpd(k) + \sum_{i=2}^{3} w_{i} \, \exp(-2 \pi^2
  \sigma_{s,i}^2 k^2) + w_{4} \, \mathrm{MTF}_{\mathrm{off}}(k).
\end{equation}  
where 
\begin{equation}
  \mathrm{MTF}_{\mathrm{off}}(k) = \lim_{\sigma \to \infty} \exp(-2 \pi^2
  \sigma^2 k^2) = \left\{
  \begin{array}{rl}
    1 & \textrm{if } k = 0,\\
    0 & \textrm{elsewhere}
  \end{array} \right.
\end{equation}
is a term which equally redistributes a fraction $w_4$ of the total intensity inside the FOV. Naturally the condition on the weights changes to $w_1 + \ldots + w_4 = 1$.\\

In Table~\ref{tab:psfpar_me} we show the best-fit parameters to the observed limb profiles at the wavelengths of SuFI. We also show the 95\% confidence intervals (CIs) of the best-fit parameters, $p_0$.

To compute the 95\% CIs, we use the bootstrap method \citep{press_numerical_1992}. This method consists of generating synthetic data sets (in this case, limb profiles) drawn from the original data set (with best-fit solution $p_0$ of dimension $n=6$). Each data set returns a set of best-fit parameters, $p_i$. The sigma of the $p_0-p_i$ distribution in the $n$ dimensional space of all parameters is the standard error $\Delta p$ (with a probability of 68.3\% to find the true value). We then scale the standard error with a factor $f\approx1.96$ to get the confidence region for a probability of 95\% (assuming that the parameter values follow a standard normal distribution).

To assess the effect of these uncertainties on the computed RMS contrasts, we deconvolve the images with two PSFs corresponding to $p_0+f.\,\Delta p$ and $p_0-f.\,\Delta p$. We then compute the corresponding RMS contrasts. The results are shown in Table~\ref{tab:RMS_CI} for solar limb fitting in the three wavelengths of 300\,nm, 313\,nm and 388\,nm, and for three $\mu$ values ($\mu = 0.4, 0.6, 1$). Each QS intensity contrast value in the table corresponds to an average over a number of level-3.1 images. This number ranges from 50 to 150 depending on the $\mu$ value and wavelength. We note that the edges resulting from apodizing the images are trimmed before computing the intensity contrast values (even if edges are still apparent in the illustrative figures throughout the chapter).

\begin{table}

\centering
 \begin{adjustwidth}{-1cm}{}

\caption{Best-fit parameters of the extended PSF along with their 95\% confidence intervals. Columns $2-4$ correspond to the limb profiles recorded in 2009. The last column corresponds to the best-fit parameters resulting from fitting the 2013 limb profile at 300\,nm (relevant for Section~\ref{SL_up}). }
\label{tab:psfpar_me}
\centering
\begin{tabular}{cccccc}

\hline
\hline
Parameter  & \multicolumn{3}{c}{Local stray light assumption} \\
\hline
  & 300 nm & 312 nm & 388 nm & 300 nm (2013 LP)\\
\hline
 $w_1$ &0.528 [0.515,0.540]&0.484 [0.473,0.494]&0.735 [0.727,0.742]& 0.544 [0.536,0.552] \\
 $w_2$&0.289 [0.279,0.299]&0.346 [0.339,0.352]& 0.158 [0.152,0.165]& 0.291 [0.278,0.304]\\
 $w_3$&0.102  [0.099,0.105]&0.099  [0.094,0.103]&0.073 [0.069,0.077]&0.068 [0.058,0.078]\\
 $w_4$&0.082  [0.081,0.083]&0.072  [0.071,0.074]& 0.034 [0.029,0.039]& 0.097 [0.079, 0.116]\\
$\sigma_{s2}$&0.720  [0.714,0.727]&0.918  [0.915,0.920]& 0.977 [0.968,0.998] & 0.956 [0.924,0.987]\\
$\sigma_{s3}$&2.822  [2.781,2.862]&3.104  [2.994,3.213]&6.993  [6.351,7.636] & 4.703 [3.778,6.629]\\
\hline
\end{tabular}

 \end{adjustwidth}

\end{table}

\begin{table}

\caption{RMS contrasts computed for three heliocentric angles. Each value corresponds to the averaged contrast of a number of images. The contrasts are given before and after the stray light correction along with the 95\% CIs in brackets. } 
\label{tab:RMS_CI}
\centering
 \begin{adjustwidth}{-2cm}{}

\begin{tabular}{cccccccc}

\hline
\hline
Heliocentric angle  & \multicolumn{2}{c}{300 nm}  & \multicolumn{2}{c}{312 nm}  & \multicolumn{2}{c}{388 nm}  \\
\hline
\hline
& before&after &before&after&before&after\\
\hline
 $\mu = 1$ & 0.209& 0.368 [0.367,0.370] & 0.197&0.382 [0.375,0.391] &0.170& 0.230 [0.227,0.232] \\
 $\mu=0.6$&  0.160& 0.278 [0.277,0.279]&0.148&0.285 [0.279,0.291]& 0.145& 0.196 [0.194,0.198]\\
 $\mu=0.4$&  0.132&0.228 [0.227,0.229]&0.114&0.215 [0.210,0.220]&0.122&0.165 [0.163,0.167]\\
 
\hline
\end{tabular}
 \end{adjustwidth}

\end{table}

\subsection{Deconvolution of the observations}
\label{deconv}
The deconvolution is applied to PD restored images, i.e., to images which have already been deconvolved with the term $\mtfpd$ in our extended MTF of Eq.~ \ref{eq_mtf_stray}. During the PD restoration process special care has been taken, in terms of frequency filtering, to avoid numerical noise amplification in the high spatial frequency range (cf. section \ref{sec:psf_pd}). By deconvolving the PD restored images with a reduced MTF, which corresponds to the MTF of Eq.~ \ref{eq_mtf_stray} divided by $\mtfpd$ (cf. Fig. \ref{fig:mtf}, right panel), we do not interfere with this previous careful noise treatment. The values of the reduced MTF stay well above the noise level across the whole spatial frequency range and thus there is no risk of any significant noise amplification from this additional deconvolution process.

\begin{figure}
\includegraphics[scale=0.36]{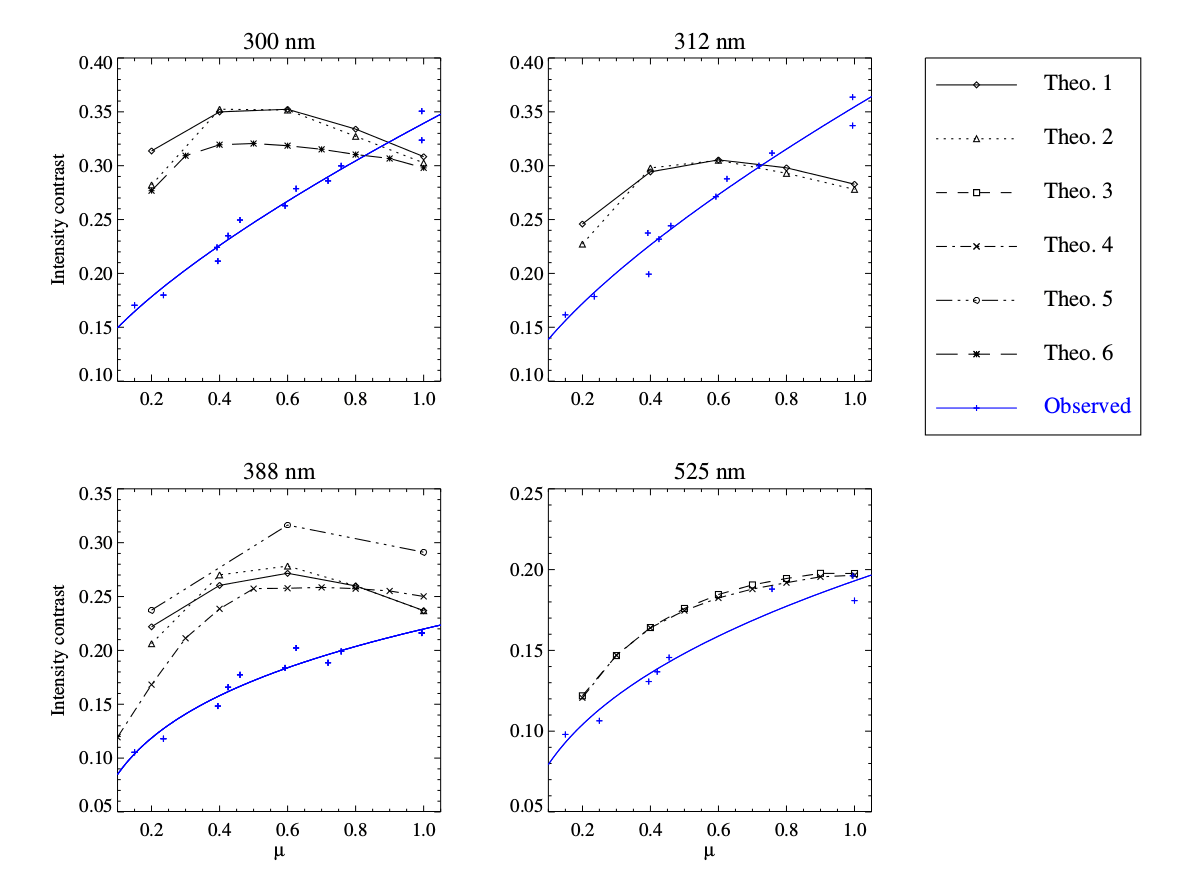}
\caption{Comparison between observed (blue) and theoretical (black) QS intensity contrasts. The observations depicted in the spectral regions 300 nm, 312 nm and 388 nm correspond to stray light corrected SuFI level-3 data. In the 525 nm region stray light corrected IMaX level-2 data are shown. The solid blue lines represent a power law fit to the respective observations. The theoretical results, denoted in the plot legend by "Theo. 1 \ldots 6" stand for different combinations of MHD and spectral synthesis computations.}
\label{fig:cvsmu_mhd}

\end{figure}

In Figure~\ref{fig:cvsmu_mhd} we plot the CLV of the observed QS intensity contrasts after dividing the Fourier transform of level-3 SuFI and IMaX data with the corresponding reduced MTFs. Plotted as well are the CLVs of QS synthetic intensity contrasts. The latter are derived from MURaM cubes by means of different spectral synthesis codes which are denoted in the legend of the figure. The details of the spectral synthesis are not relevant for this work, but one can notice that for all synthesis modes, the CLVs of the intensity contrast derived from synthetic images in the NUV result in a peak at intermediate $\mu$ values which is not reproduced by the observations.   

\section{Updating the stray light model}
\label{SL_up}

$\sunrise$ had its second science flight in 2013 June. The telescope targetted an active region (AR) NOAA 11768 (see Chapter~\ref{chapter_4}). The field of view covered a variety of magnetic features: a pore with a penumbra-like structure, a flux emergence region, pores, plage, and quiet-Sun granulation. The same scientific instruments (IMaX and SuFI) as for the first flight, with minor changes (see \cite{solanki_second_2017}) acquired images in the visible and NUV. To correct for low order aberrations, the data are phase-diversity reconstructed following the same procedure as for the $\sunrise$~I data. The same extended PSF evaluated using the limb profiles recorded during $\sunrise$~I ($\mathrm{PSF}_{\mathrm{2009}}$) is used in the deconvolution of level-3 AR data.

\begin{figure}
\centering
\includegraphics[scale=0.5]{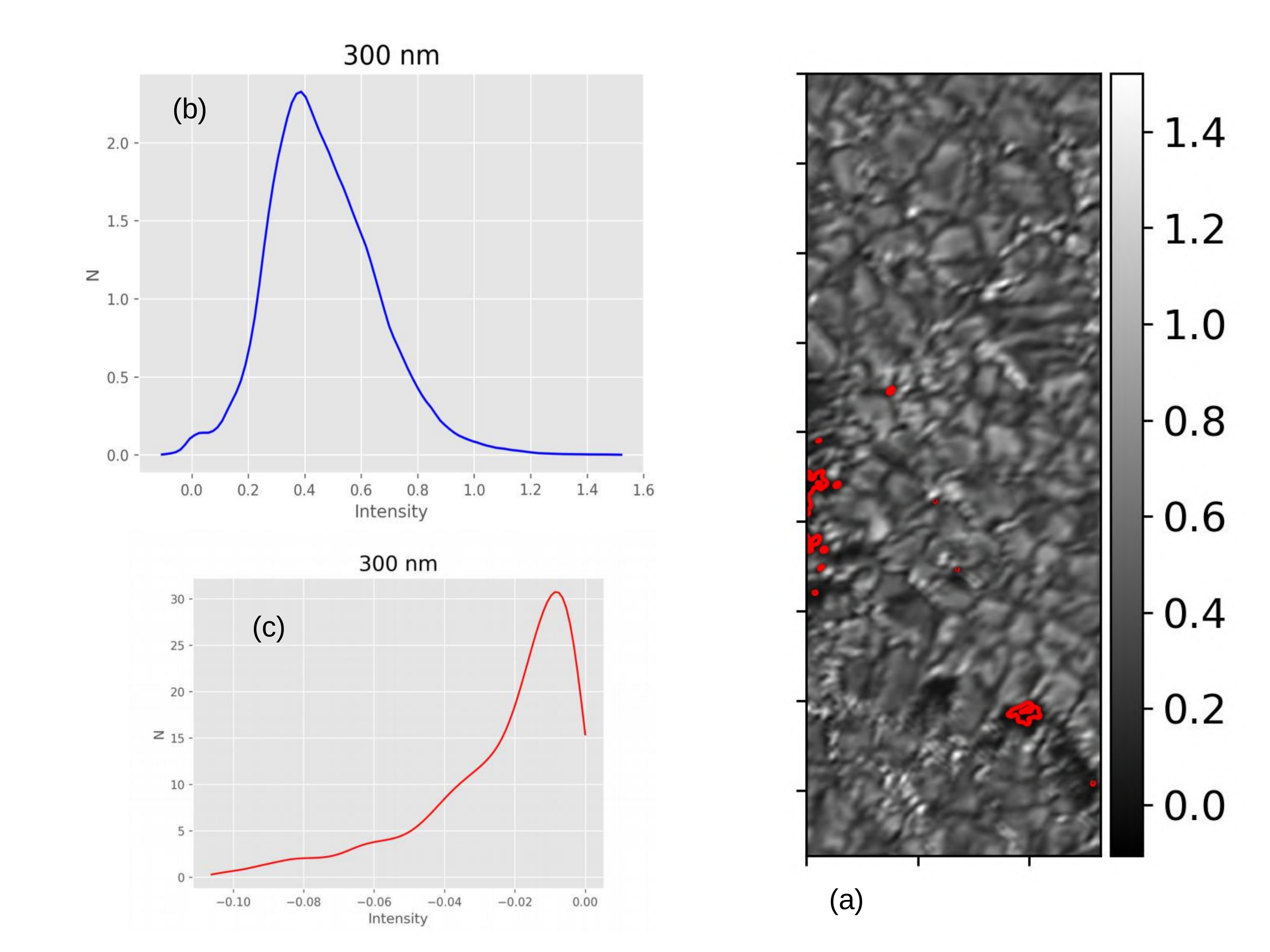}
\caption{(a) An example level-3.1 (level-3, after stray light correction) 300 nm SuFI image of the time series used in Chapter~\ref{chapter_4}. The red contours mark the pixels with negative values which belong to the interior of pores. (b) A histogram of the intensity values of all pixels in the image. (c) The histogram of the pixels with negative values (they amount to 0.6\% of all the pixels in the image). The image is normalized following the normalization method of AR SuFI data explained in Appendix~\ref{appendix_B}.}
\label{neg_int}

\end{figure}

A problem arose after correcting the $\sunrise$~II level-3 SuFI data (at disk center) using the $\mathrm{PSF}_{\mathrm{2009}}$: some pixels in the dark structures (mainly pores) end up with negative intensities (see Figure~\ref{neg_int}). The aim of this section is to revise the model developed in Section~\ref{sec:psf_limb} which was used to correct the disk center quiet-Sun data of $\sunrise$~I. 

The common SuFI wavelengths between the two flights are 300 nm and 397 nm. Limb images at 397 nm cannot be used for the stray light analysis since this window has substantial chromospheric contributions. For this reason, we restrict our model testing on the AR data acquired in the 300 nm filter (bandwidth of 4.4 nm). In the next subsections, we test the possible reasons for this puzzling finding. To avoid the confusion from alternating between the terms PSF and MTF of the same intensity distribution: \textbf{convolution} of images is done \textbf{with the PSF} and the images are \textbf{divided} \textbf{by the MTF} (check Section~2.2 of Chapter~\ref{chapter_2}).

\subsection{The Gaussian approximation for the PD PSF}
In Section~\ref{sec:psf_limb} we mentioned that the term $\mtfpd$ in the stray light model was approximated with a Gaussian function for the sake of simplicity. We test if using the real MTF captured from PD wavefront sensing in the model makes a difference in the computed parameters of the extended MTF, and therefore in the corrected AR SuFI images.

\begin{figure}[h!]
\centering
\includegraphics[scale=0.4]{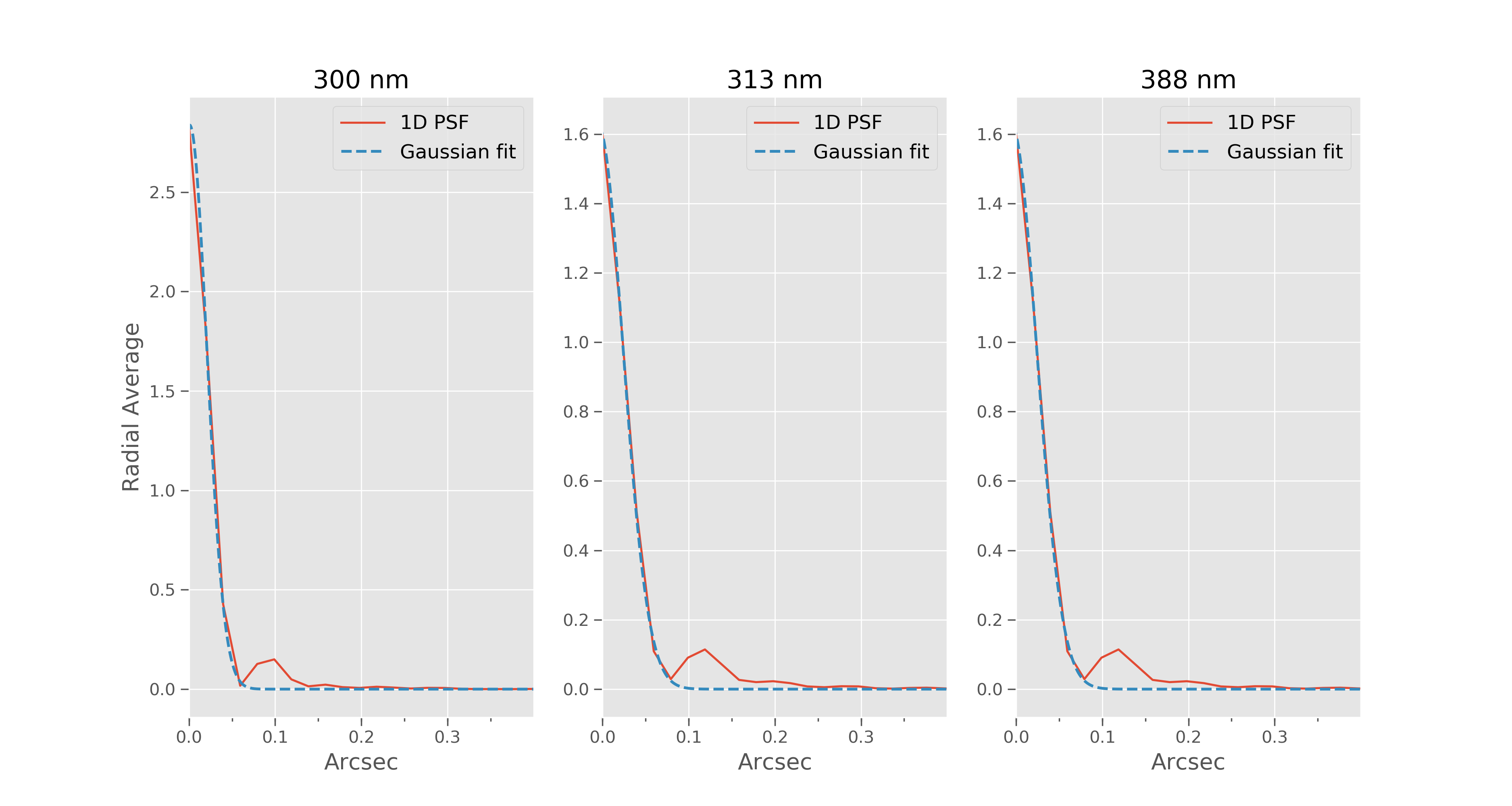}
\caption{\textit{Solid orange curves:} the azimuthal average of the 2D PSF computed from phase-diversity reconstruction in the wavelengths 300\,nm, 313\,nm and 388\,nm. \textit{Dashed blue curves:} Gaussian fits to the radial profiles. The width of the Gaussian is 0.02 arcsec for all three wavelengths. }
\label{rad_psf}
\end{figure}

We will refer to using the Gaussian approximation for the $\psfpd$, as done in Section~\ref{sec:psf_limb}, as \textit{case\,1}. Using the 1D azimuthal average of the 2D $\mathrm{PSF}_{\mathrm{PD}}$ image will be referred to as \textit{case\,2}. We show in Figure~\ref{rad_psf} the 1D radial profiles of the PD PSFs and the Gaussian fits for the three wavelengths of 300 nm, 313 nm and 388 imaged by $\sunrise$~I. 

The width of the fitted Gaussian ($\sigma_{\mathrm{PD}}$ in Equation~\ref{sigmas}) is on the order of 0\farcs02. In \textit{case~1} $\sigma_{\mathrm{PD}}$ was therefore neglected  and the intrinsic width $\sigma_{\mathrm{t}}$ was taken to be equal to $\sigma_{1}$ in Equation~\ref{sigmas} ($\sigma_{1}$ values are given in Table~\ref{cases_1_2}).
 
The updated model (with \textit{case 2}) is used now to repeat the fitting process on the 2009 limb profiles at all wavelengths. The fit is highly sensitive to the choice of the initial guess parameters, so we refer to global optimization algorithms to test for the convergence of the solution in both defined cases. The global optimization method that we use is based on the differential evolution algorithm \citep{storn_differential_1997}. After testing local and global optimization algorithms in both cases, we found that only the latter finds the deepest minimum in the solution space, which turns out to be rather similar for both cases (see Table~\ref{cases_1_2}).

\begin{table}
\caption{Best fit parameters to the 2009 limb profiles for both cases (see text for details). }
\label{cases_1_2}
\centering
\begin{tabular}{cccccccc}

\hline
\hline
Wavelength  & \multicolumn{2}{c}{300 nm}  & \multicolumn{2}{c}{312 nm}  & \multicolumn{2}{c}{388 nm}  \\
\hline
\hline
& \textit{case~1}& \textit{case~2} & \textit{case~1}&\textit{case~2}&\textit{case~1}&\textit{case~2}\\
\hline
\hline
$w_1$ & 0.528&0.528& 0.484&0.484& 0.735 & 0.733 \\
$w_2$&  0.289&0.289&0.346 & 0.346 & 0.158 & 0.158 \\
$w_3$& 0.102&0.101&  0.099& 0.098 & 0.073 & 0.066 \\
$\sigma_1$& 0.259&0.259 &0.309&0.308&  0.306&0.305\\
$\sigma_2$&0.765&0.765&0.968&0.969&1.024&1.010  \\
$\sigma_3$&2.833&2.827 &3.119&3.125 &7.011&5.842\\
 
\hline

\end{tabular}
\end{table}

According to our finding, we come to the conclusion that the Gaussian approximation for the PD PSF in the stray light model is sufficiently accurate and is not the reason for obtaining the negative intensities in the pores in the level-3.1 AR SuFI images.

\subsection{Comparing the effect of PD PSFs from $\sunrise$ I and II  }
Our second guess regarding the source of the negative intensities was the usage of the appropriate PD PSF in the expression of the extended PSF (the term $\psfpd(x)$ in Eq.~\ref{eq_strayx}). The $\psfpd(x)$ used so far to fit the limb profiles and to correct 2013 level-3 data is the one derived from wavefront sensing of the quiet-Sun data acquired in 2009. 

Considering that observational conditions and instrumental setup varied slightly between the two flights, the two PSFs (and therefore MTFs) from both flights could be different. In addition, in the correction process, level-3 images are divided by the reduced MTF which is the ratio of the extended MTF to the $\mtfpd$ to which the level-3 images should be already divided (see Section~\ref{deconv}). Hence using the 2009 $\psfpd(x)$ to deconvolve science images corrected already with the 2013 $\psfpd(x)$ could introduce a significant systematic error. We test this by incorporating the 2013 $\mathrm{MTF}_{\mathrm{PD}}$ in the expression of the extended MTF (Eq.~\ref{eq_mtf_stray}) instead of the 2009 $\mathrm{MTF}_{\mathrm{PD}}$. 

\begin{figure}
\centering
\hspace*{-2cm}\includegraphics[scale=0.55]{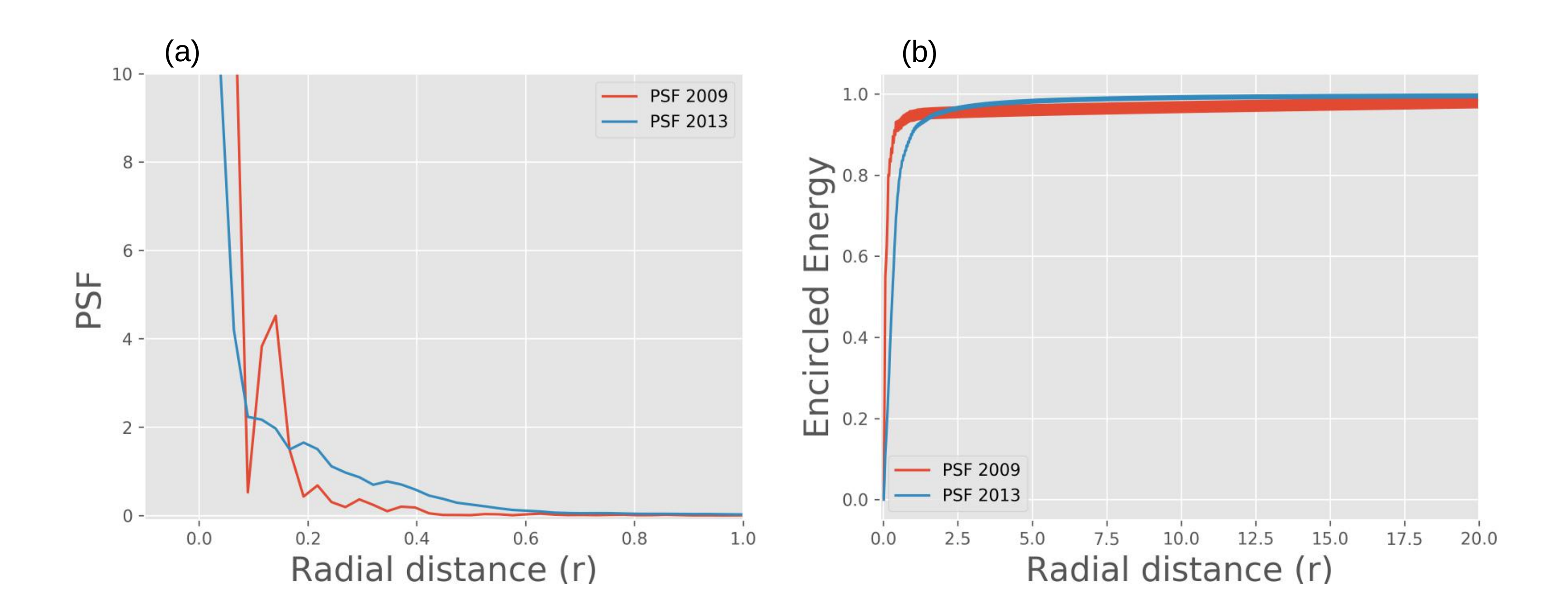}
\caption{(a) The 1D radial profile of the PD PSF resulting from phase-diversity wavefront sensing of 2009 data (red) and 2013 data (blue). (b) The encircled energy plots for both PSFs which represent the energy enclosed in the PSF within the radial distance, $r$, in arcsec.}
\label{rad_psf_ee}
\end{figure}

Employing the 2013 $\mathrm{MTF}_{\mathrm{PD}}$ in the deconvolution process had no effect on the computed RMS contrast of the image and the percent of pixels with negative intensities. This is easy to understand considering the overplot of the azimuthal averages of both PD PSFs in Figure~\ref{rad_psf_ee}. The wings do not differ by much in both cases while the cores show a distinguishable difference. According to our discussion in Section~\ref{sec:psf}, the core mainly affects the spatial resolution of the image while the wings affect the RMS contrast, hence changing the PD PSF in the expression of the extended PSF has little effect on the computed contrasts.

\subsection{Using limb profiles of $\sunrise$~II}
Now that we have eliminated the previous two possible reasons, we are sure that the parameters of the stray light components modelled with 2 Gaussians in the extended PSF are responsible for the negative pixels in the pores. Simply stated, the negative values might be a consequence of overcorrecting for the stray light in the deconvolution process (this is apparent in Figure~\ref{fig:cvsmu_mhd} where at disk center the observed 2009 intensity contrast at 300\,nm exceeds the synthesized one). This possibly stems from overestimating the wings of the extended PSF. Two factors come to mind, the function used to model the intrinsic limb profile (with a free parameter of $\sigma_{\mathrm{t}}$), and the observed limb profile used in the fitting.

\begin{figure}
\centering
\includegraphics[scale=0.8]{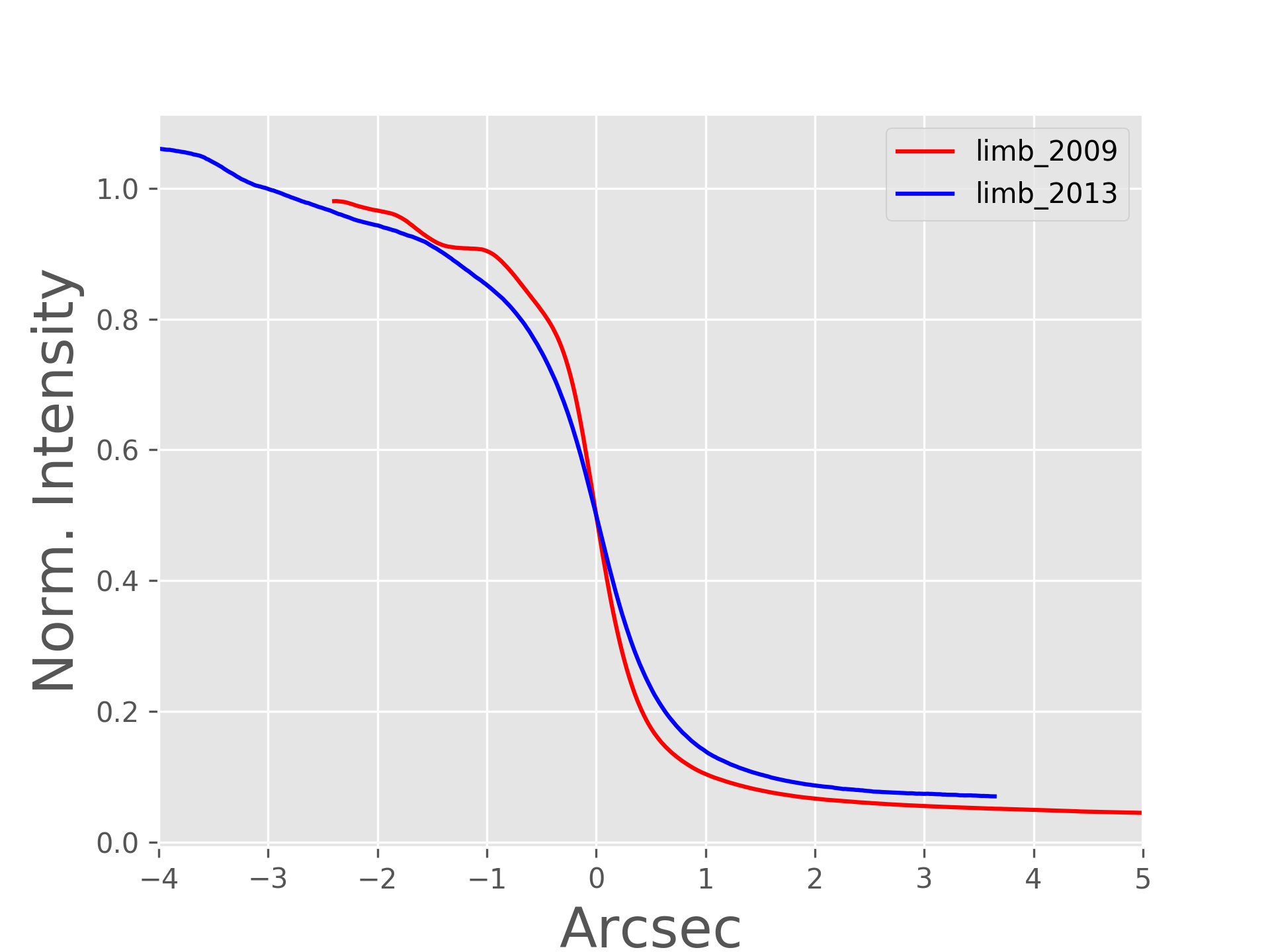}
\caption{The 1D limb profiles resulting from processing limb images recorded in 2009 during the $\sunrise$~I flight (red) and in 2013 during $\sunrise$~II (blue). The limb profiles are normalized such that the intensity at the inflection point is equal to 0.5 (see Section~\ref{sec:psf_limb} for more details). }
\label{limbs}
\end{figure}

We choose a number of good level-1 limb images recorded during $\sunrise$~II in the wavelength band of 300\,nm. We apply the same limb processing steps used for 2009 limb profiles (destretching the limb images, finding the inflection point, shifting and normalizing the 1D profile). The averaged 1D 2013 limb profile is shown in Figure~\ref{limbs} along with the 2009 limb profile for a comparison. The 2013 limb profile (blue curve in Fig.~\ref{limbs}) is broader than the 2009 profile (red curve in Fig.~\ref{limbs}), especially in the range up until 2 arcsec. 

We repeat the fitting process on the 2013 limb profiles using both local and global optimization algorithms. Both algorithms converge to the same solution shown in the last column of Table~\ref{tab:psfpar_me}.
After constructing the extended PSF based on the new parameters and deconvolving the example SuFI image of Figure~\ref{neg_int}, the negative values persisted in the pores and even increase in number (from 0.6\% with the 2009 limb profile fitting to 0.7\%). This is not surprising considering that the limb profile is broader due to higher stray light during the 2013 $\sunrise$ flight.

\subsection{The intrinsic limb profile}
The instrinsic limb profile, $l_t$, employed in the stray light model is approximated with a Heaviside function convolved with a Gaussian function of a free parameter, $\sigma_t$.
Given the narrow shape of the Heaviside function, it could be that the intrinsic broadening estimated from the limb fitting is underestimated and hence the wings of the PSF are overestimated. This can lead to an over-correction of the stray light. The shape of the adopted intrinsic limb profile in the model should be revised. For example, one can refer to synthesized limb images from MHD simulations as modelled in \cite{shelyag_spectro-polarimetric_2015}.

\subsection{The effect of residual image jitter}
\label{jitter}

The jitter, which results from telescope vibrations and residual pointing errors, acts as an additional source to image smearing. In the case of $\sunrise$, the jitter is more serious at the limb than at disk center because the correlation tracker did not lock as stably. Therefore, the limb images are more affected by the smearing due to jitter than disk center data. In fact, images taken at small $\mu$ values are still less affected by the jitter than limb images. The reason is that the optimum performance of the Correlation tracker and Wavefront Sensor (CWS) on $\sunrise$ depends on a uniform distribution of solar features. This is not the case in the limb images which display a discontinuity in features (i.e., the edge of the solar disk against a dark background). Our guess is that limb profile fitting is interpreting the jitter as an additional stray light contamination, leading to over-correcting disk center images. This is a possible reason for the QS RMS intensity contrasts at 300\,nm exceeding the values computed from radiative MHD simulations (see Figure~\ref{fig:cvsmu_mhd}). For the same reason pores in the AR images at the same wavelength can end up with negative intensities after the deconvolution.

Assuming that the jitter is the only existing degradation source that is not taken into account so far in the model, we assess its effect on our stray light estimate by constraining the fit to the observed limb profile. The constraint is such that the returned intensity contrast of a level-3.1 disk center QS image is forced to be equal to that of a synthesized image (the values of which are given in the last row of Table~\ref{tab:psfpar_cons}). The theoretical contrast computation is based on spectral synthesis in the wavelength bands of SuFI carried out by \cite{riethmuller_comparison_2014} using MURaM simulations of a QS region (average magnetic flux of 30\,G). The extended PSF that results from this constrained fit will be jitter free. 

We carry out the constrained limb profile fitting using three theoretical images (at 300\,nm, 313\,nm and 388\,nm) after resampling them to the pixel size of SuFI (0.02"/pixel). The constrained minimization is based on Sequential Quadratic Programming (SQP) algorithms, which are usually used for constrained non-linear optimization problems \citep{johansen_constrained_2004}. The best-fit parameters for all three wavelengths are listed in Table~\ref{tab:psfpar_cons}. CIs for these parameters have not been determined because the computations based on Monte Carlo simulations (the bootstrap method explained in Section~\ref{sec:psf}) are numerically too expensive.

\begin{table}[h]
\caption{Best-fit parameters of the jitter-free extended PSF for each wavelength band after applying a constraint on the fitting procedure (see text for details).}
\label{tab:psfpar_cons}
\centering
\begin{tabular}{ccccc}

\hline
\hline
Parameter  & 300 nm & 312 nm & 388 nm \\
\hline
\hline
 $w_1$ &  0.636  & 0.539&0.721 \\
 $w_2$&  0.215 &0.090&0.159\\
 $w_3$&  0.084 &0.322&0.061\\
 $w_4$&   0.063 &0.049&0.059\\
 $\sigma_t$& 0.303&0.304&0.302\\
$\sigma_{s2}$& 1.018&5.711&0.873\\
$\sigma_{s3}$&  4.970&1.070&3.583\\
\hline
forced RMS contrast & 0.300 & 0.324 & 0.308 \\
(Disk center) \\
\hline
\end{tabular}
\end{table}

\begin{figure}
\centering
\includegraphics[scale=0.8]{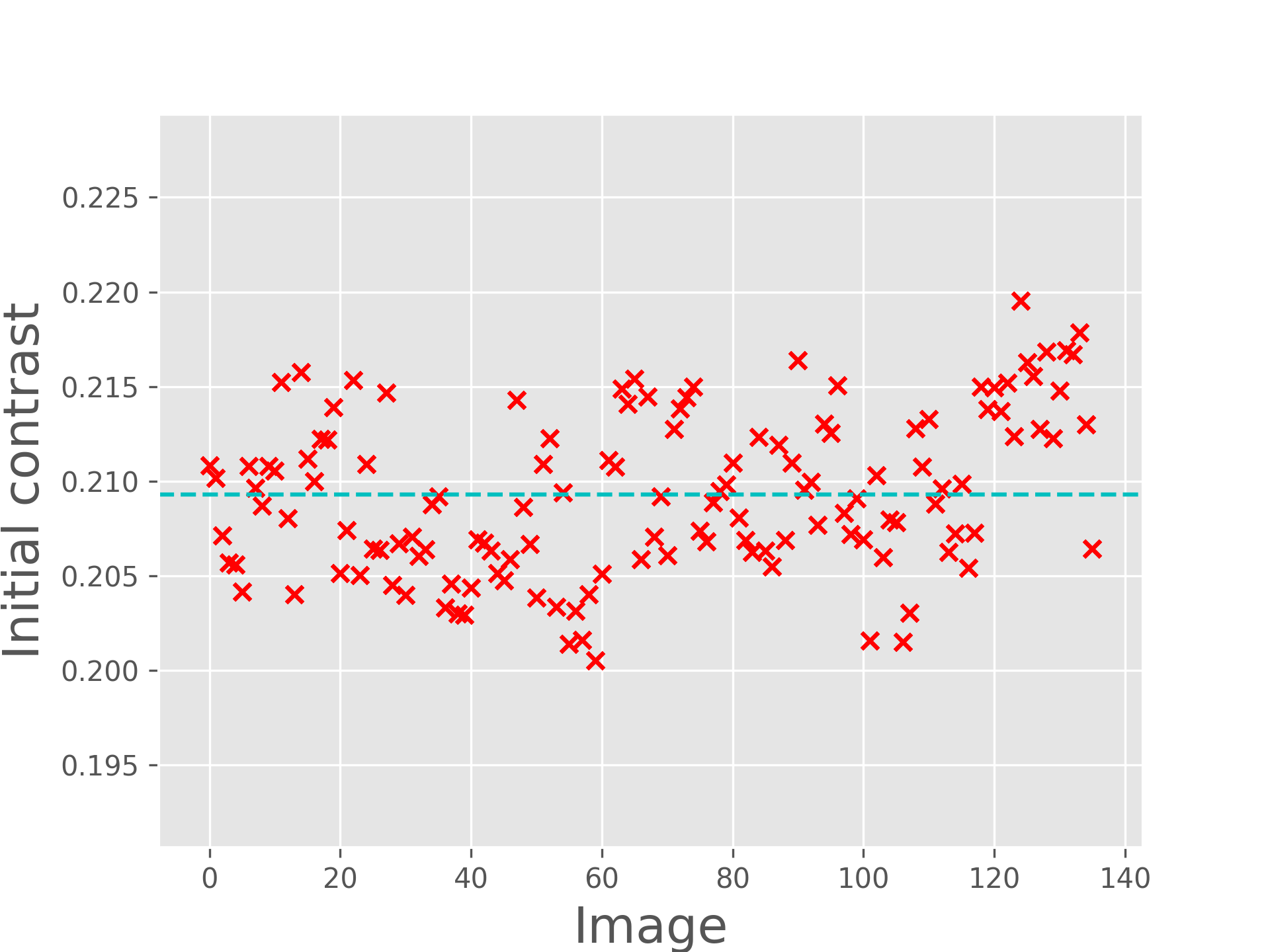}
\caption{The initial intensity contrasts of level-3 SuFI images at 300\,nm (at disk center). The red dashed line is the average value. }
\label{rms_scatter_300}
\end{figure}

From now on, we restrict our analysis to the wavelength of 300\,nm, since it is the only common wavelength between the two flights, and with which we can test the validity of the computed PSF (by inspecting whether the negative intensities in the pores of AR data are gone after the deconvolution with the extended PSF). 

The fitting process now is done on the 2009 limb profile at 300\,nm. We set a constraint such that a chosen level-3 SuFI image (2009) at disk center ends up with an intensity contrast of 0.3 after the deconvolution. We note here that the constrained fit is done using one example level-3 SuFI image. By using other images we end up with slightly different best-fit parameters since their initial intensity contrasts display a scatter (see Figure~\ref{rms_scatter_300}). We choose then an image with an initial intensity contrast close to the average value over all images of around 0.209 (cyan dashed line in Figure~\ref{rms_scatter_300}). 

\begin{figure}
\centering
\includegraphics[scale=0.65]{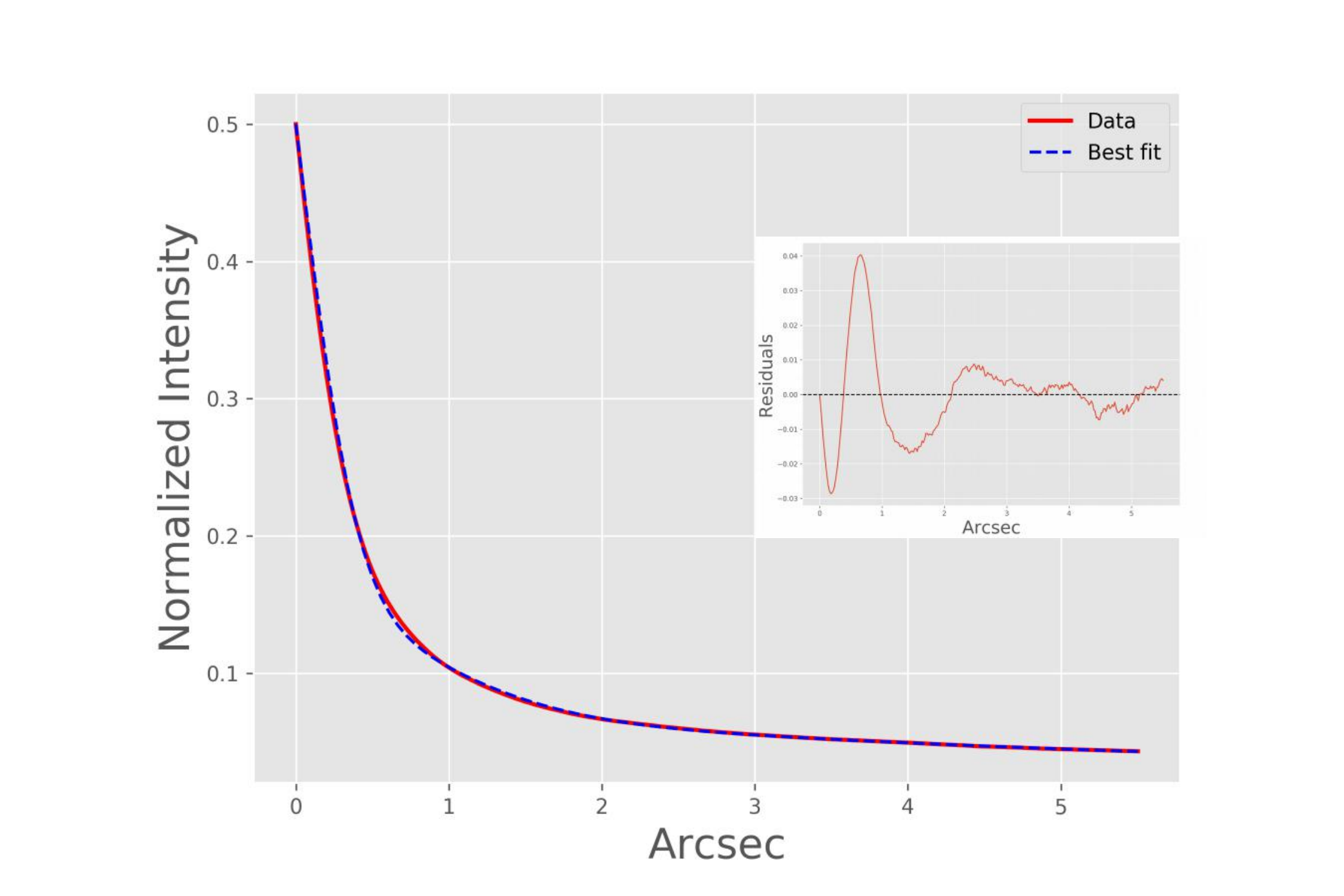}
\caption{The limb profile at 300\,nm in red. The blue dashed curve is the constrained fit (see text for details). The inlay shows the residuals (the difference between the data and best-fit profiles.) }
\label{lousy_fit}
\end{figure}

Figure~\ref{lousy_fit} shows the constrained fit to the observed limb profile. As expected, the limb profile resulting from the constrained fit shows significant deviations from the observed profile, in particular in the range up to 1 arcsec. In the context of this analysis, this difference is attributed to additional jitter at the limb, and modeled with a Gaussian term, $g_{j}(x)$. The jitter Gaussian should be convolved with the stray light Gaussians in Equation~\ref{eq_strayx}. The observed limb profile, $l(x)$ would then be modelled as follows:

\begin{equation}
  l(x) = l_{t}(x) \ast  \left(w_{1} \, \psfpd(x) + \left[\sum_{i=2}^{3} w_{i} \, g_{s,i}(x) \right] \ast g_{j}(x)\right) 
 \label{stray_jitter}
\end{equation} 

To compute the residual jitter, we fit the model of Eq.~\ref{stray_jitter} to the observed limb profile. The model now has one free parameter, $\sigma_j$, since the other parameters are fixed to the values given in Table~\ref{tab:psfpar_cons} (at 300\,nm). Unfortunately, the fit does not converge to a non-zero value. We have tested if the problem is the convergence of the solution by using a global optimization algorithm, but a test with different $\sigma_j$ values implied that the jitter cannot be modelled with a normal distribution. As a test, we have also tried modelling the jitter using the limb profiles at the other SuFI wavelengths (313\,nm and 388\,nm). In both wavelengths, the jitter could not be modeled with a normal distribution similar to the wavelength of 300\,nm. This is expected since the residual jitter is not wavelength dependent.

\begin{figure}
\center
\includegraphics[scale=0.6]{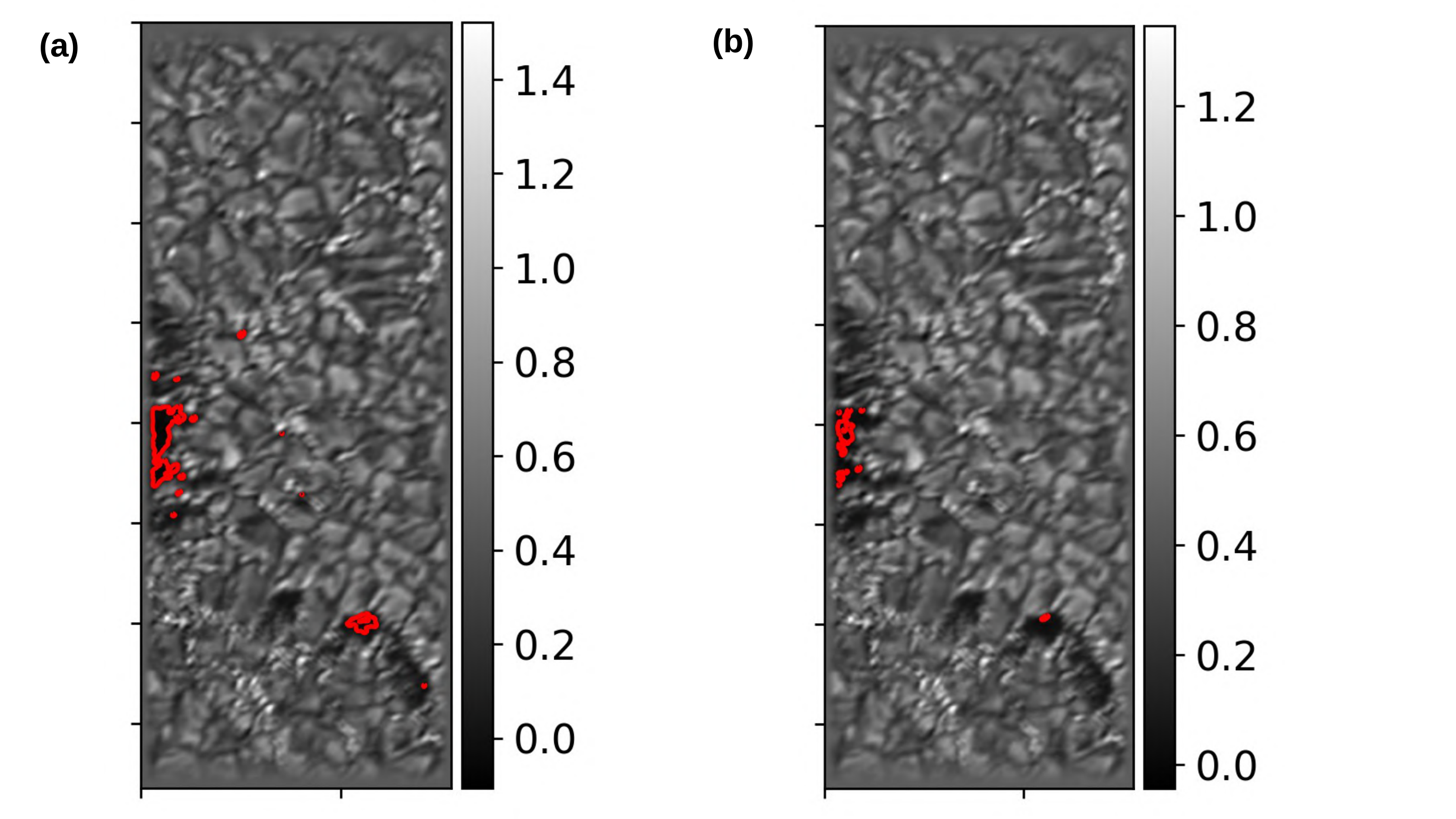}
\caption{An example level-3 SuFI image at 300\,nm (the same as in Fig.~\ref{neg_int}) and deconvolved with an extended PSF based on (a) non-constrained and (b) constrained limb fitting.}
\label{b_a_jitter}
\end{figure}

Nevertheless, to test the effect of the constrained minimization on the stray light correction of disk center data, we construct the jitter free extended PSF using the best-fit parameters in the second column of Table~\ref{tab:psfpar_cons}. Then, we use it to deconvolve the example SuFI image of Figure~\ref{neg_int}. The result is shown in panel (b) of Figure~\ref{b_a_jitter}. The red contours enclose the pixels with negative values which are reduced by around 70\% compared to the non-constrained limb fitting (panel (a)). This implies that a proper modelling of the jitter will improve the stray light correction. 

To test whether the peaking shape of the CLV of quiet Sun RMS intensity contrasts (as expected from radiative MHD simulations, see Figure~\ref{fig:cvsmu_mhd}) is reproduced upon using the jitter-free extended PSF, we repeat the deconvolution process on 2009 quiet-Sun images taken at three $\mu$ angles, 0.4, 0.6 and 1 for the three wavelength channels. The results are shown in Table~\ref{tab:RMS_cons}.

\begin{table}
\caption{The intensity contrasts of SuFI/$\sunrise$~I images at three heliocentric angles. The RMS contrasts are shown for the non-constrained (NC) limb profile fitting (on which the results of Figure~\ref{fig:cvsmu_mhd} are based), and for the constrained fit (CF) to the 2009 solar limb profiles.}
\label{tab:RMS_cons}
\centering
\begin{tabular}{cccccccc}

\hline
\hline
Heliocentric angle  & \multicolumn{2}{c}{300 nm}  & \multicolumn{2}{c}{312 nm}  & \multicolumn{2}{c}{388 nm}  \\
\hline
\hline
& NC&C &NC&C&NC&C\\
\hline
 $\mu = 1$ & 0.368& 0.318& 0.382&0.350 &0.230&0.234   \\
 $\mu=0.6$&  0.278&0.240 &0.285&0.260& 0.196&0.20\\
 $\mu=0.4$&  0.228&0.198&0.215&0.197&0.165&0.168\\
 
\hline

\end{tabular}
\end{table}

In the spectral windows 300\,nm and 313\,nm, the disk center contrasts are closer to the synhesized ones, as expected from the constraint we applied on the limb profiles. They are not exactly equal to the theoretical contrasts given in Table~\ref{tab:psfpar_cons} since the constraint, for each spectral window, has been defined on one example image (as explained earlier in this subsection). 

Going away from disk center, the contrasts (with the original stray light correction, i.e., non-constrained limb fitting) were smaller than synthesized values, now they display higher values and get closer to the synthesized ones. At 388\,nm, the contrasts were smaller than the synthesized values at all $\mu$ values (see Figure~\ref{fig:cvsmu_mhd}), but when using the new extended PSF, they lie a bit higher. The peaking shape of the observed CLV, however, could still not be reproduced with the usage of the jitter-free extended PSF.

\section{Conclusions}
\label{conclusions}
Stray light correction has a significant effect on the observed intensity contrasts. The amount and spatial distribution of stray light is estimated by analyzing solar limb images recorded during the first $\sunrise$ flight on 2009. The wavelengths studied here are restricted to the near ultra violet (SuFI data). The reason I did not include IMaX data in this chapter is because they are corrected for stray light derived from other approximations (see Chapters~\ref{chapter_3} and \ref{chapter_4}).

In Section ~\ref{sec:psf}, we have described the processing steps which led to the derivation of the 1D solar limb profiles from 2D limb images. Next, we described the model used to fit an observed limb profile. The latter is modeled as an intrinsic (i.e., solar) limb profile convolved with an extended PSF. The extended PSF comprises the PSF retrieved from phase-diversity reconstructions (on spatial scales of order of 0.1 arcsec), and two Gaussians corresponding to the stray light (which acts on the wings of the PSF). 

The limb profile fitting is carried out for three spectral windows: 300\,nm, 313\,nm, and 388\,nm. The best-fit parameters, which are used to construct the extended PSF, are shown in Table~\ref{tab:psfpar_me}. The extended PSFs are then used to correct images at all $\mu$ values and in all three wavelengths. RMS contrasts of corrected data are then computed and results are shown in Figure~\ref{fig:cvsmu_mhd} together with theoretical CLVs based on 3D atmospheric models obtained from MURaM and different spectral synthesis codes.

In all studied wavelengths, the observed RMS contrast shows a maximum at disk center ($\mu=1$), and decreases approaching the solar limb. The CLV of the RMS contrasts derived from synthetic images in the NUV results in a peak at intermediate $\mu$ values which is not reproduced by the observations.

In Section~\ref{SL_up}, we have tested the applicability of the extended PSF constructed from fitting solar limb profiles obtained during the 2009 $\sunrise$ flight to correcting active region data recorded during the second flight of $\sunrise$ in 2013. The level-3.1 corrected SuFI images ended up with negative intensities in the pores, indicating an overcorrection for stray light in these images. However, these negative intensities stay well below the noise level so that AR SuFI data can be used reliably for scientific analysis as done in Chapter~\ref{chapter_4}.

We have tested whether using the averaged 1D PD PSF in the model (instead of a Gaussian) or implementing the proper PD PSF corresponding to the 2013 data has a significant effect on the stray light correction. But we have found little effect and came to the conclusion that the error potentially originates from residual jitter, which has a $\mu$ dependent broadening effect, and which is erroneously captured by the limb profile fitting as a stray light contribution, causing over-correction of disk center images. Following this assumption, we have constrained the fit to the limb profiles from 2009 to have, after the deconvolution of an example level-3 disk center image, an RMS contrast equal to that returned from synthesized data. This constrained fit has allowed us to estimate a jitter-free extended PSF.

As shown in Section~\ref{jitter}, using this jitter-free PSF has significantly reduced but not entirely eliminated the over correction. In addition, the jitter couldn't be modeled with a Gaussian. In a future step, another mathematical function should be used to evaluate the contribution of the jitter in broadening the observed limb profiles. However, even if the broadening due to jitter is quantified at the limb, we have little information on how it behaves across the solar disk. We therefore must assume that the effect of residual jitter still contributes to an uncertainty of about $1-5$\% (depending on wavelength, cf.~Table~\ref{tab:RMS_cons}) in the present interpretation of the observed $\sunrise$ intensity contrasts. 

\chapter{Summary and outlook}
\label{chapter_6}

In this thesis, we have analysed high resolution data of two magnetically different solar regions during times of different solar activity levels. The quiet-Sun internetwork (and parts of the network) at disk center ($\mu = 0.97$) was observed by the balloon-borne observatory $\sunrise$ during its first flight on 2009. The first $\sunrise$ flight ($\sunrise$~I) occured during solar cycle minimum when the Sun showed no sign of magnetic activity, while the second flight of $\sunrise$ in June 2013 ($\sunrise$~II) allowed for observing an active region near disk center ($\mu=0.93$).
Two scientific instruments on-board the observatory and operating in different wavelength bands acquired high spatial and temporal resolution data. The spectro-polarimetric data consisting of 2D Stokes maps were collected with IMaX, an imaging magnetograph with a tunable narrow-band filter operating in the photospheric Fe\,{\sc i} line at $5250.2$\,\AA{}. The imaging data were collected with SuFI, a filtergraph with wavelength channels sampling the solar photosphere and lower chromosphere of part of the IMaX region. The brightness of the quiet-Sun data were evaluated in the SuFI wavelengths of 214\,nm, 300\,nm, 313\,nm (OH molecular band), 388\,nm (CN molecular band) and 397\,nm (core of the Ca\,{\sc ii} H line), while the brightness of the active region was measured at 300\,nm and 397\,nm (with a narrower filter than in $\sunrise$~I). Both imaging and spectropolarimetric data were corrected by means of phase-diversity reconstruction techniques for wavefront aberrations caused by the optical system (Section~\ref{intro-PD}). Data were further corrected for instrumental stray light using solar limb profiles (described in Chapter~\ref{chapter_5}). Finally, Stokes data were inverted with SPINOR to retrieve the magnetic field vector along with other physical parameters of the regions under study.

We used quasi-simultaneous time series from both, IMaX and SuFI to study the dependence of the pixel-by-pixel contrast (the intensity relative to the mean quiet-Sun level) on the longitudinal component of the field, $B_{\rm LOS}$, in solar magnetic elements. In Chapter~\ref{chapter_3} we studied this relationship in the quiet-Sun region imaged by $\sunrise$~I and we came to the following conclusions:
\begin{itemize}
\item[$\blacksquare$] The averaged contrast in the visible continuum (at 525.04\,nm) saturates at higher $B_{\rm LOS}$ values, contrary to earlier studies which reported a peak in the contrast at intermediate field strengths and a turnover at higher $B_{\rm LOS}$ values. Examples are \cite{kobel_continuum_2011} with Hinode data and \cite{lawrence_contrast_1993} with SVST data. After degrading our Stokes data to the spatial resolution of Hinode, the peak and the tunrover of the contrast were reproduced. This implies that at the high resolution of IMaX (0.15 arcsec), magnetic features in the quiet-Sun network are starting to be resolved. 
\item[$\blacksquare$] The granulation structure consisting of the magnetic field and continuum intensity distribution in the granules and intergranular lanes, is also resolved. Inspecting the shape of the scatterplot in Figure~3 of Chapter~\ref{chapter_3} in the low field region, the so-called `fishhook' pattern \citep{schnerr_brightness_2011} is very well visible. This shape marks the transition from the nearly field-free granules, via the weakly magnetized regions with an averaged negative contrast to magnetic bright concentrations in the intergranular lanes.
\item[$\blacksquare$] The contrast in the core of the IMaX line which samples heights around 400\,km above the $\tau_{500} = 1$ level \citep{jafarzadeh_inclinations_2014} increases monotonically with $B_{\rm LOS}$. The fishhook shape is less pronounced compared to the IMaX continuum contrast vs. $B_{\rm LOS}$ curve, due to the larger formation height and absence of normal granulation in this wavelength band.
\item[$\blacksquare$] The contrast in the UV (at 214\,nm, 313\,nm, and 388\,nm) is higher than in the visible continuum and reaches the highest recorded values. This is due to the higher sensitivity of the Planck function to temperature variations at shorter wavelengths and the molecular dissociation (in the lines at 313 nm and 388 nm) inside the tubes which renders them hotter than their surroundings at equal optical depths.
\item[$\blacksquare$] The photospheric magnetic field contributes to the brightening of quiet-Sun magnetic features at heights near the temperature minimum. This is indicated by the non-linear dependence of the contrast in the core of the Ca\,{\sc ii} H line on $B_{\rm LOS}$. This dependence is modelled with a logarithmic function instead of the commonly used power-law function (which works in our data only for $B_{\rm LOS}>190$\,G).
\end{itemize}

The intensity contrast variation with the magnetic field strength is revisited in the plage features spread in the active region observed by $\sunrise$~II, during a period of higher activity. Plages are composed of small-scale magnetic elements and differ from those found in network features primarly by their sizes and magnetic environment. We mention here our main findings based on the analysis of these data in Chapter~\ref{chapter_4}:
\begin{itemize}
\item[$\blacksquare$] The averaged continuum contrast around 525.04 nm of AR plages peaks at 850\,G, a value higher than what's earlier reported in e.g.\citep{narayan_small-scale_2010,kobel_continuum_2011}. The intensity then goes below the quiet-Sun level at around 1200\,G. This qualitative behaviour is seen in earlier lower spatial resolution data \citep{kobel_continuum_2011}.

\item[$\blacksquare$] The contrast in the cores of spectral lines (core of  Fe\,{\sc i} and Ca\,{\sc ii} H lines) increases with increasing $B_{\rm LOS}$ and is higher than the contrast in the lower photosphere sampled by the IMaX continuum and the wavelength band of 300\,nm imaged by SuFI (Figure~7 in Chapter~\ref{chapter_4}).

\item[$\blacksquare$] The intensity contrast vs. $B_{\rm LOS}$ relationship at all wavelength bands (except in the IMaX continuum) is modelled succesfully with a logarithmic function similar to the dependence between the two quantities found in the quiet-Sun regions of $\sunrise$~I.   

\item[$\blacksquare$] After identifying the quiet-Sun areas in the FOV of IMaX (QS-2013), we produce plots of their IMaX continuum and line core intensity contrasts against their $B_{\rm LOS}$ values and compare them to the corresponding plots of the quiet-Sun data in 2009 (QS-2009) analysed in Chapter~\ref{chapter_3}. Both curves in both wavelength bands agree qualitatively between the two flights, with the contrast in QS-2009 being larger than in QS-2013 for the same field strength. We attributed this difference to the higher magnetic activity in 2013 observations which affect the convective energy transport in the surroundings of the bright magnetic features. The saturation of the continuum contrast in kG magnetic features of QS-2013 implies that the magnetic field is at least partly resolved by IMaX in the AR observations.

\item[$\blacksquare$] The intensity contrast vs. $B_{\rm LOS}$ in quiet-Sun network (results of Chapter~\ref{chapter_3}) is compared to that of AR plage of $\sunrise$~II, and the plot is shown in Figure~11 of Chapter~\ref{chapter_4}. This comparison is restricted to the wavelengths in common between the two $\sunrise$ flights, namely IMaX (continuum and line core) and SuFI (300\,nm and 397\,nm). In all wavelength bands, the quiet Sun reaches higher contrasts than plages for high values of $B_{\rm LOS}$. The difference in the contrast is large when measured in the continuum, which is formed in the low photosphere, but decreases in higher atmospheric layers (sampled by IMaX line core and core of Ca\,{\sc ii} H). This result is in accordance with previous observational studies \citep{title_differences_1992, kobel_continuum_2011} and model atmospheres of magnetic flux tubes \citep{solanki_photospheric_1987, solanki_continuum_1992}. This points to the role of the interplay between convection and solar magnetic field in determining the contrast of magnetic elements in the lower photosphere. While the effect of convective energy transport is negligible at heights closer to the temperature minimum.

\item[$\blacksquare$] Magnetic features with different sizes coexist in the same region. These features range from small bright points (sizes of $100-200$\,km) to filamentary structures or striations (300\,km$-$400\,km) to micropores (500\,km$-$600\,km), as identified in the continuum intensity maps at 525.04 nm. The plots across such features in a direction perpendicular to the limb of the IMaX continuum intensity and $B_{\rm LOS}$ profiles are quite interesting. While the $B_{\rm LOS}$ shows a common behaviour in all features: peaking in their cores and diminishing towards their granular edges, the IMaX continuum intensity distribution exhibits 3 distinctive behaviours: (1) a maximum value in the center of the tube which decreases towards the edges (in small bright points), (2) a dip below the QS level in the center which increases above that level towards the edges (in micropores), (3) a bright part in the limb direction of the feature and a dark `lane' towards disk center (in intermediate-sized features). The latter distribution is similar to the reported profiles in the G-band observations of limb-ward faculae \citep{hirzberger_solar_2005}. 
  
\item[$\blacksquare$] The spatial photometric and magnetic distributions in magnetic features found in all IMaX images explain the scatter of data points seen in Figure~3 of Chapter~\ref{chapter_4}. We conclude that the shape of the scatterplot is not a consequence of poor spatial resolution as pointed out by studies based on radiative MHD simulations of \cite{rohrbein_is_2011} and \cite{danilovic_relation_2013}.
\end{itemize}
Based on the results and conclusions listed above, further research is to be done on the following topics. Some of these works has already been started, as described in the following subsections.

\section{Simulations of a plage region including pores}
Comparison of properties of  magnetic bright points identified in the quiet-Sun data of $\sunrise$~I to the output of radiative MURaM simulations of a quiet-Sun region was investigated by \cite{riethmuller_comparison_2014}. This comparison needs to be extended to the magnetic features in AR plages studied in Chapter~\ref{chapter_4}. With the help of MURaM, we can simulate an active region (see below) then with SPINOR in its forward computation mode, synthesize the Stokes spectra around the IMaX line, degrade them to match the resolution of our observations and use SPINOR to invert the degraded profiles and retrieve the physical parameters in the same manner as IMaX data were inverted. 

This study is tempting since the data are corrected already for wavefront aberrations and instrumental stray light so that the only degradation sources to the simulations would be the noise and jitter, plus the spatial degradation that cannot be removed by phase diversity. In addition, the magnetic field information is retrieved through the inversions of IMaX data, allowing for a direct comparison with the output of the radiative MHD simulations instead of using proxies such as the circular polarization degree as done in \cite{riethmuller_comparison_2014}. 

\begin{figure}
\centering
\hspace*{-0.5cm}\includegraphics[scale=0.12]{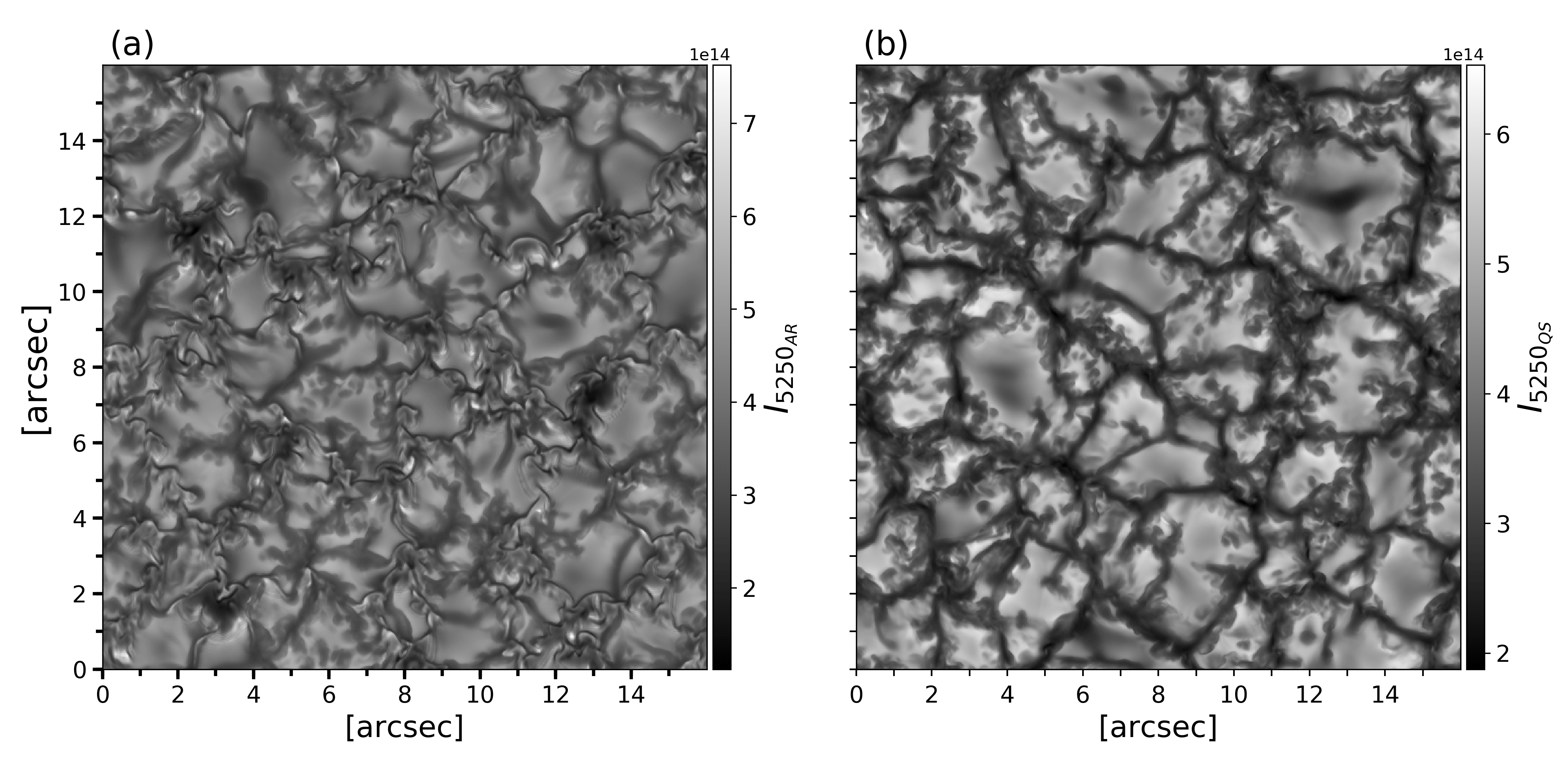}
\caption{\textit{Left panel:} The synthesized continuum brightness from a MHD snapshot corresponding to plage (400\,G vertical initial magnetic field). \textit{Right panel:} Same but for a snapshot of a hydrodynamical run (zero magnetic field). Both are computed for $\mu = 0.93$}
\label{cont_plage_qs_mhd}

\end{figure}

\begin{figure}
\centering
\hspace*{-0.8cm}\includegraphics[scale=0.12]{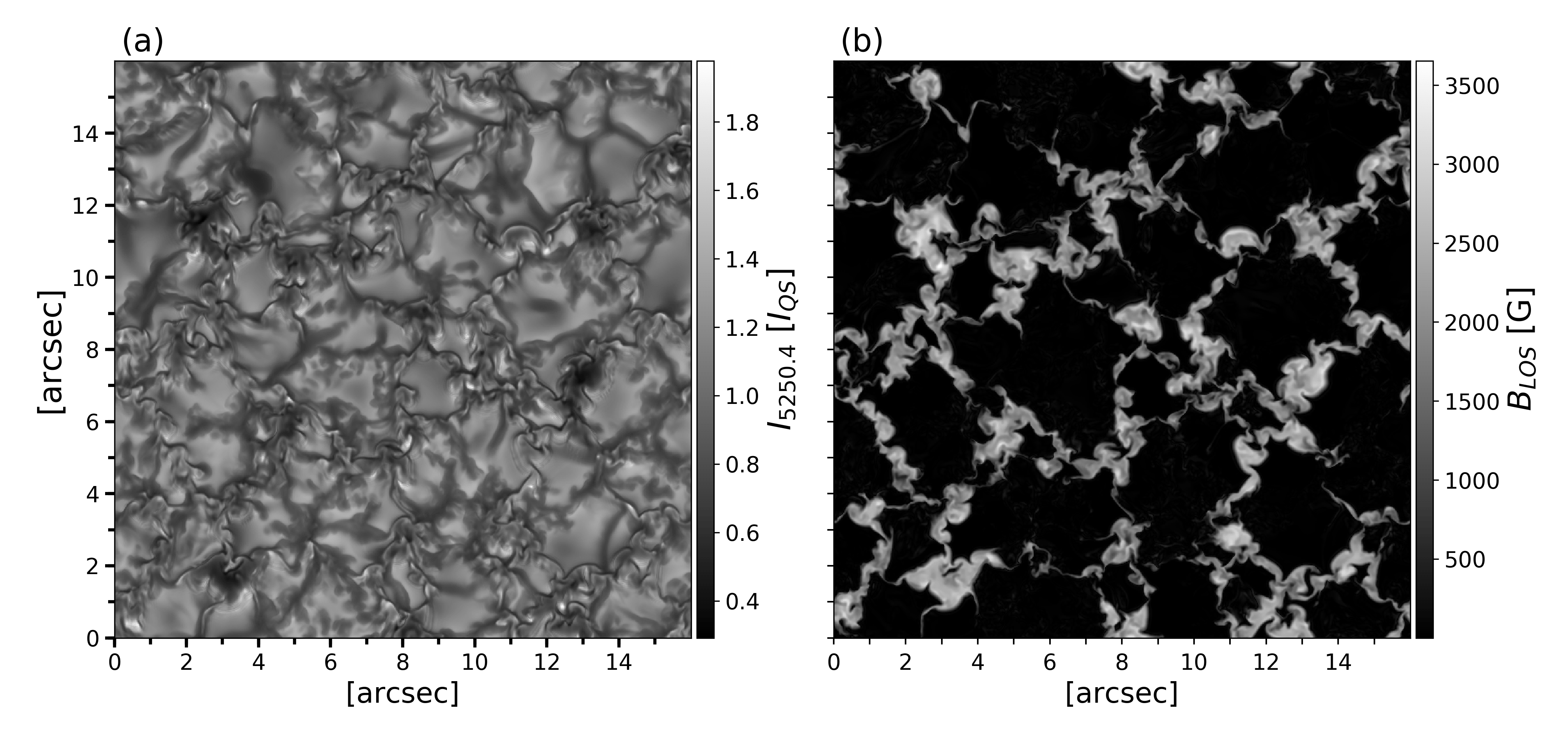}
\caption{\textit{Left panel:} The continuum intensity contrast of a plage region. \textit{Right panel:} The output magnetic field map from MuRAM at log\,$\tau=0$.}
\label{cont_B_mhd}

\end{figure}

Another motivation to carry out this study is the spatial distribution of the continuum brightness found in the different sized magnetic features observed with IMaX. As discussed in the introduction of this thesis (Section~2.1 of Chapter~\ref{chapter_1}) and in the discussion section in Chapter~\ref{chapter_4}, the simulations used so far to inspect this relationship in plages, lack pores. These simulations were used by \cite{danilovic_relation_2013} to asses the effect of the finite spatial resolution on the shape of this relationship in AR plages agreeing with the observational findings of \cite{kobel_continuum_2011}. But these conclusions contradict with our findings that plages magnetic features are resolved in IMaX. The work of \cite{danilovic_relation_2013} should be revised at the wavelength and spatial resolution of IMaX, and with MHD simulations of a solar scene containing plages and pores. Below are the properties of such simulations that I have run and are currently under study:
\begin{itemize}
\item[$\blacktriangleright$] Simulation box size: 12 Mm in both x and y horizontal directions and 3 Mm in the vertical direction (compared to 1.4 Mm in earlier simulations). Deep boxes could allow convection to produce larger magnetic structures such as pores.
\item[$\blacktriangleright$] Cell size: 20 km (0.027 arcsec) in the horizontal direction and 10 km (0.0135 arcsec) in the vertical direction.

\item[$\blacktriangleright$] Introduced unipolar vertical magnetic field strength: $B_z$ = 400\,G (compared to the usually employed 200\,G), which describes strong plage. The magnetic field is introduced to a hydrodynamical simulation and run for 5 hours to reach a thermally relaxed state (see Figure~\ref{cont_B_mhd}).

\item[$\blacktriangleright$] Grey simulations are carried out for 3 solar hours followed by non-grey radiative simulations for 2 solar hours. This allows the simulation to relax sufficiently. 

\item[$\blacktriangleright$] The optical depth unity level (at 500\,nm) reached in the non-grey simulations is on average 800\,km below the upper box boundary.

\item[$\blacktriangleright$] Stokes spectra of the Fe\,{\sc i} $5250.2$\,\AA{} line are synthesized with a forward calculation done by SPINOR. The synthesis is done in the range 5249.808\,\AA{} and 5250.808\,\AA{} with a spectral sampling of 10 m\,\AA{}. Two spectral lines are also synthesized besides the Land\'e $g=3$ Fe\,{\sc i} line $5250.2$\,\AA{}, the Co\,{\sc i} line at 5249.997\,\AA{} and the Fe\,{\sc i} line at 5250.645\,\AA{}.

\item[$\blacktriangleright$] The synthesis is done for the same heliocentric angle as the $\sunrise$~II observations ($\mu = 0.93$)

\item[$\blacktriangleright$] The synthesized continuum intensity map is computed by evaluating at each pixel the intensity at $\Delta \lambda=230$ m\,\AA{}  from the line center (close to $\Delta \lambda=270$ m\,\AA{} of the IMaX continuum point).

\item[$\blacktriangleright$] Contrast is computed by normalizing to the mean quiet-Sun level taken as the average of a synthesized continuum image from a hydrodynamical simulation (with zero magnetic field). It is also computed for $\mu = 0.93$ (see Figure~\ref{cont_plage_qs_mhd}).

\end{itemize}

By simple inspection of both, the intensity contrast and LOS magnetic field maps, one can notice the difference in the appearences of magnetic features compared to our observed data. For example, the active region observed by $\sunrise$~II contains a pore with a penumbra-like appendage and a flux emergence region, which are completely missing from our simulations of "plage" regions. Therefore, the intensity distribution seen across the IMaX features might not be reproduced by simulations, given that inclined magnetic fields are hard to be produced without the existence of large magnetic concentrations. 
\section{Center-to-Limb Variation (CLV) study of contrast}
Our study on the brightness of solar magnetic elements in Chapters~\ref{chapter_3} and \ref{chapter_4} was restricted to data observed at/near the center of the solar disk. This allows for studying the thermal structure of the deep interiors and right above the flux tubes since at disk center, the line of sight is parallel to the solar surface normal and magnetic flux tubes are vertical (due to their buoyancy). 
Nevertheless, an extension of this study to cover all solar disk positions is needed to constrain the center-to-limb dependence of the intensity contrast of network and faculae on wavelength and magnetic field strength. Studies based on full-disc observations were done, but at low spatial resolution of 1 to 4 arcsec  \citep{ortiz_intensity_2002, yeo_intensity_2013}. Such studies are missing the contribution from the granulation, which is succesfully spatially resolved with $\sunrise$.

Moreover, such CLV studies in the UV are not yet available, and $\sunrise$ delivered unprecedented data in the NUV down to 200\,nm, for both, the quiet and active Sun. Such a study is needed to improve solar irradiance modelling, especially in the UV part of the spectrum: by referring to high resolution observations of UV intensity contrasts in different magnetic regions, during times of different magnetic activity levels, and covering a wide range of $\mu$ values, one can test the reliability of the modelled UV spectrum. 

Finally, a CLV study of the statistical properties of solar magnetic elements (as explained in the previous subsection) seems promising to constrain theoretical models of the 3D geometrical structure of these elements. Off disk center data are available from both $\sunrise$ flights. The first steps would be to correct them for wavefront aberrations and for stray light after improving the model described in chapter~\ref{chapter_5} (by modelling the jitter), and to invert the corresponding polarimetric data. This CLV study is tempting, and it will reveal, at the high spatial resolution of our data, new information on the magneto-convective properties, height dependence of the temperature inhomogeneities, and geometry of small-scale magnetic elements on the solar surface.

\section{Ca\,{\sc ii} H brightness and inclined magnetic field lines}
In addition to probing the photospheric brightness, we have focused in Chapters~\ref{chapter_3} and \ref{chapter_4} on the emission from the lower chromosphere sampled by the Ca\,{\sc ii} H line core and imaged by SuFI. We have inspected the pixel-by-pixel correspondence of the integrated brightness across this wavelength band with the strength of the longitudinal magnetic field in the quiet-Sun internetwork/network and active region faculae at disk center. However, at the fine spatial resolution of SuFI and IMaX, care has to be taken when associating IMaX photospheric magnetograms (originating from a mean height of 100\,km above the $\tau_{500}$ = 1 level) with the filtergrams of the Ca\,{\sc ii} H line core which has a considerable contribution from the lower chromosphere and with mean formation height around 500\,km (with the exact value depending on the solar region and employed filter). The pixel-by-pixel correspondence (scatterplots shown in Figure~8 in Chapter~\ref{chapter_3} and Figure~6 in Chapter~\ref{chapter_4}) might not be valid at some locations since the brightening of a certain pixel in the calcium image might not be caused by the magnetic field line whose footpoint is located at the same pixel position in the corresponding IMaX magnetogram. That is due to two main factors: the expansion of magnetic features with height (in both QS and active region), and the inclination of the magnetic field vector at some locations, which in the case of the AR data of $\sunrise$~II, results from bending of magnetic field lines in magnetic elements close to pores, forming magnetic canopies. These structures are identified as `slender fibrils' in the SuFI calcium images and they generally lie over regions with weak magnetic fields (see Figure~2 in \cite{jafarzadeh_slender_2017}). Note that they could also emanate from pores (or micropores) and extend to nearby plages magnetic elements, so that the Ca\,{\sc ii} H brightness of the latter has a contribution from both, the corresponding field line and nearby inclined fields. 

\begin{figure}
\centering
\hspace*{-1cm}\includegraphics[scale=0.15]{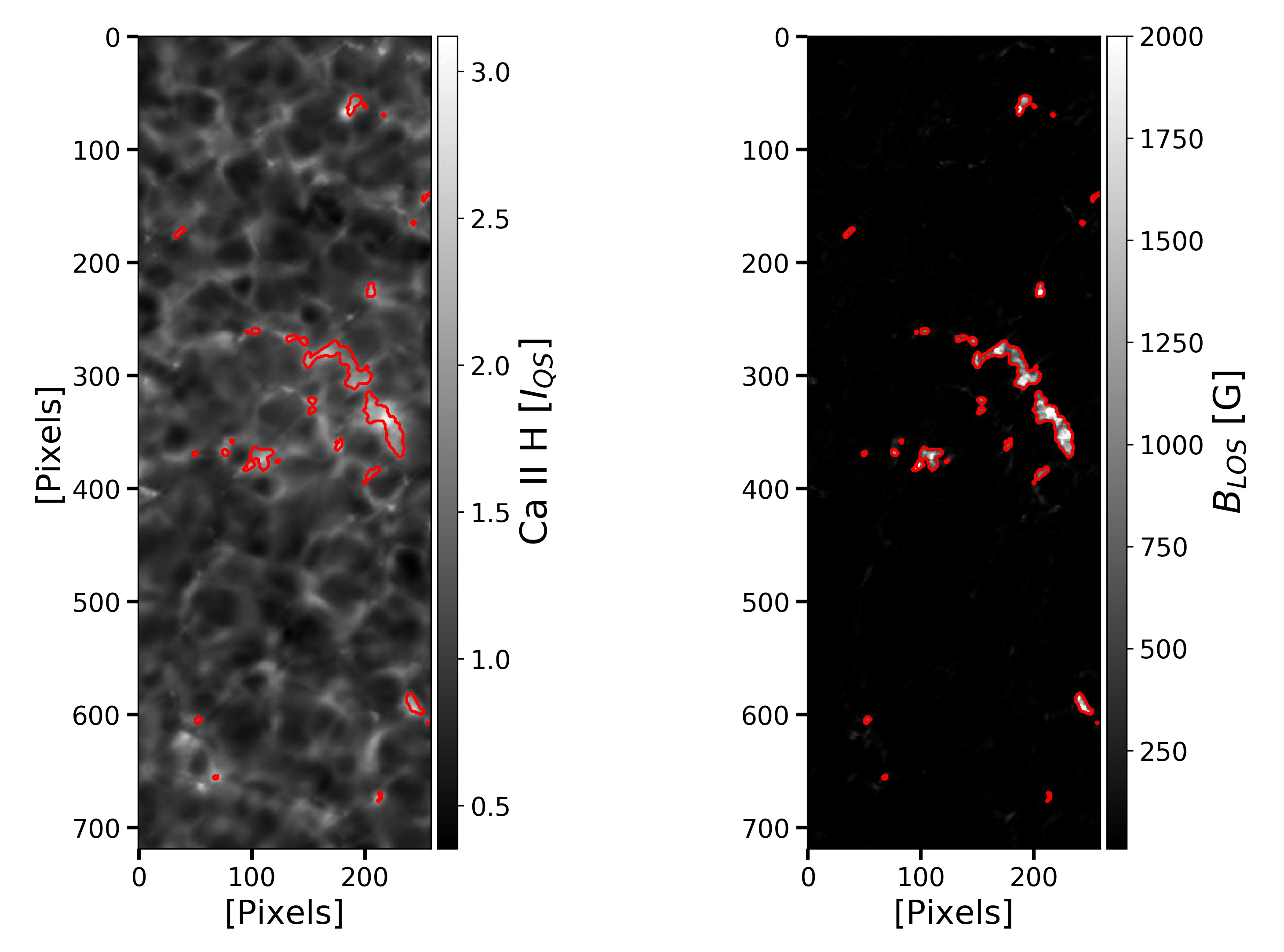}
\caption{Ca\,{\sc ii} H contrast image (left panel) with the co-spatial and co-temporal usigned LOS magnetic field map derived from IMaX data inversions (right panel). The maps are taken from the quiet-Sun time series studied in Chapter~\ref{chapter_3}. The red contours overlaid on both images enclose pixels with $B_{\rm LOS} > 400$\,G and $B_{\rm LOS} <  2500$\,G.  }
\label{cal_contours_qs}
\end{figure}

\begin{figure}
\centering
\hspace*{-1cm}\includegraphics[scale=0.15]{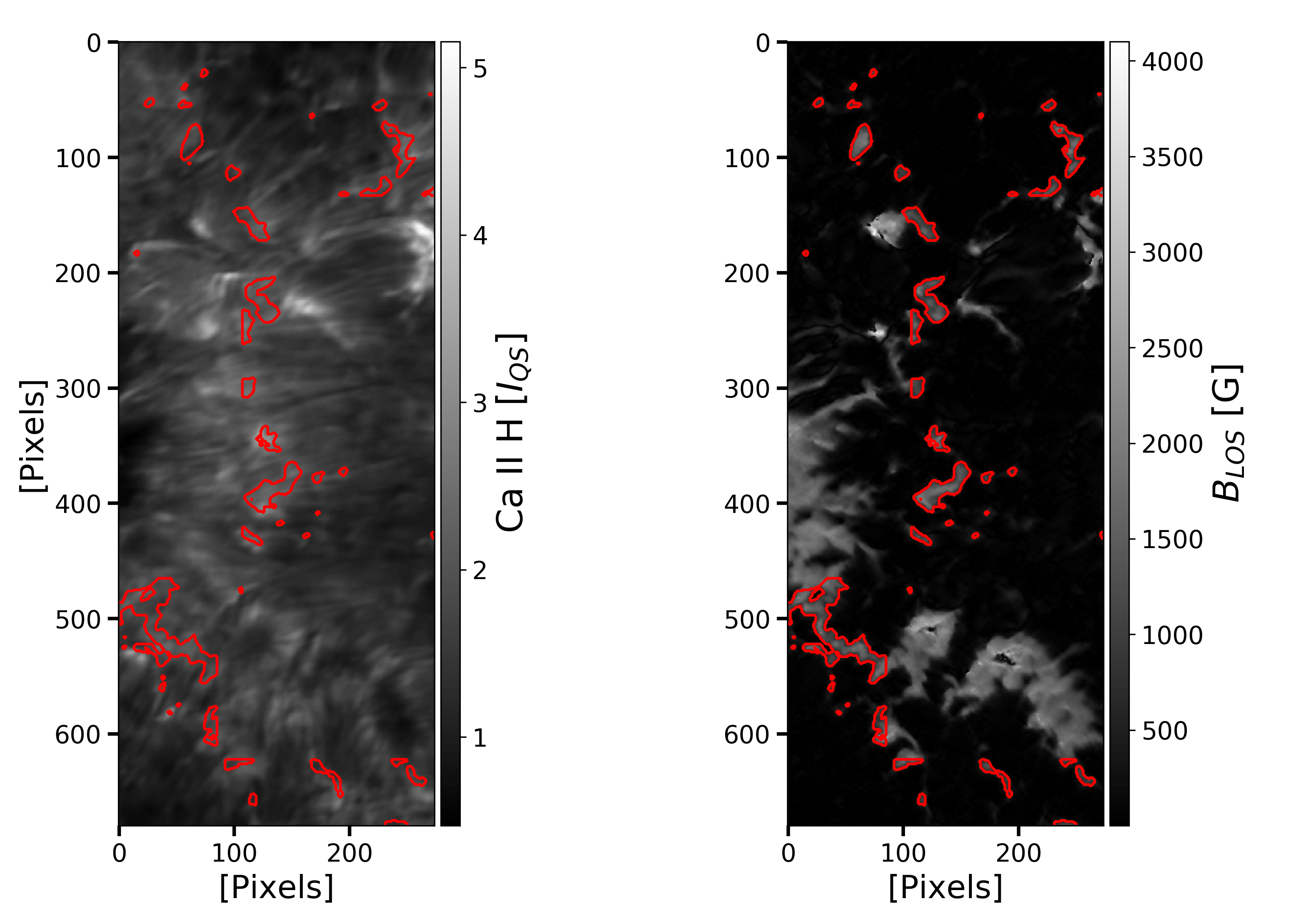}
\caption{Same as Figure~\ref{cal_contours_qs} but for the maps taken from the active region time series studied in Chapter~\ref{chapter_4}. The red contours overlaid on both images enclose pixels with $B_{\rm LOS} > 400$\,G and $B_{\rm LOS} <  2500$\,G, excluding the contribution from pores.}
\label{cal_contours_AR}
\end{figure}

In the quiet-Sun data analysed in Chapter~\ref{chapter_3}, the picture is less complicated since fibrils are non-existent and the kG magnetic field is considered to be mainly vertical in the absence of large magnetic features (pores and micropres) that form with higher average magnetic flux. Although some magnetic bright points seem to be shifted with respect to the corresponding magnetic flux concentration as can be seen in Figure~\ref{cal_contours_qs}, where I show the contours enclosing pixels with $B_{\rm LOS} > 400$\,G and $B_{\rm LOS} <  2500$\,G. The chromospheric bright point on the top right of the SuFI image is slightly off the contour, indicating a possible inclination of the magnetic field lines originating from the same contoured flux concentration in the magnetic field map. Figure~\ref{cal_contours_AR} shows an example image from the AR data analysed in Chapter~\ref{chapter_4} in the Ca\,{\sc ii} H band and the corresponding $B_{\rm LOS}$ map. Magnetic bright points (or grains) or clumps of them are hard to be identified since fibrils dominate in the image. 

The effect of inclined magnetic field lines could be disregarded for the pixel-by-pixel correspondence of the IMaX continuum and 300 nm contrasts with $B_{\rm LOS}$ in plage features due to the smaller height difference between the layers at which these wavelengths form and the height at which the IMaX line senses the magnetic field.

To derive the right dependence and take the topology of the magnetic field lines into account, we can proceed as follows: with the assumption that the mean height difference between IMaX magnetograms and calcium images is around 450 km as computed by \cite{jafarzadeh_structure_2013}, we can compute for each pixel of position (1) on the magnetograms the position (2) of the mganetic field vector at the height of formation of Ca\,{\sc ii} H, using both the inclination $\gamma$ and azimuthal angle $\phi$ computed from SPINOR inversions of IMaX spectropolarimetric data. The brightness at pixel position (2) is associated with the magnetic field value at pixel position (1). However, this method has three main drawbacks:
\begin{itemize}
\item[$\bullet$] Because of the 180 ambiguity, using the azimuthal information to compute the exact position of the calcium brightening associated with the magnetic field line is not reliable.

\item[$\bullet$] It is not known exactly what happens to the magnetic field line once it emerges out of the photosphere. It can return to the solar surface as a low-lying loop and never reach the layer at which the calcium line core is formed.

\item[$\bullet$] For weaker signals noise leads to strongly exaggerated inclinations \citep{borrero_inferring_2011, jafarzadeh_inclinations_2014}. 
\end{itemize}
Another method would be to refer to the non-linear force-free and magneto-static field extrapolations above the quiet Sun internetwork and active region done by \cite{wiegelmann_evolution_2013} and \cite{wiegelmann_magneto-static_2017}, respectively. The magnetic field vector is extrapolated to the corona using IMaX photospheric measurements of the magnetic field as a boundary condition. With such extrapolations, we can estimate at each pixel the location of the magnetic field vector and track its trajectory at the height of formation of Ca\,{\sc ii} H line core. The scatterplot can be reproduced with the new associated calcium brightness-$B_{\rm LOS}$ values. Although, the problem of the unreliable information on the magnetic inclination for weak fields persists. 
\section{Sunrise III} 
With the promised third flight of $\sunrise$, new (and upgraded) instruments will be installed on-board, including the Sunrise UV Spectropolarimteric Imager (SUSI). This instrument will be recording Stokes data of about 150 chromospheric lines, allowing to probe not only the photospheric magnetic field, but that of the chromosphere. The flight is scheduled to be launched on June 2021, in a period of low solar activity. Therefore, we can learn more about the dynamics of magnetic field concentrations in small-scale magnetic elements by combining simultaneous obeservations of the brightness and magnetic field configuration in solar heights ranging from the lower photosphere to the middle chromosphere.

\bibliography{MyLibrary}

\newpage

\appendix

\thispagestyle{empty}
\vspace*{2.2cm}
{\Huge\bf Appendix}
\chapter{Non Parametric Regression}
\label{appendix_A}

\section{Kernel Smoothing}

Kernel smoothing is a non-parametric regression technique where a non-linear relationship between two quantities, in our case the LOS component of the magnetic field $B_{\rm LOS}$, and the corresponding contrast value $C$ is approximated locally. In the following we write $B$ for $B_{\rm LOS}$ for simplicity. At each field value $B_0$ we want to find a real valued function $\hat{m}_h$ to compute the corresponding contrast value, $\hat{m}_h(B_0)$ or the conditional expectation of $C$ given $B_0$, i.e., $E[C|B_0]$, which is the outcome of the smoothing procedure, and is called \textit{the smooth} \cite{tukey_exploratory_1977}. The curve joining the smooth values at each $B_0$ in our data set is called the non-parametric regression or NPR curve. To do this, the function $\hat{m}_h$ is estimated at the $n$ neighboring data points,  $\{B_i,C_i\}$, falling within a bandwidth $h$ around $B_0$, with $i$ ranging from $1$ to $n$, and weighted by a kernel density function. The latter is defined for $B_0$ as:
\begin{equation}
\hat{f}(B_0) = \frac{1}{n h} \Sigma_{i=1}^{n} \mathcal{K}(\frac{B_0-B_i}{h})
\end{equation}

Here $\mathcal{K}$ is a kernel centered at $B_0$, giving the most weight to those $B_i$ nearest $B_0$ and the least weight to points that are furthest away. The shape of the kernel is determined by the type of the used kernel (often Gaussian), and by the magnitude of the bandwidth or \textit{smoothing parameter} $h$.
\\

Following a derivation that can be found in \cite{nadaraya_estimating_1964} and \cite{watson_smooth_1964}, the following expression for $\hat{m}_h(B_0)$ is found:
\begin{equation}
\hat{m}_h(B_0) = \frac{n^{-1}\Sigma_{i=1}^{n} \mathcal{K}_{h}(B_0-B_i)C_i}{n^{-1} \Sigma_{i=1}^{n}\mathcal{K}_{h}(B_0-B_i)},
\label{SA}
\end{equation}

which is called the \textit{Nadaraya-Watson estimator}.
\\
\subsection{Local averaging}

For local averaging, we compute at each $B_0$ a weighted average of all the $n$ data points $\{B_i,C_i\}$ that fall within the bandwidth $h$, with $i$ ranging from $1$ to $n$. Hence, the process of kernel smoothing defines a set of weights $\{W_{hi}\}_{i=1}^{n}$ for each $B_0$ and defines the function $\hat{m}_h$ as:

\begin{equation}
\hat{m}_h(B_0) = \frac{1}{n} \Sigma_{i=1}^{n} W_{hi} (B_0) C_i \\.
\label{weights}
\end{equation}

Comparing Eq.~\ref{weights} with Eq.~\ref{SA}, the weight sequence is then defined by:

\begin{equation}
W_{hi} (B_0) = \frac{\mathcal{K}(\frac{B_0-B_i}{h})}{n^{-1} \Sigma_{i=1}^{n}\mathcal{K}(\frac{B_0-B_i}{h})}
\end{equation}

\subsection{Local polynomial smoothing}
Apart from local averaging, the second kernel regression method we tested is local-polynomial smoothing. There the set of $n$ data points $\{B_i,C_i\}$ around each field strength value $B_0$ are fit with a local polynomial of degree $q$:

\begin{equation}
\hat{m}_h(B) = a_0-a_1(B-B_0)-...-a_q\frac{(B-B_0)^q}{q!}\\.
\label{taylor}
\end{equation} 
In this case, the best-fit parameters ($a_0, a_1,...,a_q$) are computed via least-squares minimization techniques, i.e., the parameters that minimize the following function: 

\begin{equation}
\Sigma_{i=1}^{n} \mathcal{K}(\frac{B_0-B_i}{h})(C_i-a_0-a_1(B_0-B_i)-...-a_q\frac{(B_0-B_i)^q}{q!})^2 \\.
\label{LPq}
\end{equation}

The smooth is the value of the fit at $B_0$, i.e., $\hat{m}_h(B_0)$, which according to Eq.~\ref{taylor} is simply $a_0$. 

This procedure is applied at each $B_0$ value we have in our data, and the curve joining the smooth values $\hat{m}_h(B_0)$ is the NPR curve.

\subsection{Tests using non-parametric regression}
\label{npr_tests}
As mentioned in the main text, these techniques are applied to the various scatterplots shown in the paper to test the validity of our binning method. In Fig.~\ref{bc_smooth} we showed the scatterplot of the IMaX continuum contrast vs. $B_{\rm LOS}$ with the NPR curve plotted in green. It should be mentioned here that using a local averaging (Eq.~\ref{SA}) or a local-polynomial fit (Eq.~\ref{taylor}) returns the same result, except at the boundaries, where the curve resulting from the local-polynomial fit is smoother. The reason is the non-equal number of data points around $B$ when the latter is close to the boundary, which leads to a bias at the boundary upon locally averaging the contrast values there, meaning that the computed average will be larger than one would get if data points were symmetrically distributed around $B_0$. Using a high order polynomial reduces this bias at the boundaries, but increases the variance. 
We have tested different orders and saw that linear, quadratic and cubic polynomials give very similar results. For all the scatterplots analysed here, we only show the cubic local-polynomial ($q=3$) regression curve with a bandwidth of 6\,G\footnote{The bandwidth is taken as the square root of the covariance matrix of the Kernel used.}.
\\

\begin{figure*}
\centering
\includegraphics[width=\textwidth]{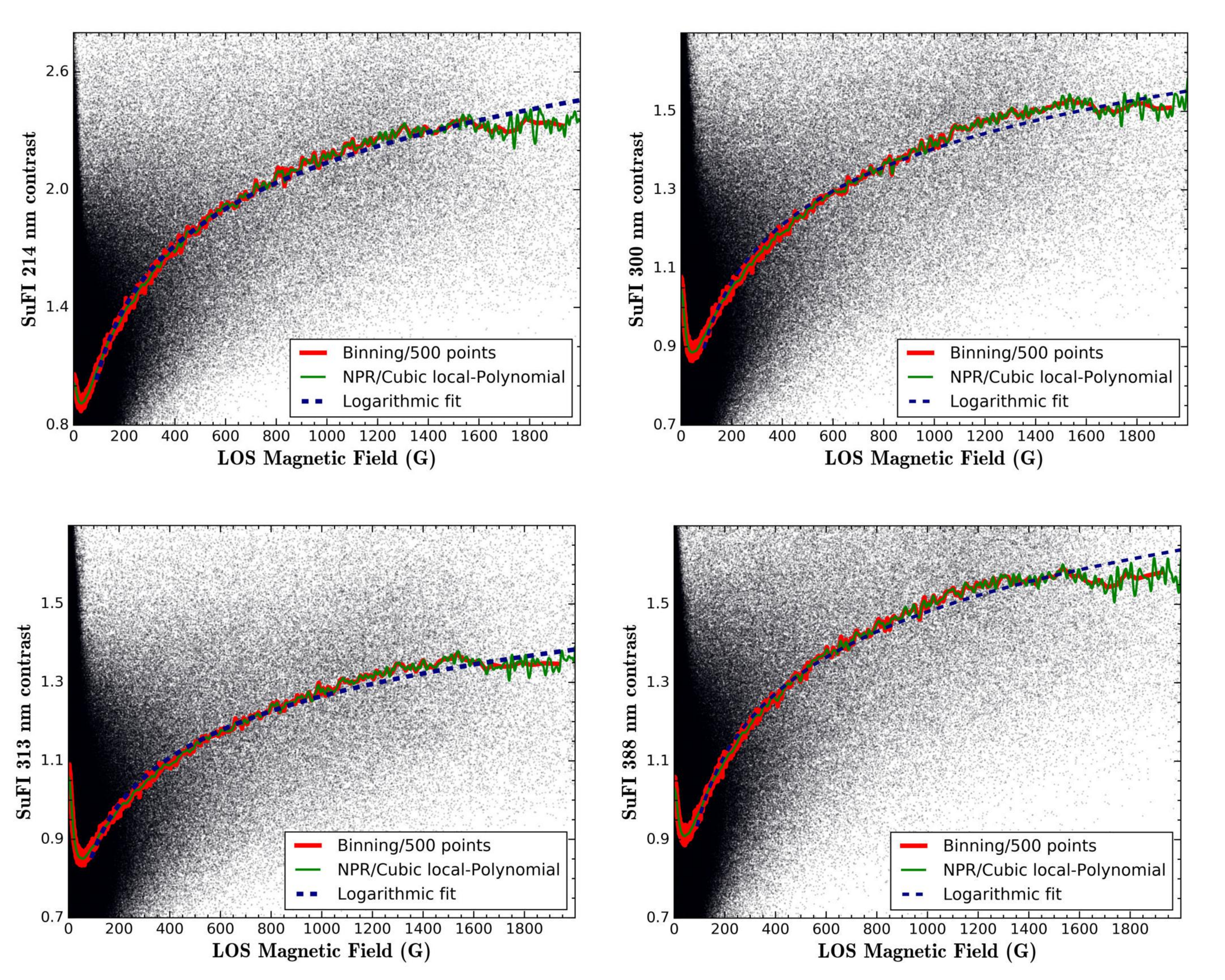}
\caption{Same scatterplots as in Fig.~\ref{sufi} of the NUV contrast vs. $B_{\rm LOS}$. The binned values are plotted in red, the NPR curves in green, and the logarithmic fits in dashed blue, starting at 90\,G.}
\label{sufi_nonparam}
\end{figure*}

Figure \ref{sufi_nonparam} shows the scatterplots analysed in Sect.~\ref{uv_vs_B}, for the NUV contrast vs. $B_{\rm LOS}$ at 214 nm, 300 nm, 313 nm, and 388 nm. The binned data points are plotted in red, the NPR curves in green, and the logarithmic fits (starting from 90\,G) in dashed blue.
The non-parametric regression curves again agree almost perfectly with the binned data (although the NPR curves are smoother, with less scatter than the binned values at small $B_{\rm LOS}$ values), and they agree very well with the logarithmic fits for almost all magnetic field values above 90\,G.\\

\begin{figure}
\centering
\includegraphics[scale=0.2]{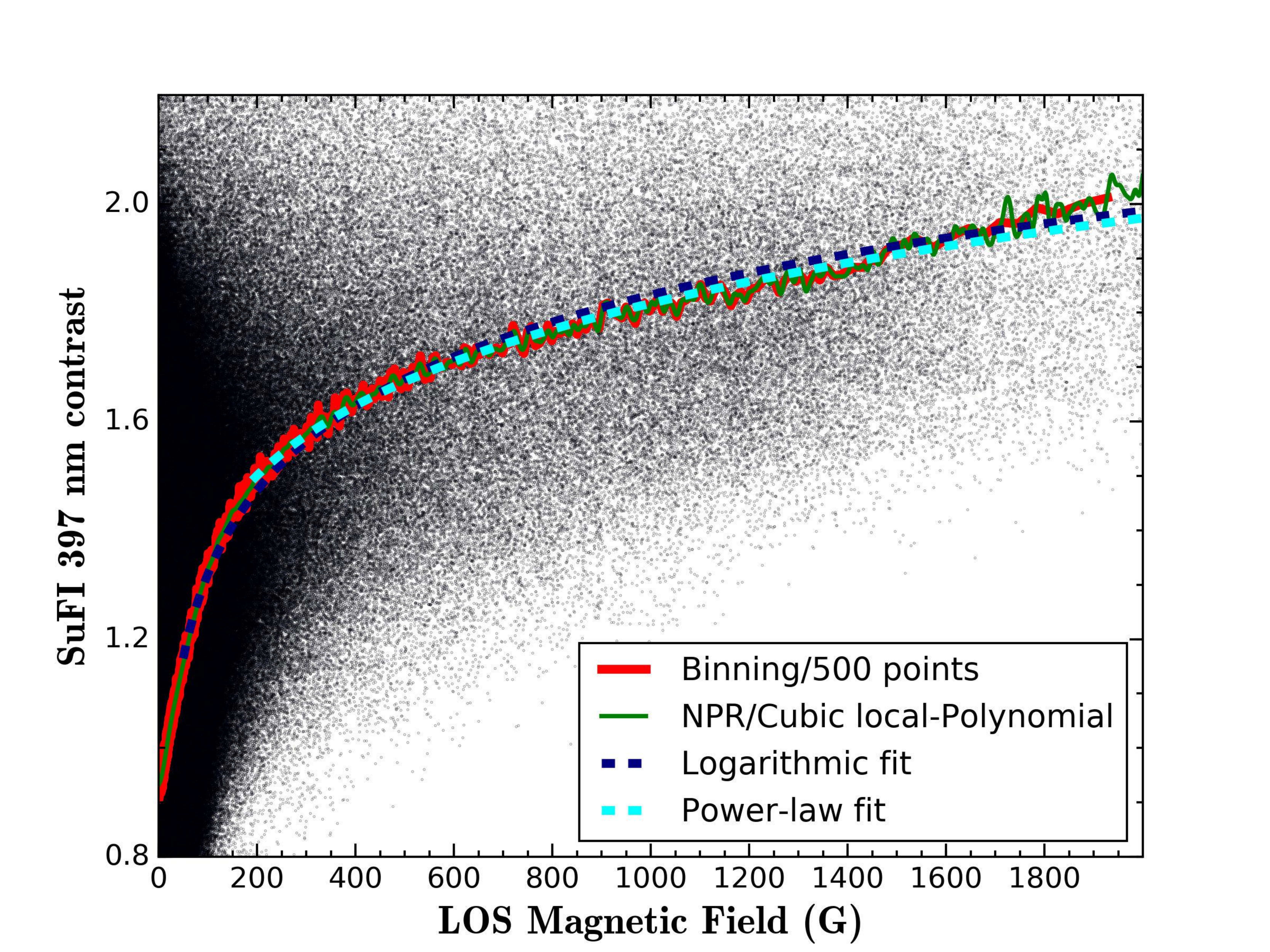}
\caption{Same scatterplot as Fig.~\ref{b_ca} of the Ca\,{\sc ii} H contrast vs. $B_{\rm LOS}$. The red curve is the binned contrast values. The NPR curve is plotted in green, with a bandwidth of $h$=6\,G, the dark blue dashed curve is the logarithmic fit to the binned curve lying above $B_{\rm LOS}$ = 50\,G, and the light blue dashed curve is the power-law fit starting at 190\,G. }
\label{b_ca_smooth}
\end{figure}

Finally, we show in Fig.~\ref{b_ca_smooth} the scatterplot of SuFI 397 nm Ca\,{\sc ii} H contrast vs. $B_{\rm LOS}$. The NPR curve is overplotted, along with both, the logarithmic fit starting at 50\,G, and the power-law fit starting at 190\,G. In addition to the similarity between the binned values plotted in red and the NPR curve (in green), it can be inferred from this plot that both parametric model fits reliably represent the variation of the Ca\,{\sc ii} H contrast vs. $B_{\rm LOS}$, since they also agree with the non-parametric regression curve for almost all values of $B_{\rm LOS}$ larger than the threshold above which the fits are applied.\\                                                                                                                                                                                                                                                                                                                                                                                                                                                                                                                                                                                                                                                                                                                                                                                                                                              

\chapter{The quiet Sun's mean intensity in SuFI data}
\label{appendix_B}

We use the dark corrected (level~$0.1$) data taken over nearly 4 hours (from June 12, 22:15:37 UT to June 13, 01:58:00 UT) to visualize the day-to-night cycles of the mean quiet-Sun intensity at 300\,nm and 397\,nm. Quiet Sun images are acquired for more than an hour at the beginning of the series, just prior to the 1\,h spent observing the active region, and in three intervals after the end of the AR observations. 

\begin{figure*}[ht!]
\begin{center}
\hspace*{-3cm}\includegraphics[scale=0.56]{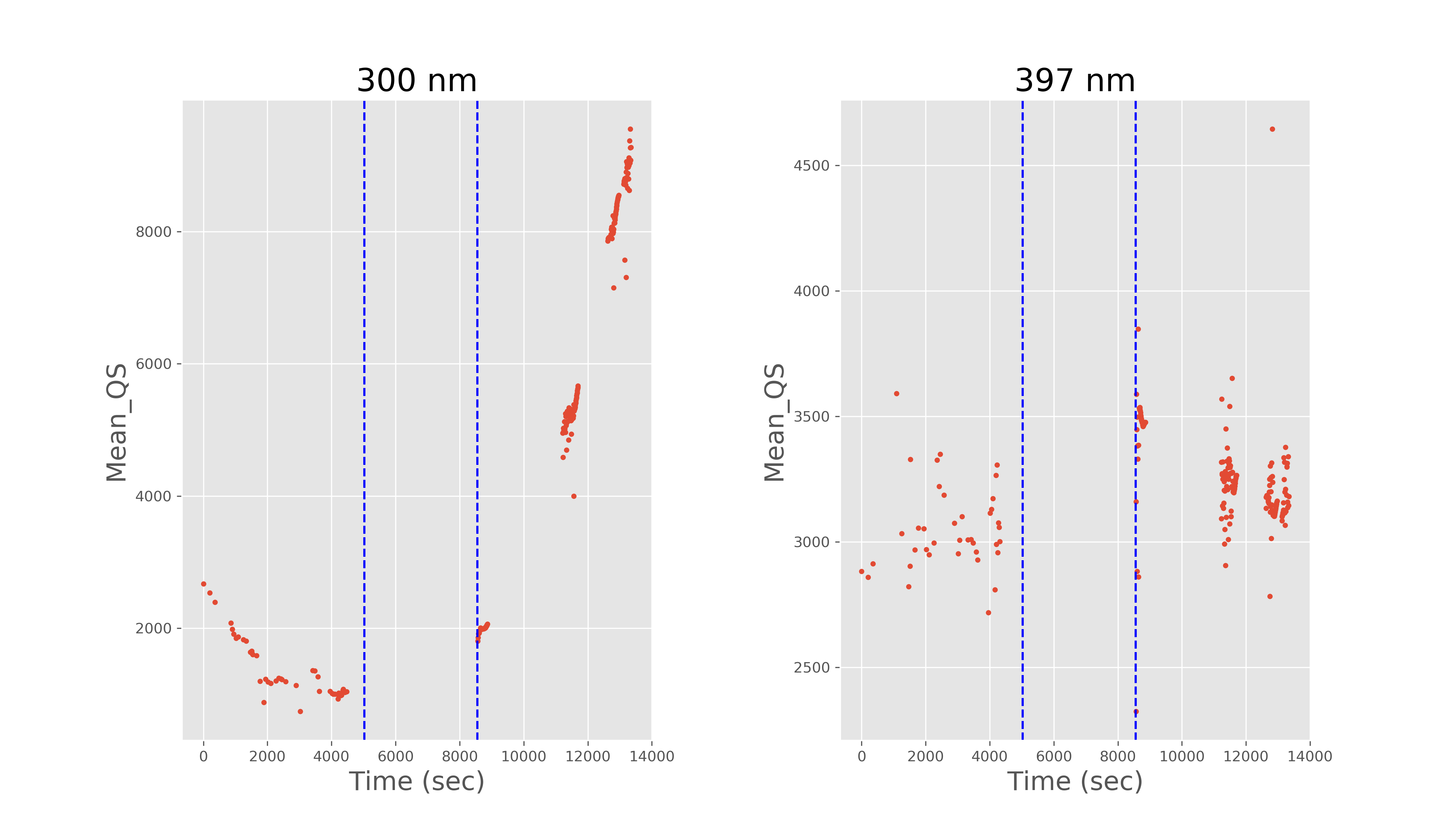}
\caption{Part of the day-to-night cycle of the photon flux at 300\,nm and 397\,nm for quiet-Sun images taken at disk center.
}
\end{center}
\label{cycle}
\end{figure*}

We plot for both wavelengths the averaged level~$0.1$ QS intensity ($I$) versus time ($t$), with the time at which the first image of the time series was recorded being referred to as $t=0$. The day-night cycle is clearly visible in the left panel of Figure~\ref{cycle}. The dashed blue lines delimit the 1 hour observation of the active region, which was partly analysed in this work.

For fitting the day-to-night cycles we use the Beer-Lambert law, which gives the amount of flux (of original value $I_0$) absorbed by a medium with a Rayleigh airmass of $m$: 

\begin{equation}
I = I_0  \times e^{- m\tau_{\lambda}}.
\label{beer}
\end{equation}

$\tau_{\lambda}$ is the optical depth of the terrestrial atmosphere at a given wavelength $\lambda$. The Rayleigh airmass depends on the Sun's elevation angle, $\phi$ \citep{pickering_southern_2002}, and for a spherically-symmetric atmosphere it is given by:

\begin{equation}
m = \frac{1}{sin(\phi +\frac{244}{165 + 47\phi^{1.1}})}.
\end{equation} 

\begin{table}[ht!]
\caption{Best-fit parameters at each wavelength obtained by using Eq.~\ref{beer} and applied to $I$ vs.~$m$ scatterplots. }
\label{fit_params}
\centering
\begin{tabular}{c c c}
\hline \hline

Wavelength (nm) &$I_0$ & $\tau_{\lambda}$  \\
\hline
300 &  188579.7 & 0.3 \\
\hline
397 & 3198.101 &  -0.0003 \\
\hline
\hline
\end{tabular}

\end{table}

\begin{figure*}[ht!]
\centering
\hspace*{-2cm}\includegraphics[scale=0.1]{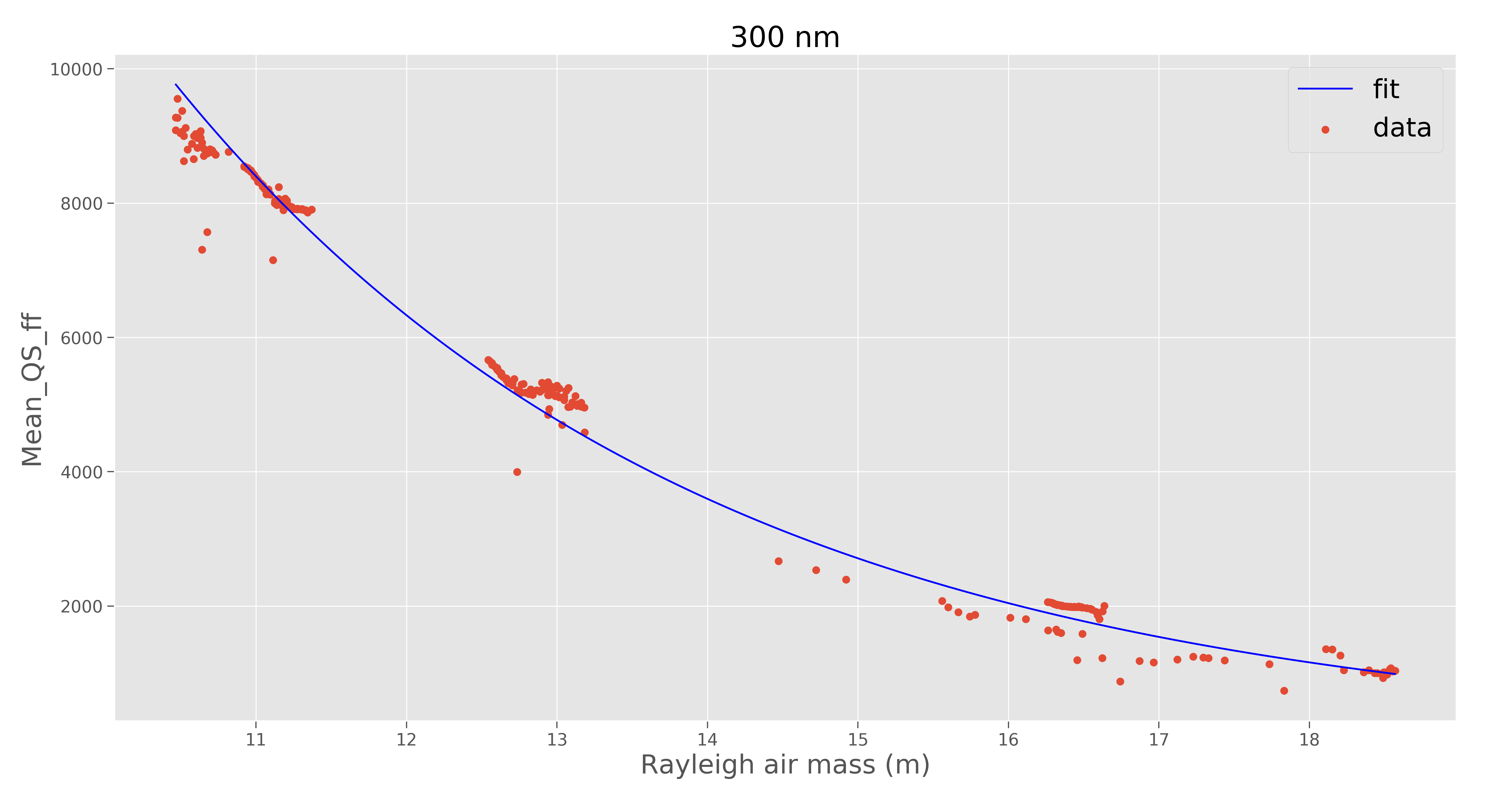}
\caption{Red data points: the 300\,nm quiet-Sun photon flux vs. $m$, the air mass factor during the 4 hours observing period. The best-fit blue curve was calculated according to Equation~\ref{beer} with the parameters given in Table~A1.}
\label{I_m_300}
\end{figure*}

\begin{figure*}[ht!]
\centering
\hspace*{-2cm}\includegraphics[scale=0.1]{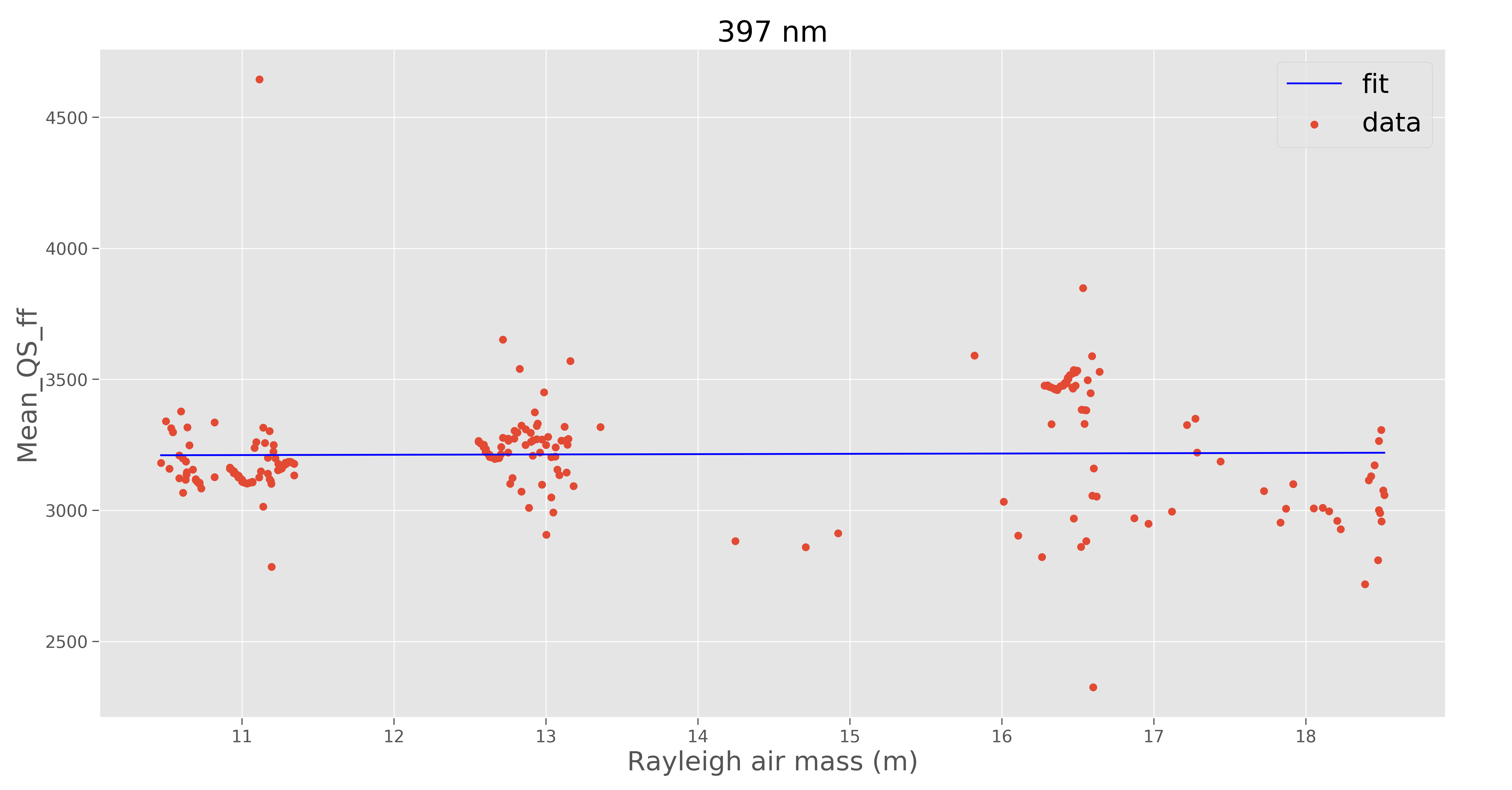}
\caption{Same as Figure~\ref{I_m_300}, but for the 397\,nm data.}
\label{I_m_397}
\end{figure*}

For each wavelength, $I_0$ and $\tau_{\lambda}$ are determined from a fit to the $I$ vs.~$m$ data based on Eq.~\ref{beer}. They are shown in Table~A1. Plots of $I$ versus $m$ and the corresponding fits are shown in Figures~\ref{I_m_300} and \ref{I_m_397} for 300\,nm and\,397 nm, respectively.
As expected, the day-to-night cycle variation is negligible at 397 nm since the average quiet-Sun intensity does not vary with time (or elevation angle). Therefore, the average of the flatfields is the average QS intensity. 

At 300\,nm, we use the best-fit parameters to evaluate $I$ at the airmass values of our AR data. The $I$ values determined in this way are the true mean quiet-Sun intensities. We normalize each image in our level 3.1 data with the corresponding evaluated mean after restoring the original flux. The latter is simply the product of every level~$3.1$ pixel value with the averaged flatfield to which our data were normalized (assuming the the straylight-correction and phase-diversity reconstruction do not affect the mean value of the image).\\

\chapter*{Publications\markboth{Publications}{Publications}}
\addcontentsline{toc}{chapter}{Publications}
\begin{flushleft}

{\bf\large Refereed publications}\\
\begin{itemize}
\item Kahil, F., Riethm\"{u}ller, T. L., and Solanki, S. K., 2017, "Brightness of Solar Magnetic Elements As a Function of Magnetic Flux at High Spatial Resolution", The Astrophysical Journal Supplement Series, 229(1), 12  
\item Kahil, F., Riethm\"{u}ller, T. L., and Solanki, S. K., 2019, "Intensity contrast of solar plage as a function of magnetic flux at high spatial resolution", Astronomy and Astrophysics, 621, A78  \\
\end{itemize}
\end{flushleft}

\begin{flushleft}
{\bf\large Conference Contributions} \\

\begin{enumerate}
\item On the contrast of solar magnetic elements observed by SUNRISE, Hinode-12: The many Suns, Granada, Spain, September 10 $-$ 13, 2018 (Poster)
\item On the contrast of solar magnetic elements in the quiet Sun and active region plage, International Astronomical Union General Meeting, Vienna, Austria, August 20 $-$ 31, 2018 (Poster)
\item Photometric and Magnetic Properties of Solar Plage Observed by SUNRISE, Annual Meeting of the German Astronomical Society, G\"{o}ttingen, Germany, September 18 $-$ 22, 2017 (Talk)
\item Solar Magnetic Elements at High Spatial Resolution, 15th European Solar Physics Meeting, Budapest, Hungary, September 04 $-$ 08, 2017 (Poster)
\item Brightness of solar magnetic elements as a function of magnetic flux at high spatial resolution, Solar Polarization Workshop 8, Florence, Italy, September 12 $-$ 16, 2016 (Talk)
\item Longitudinal magnetic field computation in the quiet Sun: Inversions vs. Centre-of-Gravity method, 10th Sunrise Science Meeting, G\"{o}ttingen, Germany, May 03 $-$ 04, 2016 (Talk)
\item On the Contrast-Magnetic field relation in the quiet Sun, 9th Sunrise Science Meeting, G\"{o}ttingen, Germany, September 28 $-$ 29, 2015 (Talk)
\end{enumerate}

\end{flushleft}

\chapter*{Acknowledgements\markboth{Acknowledgements}{Acknowledgements}}
\addcontentsline{toc}{chapter}{Acknowledgements}


First, I would like to thank my supervisor, Prof.~Dr.~Sami Solanki. It has been a pleasure to have such a distinguished scientist as a guide and mentor. His passion, enthusiasm, and wide knowledge have inspired me to become a better hardworking scientist. I want to thank him for revising my scientific papers and guiding me to be a better scientific writer, for effectively revising the chapters of my thesis, and for being a great teacher who can deliver the hardest concepts in a clear and understandable, yet a rich way.

I want to also thank my supervisor, Dr.~Tino Riethm\"uller for his help throughout my project with the Sunrise data. He has been patient all the way and always available for revising my scientific papers, conference abstracts, and presentations.

Special thanks for Dr.~Alex Feller for his supervision on the stray-light project, and for the knowledge I have gained about this topic. He has been patient with me, and gladly available for any question or discussion.

Special thanks to Dr.~Andreas Lagg for his guidance during my first days in the institute and for his constant help and fruitful discussions throughout my PhD years.

I would like to thank all the postdocs and senior scientists of the Solar Lower Atmosphere and Magnetism (SLAM) research group for their helpful comments and suggestions regarding any complication in my work. It has been an honor to work with such great minds.

Special thanks to Dr.~Lakshmi Pradeep Chitta, Dr.~Ivan Milic, Dr.~Kok Leng Yeo for the helpful discussions. A big thanks to Dr.~Anusha Bhasari, who has been an amazing big sister and colleague. 

I would like to thank the International Max Planck School for Solar System Science (IMPRS) for funding my project and the school coordinator, Dr. Sonja Schuch, for being patient all the way and guiding me during my studies and teaching periods in the school. 

I want to thank Ines Dominitzki for her help (with a big smile on her face) with every single matter that made my stay in the insitute more comfortable. And to the secretaries, Johanna Wagner-Farssi, Sibylla Siebert-Rust, Grit Kolleck and Tanja Macke, thank you for your help in administrative matters.

To Ankit, Theo, David, Holy, Ivan, Sebastian, Eliana, Paul-Louis, Sihane, Mayukh, Sabrina, Sudharshan, Ameya, L\'ea, Lujane, thank you for the good times. Special thanks to Franzi for being an amazing friend, and for being patient whenever I annoy her with my German practice. Danke, dass du die fantastischste Person ist!

Alessandro, you have been a friend and a big brother right from the beginning of this journey, I wish you were here to witness the end of it. Thank you, wherever you are.

Finally, I want to thank my family for giving up on so much to help me get where I am today. I owe you the world.\\

\chapter*{Curriculum vitae\markboth{Curriculum vitae}{Curriculum vitae}}
\addcontentsline{toc}{chapter}{Curriculum vitae}
\begin{flushleft}

{\bf\large Personal Information}\\

\begin{itemize}
\item Name: Fatima Kahil
\item Date of birth: March 6, 1992
\item Nationality: Lebanese
\item Languages: Arabic, English, French, German\\
\end{itemize}
\end{flushleft}

\begin{flushleft}

{\bf\large Education}\\

\begin{itemize}
\item PhD., Solar Physics (2015$-$2018) 
\begin{itemize}
\item[$\blacksquare$] International Max Planck Institute for Solar System Science at the University of G\"ottingen, G\"ottingen, Germany
\item[$\blacksquare$] Dissertation: Brightness Contrast of Solar Magnetic Elements Observed by Sunrise 
\item[$\blacksquare$] Supervisors: Prof. Dr. Laurent Gizon, Prof. Dr. Sami Solanki, Dr. Tino Riethm\"uller
\end{itemize}

\item MSc., Astronomy and Astrophysics (2012$-$2014)

\begin{itemize}
\item[$\blacksquare$] Notre Dame University, Beirut, Lebanon
\item[$\blacksquare$] Universit\'{e} Saint Joseph, Beirut, Lebanon
\item[$\blacksquare$] Dissertation: MESSENGER Spectroscopic Observations of Mercury's Sodium Exosphere
\item[$\blacksquare$] Supervisor: Dr. Nelly Mouawad

\end{itemize}

\item B.S., Fundamental Physics (2009$-$2012)
\begin{itemize}

\item[$\blacksquare$] Lebanese University, Faculty of Physics, Nabatieh, Lebanon
\end{itemize}

\end{itemize}

\end{flushleft}

\end{document}